\pdfoutput=1

\newcommand*{\ATLASLATEXPATH}{latex/}

\documentclass[cernpreprint,texlive=2016,txfonts,UKenglish]{\ATLASLATEXPATH atlasdoc}

\usepackage{\ATLASLATEXPATH atlaspackage}
\usepackage{\ATLASLATEXPATH atlasbiblatex}
\usepackage{\ATLASLATEXPATH atlascontribute}
\usepackage{\ATLASLATEXPATH atlasphysics}
\usepackage{\ATLASLATEXPATH atlascover}

\usepackage{PaperZ3D-defs}
\usepackage[utf8]{inputenc}
\usepackage{rotating}
\usepackage{verbatim}
\usepackage{tabularx}
\newcolumntype{L}[1]{>{\raggedleft\arraybackslash}p{#1}}
\newcolumntype{R}[1]{>{\raggedright\arraybackslash}p{#1}}
\newcolumntype{C}[1]{>{\centering\arraybackslash}p{#1}}

\graphicspath{{logos/}{figures/}}

\AtlasTitle{Measurement of the Drell--Yan triple-differential cross section in {\pp} collisions at $\rts = 8$~{\TeV}}

\author{The ATLAS Collaboration}

\date{\today}

\AtlasRefCode{STDM-2016-04-002}
\AtlasNote{ATL-COM-PHYS-2015-1575}
\PreprintIdNumber{CERN-EP-2017-152}

\AtlasDate{\today}

\arXivId{1710.05167}

\HepDataRecord{10.17182/hepdata.77492}

\AtlasJournal{JHEP}
\AtlasJournalRef{JHEP 12 (2017) 059}
\AtlasDOI{10.1007/JHEP12(2017)059}

\AtlasAbstract{%
This paper presents a measurement of the triple-differential cross section
for the Drell--Yan process $Z/\gamma^*\rightarrow \ell^+\ell^-$ where
$\ell$ is an electron or a muon.  The measurement is performed for invariant
masses of the lepton pairs, $m_{\ell\ell}$, between $46$ and $200$~{\GeV}
using a sample of $20.2$~fb$^{-1}$ of $pp$ collisions data at a centre-of-mass
energy of $\sqrt{s}=8$~{\TeV} collected by the ATLAS detector at the LHC in
2012. The data are presented in bins of invariant mass, absolute dilepton
rapidity, $|y_{\ell\ell}|$, and the angular variable $\cos\theta^{*}$ between
the outgoing lepton and the incoming quark in the Collins--Soper frame. The
measurements are performed in the range $|y_{\ell\ell}|<2.4$ in the muon
channel, and extended to $|y_{\ell\ell}|<3.6$ in the electron channel. The
cross sections are used to determine the $Z$ boson forward-backward
asymmetry as a function of $|y_{\ell\ell}|$ and $m_{\ell\ell}$. The
measurements achieve high-precision, below the percent level in the pole
region, excluding the uncertainty in the integrated luminosity, and are in
agreement with predictions. These precision data are sensitive to the parton
distribution functions and the effective weak mixing angle.
}

\AtlasCoverSupportingNote{Supporting documentation}{https://cds.cern.ch/record/2117171}

\AtlasCoverAnalysisTeam{Lewis~Armitage, Sasha~Glazov, Richard~Keeler, Tony~Kwan,
Eram~Rizvi, Andrey~Sapronov, Nenad~Vranjes, Elena~Yatsenko}

\AtlasCoverEdBoardMember{Maarten~Boonekamp~(chair), Mykhailo~Lisovyi, Nicolas~Morange}

\AtlasCoverEgroupEditors{atlas-stdm-2016-04-editors@cern.ch}

\AtlasCoverEgroupEdBoard{atlas-stdm-2016-04-editorial-board@cern.ch}

\hypersetup{pdftitle={ATLAS document},pdfauthor={The ATLAS Collaboration}}

\begin{document}

\maketitle

\tableofcontents

\clearpage

\clearpage
\section{Introduction}
\label{sec:Introduction}

In the Drell--Yan process ~\cite{Drell:1970wh,Drell:1971ap}
$q\bar{q}\rightarrow Z/\gamma^* \rightarrow \ell^+\ell^-$, parity
violation in the neutral weak coupling of the mediator to fermions induces a
forward-backward asymmetry, $A_{\textrm{FB}}$, in the decay angle
distribution of the outgoing lepton ($\ell^-$) relative to the incoming quark
direction as measured in the dilepton rest frame.  This decay angle 
depends on the sine of the weak mixing angle, $\sin^2{\theta_{\textrm {W}}}$,
which enters in the fermionic vector couplings to the $Z$ boson. At
leading order in electroweak (EW) theory it is given by
$\sin^2{\theta_{\textrm {W}}} = 1-m_W^2/m_Z^2$, where $m_W$ and $m_Z$
are the $W$ and $Z$ boson masses, respectively. Higher-order loop
corrections modify this relation depending on the
renormalisation scheme used, and so experimental measurements are
often given in terms of the sine of the effective weak mixing angle,
$\sin^2\theta^{\textrm{eff}}$~\cite{LEP-EWWG:2006}. High-precision cross-section
measurements sensitive to the asymmetry, and therefore to the
effective weak mixing angle, provide a testing ground
for EW theory and could offer some insight into physics beyond the Standard
Model (SM).

Previous measurements by ATLAS and CMS of the Drell--Yan (DY) process
include measurements of fiducial cross sections~\cite{STDM-2015-03,STDM-2016-02,CMS-EWK-10-005,CMS-SMP-12-011},
and one-dimensional differential cross sections as a function of
rapidity~\cite{STDM-2011-06,CMS-EWK-10-010}, transverse
momentum~\cite{STDM-2012-06,STDM-2012-23,CMS-EWK-10-010,CMS-SMP-14-012},
and invariant mass~\cite{STDM-2012-10,STDM-2011-41,CMS-EWK-10-007}. 
Double-differential cross-section measurements as a function of
invariant mass and either rapidity or
transverse momentum~\cite{STDM-2014-12,STDM-2014-06,STDM-2012-20,CMS-SMP-13-013,CMS-SMP-14-003,CMS-SMP-13-003}
have also been published, as well as $Z$ boson polarisation
coefficients~\cite{STDM-2014-10,CMS-SMP-13-010} and the
forward-backward asymmetry~\cite{STDM-2011-34,CMS-EWK-11-004}.
Extraction of the effective weak mixing angle in leptonic $Z$ boson decays,
$\sin^2\theta^{\textrm{eff}}_{\textrm{lept}}$, from $A_{\textrm{FB}}$
measurements has been performed by ATLAS using 5 fb$^{-1}$ of
proton-proton collision data at $\sqrt{s}=$ 7 {\TeV}~\cite{STDM-2011-34} --
a result in which the largest contribution to the uncertainty was due to limited
knowledge of the parton distribution functions (PDFs) of the proton. 

A complete description of the Drell--Yan cross section to all orders
in  quantum chromodynamics (QCD) depends on five kinematic variables of the Born-level leptons, namely $m_{\ell\ell}$, the
invariant mass of the lepton pair; $y_{\ell\ell}$, the rapidity of the
dilepton system; $\theta$ and $\phi$, the lepton decay angles in the
rest frame of the two incident quarks; and $p_{\textrm{T},Z}$, the
transverse momentum of the vector boson. In this paper, measurements of
the triple-differential Drell--Yan cross section,
$\mathrm{d}^3\sigma/\mathrm{d}m_{\ell\ell}\mathrm{d}|y_{\ell\ell}|\mathrm{d}\textrm{cos}\theta^*$,
are reported as a function of $m_{\ell\ell}$, $|y_{\ell\ell}|$, and
$\cos{\theta^*}$, where the lepton decay angle is defined in the
Collins--Soper (CS) reference frame~\cite{Collins-Soper:1977}. These
cross-section measurements are designed to be simultaneously sensitive
to $\sin^2\theta^{\textrm{eff}}_{\textrm{lept}}$ and to the PDFs,
therefore allowing a coherent determination of both. A simultaneous
extraction has the potential to reduce the PDF-induced uncertainty in the
extracted value of the effective weak mixing angle.

At leading order (LO) in perturbative electroweak and 
QCD theory, the Drell--Yan triple-differential cross
section can be written as
\begin{equation}
\frac{\mathrm{d}^3\sigma}{\mathrm{d}m_{\ell\ell}\mathrm{d}y_{\ell\ell}\mathrm{d}\cos\theta^*} = 
	\frac{\pi\alpha^2}{3m_{\ell\ell}s} \sum_{q}P_q
  	\left[f_q(x_1,Q^2)f_{\bar{q}}(x_2,Q^2) + (q\leftrightarrow\bar{q})\right],
\label{eq:DY-PDF}
\end{equation}
where $s$ is the squared proton-proton ($pp$) centre-of-mass energy; the incoming
parton momentum fractions are
\(x_{1,2} = (m_{\ell\ell}/\sqrt{s})\textrm{e}^{\pm y_{\ell\ell}}\); and
$f_q(x_1,Q^2)$ are the PDFs for parton flavour $q$. Here, $Q^2$ is the
four-momentum transfer squared and is set to the dilepton
centre-of-mass energy, $m_{\ell\ell}$, which is equal to the partonic
centre-of-mass energy. The $q\leftrightarrow\bar{q}$ term
accounts for the case in which the parent protons of the $q$ and
$\bar{q}$ are interchanged. The function $P_q$ in
equation~(\ref{eq:DY-PDF}) is given by
\begin{equation}
\begin{split}
  P_q &=  e_{\ell}^2e_q^2(1+\cos^2\theta^*) \\
      &+  e_{\ell}e_q\frac{2m_{\ell\ell}^2(m_{\ell\ell}^2-m_Z^2)}{\sin^2\theta_{\textrm{W}}\cos^2\theta_{\textrm{W}} \label{eq:EWsigma}
          \big[(m_{\ell\ell}^2-m_Z^2)^2+\Gamma_Z^2m_Z^2\big]}
          \big[v_{\ell}v_q(1+\cos^2\theta^*)+2a_{\ell}a_q\cos\theta^*\big]  \\
      &+  \frac{m_{\ell\ell}^4}{\sin^4\theta_{\textrm{W}}\cos^4\theta_{\textrm{W}}
          \big[(m_{\ell\ell}^2-m_Z^2)^2+\Gamma_Z^2m_Z^2\big]}
          \big[(a_{\ell}^2+v_{\ell}^2)(a_q^2+v_q^2)(1+\cos^2\theta^*)+8a_{\ell}v_{\ell}a_qv_q\cos\theta^*\big]. 
\end{split}
\end{equation}
In this relation $m_Z$ and $\Gamma_Z$ are the $Z$ boson mass and width, respectively;
$e_{\ell}$ and $e_q$ are the lepton and quark electric charges;  and
$v_{\ell}=-\frac{1}{4}+\sin^2\theta_{\textrm W}$,
$a_{\ell}=-\frac{1}{4}$, 
$v_q=\frac{1}{2}I^3_q-e_q\sin^2\theta_{\textrm W}$, and
$a_q=\frac{1}{2}I^3_q$ are the vector and axial-vector lepton and
quark couplings, respectively where $I^3_q$ is the third
component of the weak isospin.

The first term in equation~(\ref{eq:EWsigma}) corresponds to pure virtual photon,
$\gamma^*$, exchange in the scattering process, the second corresponds
to the interference of $\gamma^*$ and $Z$ exchange, and the last term
corresponds to pure $Z$ exchange. Thus the DY invariant mass spectrum
is characterized by a $1/m_{\ell\ell}^2$ fall-off from $\gamma^*$ exchange
contribution, an $m_{\ell\ell}$-dependent Breit--Wigner peaking at the mass
of the $Z$ boson, and a $Z/\gamma^*$ interference contribution which
changes sign from negative to positive as $m_{\ell\ell}$ increases
across the $m_Z$ threshold. 

The terms which are linear in $\cos\theta^*$ induce the
forward-backward asymmetry. The largest contribution comes from the
interference term, except at $m_{\ell\ell}=m_Z$ where the interference
term is zero, and only the $Z$ exchange term contributes to the asymmetry. The
resulting asymmetry is, however, numerically small due to the small value of
$v_{\ell}$. The net effect is an asymmetry which is negative for
$m_{\ell\ell}<m_Z$ and increases, becoming positive for
$m_{\ell\ell}>m_Z$. The point of zero asymmetry occurs slightly below
$m_{\ell\ell}=m_Z$.

The forward-backward asymmetry varies with $|y_{\ell\ell}|$.
The incoming quark direction can only be determined probabilistically:
for increasing $|y_{\ell\ell}|$ the momentum fraction of one parton
reaches larger $x$ where the valence quark PDFs dominate because the valence quarks typically
carry more momentum than the antiquarks. Therefore, the $Z/\gamma^*$
is more likely to be boosted in the quark direction. Conversely, at
small boson rapidity, $|y_{\ell\ell}|\sim 0$, it becomes almost impossible to
identify the direction of the quark since the quark and antiquark
have nearly equal momenta.

The sensitivity of the cross section to the PDFs arises
primarily from its dependence on $y_{\ell\ell}$ (and therefore $x_1$ and $x_2$) in
equation~(\ref{eq:DY-PDF}). Further sensitivity is gained by analysing
the cross section in the $m_{\ell\ell}$ dimension, since in the $Z$
resonance peak the partons couple through the weak interaction
and off-peak the electric couplings to the $\gamma^*$
dominate. Therefore, the relative contributions of up-type and
down-type quarks vary with $m_{\ell\ell}$. Finally, the $\cos\theta^*$
dependence of the cross section provides sensitivity to terms
containing $a_{\ell}a_q$ and $v_{\ell}v_qa_{\ell}a_q$ in
equation~(\ref{eq:EWsigma}). Three different combinations of
couplings to the incident quarks contribute to the LO cross
section. The magnitude of the asymmetry is proportional to the valence
quark PDFs and offers direct sensitivity to the corresponding PDF component.

The full five-dimensional cross section can also be decomposed into harmonic polynomials for
the lepton decay angle scattering amplitudes and their corresponding
coefficients $A_{0-7}$~\cite{STDM-2014-10}. Higher-order QCD
corrections to the LO $q\qbar$ process involve
$qg+\qbar g$ terms at next-to-leading order (NLO), and $gg$ terms at
next-to-next-to-leading order (NNLO). These higher-order terms modify
the decay angle dependence of the cross section. Measuring the
$|\cos\theta^*|$ distribution provides additional sensitivity to the
gluon versus sea-quark PDFs and is related to the measurements
of the angular coefficients as a function of the $Z$ boson transverse
momentum~\cite{STDM-2014-10,CMS-SMP-13-010}.

Initial-state QCD radiation can introduce a
non-zero transverse momentum for the final-state lepton pair, leading
to quark directions which may no longer be aligned with the incident
proton directions.  Hence, in this paper, the decay angle is measured
in the CS reference frame~\cite{Collins-Soper:1977} in
which the decay angle is measured from an axis symmetric with respect
to the two incoming partons. The decay angle in the CS frame ($\theta^*$)
is given by
\begin{equation}
\cos{\theta^*} = \frac{p_{\textrm z,\ell\ell}}{\mll |p_{\textrm z,\ell\ell}|}\frac{p_1^+p_2^- -
  p_1^-p_2^+}{\sqrt{\mll^2 + p_{\textrm T,\ell\ell}^2}} ~,
\nonumber
\end{equation}
where $p_i^{\pm}=E_i\pm p_{{\textrm z,}i}$ and $i=1$ corresponds to
the negatively-charged lepton and $i=2$ to the positively-charged
antilepton. Here, $E$ and $p_{\textrm z}$ are the energy and
longitudinal $z$-components of the leptonic four-momentum,
respectively; $p_{\textrm z,\ell\ell}$ is the dilepton
$z$-component of the momentum; and $p_{\textrm T,\ell\ell}$ the
dilepton transverse momentum.

The triple-differential cross sections are measured using
20.2~fb$^{-1}$ of $pp$ collision data at $\sqrt{s}=8$~{\TeV}.  The
measurements are performed in the electron and muon decay channels for
$|y_{\ell\ell}| < 2.4$. The electron channel analysis is extended to high rapidity in the region
$1.2 < |y_{\ell\ell}|< 3.6$. The measured cross sections cover the kinematic
range $46 < m_{\ell\ell} < 200$~{\GeV}, $0 < |y_{\ell\ell}|< 3.6$,
and $-1 < \cos\theta^*< +1$. For convenience the notation
\begin{equation}
\d3s\equiv 
\frac{\mathrm{d}^3\sigma}{\mathrm{d}m_{\ell\ell}\mathrm{d}|y_{\ell\ell}|\mathrm{d}\cos\theta^*}
\nonumber
\end{equation}
is used. The cross sections are classified as either \textit{forward}
($\cos\theta^* > 0$) or \textit{backward} ($\cos\theta^* < 0$) and
used to obtain an experimental measurement of $A_{\textrm{FB}}$
differentially in $m_{\ell\ell}$ and $|y_{\ell\ell}|$:
\begin{equation}
A_{\textrm{FB}} = \frac{\mathrm{d}^3\sigma(\cos{\theta^*}>0) -
  \mathrm{d}^3\sigma(\cos{\theta^*}<0) }{\mathrm{d}^3\sigma(\cos{\theta^*}>0) +
  \mathrm{d}^3\sigma(\cos{\theta^*}<0)} ~.
\label{eq:afb}
\end{equation}

\FloatBarrier

\section{ATLAS detector}
\label{sec:detector}

The \mbox{ATLAS} detector~\cite{PERF-2007-01} consists of an inner
tracking detector (ID) surrounded by a thin superconducting solenoid,
electromagnetic and hadronic calorimeters, and a muon spectrometer
(MS).  Charged particles in the pseudorapidity\footnote{ATLAS uses a
right-handed coordinate system with its origin at the nominal
interaction point in the centre of the detector and the $z$-axis
along the beam pipe. The $x$-axis points from the interaction point
to the centre of the LHC ring, and the $y$-axis points
upward. Cylindrical coordinates $(r,\phi)$ are used in the
transverse plane, $\phi$ being the azimuthal angle around the beam
pipe. The pseudorapidity is defined in terms of the polar angle
$\theta$ as $\eta=-\ln\tan(\theta/2)$.}  range $|\eta| < 2.5$ are
reconstructed with the ID, which consists of layers of silicon pixel
and microstrip detectors and a straw-tube transition-radiation tracker
having a coverage of $|\eta| < 2.0$.  The ID is immersed in a 2~T
magnetic field provided by the solenoid.  The latter is surrounded by
a hermetic calorimeter that covers $|\eta| < 4.9$ and provides
three-dimensional reconstruction of particle showers.  The
electromagnetic calorimeter is a liquid-argon sampling calorimeter,
which uses lead absorbers for $|\eta| < 3.2$.  The hadronic sampling
calorimeter uses plastic scintillator tiles as the active material and
steel absorbers in the region $|\eta| < 1.7$.  In the region $1.5 <
|\eta| < 3.2$, liquid argon is used as the active material, with copper
absorbers. A forward calorimeter covers the range $3.2 < |\eta| <
4.9$ which also uses liquid argon as the active material, and copper and
tungsten absorbers for the EM and hadronic sections of the subdetector,
respectively.

Outside the calorimeters, air-core toroids supply the
magnetic field for the MS.  There, three layers of precision
chambers allow the accurate measurement of muon track curvature in the
region $|\eta| < 2.7$.  The majority of these precision chambers is
composed of drift tubes, while cathode-strip chambers provide coverage
in the inner layers of the forward region $2.0 < |\eta| < 2.7$.
The muon trigger in the range $|\eta| < 2.4$ uses resistive-plate chambers in the central region 
and thin-gap chambers in the forward region.  A three-level trigger
system~\cite{PERF-2011-02} selects events to be recorded for offline
analysis.
\FloatBarrier

\section{Simulated event samples}
\label{sec:MC}

Monte Carlo (MC) simulation samples are used to model the expected
signal and background yields, with the exception of certain
data-driven background estimates.  The MC samples are normalised using
the highest-order cross-section predictions available in perturbation
theory.

The DY process was generated at NLO using
\powheg-Box (referred to as \powheg \;in the following)~\cite{Nason:2004rx,Frixione:2007vw,Alioli:2008gx,Alioli:2010xd}
and the CT10 PDF set~\cite{CT10}, with \pythia~8~\cite{pythia8} to model
parton showering, hadronisation, and the underlying event (UEPS).
The $Z/\gamma^* \rightarrow \ell^+\ell^-$ differential cross section as a function of mass has
been calculated at NNLO in perturbative QCD (pQCD) using
\fewz~3.1~\cite{FEWZnnlo,FEWZ2,FEWZ4} with the MSTW2008NNLO
PDF set~\cite{mstw}. The renormalisation, $\mu_{\textrm{r}}$, and
factorisation, $\mu_{\textrm{f}}$, scales were both
set equal to $m_{\ell\ell}$. The calculation includes NLO EW
corrections beyond final-state photon radiation (FSR) using the
$G_{\mu}$ EW scheme~\cite{Hollik:1988ii}. A mass-dependent
$K$-factor used to scale the $Z/\gamma^{\ast}  \rightarrow \ell^+\ell^-$ MC sample is
obtained from the ratio of the calculated total NNLO pQCD
cross section with the additional EW corrections, to the total cross
section from the \powheg\ sample.  This one-dimensional (and therefore
partial) NNLO $K$-factor is found to vary from 1.035 at the lowest invariant
mass values considered in this analysis to 1.025 at the highest.
This factor also improves the modelling of the $Z$ boson
lineshape. The DY production of $\tau$ pairs was modelled using
\powheg\ in the same way as the signal simulation.

The scattering amplitude coefficients describing the distributions of
lepton decay angles are known to be not accurately modelled in \powheg\,
particularly $A_0$ at low $p_{\textrm{T},Z}$~\cite{STDM-2014-10}. For this
reason, the signal MC events are reweighted as a function of $p_{\textrm{T}, Z}$
and $y_{\ell\ell}$ to improve their modelling. These weights were
calculated using the cross-section calculator DYNNLO~\cite{Catani:2009sm}.

The photon-induced process, $\gamma \gamma \to \ell\ell$, is simulated
at LO using \pythia~8 and the MRST2004qed PDF set~\cite{MRST2004QED}.
The expected yield for this process also accounts for NLO QED/EW corrections
from references~\cite{Bardin:2012jk,Bondarenko:2013nu}, which decrease the
yield by approximately 30\%.

The production of top quark pairs with prompt isolated leptons
from electroweak boson decays constitutes a dominant background. It is
estimated at NLO in QCD using \powheg\ and the CT10 PDF set, with
\pythia~6~\cite{Sjostrand:2006za} for UEPS. The \ttbar\ sample is normalized using a cross
section calculated at NNLO in QCD including resummation effects
\cite{Cacciari:2011,Baernreuther:2012ws,Czakon:2012zr,Czakon:2012pz,Czakon:2013goa,TopPP:2011}.
Small compared to the \ttbar\ contribution, single-top production
in association with a $W$ boson ($Wt$) is also modelled by \powheg\
and the CT10 PDF set, with \pythia~6 for UEPS. Both the \ttbar\ and $Wt$ contributions are summed
and collectively referred to as the top quark background.

Further small background contributions are due to diboson ($WW$,
$WZ$ and $ZZ$) production with decays to final states with at least
two leptons.  The diboson processes were generated at LO
with \herwig, using the CTEQ6L1 PDF set~\cite{Pumplin:2002vw}.  The
samples are scaled to NLO calculations~\cite{Campbell:1999mcfm,Campbell:2011bn}
or to ATLAS measurements as described in reference~\cite{STDM-2014-06}.
Additionally, the background arising from $W$ boson production in association
with jets ($W$+jets) is studied with MC samples generated with \powheg\ under
identical conditions as the DY signal samples.

All MC samples used in the analysis include the effects of QED FSR, multiple
interactions per bunch crossing (``pile-up''), and detector simulation. QED FSR
was simulated using \photos~\cite{fsr_ref}, while the effects of pile-up were
accounted for by overlaying simulated minimum-bias events~\cite{SOFT-2010-01}
generated with \pythia 8~\cite{pythia8}. The interactions of particles with the
detector were modelled using a full \mbox{ATLAS} detector simulation~\cite{SOFT-2010-01}
based on \geant4~\cite{geant}. Finally, several corrections are applied to the
simulated samples, accounting for differences between data and
simulation in the lepton trigger, reconstruction, identification, and
isolation efficiencies as well as lepton resolution and muon momentum
scale~\cite{PERF-2013-05,PERF-2013-03,PERF-2016-01,PERF-2014-05,TRIG-2012-03,TRIG-2012-03}.
The electron energy scale corrections are applied to the data.

An overview of the simulated event samples is given in table~\ref{tab:mc}. 

\begin{table}[htp]
\centering
\footnotesize
\begin{tabular}{c@{\hskip 1.2cm}c@{\hskip 1cm}c@{\hskip 1cm}c@{\hskip 1cm}c@{\hskip 1cm}}
\toprule
Process          										& Generator    					& Parton shower \&		& Generator 			& Model parameters\\
			          										& 				    					& underlying event		& PDF 					& (``Tune'')\\
\midrule
$Z/\gamma^*\rightarrow\ell\ell$		& \powheg~v1(r1556) 		& \pythia~8.162			& CT10					& AU2~\cite{ATL-PHYS-PUB-2012-003}\\
$Z/\gamma^*\rightarrow\tau\tau$	& \powheg~v1(r1556) 		& \pythia~8.162			& CT10					& AU2\\
$\gamma \gamma \to \ell\ell$					& \pythia~8.170	        	& \pythia~8.170			& MRST2004qed	& 4C~\cite{Corke:2010yf}\\
\ttbar													& \powheg~v1(r1556) 		& \pythia~6.427.2		& CT10					& AUET2 \cite{ATL-PHYS-PUB-2011-008}\\
$Wt$													& \powheg~v1(r1556) 		& \pythia~6.427.2		& CT10					& AUET2\\
Diboson												& \herwig~6.520				& \herwig~6.520  		& CTEQ6L1			& AUET2\\
$W\rightarrow\ell\nu$	        			& \powheg~v1(r1556) 		& \pythia~8.162			& CT10					& AU2\\
\bottomrule
\end{tabular}
\caption{Overview of the Monte Carlo samples used in this analysis.}
\label{tab:mc}
\end{table}

\FloatBarrier

\section{Event selection}
\label{sec:selection}

Events are required to have been recorded during stable beam condition
periods and must pass detector and data-quality requirements. This
corresponds to an integrated luminosity of $20.2$~fb$^{-1}$ for the
muon channel.  Small losses in the data processing chain lead to an
integrated luminosity of $20.1$~fb$^{-1}$ for the electron channel.
Due to differences in the detector response to electrons and muons the
selection is optimised separately for each channel and is described in
the following.

\subsection{Central rapidity electron channel}
\label{sec:sel_cen_elec}

The electron data were collected using a dilepton trigger which uses
calorimetric and tracking information to identify compact electromagnetic
energy depositions. Identification algorithms use calorimeter shower shape
information and the energy deposited in the vicinity of the electron candidates
to find candidate electron pairs with a minimum transverse energy of
$12$~{\GeV} for both the leading and subleading electron. 

Electrons are reconstructed by clustering energy deposits in the
electromagnetic calorimeter using a sliding-window algorithm. These
clusters are then matched to tracks reconstructed in the inner
detector. The calorimeter provides the energy measurement and the
track is used to determine the angular information of the electron
trajectory. An energy scale correction determined from $Z\to e^+e^-$,
$W\to e \nu$, and $J/\psi\to e^+e^-$ decays~\cite{PERF-2013-05} is
applied to data. Central electron candidates are required to have
$|\eta^e|<2.4$. Furthermore, candidates reconstructed within the
transition region between the barrel and endcap calorimeters,
$1.37<|\eta^e|<1.52$, are excluded from the measurement. Each
candidate is required to satisfy the ``medium'' electron
identification~\cite{PERF-2013-03,PERF-2016-01}
criteria, based on calorimetric shower shapes and track parameters.
To ensure the selected electrons are on the efficiency plateau of the trigger,
electrons are required to have $\et^e>20~\GeV$.
Candidate events are required to have exactly one pair of oppositely-charged
electrons and their invariant mass is required to be in the range
$46 <  m_{ee} < 200~\GeV$.

\subsection{High rapidity electron channel}
\label{sec:sel_fwd_elec}
In this channel, the rapidity range of the measurement is extended by
selecting one central electron and one forward electron. Forward electrons
are defined as having pseudorapidities in the range $2.5 < |\eta^e| < 4.9$,
reconstructed by the endcap or forward calorimeters. The data were collected using
two single-electron triggers in the central calorimeter region with
$\et^e>24~\GeV$ or $\et^e>60~\GeV$. The lower-threshold trigger
has additional criteria for the shower shape and energy deposited in
the vicinity of the electron candidate. The reconstructed central electrons
are required to have $\et^e>25~\GeV$, $\left|\eta^e\right|<2.4$,
and must satisfy the ``tight'' identification criteria. Electrons in the
calorimeter transition regions $1.37<\left|\eta^e\right|<1.52$ are
rejected.  Leptons produced in the Drell--Yan process are expected to
be well isolated from other particles not associated with the lepton.
This provides a good discriminant against the multijet background
arising from the semileptonic decays of heavy quarks or hadrons
faking electrons. The track isolation is defined as the scalar sum
of the transverse momenta, $\sum p_{\textrm T}$, of the additional tracks
contained in a cone of size $\Delta R=\sqrt{(\Delta\phi)^2+(\Delta\eta)^2}=0.2$
around the electron (omitting the contribution from the electron
track). Central electrons are required to have a track isolation less
than 14\% of $\et^e$.

The forward electron is required to satisfy ``tight'' identification
criteria, $\et^e>20~\GeV$, and $2.5<\left|\eta^e\right|<4.9$, excluding
the transition region between the endcap and forward calorimeters,
$3.00<\left|\eta^e\right|<3.35$. Due to insufficient accuracy in the modelling
of the material in front of the endcap calorimeter, forward electrons in the region
$2.70<|\eta^e|<2.80$ are also rejected.

A dedicated calibration procedure is performed for the forward
electrons. Energy scale and Gaussian resolution corrections are
derived in bins of $\eta^e$ by comparing the peak position and the width
of the $m_{ee}$ distributions in data and simulation. The scale and
resolution corrections are the values that bring the peak regions,
$80<m_{ee}<100$~{\GeV}, of the data and simulation into the best
agreement.

No isolation criteria are applied to the forward electron and due to
the absence of tracking information in the forward region, no charge
requirements are placed on the selected electron pair. Lastly, events
in the high rapidity electron channel are required to have exactly one
central-forward pair of electrons with an invariant mass in the range
$66 < m_{ee} < 150~\GeV$. Events with more than one possible
central-forward pair are not used in this measurement channel.

\subsection{Central rapidity muon channel}
\label{sec:sel_muon}

Candidate events in the muon channel were collected using two sets of
triggers with the set of triggers used depending on the $p^{\mu}_{\textrm{T}}$
of the muon with the larger transverse momentum. For
$p^{\mu}_{\textrm T} > 25$~{\GeV}, two single-muon triggers are used, with
transverse momentum thresholds of $24$~{\GeV} and $36$~{\GeV}. The
low-threshold trigger requires the muon to be isolated. This combination of
triggers collected the majority of the events in the data sample.
For $p^{\mu}_{\textrm T}<25$~{\GeV}, a dimuon trigger is used which
requires two muons with transverse momentum thresholds of $18$~{\GeV}
for one muon and $8$~{\GeV} for the other.

Muons are identified by tracks reconstructed in the muon spectrometer
matched to tracks reconstructed in the inner detector, and are required
to have $p^{\mu}_{\textrm T} > 20$ {\GeV} and $|\eta^{\mu}| < 2.4$.
Additionally, they must satisfy identification criteria based on 
the number of hits in the inner detector and muon spectrometer, and
on the consistency between the charge and momentum measurements
in both systems~\cite{PERF-2014-05}. Backgrounds from multijet events
are efficiently suppressed by imposing an isolation condition requiring
that the sum of the transverse momentum, $\sum p_{\textrm{T}}$, of the
tracks contained in a cone of size $\Delta R=0.2$ around the muon (omitting
the contribution from the muon track) to be less than 10\% of
$p^{\mu}_{\textrm T}$. A small contribution of cosmic muons is removed by
requiring the magnitude of the longitudinal impact
parameter with respect to the primary interaction vertex $z_0$ to be less than $10$~mm.
Events are selected if they contain exactly two oppositely-charged muons
satisfying the isolation and impact parameter requirements. Finally, the
dilepton invariant mass must be in the range $46 < m_{\mu\mu} < 200~\GeV$.

In order to minimise the influence of residual misalignments between
the ID and MS, muon kinematic variables are measured using the ID
only. A small residual $\eta^{\mu}$- and charge-dependent bias in the
muon momentum was observed, most likely arising from residual
rotational misalignments of the inner detector. Such ID misalignments
bias the measurement of the muon track sagitta and have an opposite
effect on the momentum of positively- and negatively-charged
muons. Hence, the reconstructed invariant mass or rapidity of muon
pairs are not affected, in contrast to measurements of $\cos\theta^*$
which are charge-dependent. These residual inner detector misalignments
are corrected for based on two methods, one which uses $Z\rightarrow\mu^+\mu^-$
events, and another using $Z\rightarrow e^+e^-$ events as described in
reference~\cite{STDM-2014-18}. Together with a $\chi^2$ minimisation technique,
the dimuon data sample is used to determine the corrections binned finely,
which are however insensitive to the $\eta$-independent component of the
track curvature bias. This bias is corrected for using dielectron data by
comparing the ratio of the calorimeter energy to the track momentum
for electrons and positrons.

\subsection{Measurement bins}
\label{sec:binning}
The measurement bins are chosen by taking into consideration several competing
demands on the analysis such as its sensitivity to the underlying physics,
the statistical precision in each bin, and detector resolution effects particularly
in the $m_{\ell\ell}$ dimension. The binning must also match those used in recent
ATLAS cross section measurements~\cite{STDM-2012-10,STDM-2012-20}.

The measurement is performed in seven bins of $m_{\ell\ell}$ from
$46$~{\GeV} to $200$~{\GeV} with edges set at $66$, $80$, $91$, $102$, $116$,
and $150$~{\GeV}; $12$ equidistant bins of $|y_{\ell\ell}|$ from $0$ to $2.4$;
and bins of $\cos\theta^*$ from $-1$ to $+1$, separated at $-0.7$, $-0.4$,
$0.0$, $+0.4$, $+0.7$ giving 6 bins. In total, $504$ measurement
bins are used for the central rapidity electron and muon channel measurements.

For the high rapidity electron channel the measurement is restricted to the $5$
invariant mass bins in the region $66<m_{\ell\ell}<150$~{\GeV}. The $|y_{\ell\ell}|$
region measured in this channel ranges from $1.2$ to $3.6$ in $5$ bins
with boundaries at $1.6$, $2.0$, $2.4$, $2.8$. The $\cos\theta^*$ binning
is identical to the binning of the central analyses. A total of $150$ measurement
bins is used in this channel.

\FloatBarrier

\section{Background estimation}
\label{sec:bg}

The background from processes with two isolated final-state leptons of
the same flavour is estimated using MC simulation. The processes with
non-negligible contributions are $Z/\gamma^* \rightarrow \tau\tau$,
diboson ($WW$, $WZ$ and $ZZ$), and photon-induced dilepton production
-- together termed the \textit{electroweak} background sources. The
top quark background arising from {\ttbar} and $Wt$ production is also
estimated using MC simulation. The samples used for these estimates
are listed in table~\ref{tab:mc}.

Background contributions from events where at least one final state
jet satisfies the electron or muon selection criteria, hereafter
referred to as the \textit{fake lepton} background, are determined using
a combination of data-driven methods and MC simulation. By far the largest
contribution to the fake lepton background comes from light- and heavy-flavour
multijet production, referred to as the \textit{multijet} background, which is
determined from data. Descriptions on the fake background estimations used
in each of the three channels are given in the following subsections.

\subsection{Fake lepton background estimation in the central rapidity electron channel}
\label{sec:bg_elec_data_driven}

To separate the signal from the multijet background, the analysis
relies on the electron relative transverse energy isolation
distribution ($I^e$). This is a good discriminant for the multijet
contribution, which has larger values of $I^e$ than the signal
process. It is defined as the ratio of the summed calorimetric
transverse energy contained in a cone of size $\Delta R=0.2$ around
the electron to the electron transverse energy:
$I^e=\sum E_{\textrm{T}}(\Delta R=0.2)/E^{e}_{\textrm T}$. The smaller
of the $I^e$ values of the two electron candidates is chosen to represent
each event, as it was found to provide optimal discrimination.

The multijet fraction is then estimated from data by fitting this
distribution using a template method. The background template is
selected with inverted electron identification requirements and the
signal, electroweak, and $W+$jet templates are taken from
simulation. The non-isolated sample where the smaller $I^e$ of the two
electrons exceeds a certain value is found to be dominated by multijet
background and is used to adjust the normalization of the background
template, taking into account the small signal contamination. Since
the multijet background is not expected to exhibit any parity
violating effects and the $\cos\theta^*$ background templates in data
were found not to show any asymmetry about $\cos\theta^* = 0$, the
method is symmetrised in bins of $|\cos\theta^*|$, resulting in a doubling
of the sample sizes and therefore more stable results.

The multijet contribution is found to be largest at low $m_{ee}$ and
also at large $|\cos\theta^*|$ for $|y_{ee}|\sim0$, where it reaches
15\% of the expected number of signal events. In the pole region,
$80< m_{ee} < 102$~{\GeV}, the contribution is less than 0.1\%.
 
The contribution of $W+$jet production to the fake lepton background
is estimated from MC simulation. It is small compared to the multijet
background for all kinematic regions, and therefore does not introduce
any significant charge asymmetry.

\subsection{Fake lepton background estimation in the high rapidity electron channel}
\label{sec:bg_fwd_elec_data_driven}

The multijet background in the high rapidity electron channel is estimated
using a template method similar to the one used in the central electron channel
with, however, some small adjustments. The isolation variable is used for the
normalisation of the multijet background only for the mass bins in the range
$80<m_{ee}<102$~{\GeV}. The size of the isolation cone in this case is
increased to $\Delta R =0.3$, which was found to improve the stability of
the fits. For the off-peak mass bins, the transverse energy of the forward
electron is used as an alternative discriminating variable, where the
multijet background contributes mostly at low $E_{\textrm{T}}$. This
decreases the statistical uncertainty of the estimation and reduces its
dependence on the $W+$jet background modelling, as discussed below.

The multijet background is the dominant contribution to the background
in this measurement channel and is typically about 5--10\% of the
expected signal, but increases rapidly at large $|\cos\theta^*|$. It can
be as large as 30--60\% in some bins at large
$|y_{ee}|$ where the $|A_{\textrm{FB}}|$ is large and the signal cross
section is suppressed, i.e. $\cos\theta^*<0$ for $m_{ee}>m_Z$.

The $W+$jet background is estimated using MC simulation. As was the
case in the central electron analysis, it is found to be small under the
peak of the $Z$ resonance. It is found to be more
significant off peak, reaching 30\% of the fake lepton background. 

\subsection{Fake lepton background estimation in the central rapidity muon channel}
\label{sec:bg_muon_data_driven}

The multijet background remaining after event selection in the muon
channel is largely due to heavy-flavour $b$- and $c$-quark decays,
and is estimated in two steps. First, the shape as a function of
$|y_{\mu\mu}|$ and $|\cos\theta^*|$ is estimated in each
$m_{\mu\mu}$ bin. Next its overall normalisation is then determined
in each invariant mass region.

Three orthogonal control regions with inverted muon isolation
requirements defined by $I^{\mu}=\sum p_{\textrm T}(\Delta
R=0.2)/p^{\mu}_{\textrm T}>0.1$, and/or inverted muon pair
charge requirements are used to determine the multijet
background.  In each control region the contamination from
signal and electroweak background is subtracted using simulation.

A comparison of the shape of the $I^{\mu}$ distributions for muons in
events with same-charge and opposite-charge muon pairs shows a small
linear deviation from unity of up to +10\% when extrapolated into the
isolated signal region $I^{\mu}<0.1$. This is found to be independent
of $m_{\mu\mu}$, and is accounted for in the extrapolation. The
$|y_{\mu\mu}|$ and $|\cos\theta^*|$ dependence of the background
in each $m_{\mu\mu}$ bin is obtained in the multijet enriched data
control region in which pairs of same-charge and opposite-charge
muons satisfy $I^{\mu}>0.1$. Finally, the resulting $|y_{\mu\mu}|$ and
$|\cos\theta^*|$ spectra are normalised in the signal region using the
constraint that the yield ratio of opposite-charge to same-charge muon
pairs is similar in the isolated and non-isolated control regions.

This method does not account for a potential $W$+jets background
contribution. This component is estimated from simulation and found
to be negligible. 

The estimated fake lepton background contribution in the muon channel is
everywhere smaller than its contribution in the central electron channel, and
never more than 5\% of the expected signal yield.

\subsection{Top quark and electroweak backgrounds}
\label{sec:bg_top_ew}

These sources of background arise from QCD and EW processes in which
two prompt isolated leptons are produced. Their contributions are estimated
using MC simulation.

Background events from top quark processes increase with
$m_{\ell\ell}$ and are typically below 2\% of the expected signal
yields. The contribution is largest at the extremes of $\cos\theta^*$
where it can reach 10--20\% of the expected signal in the central
channels. At high rapidity, this background source is typically below
5\% everywhere.

The diboson background increases with invariant mass
and reaches about 6\% of the expected signal yield at large
$|\cos\theta^*|$ in both the central electron and muon channels. In
the high rapidity electron channel it reaches about 3\% at moderate
$|y_{\ell\ell}|$.

The background from $Z\rightarrow\tau\tau$ is significant only at
low $m_{\ell\ell}$, where it can reach 7\% in the central rapidity channels and
3\% in the high rapidity channel.

Photon-induced production of dilepton pairs gives a small background
contribution of 2\% or less in all channels. However, for large values of
$m_{\ell\ell}$, this contribution can reach about 5\%.

\FloatBarrier

\section{Cross-section measurement}
\label{sec:methodology}

As defined in section~\ref{sec:binning}, the binning scheme used for
the triple-differential measurements consists of 504 bins for the central
rapidity electron and muon channels, and 150 bins in the high rapidity
electron channel. The Drell--Yan cross section is measured in the
central rapidity channels within the fiducial region defined
by $p_{\textrm T}^{\ell}>20$~{\GeV}, $|\eta^{\ell}|<2.4$, and
$46<m_{\ell\ell}<200$~{\GeV}. In the high rapidity electron channel the
fiducial region of the measurement is defined by
$p_{\textrm{T}}^{\ell}>25$~{\GeV} and $|\eta^{\ell}|<2.4$ for the central electron,
$p_{\textrm {T}}^{\ell}>20$~{\GeV} and $2.5<|\eta^{\ell}|<4.9$ for the
forward electron, and $66<m_{\ell\ell}<150$~{\GeV}. 

The cross-section results are first unfolded to the ``dressed''-level,
defined at the particle-level using leptons after FSR recombined with
radiated photons within a cone of $\Delta R = 0.1$.  The unfolded data
are then corrected to the Born-level, before final-state QED radiation
at the particle-level, using a correction factor obtained from the
\powheg\ MC sample. This procedure neglects the bin migrations
between the dressed- and Born-level kinematics, an approximation
which was verified to have a negligible impact on the central values
and uncertainties of the results presented in this paper.

The triple-differential cross section is calculated as
\begin{equation} 
 \frac{{\textrm{d}^3} \sigma} {{\textrm{d}} m_{\ell\ell} \, {{\textrm{d}}|y_{\ell\ell}|}  \, {\textrm{d}}\cos\theta^*} \Bigg|_{l,m,n}
= \mathcal{M}_{ijk}^{lmn}\,\cdot\,
\frac{N^{\textrm{data}}_{ijk} - N^{\textrm{bkg}}_{ijk} }{\mathcal{L}_{\textrm{int}}} \, \frac{1}
{\Delta_{m_{\ell\ell}} \,\cdot\, 2  \Delta_{|y_{\ell\ell}|}  \,\cdot\,   \Delta_{\cos\theta^*}  }\,\,\,,
\label{eq:xsec}
\end{equation}
where $i,j,k$ are the bin indices for reconstructed final-state kinematics;
$l,m,n$ are the bin indices for the generator-level kinematics; and
$\mathcal{L}_{\textrm{int}}$ is the integrated luminosity of the data set.
Quantity $N^{\textrm{data}}$ is the number of candidate signal events
observed in a given bin of width $\Delta_{m_{\ell\ell}}$, $\Delta_{|y_{\ell\ell}|}$,
and $\Delta_{\cos\theta^*}$, while $N^{\textrm{bkg}}$ is the number of
background events in the same bin. The factor of two in the denominator
accounts for the modulus in the rapidity bin width. Integrated single- and
double-differential cross sections are measured by summing over the
corresponding indices of equation~(\ref{eq:xsec}).

The factor $\mathcal{M}$ is the inverted response matrix and takes
into account the efficiency of the signal selection and bin migration
effects. It gives the probability that a selected event reconstructed in
some measurement bin was originally generated in a given fiducial
(generator-level) bin. The factor $\mathcal{M}$ is obtained from the Drell--Yan
signal samples after correcting for differences in the reconstruction,
identification, trigger, and isolation efficiencies between data and
simulation, as well as for momentum scale and resolution mismodelling
effects. It also accounts for events originally outside of
the fiducial selection that migrate into the reconstructed event
sample.  Finally, $\mathcal{M}$ also includes extrapolations over the
regions that are excluded from the electron selection
($1.37<|\eta^e|<1.52$, 
$2.70<\left|\eta^e\right|<2.80$,
and
$3.00<\left|\eta^e\right|<3.35$
).

\begin{figure}[htp!]
\centering
\includegraphics[width=0.43\textwidth]{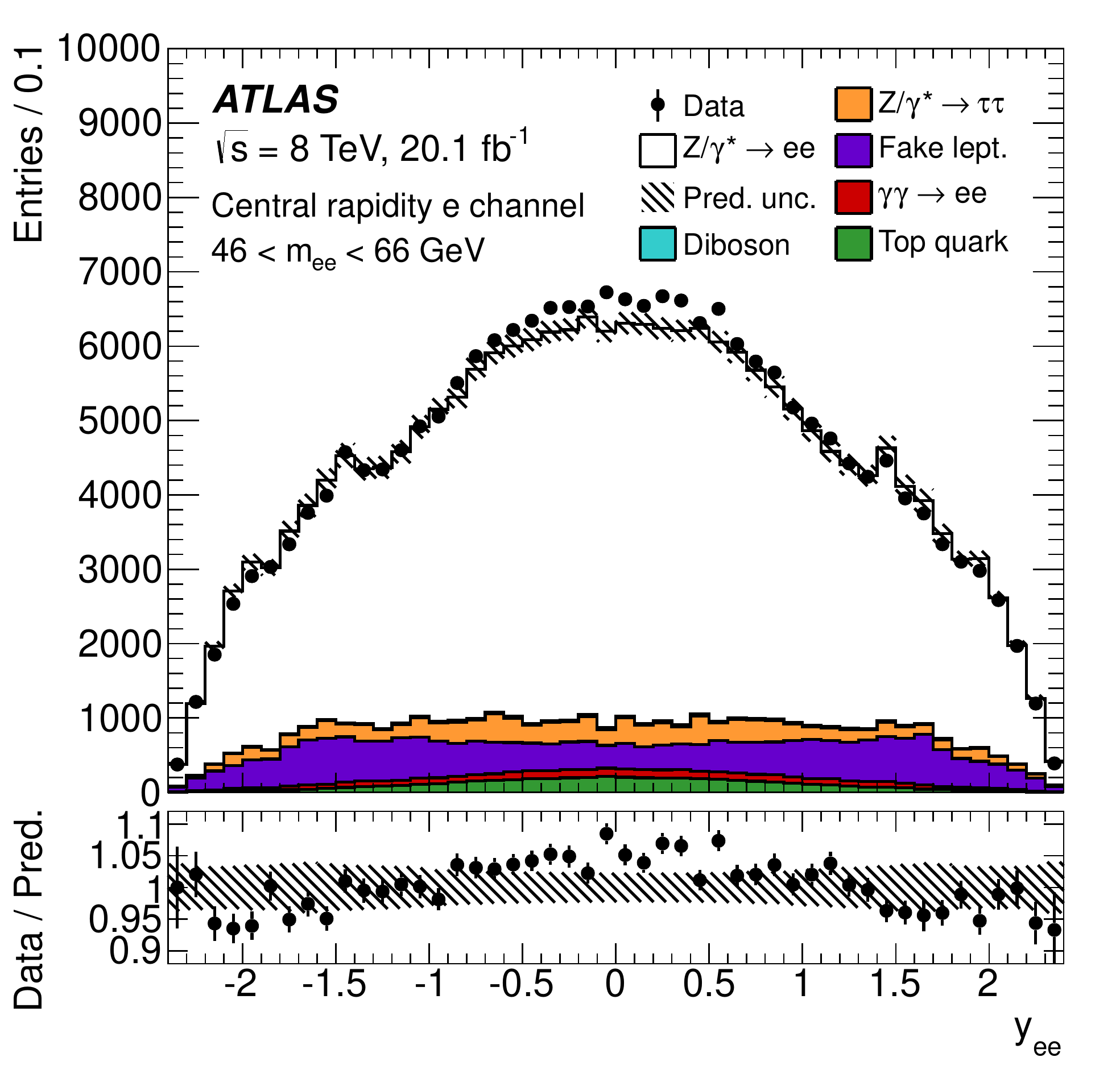}~
\includegraphics[width=0.43\textwidth]{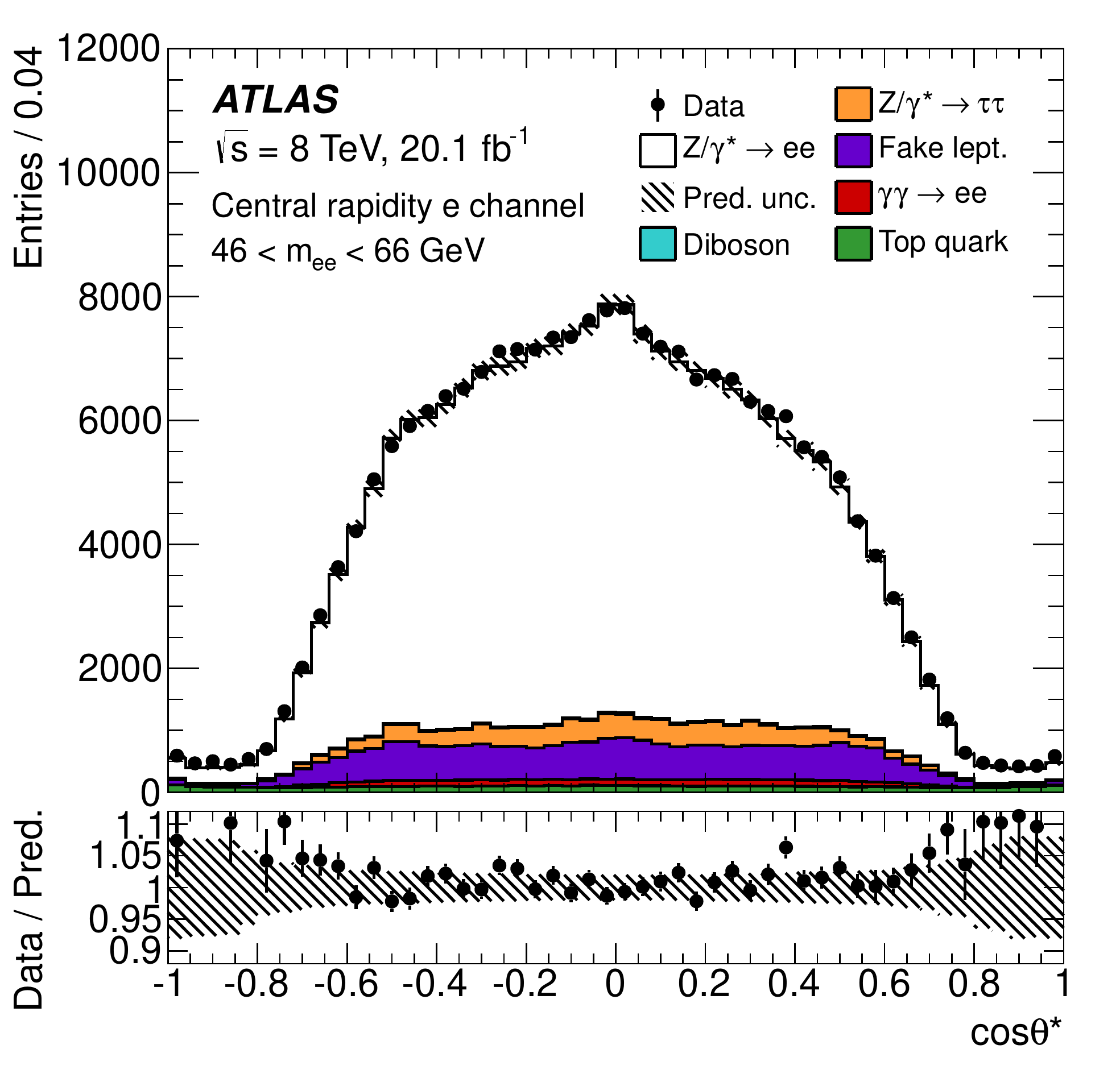}\\
\includegraphics[width=0.43\textwidth]{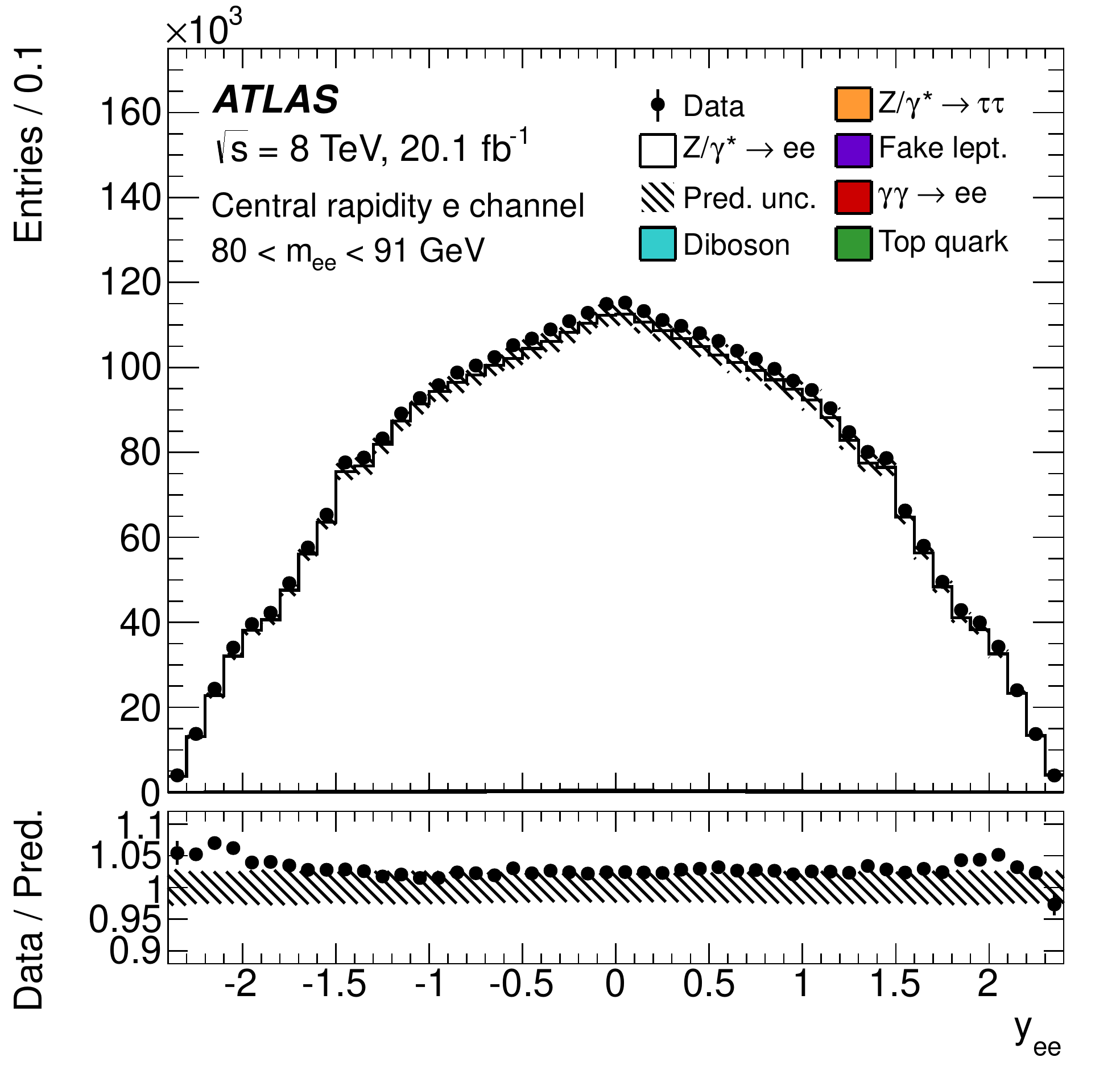}~
\includegraphics[width=0.43\textwidth]{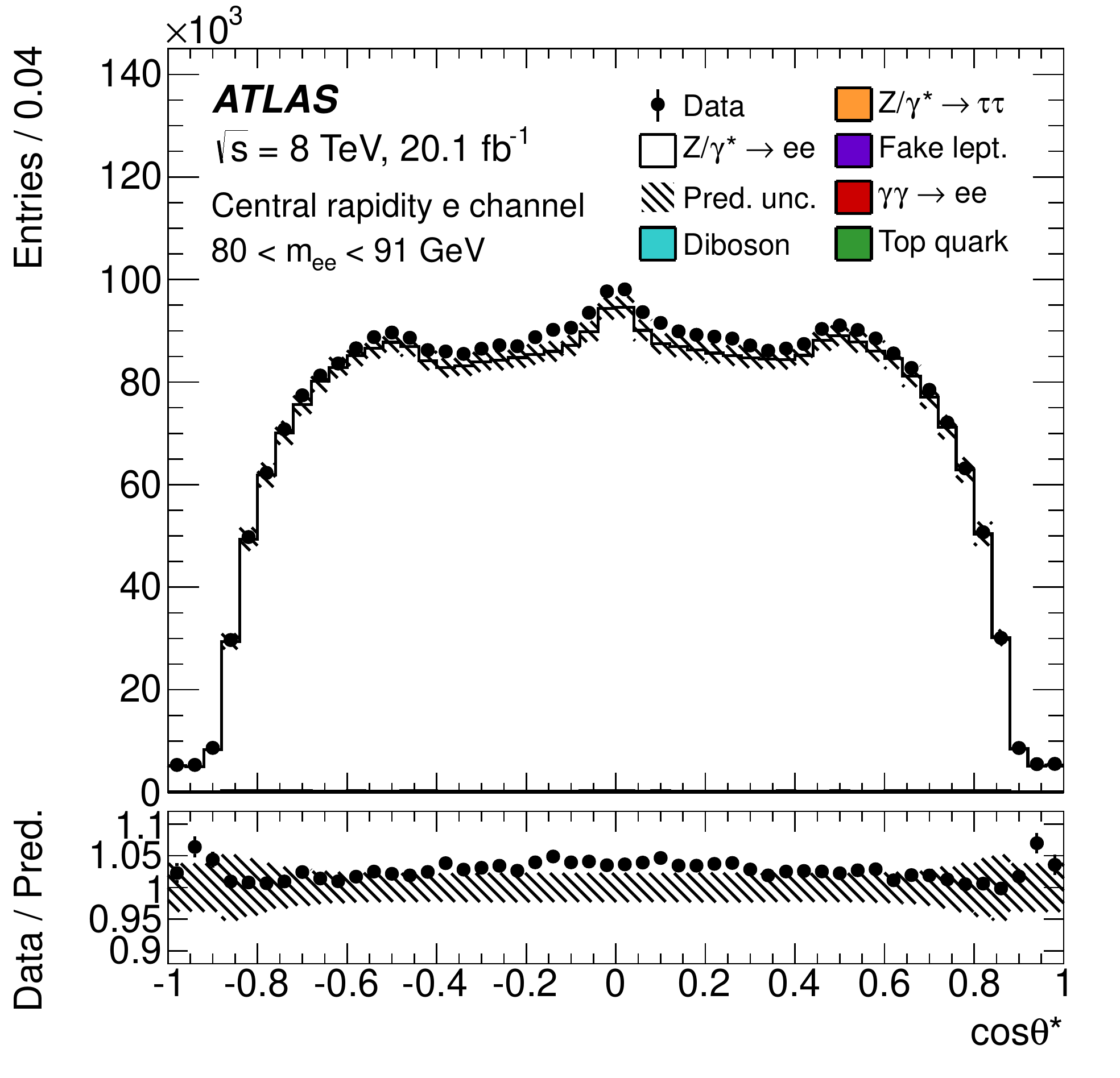}\\
\includegraphics[width=0.43\textwidth]{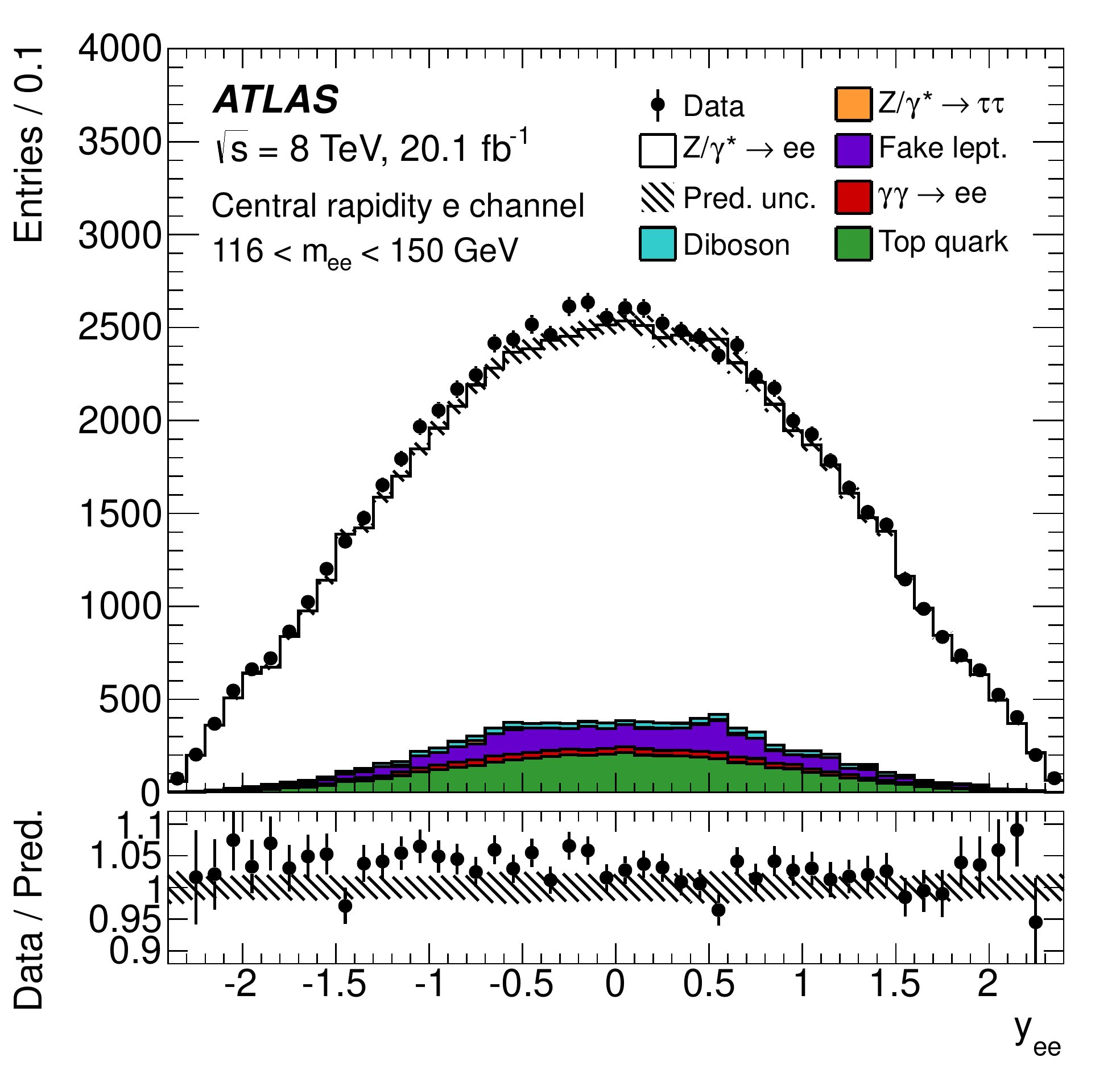}~
\includegraphics[width=0.43\textwidth]{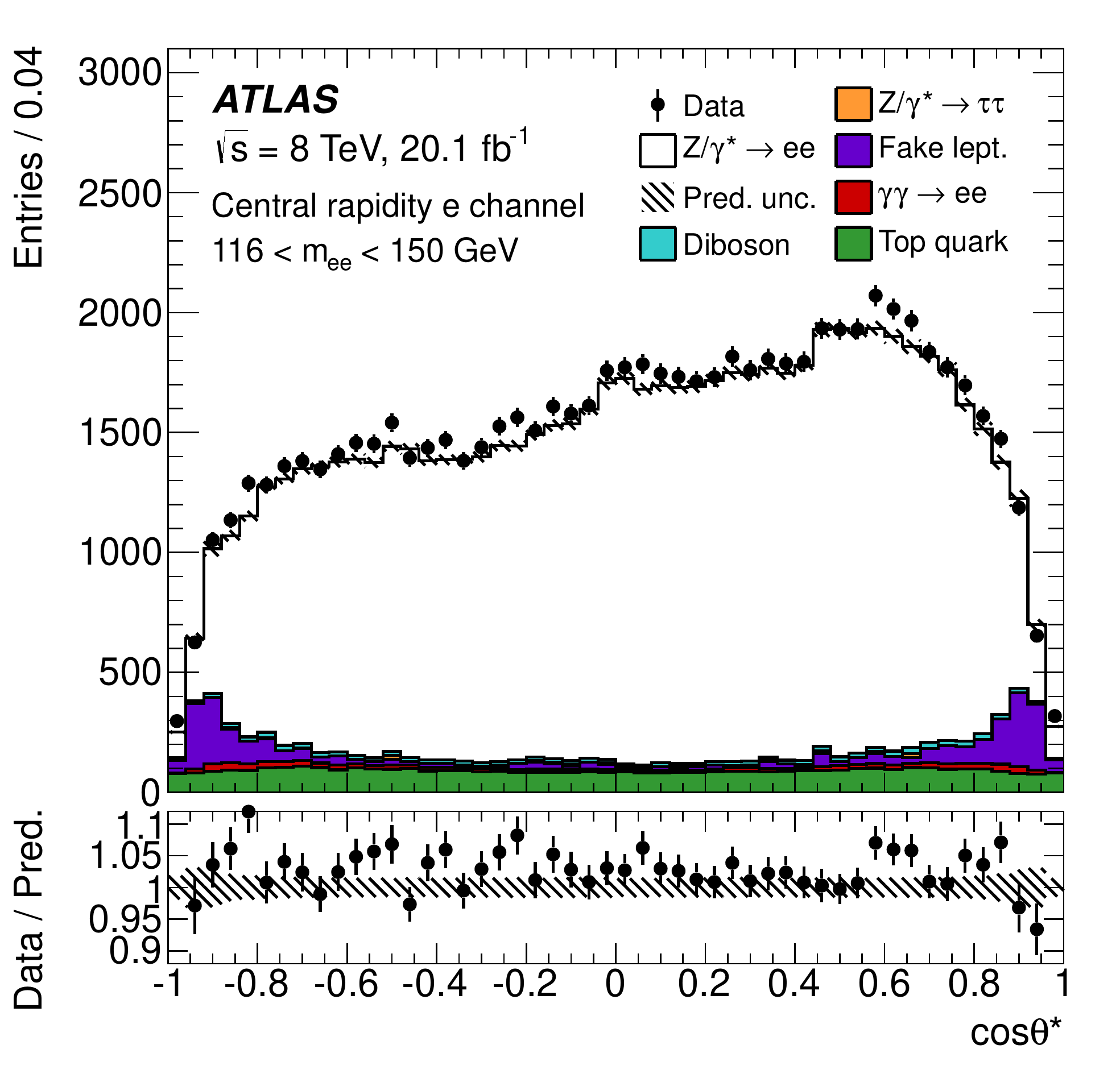}\\
\caption{Distributions of dilepton rapidity (left) and $\cos\theta^*$ (right)
	in the central rapidity electron channel for $m_{ee}$ bins
	46--66~{\GeV} (top row), 80--91~{\GeV} (middle), and
	116--150~{\GeV} (bottom). The data (solid markers)
	and the prediction (stacked histogram) are shown after event
	selection. The lower panels in each plot show the ratio of data
	to prediction. The error bars represent the data statistical
	uncertainty while the hatched band represents the systematic
	uncertainty in the prediction.}
\label{fig:zcc_control}
\end{figure}

\begin{figure}[htp!]
\centering
\includegraphics[width=0.43\textwidth]{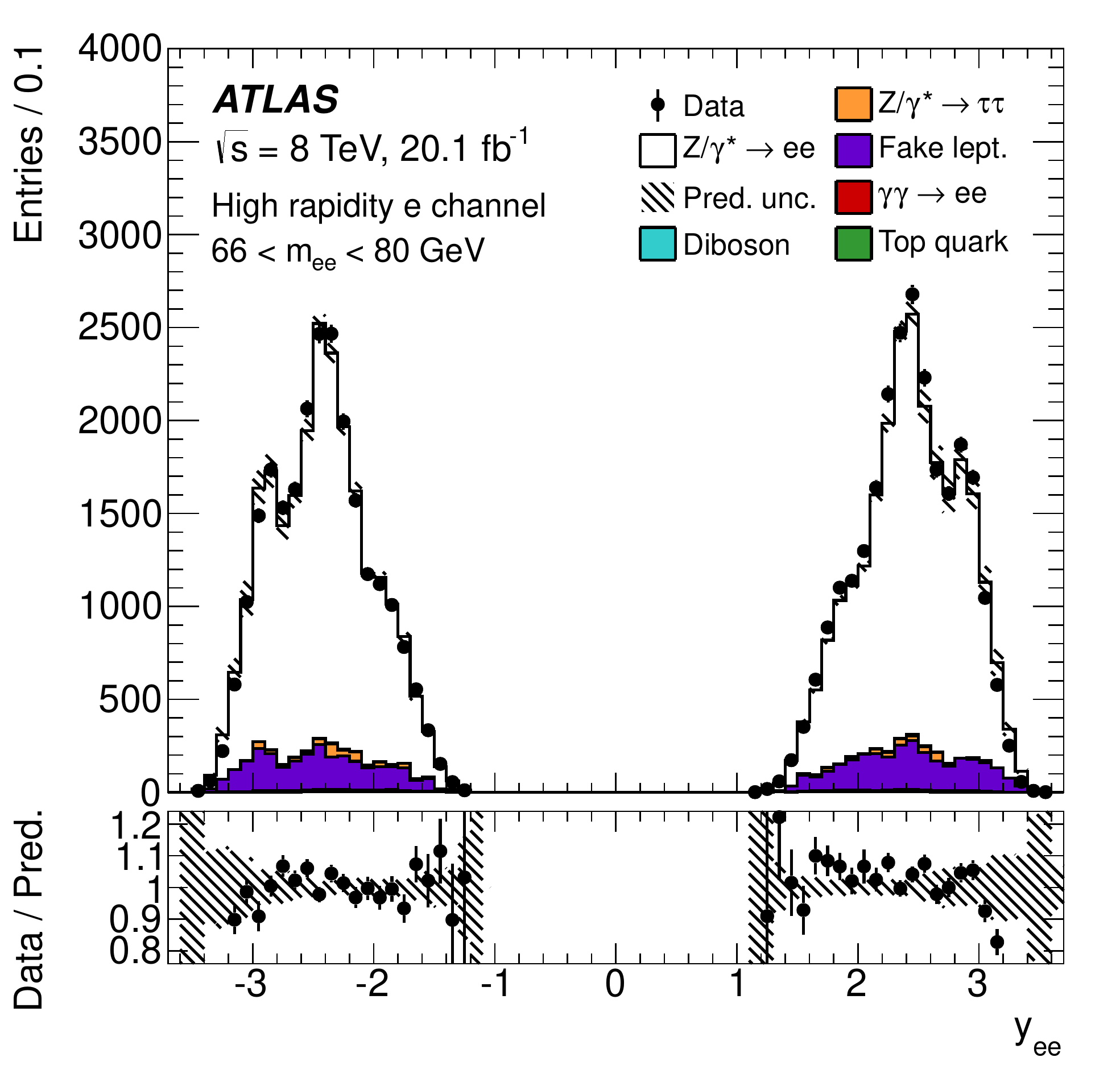}~
\includegraphics[width=0.43\textwidth]{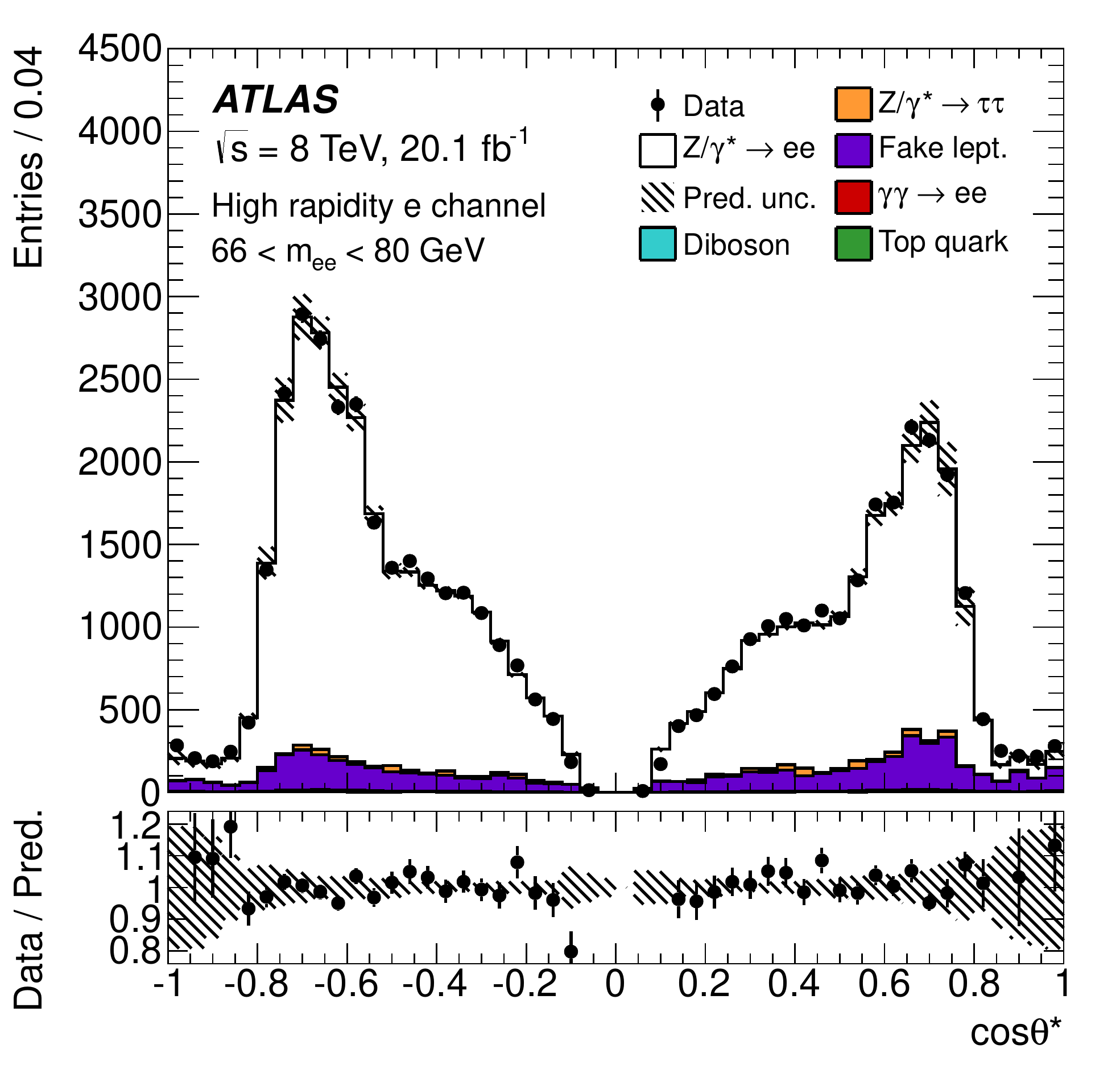}\\
\includegraphics[width=0.43\textwidth]{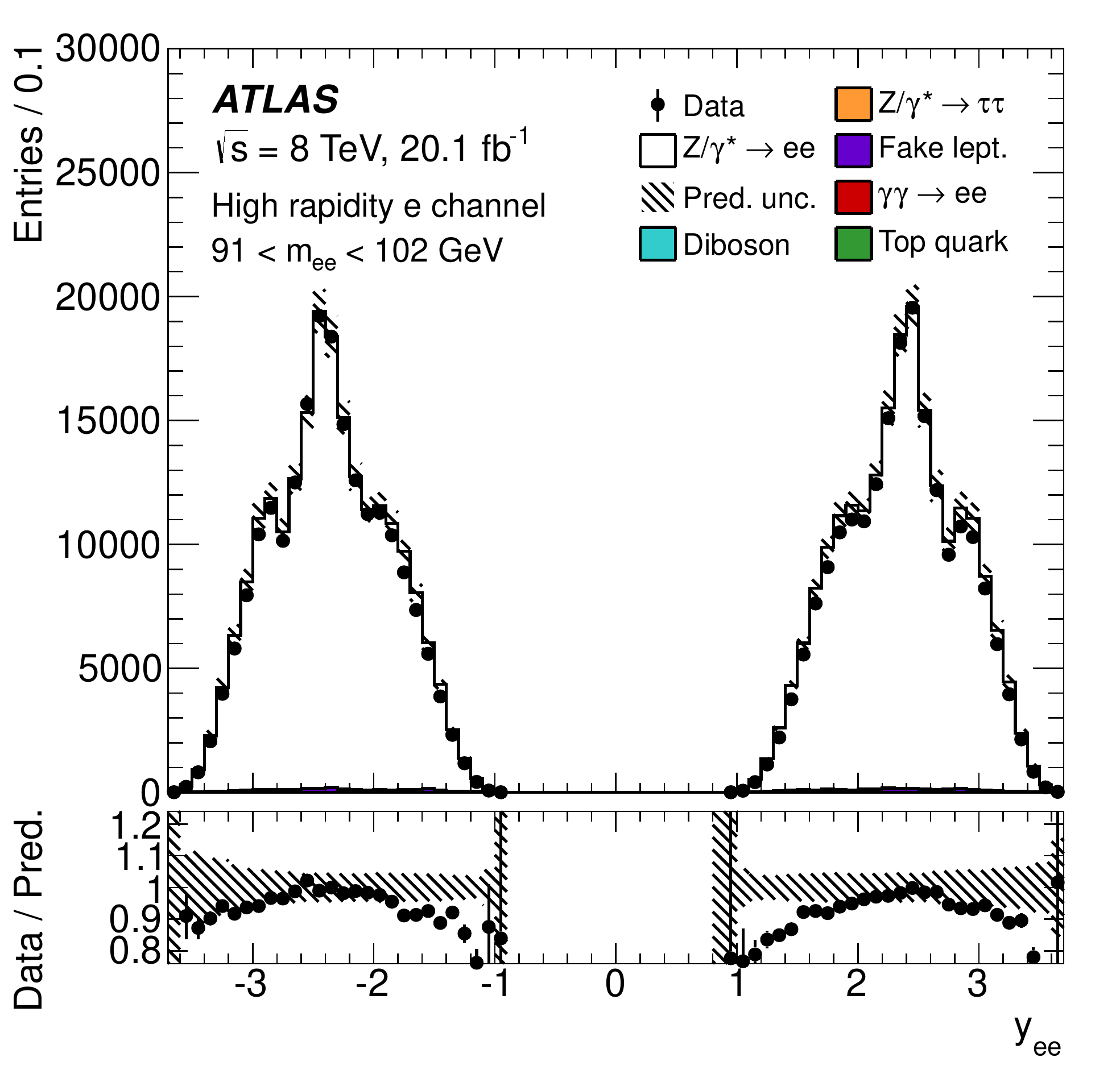}~
\includegraphics[width=0.43\textwidth]{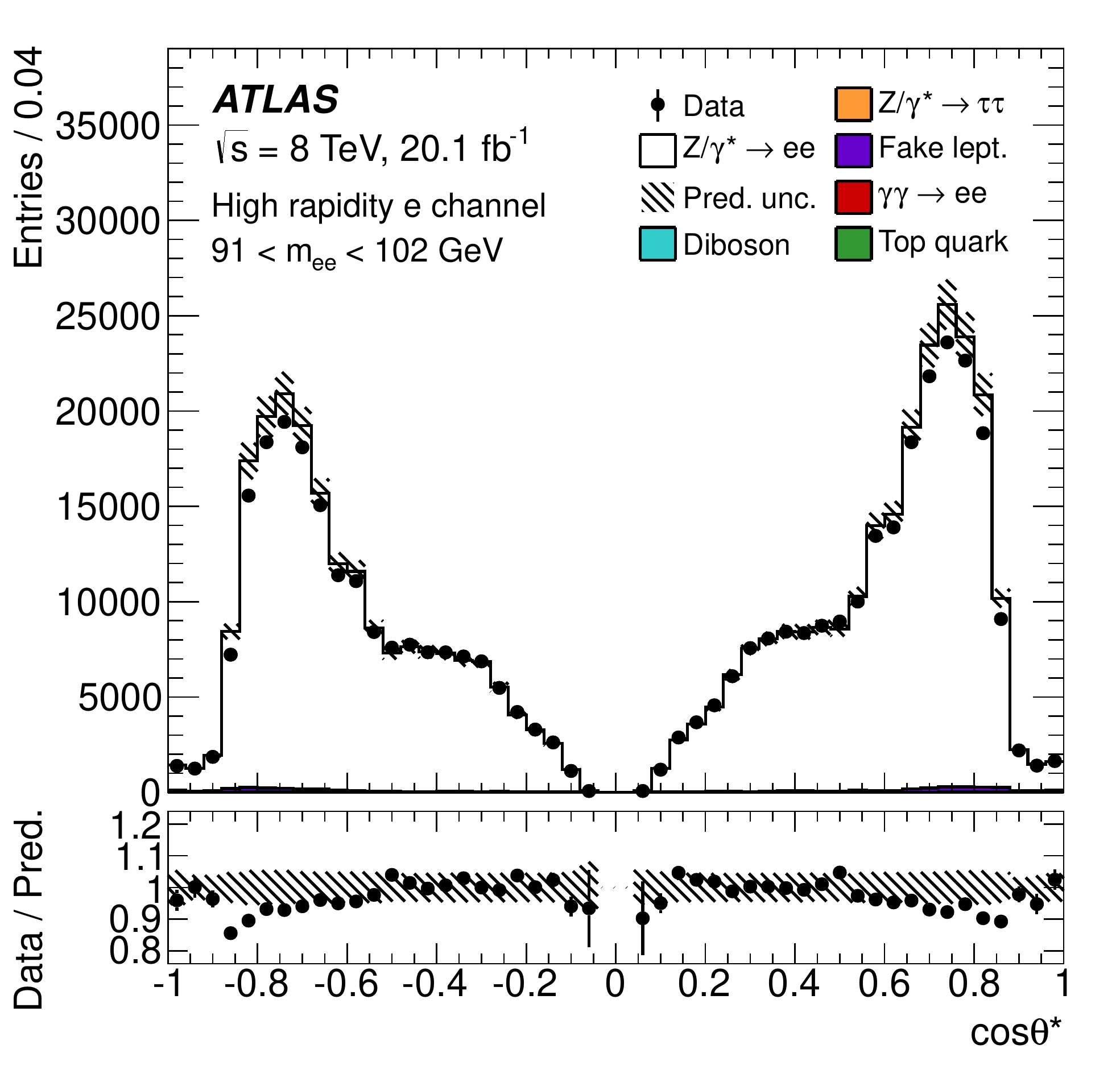}\\
\includegraphics[width=0.43\textwidth]{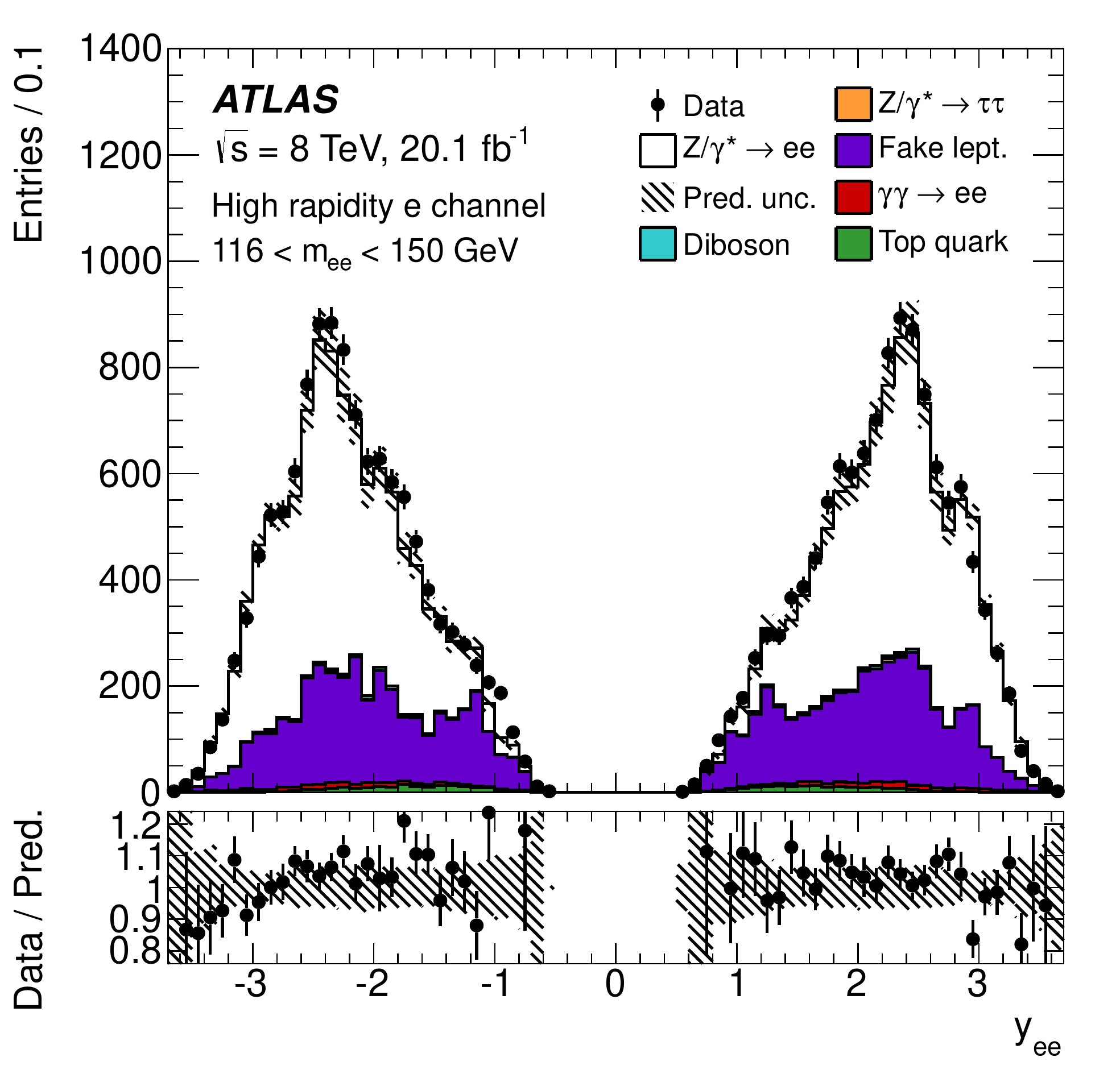}~
\includegraphics[width=0.43\textwidth]{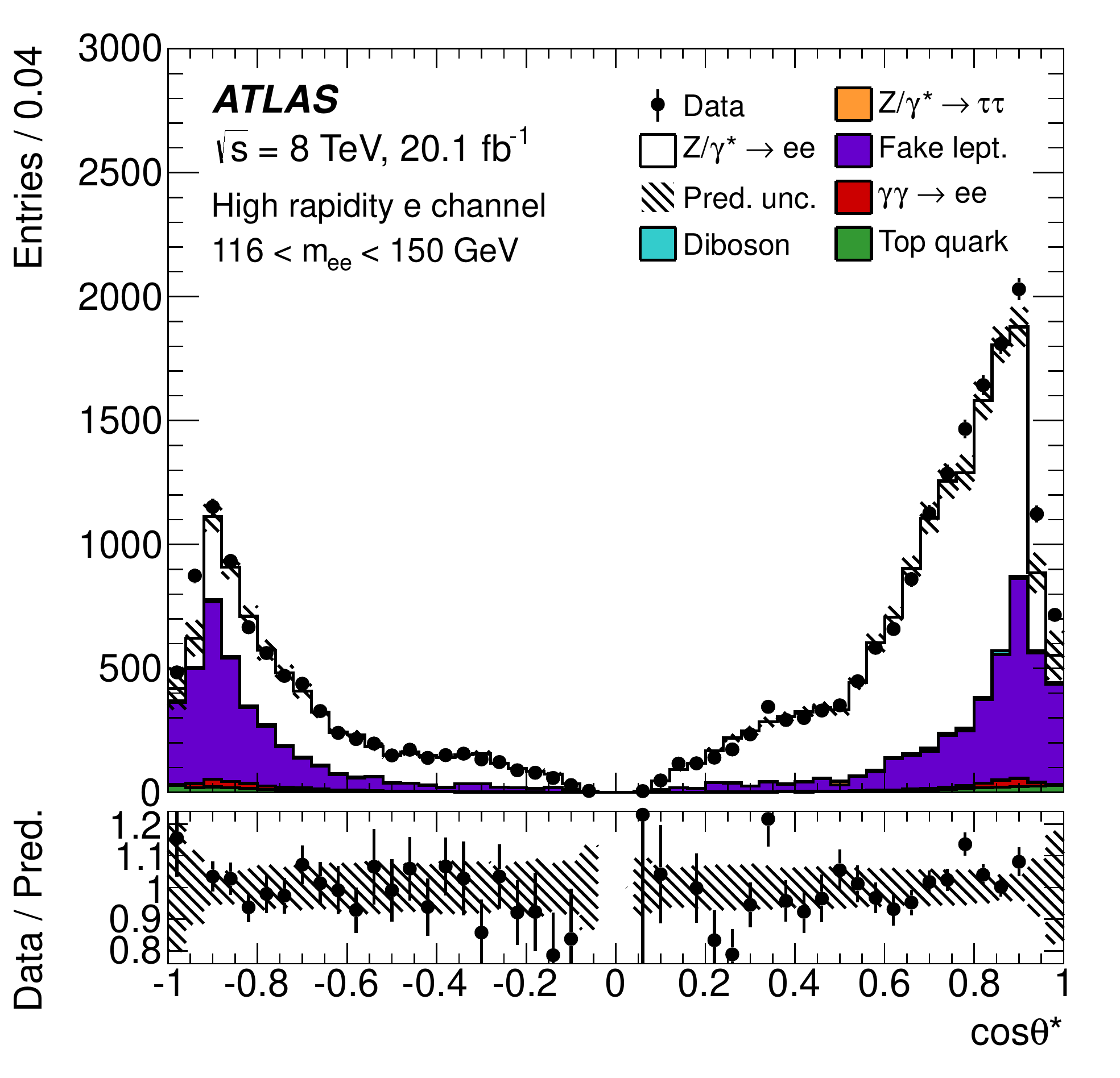}\\
\caption{Distributions of dilepton rapidity (left) and $\cos\theta^*$ (right)
	in the high rapidity electron channel for $m_{ee}$ bins
	66--80~{\GeV} (top row), 91--102~{\GeV} (middle), and
	116--150~{\GeV} (bottom). The data (solid markers)
	and the prediction (stacked histogram) are shown after event
	selection. The lower panels in each plot show the ratio of data
	to prediction. The error bars represent the data statistical
	uncertainty while the hatched band represents the systematic
	uncertainty in the prediction.}
\label{fig:zcf_control}
\end{figure}

\begin{figure}[htp!]
\centering
\includegraphics[width=0.43\textwidth]{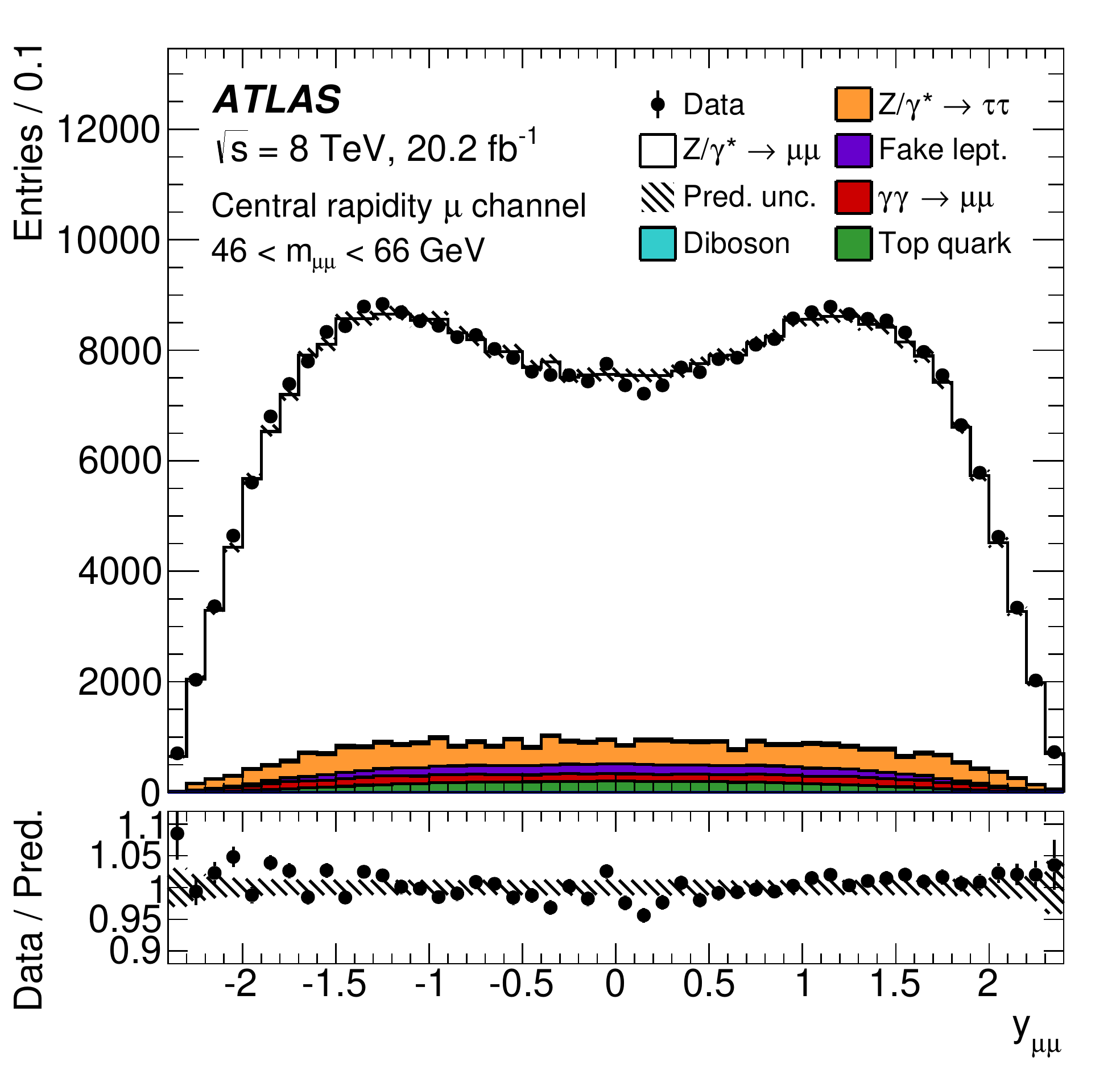}~
\includegraphics[width=0.43\textwidth]{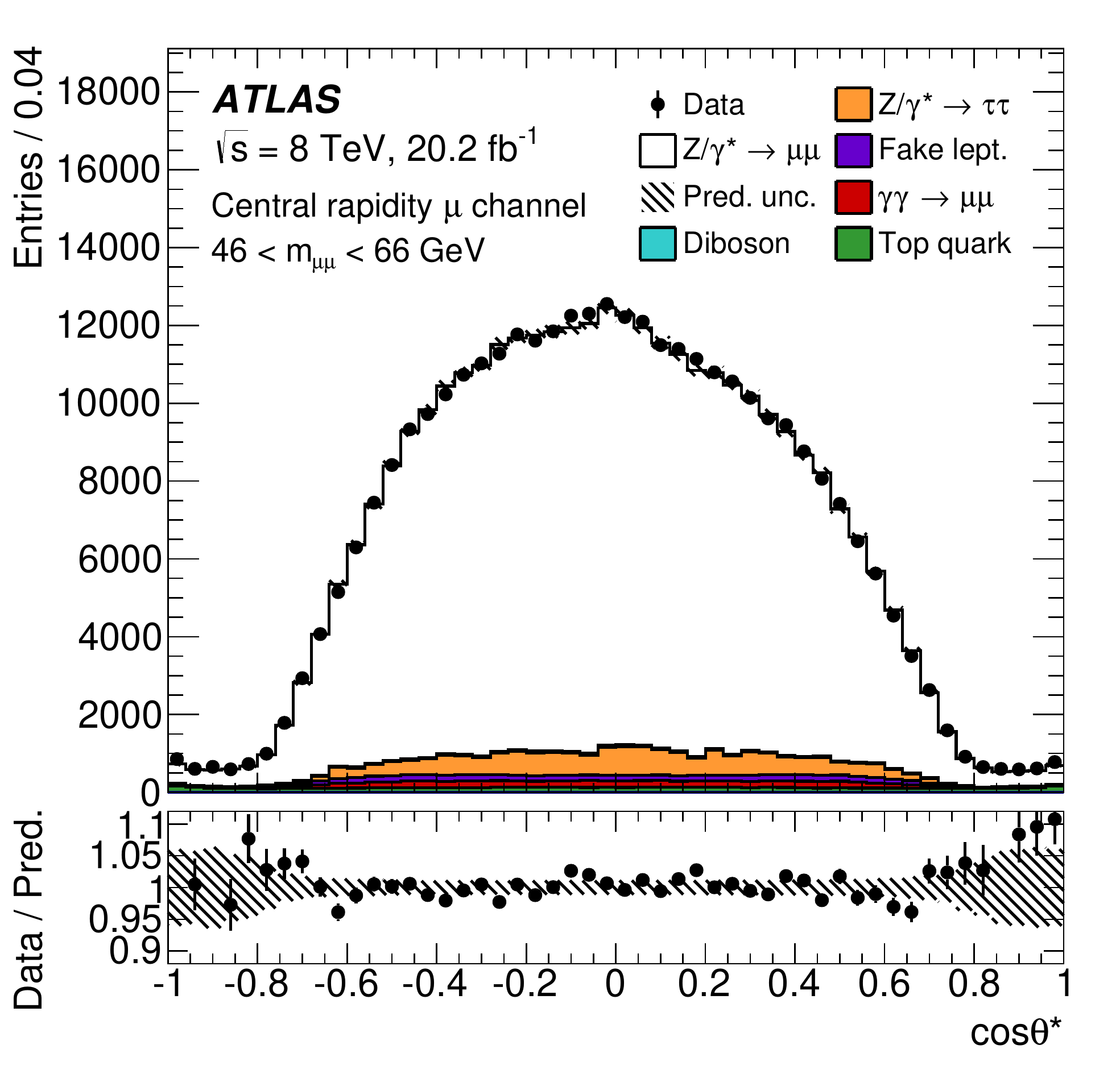}\\
\includegraphics[width=0.43\textwidth]{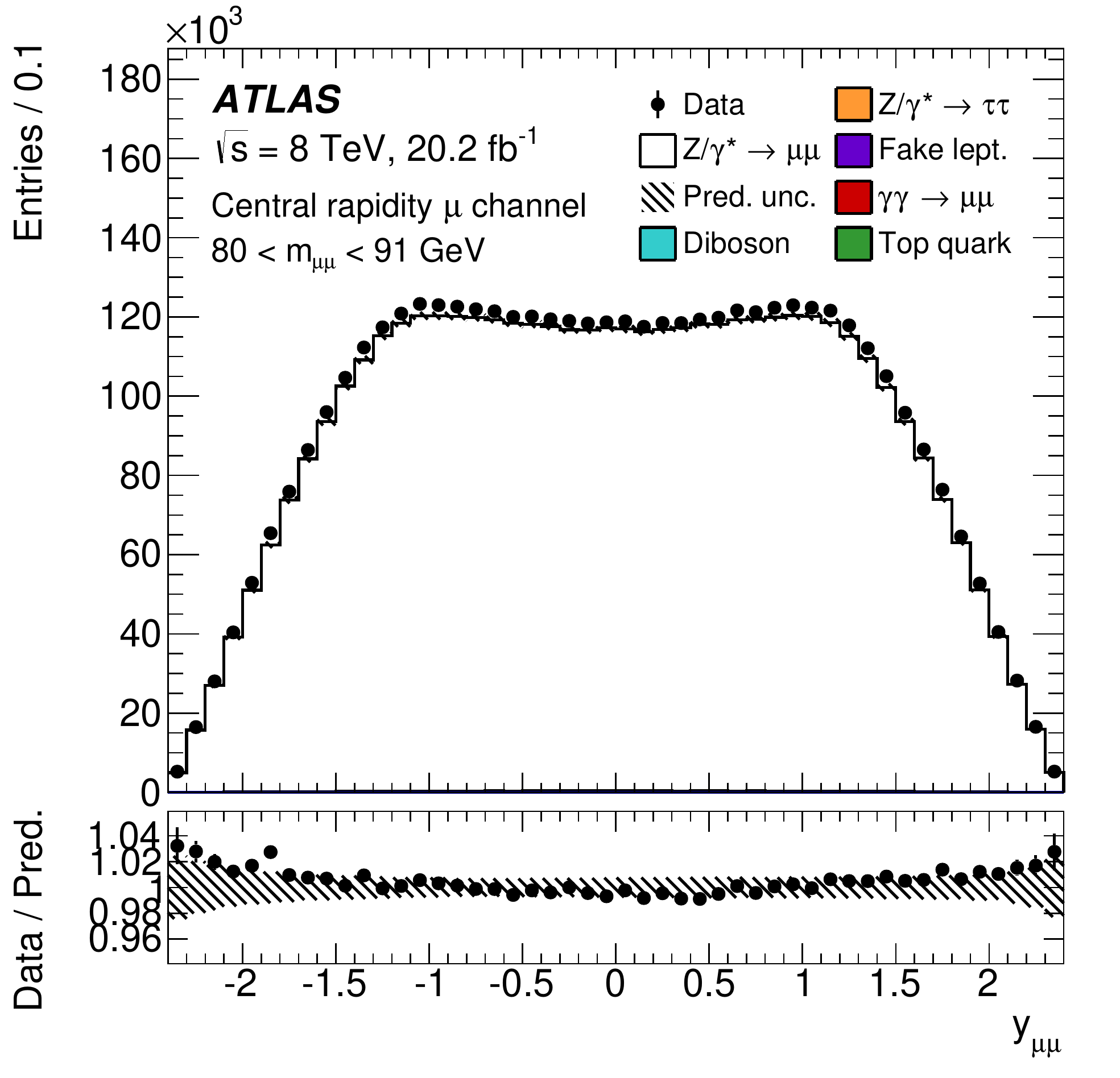}~
\includegraphics[width=0.43\textwidth]{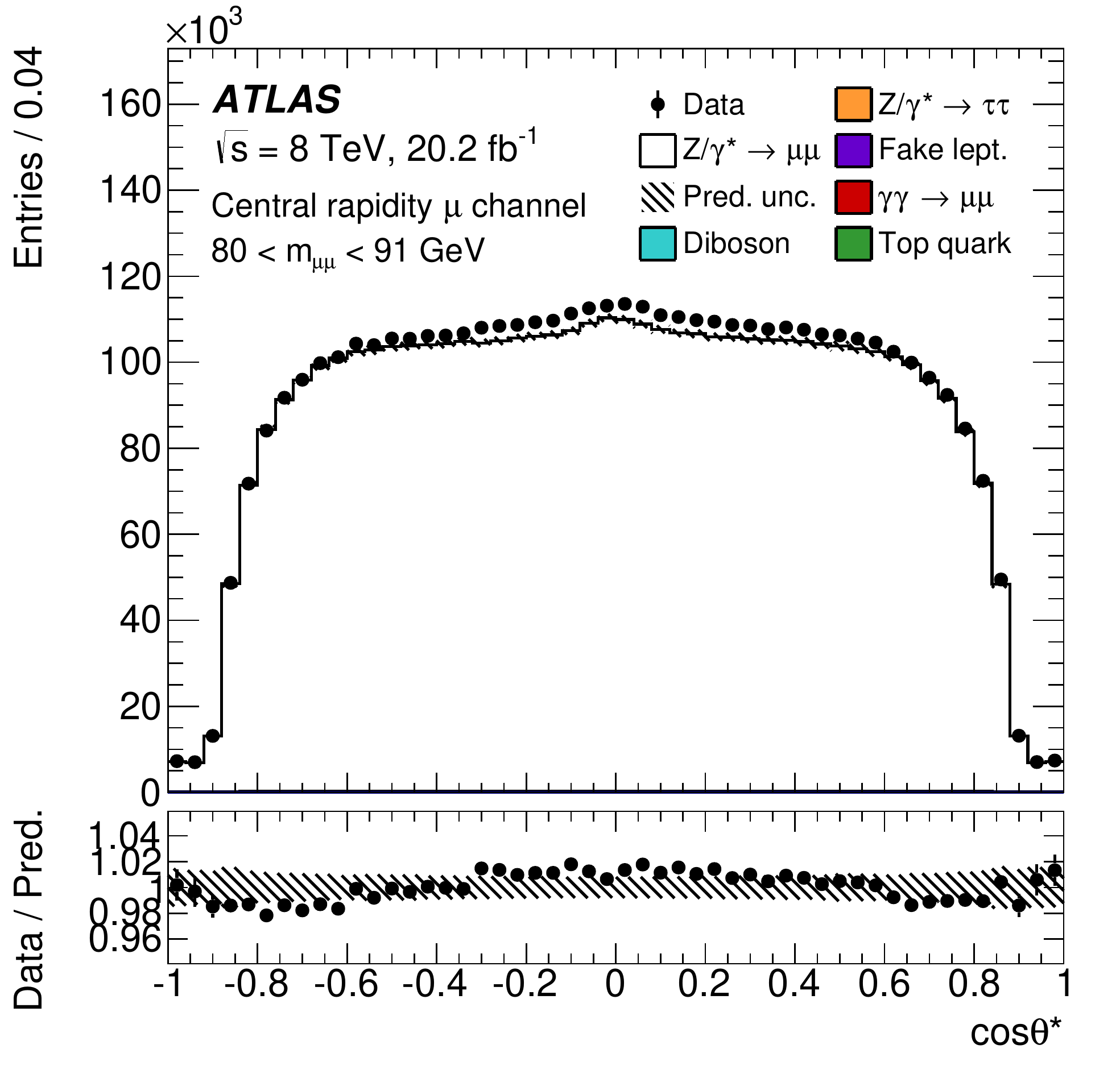}\\
\includegraphics[width=0.43\textwidth]{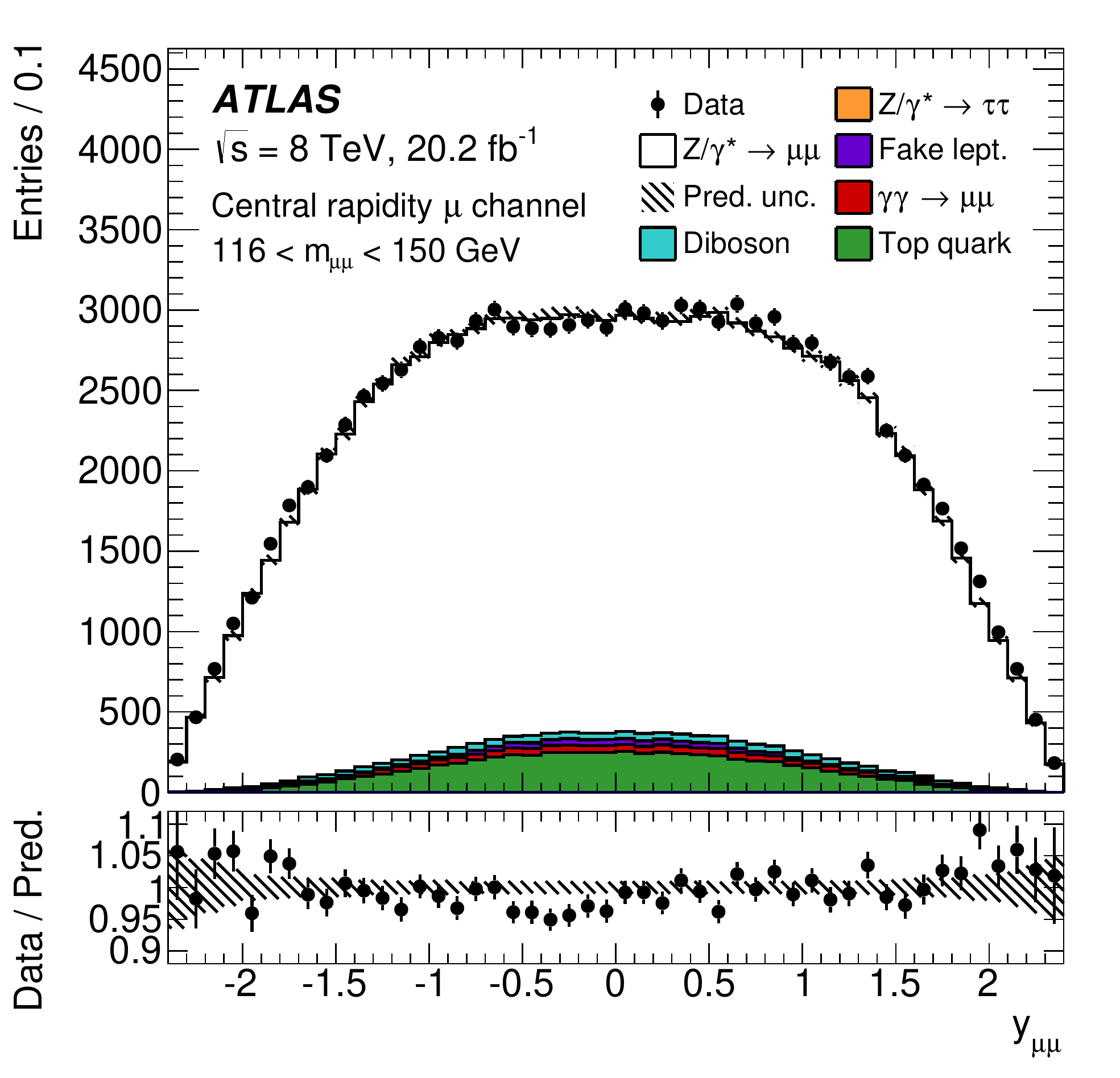}~
\includegraphics[width=0.43\textwidth]{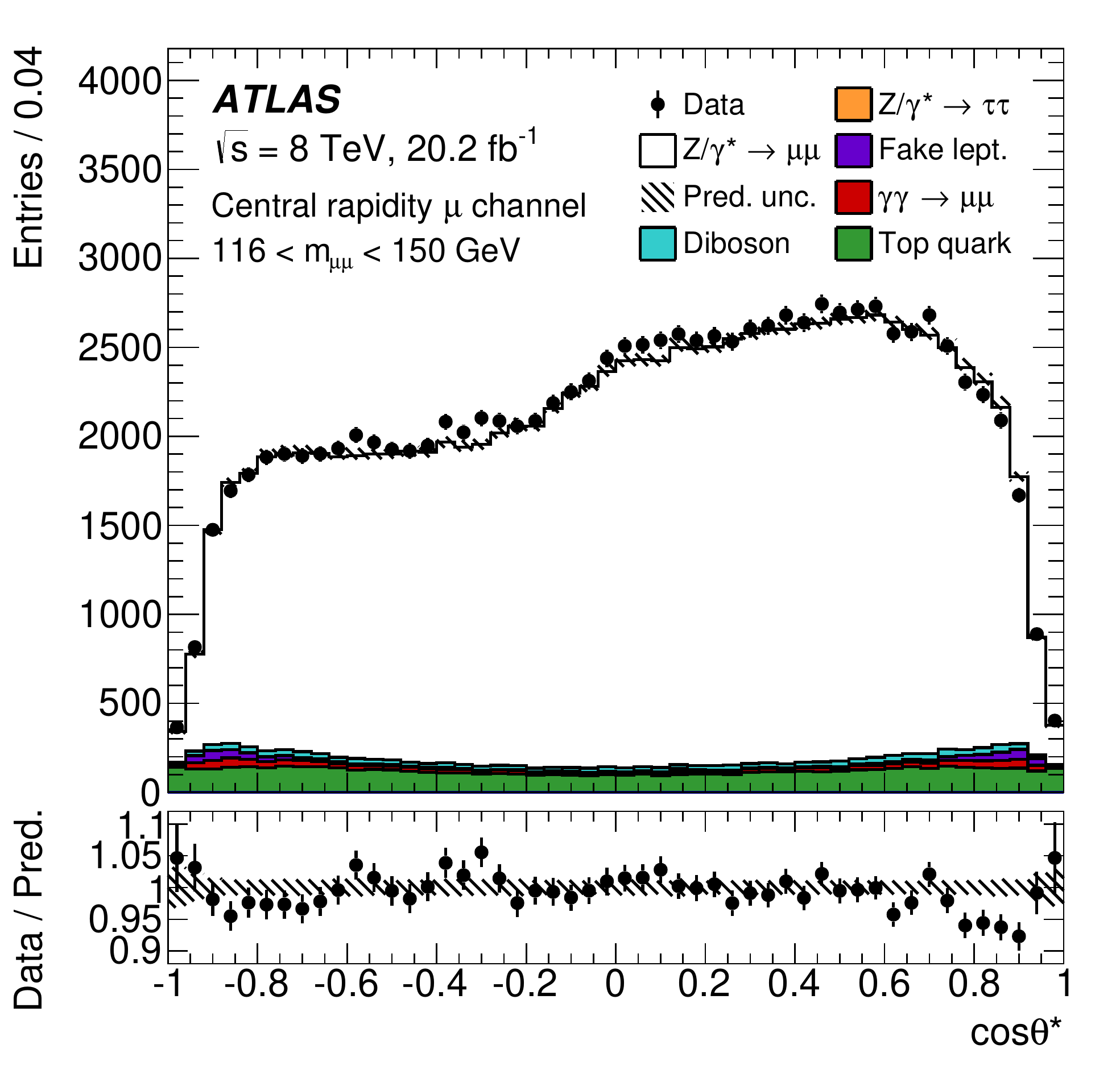}\\
\caption{Distributions of dilepton rapidity (left) and $\cos\theta^*$ (right)
	in the central rapidity muon channel for $m_{\mu\mu}$ bins
	46--66~{\GeV} (top row), 80--91~{\GeV} (middle), and
	116--150~{\GeV} (bottom). The data (solid markers)
	and the prediction (stacked histogram) are shown after event
	selection. The lower panels in each plot show the ratio of data
	to prediction. The error bars represent the data statistical
	uncertainty while the hatched band represents the systematic
	uncertainty in the prediction.}
\label{fig:zmm_control}
\end{figure}

\begin{figure}[htp!]
\centering
\includegraphics[width=0.43\textwidth]{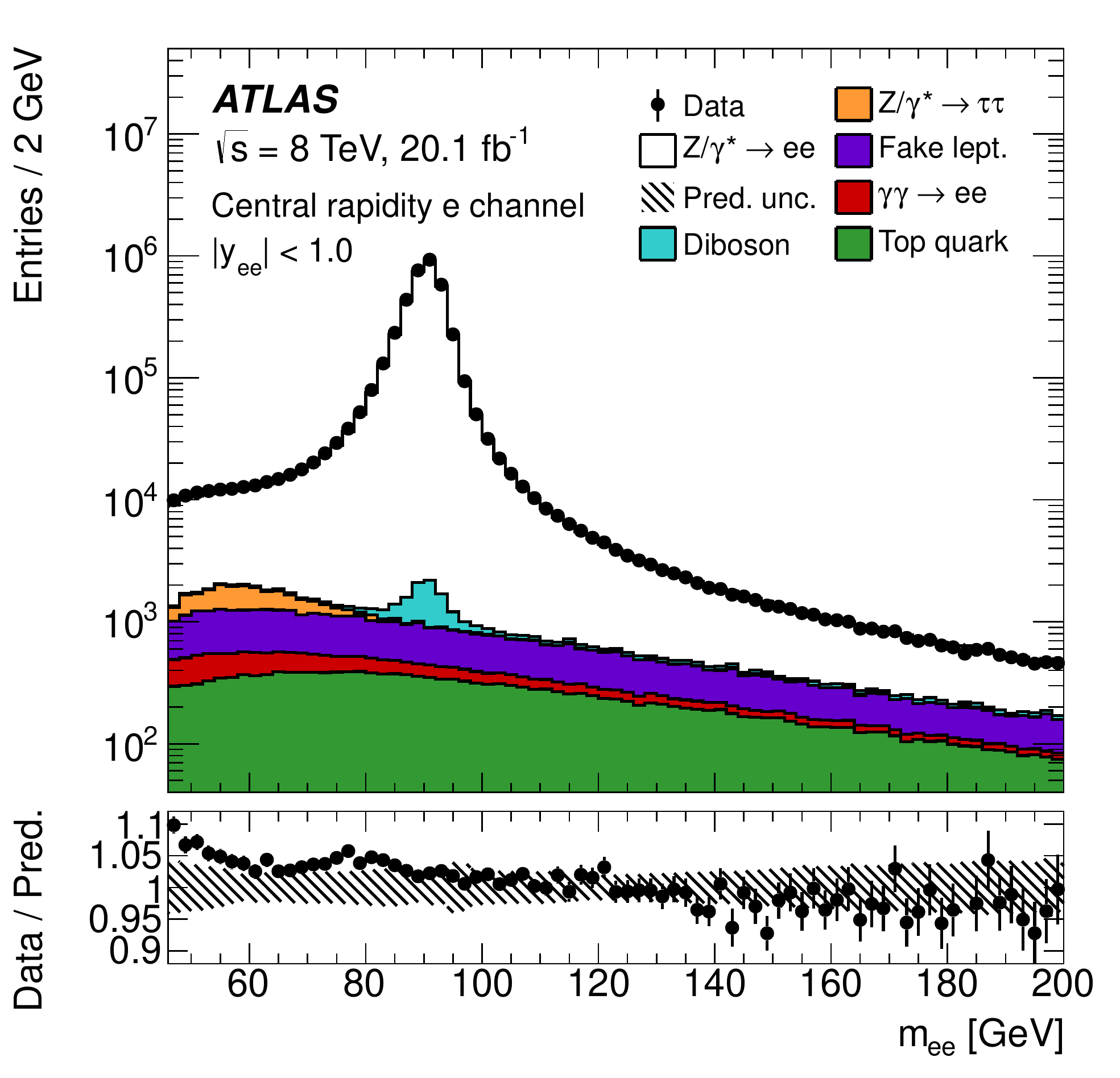}~
\includegraphics[width=0.43\textwidth]{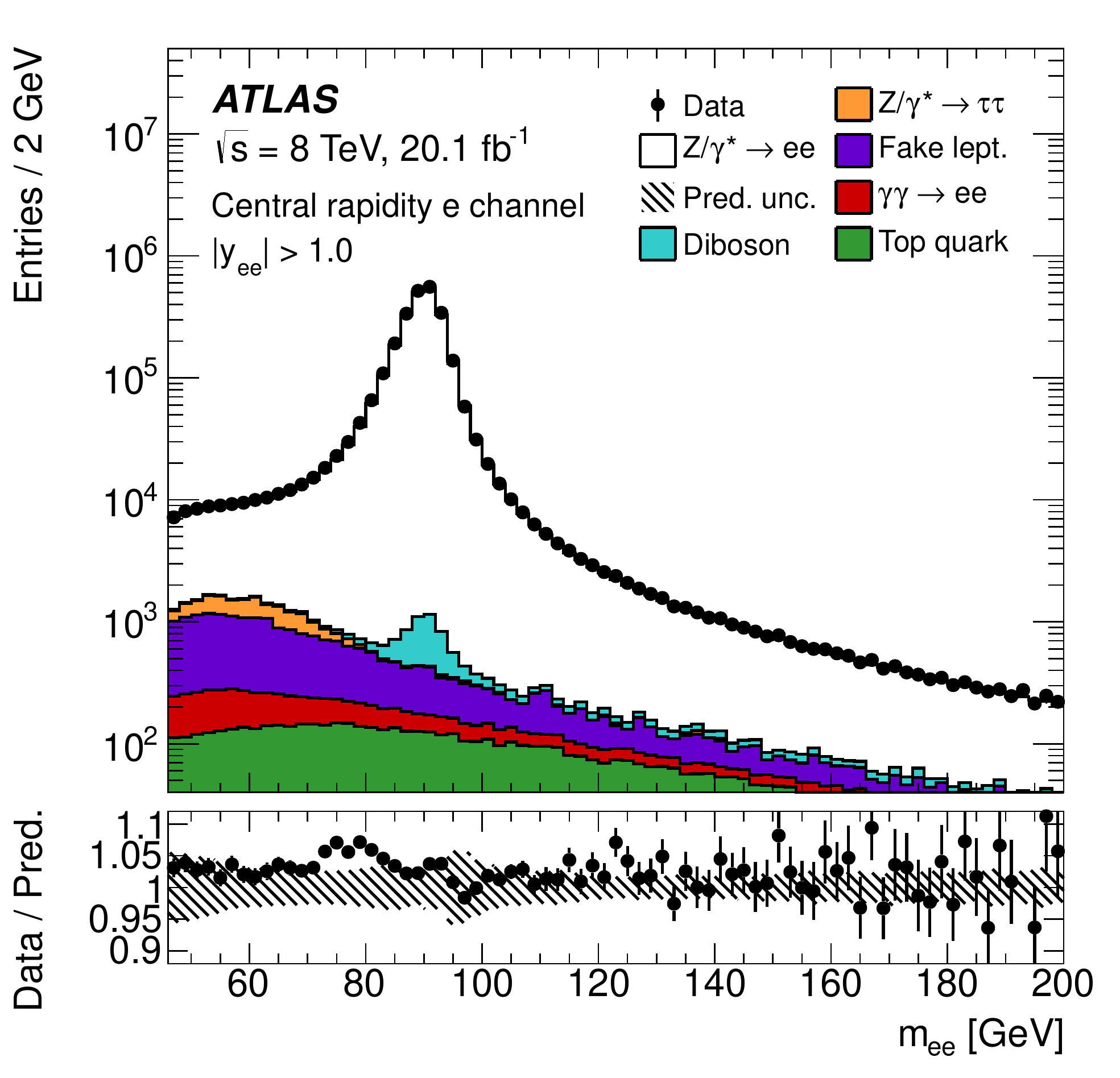}\\
\includegraphics[width=0.43\textwidth]{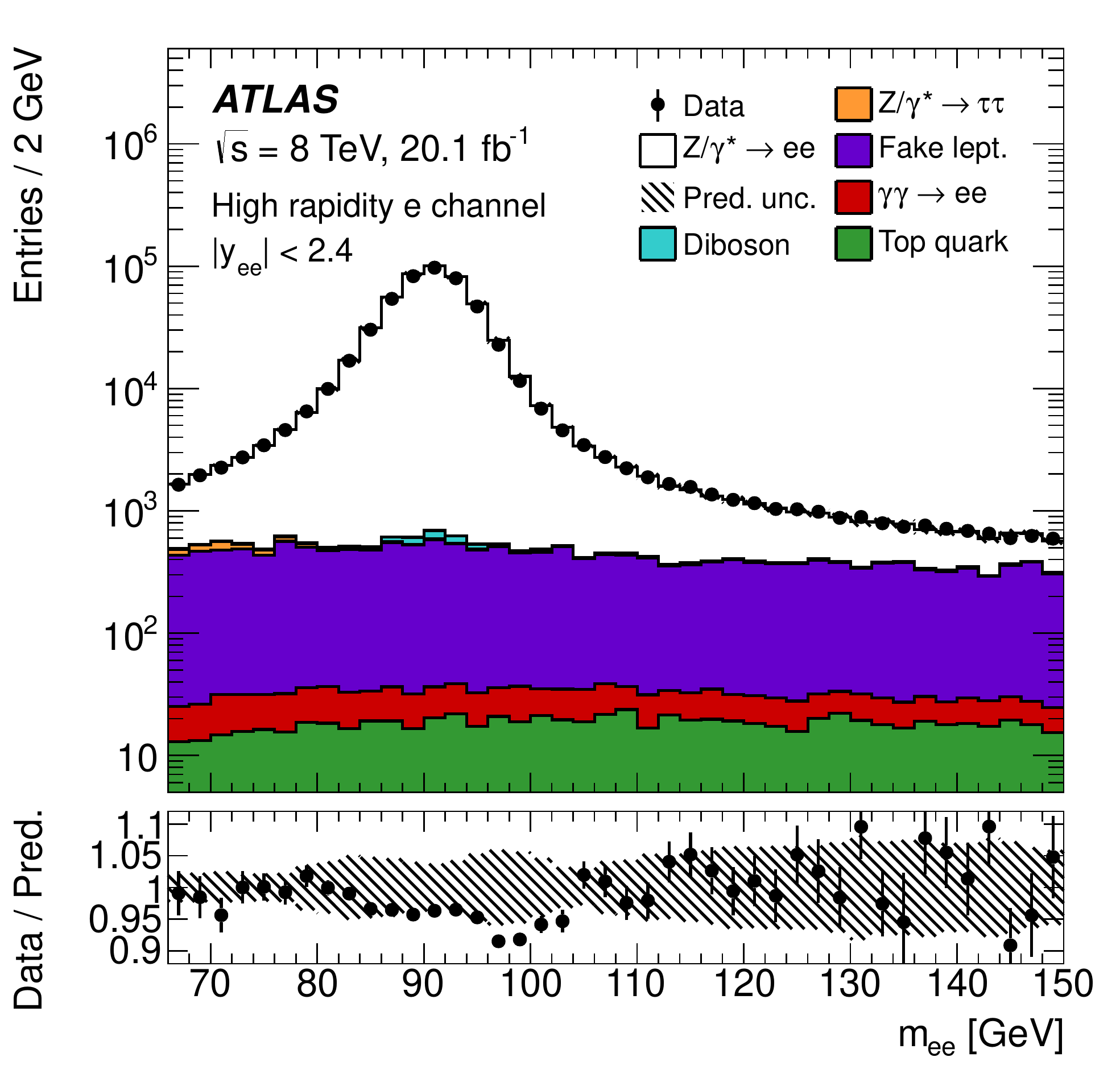}~
\includegraphics[width=0.43\textwidth]{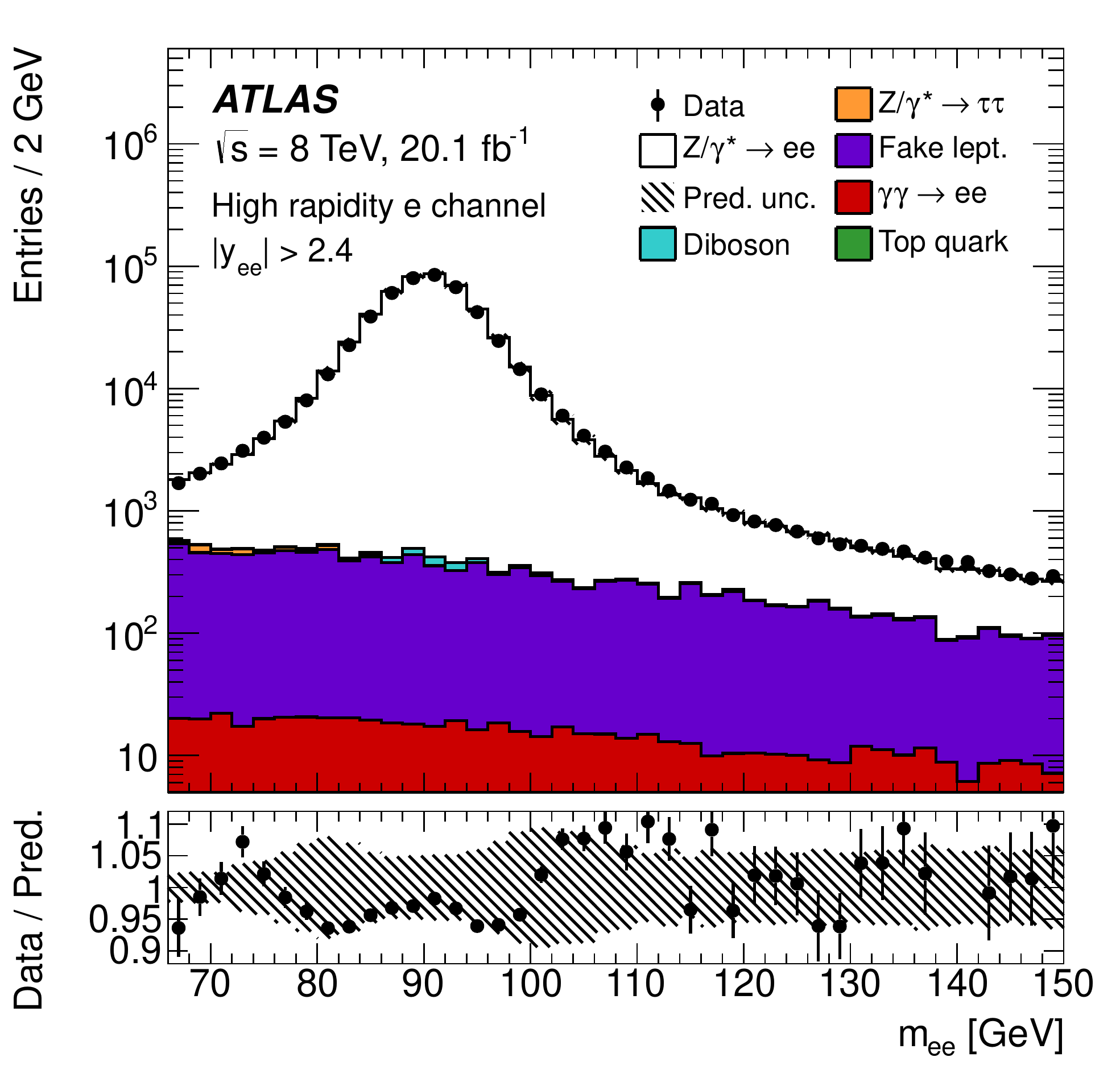}\\
\includegraphics[width=0.43\textwidth]{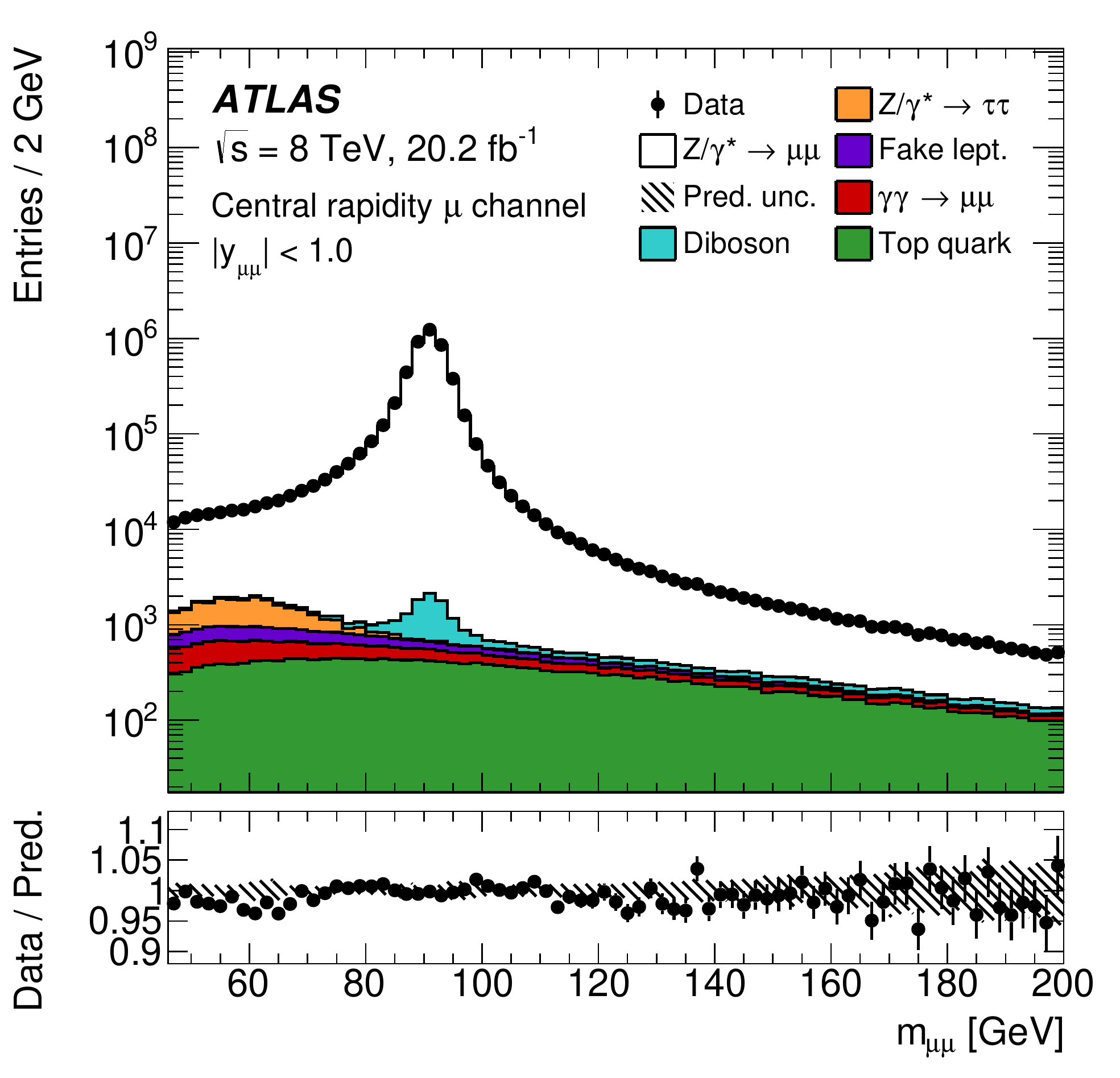}~
\includegraphics[width=0.43\textwidth]{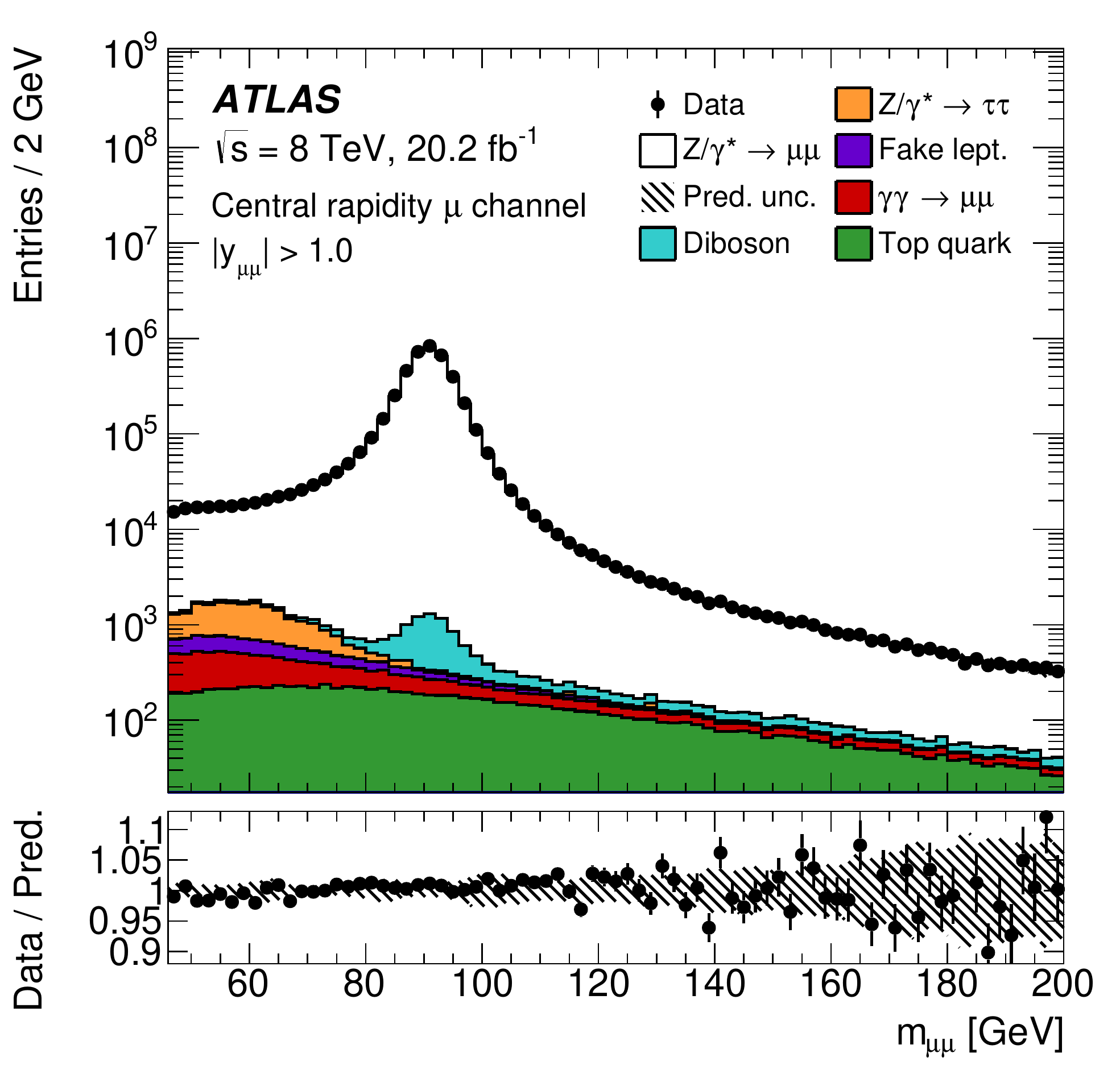}\\
\caption{Distributions of invariant mass for all three measurements:
	the central rapidity electron (top row), the high rapidity electron
	channel (middle), and the central rapidity muon (bottom) channels.
	For the central measurements, the distributions are plotted for
	$|y_{\ell\ell}|<1.0$ (left) and $|y_{\ell\ell}|>1.0$ (right) while
	for the high rapidity measurement, regions $|y_{ee}|<2.4$ (left)
	and $|y_{ee}|>2.4$ (right) are shown. The data (solid markers)
	and the prediction (stacked histogram) are shown after event
	selection. The lower panels in each plot show the ratio of data
	to prediction. The error bars represent the data statistical
	uncertainty while the hatched band represents the systematic
	uncertainty in the prediction.}
\label{fig:z_control_m}
\end{figure}

The quality of the simulation and its ability to describe the data are
checked in figures~\ref{fig:zcc_control}--\ref{fig:z_control_m},
comparing data and prediction for the $y_{\ell\ell}$, $\cos\theta^*$,
and $m_{\ell\ell}$ distributions in selected regions of the measured
kinematic range, as indicated in the figure captions.  The expected
number of events is calculated as the sum of expected signal and
background yields. Acceptable agreement is found in all channels,
given that the simulation is only accurate to NLO for the observables
shown in figures~\ref{fig:zcc_control}--\ref{fig:zmm_control}, and
to NNLO accuracy for the $m_{\ell\ell}$ distribution shown in figure~\ref{fig:z_control_m}.

The background-subtracted data are unfolded to fiducial cross
sections using the inverse of the response matrix obtained using an iterative
Bayesian unfolding method~\cite{DAgostini:1994zf} in which the prior
is improved at each iteration. When using such
methods the statistical and systematic uncertainties (discussed in
section~\ref{sec:sys}) increase with each unfolding iteration, while
the residual bias from the initial prior decreases. A balance between
these two competing effects must be struck when deciding on the number
of iterations to be used to unfold the measurement. Only small changes to the
prior are expected, however, since the lineshape of the $Z$ boson resonance
and the PDFs are known to high-precision. Moreover, the prior (\powheg) is
enhanced using QCD and EW corrections and describes the data within
experimental uncertainties. An optimum was found using two
iterations in this analysis.

Finally, measurement bins which are predicted by signal MC simulation to
have fewer than 25 signal events are expected to have large statistical
uncertainties and therefore these bins are removed from the analysis.
Approximately 50 bins are discarded in each of the central electron and muon
channels. They typically lie at large $|y_{\ell\ell}|$ and large $|\cos\theta^*|$.
In the high rapidity electron channel, 27 bins are removed, all corresponding
to small $|\cos\theta^*|$. In all cases the discarded bins correspond
to ones for which the signal prediction at LO in QCD is consistent with
zero.

\FloatBarrier

\section{Measurement uncertainties}
\label{sec:sys}

The uncertainties in the measurements are discussed separately
starting with the sources relevant to both
electron channels, then the sources only appearing in the high rapidity
electron channel. Next, sources of uncertainty specific to the muon
channel are given followed by the sources common to all three measurements.
Uncertainties due to statistical sources from both the data and
MC samples, the modelling of the energy and momentum response to
leptons, lepton selection efficiencies, background subtraction, and
theoretical uncertainties are covered in this section. Each source is classified as being correlated
or uncorrelated between measurement bins in a single channel. The
sources are propagated using one of three techniques: the bootstrap method~\cite{efron1979}, the
pseudo-experiment method, or the offset method.

\subsection{Statistical uncertainties}
The impact of the statistical uncertainty in the number of events in the data
and MC simulations on the cross-section measurement is quantified using the
bootstrap method, a statistical resampling technique in which each event is
reweighted with a random number drawn from a Poisson distribution with a
mean of unity. This reweighting procedure is done 1000 times producing 1000
replicas of the measurement. All replicas are then unfolded and the uncertainty
is taken as the standard deviation of the measured cross sections. In the case of the
signal MC sample the bootstrap replicas are used to produce an ensemble of 1000 response
matrices which are used to unfold the measurement. The standard deviation of the
unfolded cross sections is used as the signal MC statistical uncertainty.

\subsection{Systematic uncertainties}
The pseudo-experiment method is used for correction factors determined
in bins of lepton kinematics, typically $\eta$ and transverse
energy/momentum. These correction factors have statistical and
systematic uncertainties which are fluctuated randomly using 1000
pseudo-experiments according to a Gaussian distribution whose
mean and standard deviation are set to the value and uncertainty of
the correction factor, respectively. For correlated sources, a
single set of varied correction factors is used for all measurement
bins, whereas for uncorrelated sources the random shifts are applied
separately for each bin. The uncertainties are propagated via the
unfolding procedure yielding 1000 cross-section results which are used
to determine a covariance matrix.

In the offset method the correction factor values from each source are
coherently shifted upwards and downwards by one standard deviation and
the measurement is remade using the varied values. The uncertainty is taken
as half the difference between the two unfolded measurements.

\subsection{Central and high rapidity electron channels}
The systematic uncertainties in the cross section that are unique to
the electron channels are dominated by the uncertainties in the
electron energy scale, and the electron reconstruction and
identification efficiency uncertainties. In addition, a large contribution
to the uncertainty arises from the electron energy resolution
uncertainty in the two neighbouring $m_{ee}$ bins at the $Z$-peak,
$80<m_{ee}<91$~{\GeV} and $91<m_{ee}<102$~{\GeV}.

\subsubsection{Energy scale and resolution}
The electron energy scale and resolution and their corresponding
uncertainties are determined using $Z\to e^+e^-$, $W\to e \nu$, and
$J/\psi\to e^+e^-$ decays.  The uncertainty in the energy scale is
separated into a statistical component and 14 uncorrelated
systematic sources. Some of these sources are split into fine $\eta^e$
bins, while others are coarsely binned into barrel and endcap
regions as described in reference~\cite{PERF-2013-05}. These
sources are found to be strongly anti-correlated between the regions
$m_{ee}<m_Z$ and $m_{ee}>m_Z$. The statistical
uncertainty in the energy scale is found to be negligible. Adding the
effects of the 14 sources of uncertainty in the energy scale in
quadrature after propagating to the measured cross sections, the
combined uncertainty is 1--2\% for the mass bins $80<m_{ee}<91$~{\GeV} and
$91<m_{ee}<102$~{\GeV}, but is less than 1\% at low and high
$m_{ee}$. However, in the integrated $m_{ee}$ cross-section measurement the
effect of these sources is strongly reduced as a result of the
anti-correlation between these two $m_{ee}$ bins. 

The uncertainty in the energy resolution is separated into seven
uncorrelated systematic sources which are propagated to the
cross-section measurements individually. This combined uncertainty is
typically 0.1--0.5\% except in the invariant mass regions neighbouring
the $Z$-peak where it reaches 1\%.

\subsubsection{Reconstruction and identification efficiencies}
The reconstruction and identification efficiencies of electrons are
determined from data using various tag-and-probe methods in $Z$ and
$J/\psi$ decays, following the prescription in
reference~\cite{PERF-2013-03} with certain improvements and adjustments for
the 2012 conditions~\cite{ATLAS-CONF-2014-032}.
The uncertainties arise from variations in the tag-and-probe selection
and the background subtraction methods. The correlated systematic
uncertainty is taken from the RMS of all variations, separately for
the reconstruction and identification efficiency sources, and
propagated using the pseudo-experiment method.

The influence of the identification efficiency uncertainty is found to
be 0.2--0.4\% increasing for larger $|\cos\theta^*|$, and up to 2\%
at low $m_{ee}$. The reconstruction efficiency uncertainty
translates into a variation of the measured cross section which is
generally below 0.2\% but as large as 0.4\% at low $m_{ee}$.

\subsubsection{Trigger efficiency}
The trigger efficiency is measured in both the data and MC simulation
using a tag-and-probe method in $Z\to e^+e^-$ decays and is composed of
a statistical uncorrelated component which is small, and a correlated
piece which is propagated using the pseudo-experiment method. The
resulting uncertainty in the cross section amounts to approximately $0.5\%$ at low
$m_{ee}$ but decreases to approximately $0.1\%$ for  $m_{ee}>116$~{\GeV}.

\subsubsection{Charge misidentification}
The electron charge is determined from the sign of the curvature of
the associated ID track. Bremsstrahlung radiation and subsequent
conversion of the radiated photons can lead to misidentification of
the charge. This is measured in $Z$ boson decays in which one
lepton has an incorrectly reconstructed charge. Such events are
selected by requiring the electron pair to possess the same electric
charge and an invariant mass to be near $m_Z$, consistent with a
$Z$ boson decay. The resulting correlated uncertainty is propagated
with the offset method and found to be less than 0.2\% everywhere.

\subsubsection{Multijet background}
Uncertainties in the multijet  estimation arise from the sample size used in the method, the
subtracted signal and EW contamination, the shape of the multijet
distribution, and the range of the isolation distribution used. The
subtracted top quark and diboson contamination is varied coherently
within the theoretical cross-section uncertainties. The subtracted signal contamination is varied
by $\pm5$\%.  The shape of the multijet distribution is varied by
relaxing the same-sign charge requirement in the case of the central
electron channel, and using the transverse energy $E_{\textrm T}^e$ of
the forward electron as an alternative discriminant in the high
rapidity electron channel. The range of the isolation distribution
used is varied by $\pm15\%$.

The variations made to account for systematic uncertainties in the
method lead to changes in the estimated multijet yield in the central
electron channel. The variations in the multijet yields range from about
10\% at low $m_{ee}$ and $\cos\theta^*\sim 0$, to more than 100\% in
regions where the nominal multijet yield is small, e.g. at large
$|\cos\theta^*|$ and high $m_{ee}$.

The uncorrelated statistical component is propagated to the measured
cross sections with the bootstrap replica method. The remaining two
correlated components are propagated with the offset method, which
when summed in quadrature amount to a measurement uncertainty of less
than 0.1\% of the cross section, except at low $m_{ee}$ and large $|\cos\theta^*|$ where it
grows to almost 1\% in the central electron channel.

In the high rapidity channel the multijet yields range from 15\% to
more than 100\% due to systematic uncertainties in the method. At small
$\cos\theta^*$ and high invariant masses where the signal contribution
is suppressed, the expected multijet background can be very large, as noted
in section~\ref{sec:bg_fwd_elec_data_driven}. Here, the systematic
uncertainty in the multijet background is 20--70\%
depending on $|y_{ee}|$, resulting in a measurement uncertainty of
30\% or greater when propagated to the triple-differential cross section.

\subsection{High rapidity electron channel}
\label{sec:sys_high_rap}
The high rapidity electron analysis differs from the central electron channel
measurement by requiring one electron to be in the forward region
$2.5 < |\eta^e| < 4.9$ where there is no tracking system,
which leads to larger background contamination. This is compensated for
by the addition of an isolation requirement on the central electron, and
more restrictive identification requirements (see section~\ref{sec:sel_fwd_elec})
on the central and forward electrons. The technique used to calibrate the forward calorimeters is also different, and the impact of
potential charge misidentification is different. Since the charge can be measured only for the central
electron, the impact of misidentification is to swap the sign of $\cos\theta^*$. Each of these leads to
additional sources of systematic uncertainty which are discussed in
the following.

The energy scale and resolution corrections for forward electrons lead
to correlated sources of uncertainty propagated using the offset
method. They arise from changes in the event selection used to perform
the calibration as well as variations of the methodology. The influence
of the scale uncertainty on the measurement is about 1\% but can reach
5\% at high $|\cos\theta^*|$. The resolution uncertainty amounts to
0.1--0.3\% increasing to 3--5\% at large $|\cos\theta^*|$ and off-peak
mass bins.

The uncertainty in the cross-section measurement due to the
identification efficiency of forward electrons
is considered to be correlated across the measurement bins and is
estimated using the pseudo-experiment method. It amounts to about 1\%
uncertainty in the cross section.

The efficiency of the isolation selection for central electrons
is derived using a tag-and-probe method in central $Z\to e^+e^-$ decays and
is well described by the simulation. The resulting
uncertainty in the cross section is negligible.

To verify that the modelling of the $W+$jet background does not affect
the estimation of the total fake lepton background in the high rapidity
channel, its normalisation is varied by $60\%$
(as motivated by reference~\cite{STDM-2012-20}) and the
fit of the multijet background is repeated. Since the shape of the
$E_{\textrm T}$ distribution is similar for the $W+$jet and
multijet backgrounds, the total fake lepton background remains almost
invariant for the off-peak regions while for the peak mass bins the
variation is small compared to the multijet background uncertainty. 

\subsection{Central rapidity muon channel}
Uncertainties related to the muon momentum scale and resolution, and
the efficiencies of the muon trigger, reconstruction, and isolation
and impact parameter selections are all studied using
$Z\rightarrow\mu^+\mu^-$ events, and in some cases $J/\psi\to
\mu^+\mu^-$ events are also used. The efficiencies are determined using a
tag-and-probe method. The largest contributions to the systematic
uncertainty in the measurements typically arise from the
reconstruction efficiency and isolation efficiency modelling, and from
the muon momentum scale calibration.

\subsubsection{Momentum scale and resolution}
Corrections to the muon momentum scale and resolution are obtained
from fits to the $Z\rightarrow\mu^+\mu^-$ and $J/\psi\to \mu^+\mu^-$
lineshapes with scale and resolution parameters derived in local detector
regions~\cite{PERF-2014-05}. These sources are separated into 12 correlated
components for the resolution in fine $\eta^{\mu}$ bins and one
correlated component for the momentum scale.  Uncertainties in the momentum scale arising
from the methodology, and uncertainties in the ID material simulation, muon
angle reconstruction, and  alignment are propagated using the offset method. They
result in a systematic uncertainty correlated in $\eta^{\mu}$ bins of
the measured cross sections of typically $0.3\%$, increasing for
larger $|y_{\mu\mu}|$, $|\cos\theta^*|$, and $m_{\mu\mu}$ to
$2\%$. The correlated resolution uncertainty has a small influence on the
measurement and is also propagated with the offset method. 

The influence of residual misalignments is estimated from two
sources. The first arises from the statistical uncertainty of the
alignment corrections derived using $Z\rightarrow \mu^+\mu^-$ data and is
considered uncorrelated. This component is propagated to the cross
section using the pseudo-experiment method, and is separated into 84
uncorrelated components. The second source accounts for biases in the
correction method, and is defined as the difference between the corrections
derived for data and simulation in bins of $\eta^{\mu}$. This uncertainty
is separated into 40 correlated components. After propagating this correlated
source to the cross section using the pseudo-experiment method, the
resulting uncertainty is found to be about 0.2\%, increasing significantly
with $|\cos\theta^*|$ at large $|y_{\mu\mu}|$.

\subsubsection{Reconstruction efficiency}
The uncertainty due to the muon reconstruction efficiency is parameterised as
a function of $\eta^{\mu}$ and $p_{\textrm T}^{\mu}$~\cite{PERF-2014-05}
and is decomposed into correlated and uncorrelated parts. The uncertainty is
propagated to the cross section using the offset and pseudo-experiment
methods for the correlated and uncorrelated components, respectively.
The correlated component has an uncertainty of 0.1\%, which corresponds
to an uncertainty in the measured cross section of 0.2--0.4\%.

\subsubsection{Trigger efficiency}
The efficiency corrections for single-muon and dimuon triggers are
obtained using the tag-and-probe method as described in
reference~\cite{TRIG-2012-03}.  They are parameterised in terms of muon
pseudorapidity $\eta^{\mu}$, azimuthal angle $\phi^{\mu}$, and
electric charge. The correlated uncertainty components arise from the
background contamination, a possible residual dependence on muon
$p_{\textrm T}^{\mu}$, and an uncertainty based on the event topology,
which are propagated using the offset method. The uncorrelated
statistical uncertainty is propagated to the cross section using the
pseudo-experiment method. Events selected with the single-muon
triggers ($p_{\textrm T}^{\mu}>25$~{\GeV}) cover most of the kinematic
range of the measurement, whereas the dimuon triggers supplement the
selection at low $m_{\mu\mu}$ and have somewhat larger
uncertainties. This translates into a correlated uncertainty in the
measured cross section which is typically $0.1\%$ where the single-muon
triggers are used, and can reach 0.6\% at large $|\cos\theta^*|$ in
the lowest $m_{\mu\mu}$ bin.

\subsubsection{Isolation and impact parameter efficiency}
Muon isolation and impact parameter selection
efficiencies give rise to additional systematic uncertainties and
are estimated together. The sources considered include the remaining
background contamination, the residual variation in $\eta^{\mu}$, and
a possible bias from the event topology estimated by varying the
azimuthal opening angle between the two muons used in the
tag-and-probe method.  The resulting correlated cross-section
uncertainty determined with the pseudo-experiment method is found to
be typically $0.2\%$, rising to $0.5\%$ at high $m_{\mu\mu}$.

\subsubsection{Multijet background}
The uncertainty in the multijet background estimate comes from
several sources. The uncorrelated statistical uncertainty of the
control regions is propagated using the bootstrap replica method and
can be significant, in particular from the isolated same-charge control
sample. The subtracted top quark and diboson contamination in the
control regions is varied coherently within the theoretical cross-section
uncertainties given in section~\ref{sec:MC}.  The subtracted
signal contamination is varied by $\pm5\%$.  The correlated
uncertainty in the shape of the $|y_{\mu\mu}|$ and $|\cos\theta^*|$
spectra is determined from the RMS of these distributions in five
regions of increasing non-isolation of the muon pairs obtained from
the control regions. The final contribution comes from the fit
extrapolation of the background estimate into the signal region and is
assessed by varying the range of the fit. Systematic components lead to changes in
the multijet yields of 15\% to 30\% of the expected signal
contribution. This is largest in the regions of large
$|\cos\theta^*|$. The variations can be up to 60\% for large
$|\cos\theta^*|$ and large $|y_{\ell\ell}|$.

Both the shape and extrapolation uncertainties are propagated to
the cross section using the offset method and dominate the total
uncertainty.  The combined uncertainty in the background estimate
when propagated to the cross-section measurement is below $0.1\%$
in all measurement bins except in the lowest $m_{\mu\mu}$ bin where it
reaches 1\% at large $|\cos\theta^*|$ and small $|y_{\mu\mu}|$.

\subsection{Systematic uncertainties common to all channels}
\label{sec:sysCommon}
The systematic uncertainties common to all three channels are derived
using identical methods. With the exception of the statistical uncertainties
arising from the MC samples used, which are uncorrelated between the measurement
channels, common systematic uncertainties are assumed to be fully
correlated between the channels. The dominant common uncertainty
is the uncertainty in the luminosity measurement.

\subsubsection{Top, diboson, $W+$jet, $Z/\gamma^*\rightarrow\tau\tau$, and photon-induced background normalisation}
The normalisation uncertainties considered for these background
sources arise from variations in the PDFs, $\alpha_S$, and the QCD
scales used in the theoretical predictions. The normalisation
uncertainty in the top quark background, which is dominated by
\ttbar~production, is taken to be 6\% following the PDF4LHC
prescription~\cite{Botje:2011sn}. The uncertainty includes scale
and $\alpha_S$ variations and also takes into account the uncertainty
in the top-quark mass. Diboson ($WW$, $WZ$ and $ZZ$) production
is another important background source for which the normalisation
uncertainties are about $10\%$. See reference~\cite{STDM-2014-06}
for additional information on the normalisation uncertainties of the
various Monte Carlo samples used.

The background contributions from $W+$jet processes are assigned
a normalisation uncertainty of 5\% for the central rapidity measurements.
For the high rapidity electron channel, where $W+$jet is a dominant
background, a variation of 60\% is considered (see section~\ref{sec:sys_high_rap}).

The background contribution from $Z/\gamma^*\rightarrow\tau\tau$
decays is assigned a normalisation uncertainty of 5\%. The photon-induced
background is assigned an uncertainty of 40\%, derived by calculating the
photon-induced contribution in a constituent and a current mass scheme for the
quark~\cite{MRST2004QED}, and taking the magnitude of the difference
between either scheme and their average~\cite{STDM-2012-10}. In all
cases the normalisation uncertainties are propagated to the final cross
sections using the offset method.

\subsubsection{Unfolding bias}
The simulation used as an initial prior in the unfolding process could
lead to a potential bias in the measured cross sections.  This
potential bias is quantified by varying the predictions within
theoretical uncertainties. The PDF bias is probed using signal MC events
reweighted to each of the 26 different eigenvector variations of the
CT10 PDF set in the determination of $\mathcal{M}$. For each variation
the change in the unfolded cross section is found to be much smaller
than the change in the predicted cross section using each eigenvector
PDF set. Changing the PDF set can alter the predicted cross section by up
to a few percent but the influence on the unfolded result is less
than 0.1\%.  Furthermore, the change in the unfolded result, using one
to five iterations of unfolding, is much smaller than the total
uncertainty in the data. This study is repeated by reweighting the
signal MC events to different values of the scattering amplitude coefficient
$A_4=\frac{8}{3} A_{\textrm{FB}}$, which is proportional to
$\sin^2\theta_W$. A variation of $\pm0.01$ is used, corresponding to
a maximum change of $0.5\%$ in the cross-section prior, which results in
a change in the unfolded cross section of less than 0.1\%.  These
studies show that potential biases are small for five iterations or
less.

A potential overestimate or underestimate of the statistical and systematic
uncertainties of the measurement due to the chosen number of unfolding
iterations is also studied. Tests of the statistical uncertainty are performed
using pseudo-data generated using an alternative PDF. Ultimately, two unfolding
iterations are used for the final cross-section determination. This number has
a negligible bias due to the initial prior and produces a negligible
bias in the data statistical and systematic uncertainties.

\subsubsection{MC modelling}
The $Z$ boson $p_{\textrm{T}}$ distribution is not well modelled in
MC simulation and could influence the measurement. The potential
bias is estimated by reweighting the signal MC events to the observed
data spectrum at reconstruction-level.  This reweighted MC sample is
used to unfold the cross section and the difference to the nominal
measurement is taken as the uncertainty, which is typically below
0.1\%, rising to about 1\% at large $|\cos\theta^*|$ and large
$|y_{\ell\ell}|$.

Adjustments to the reweighting of the scattering amplitude coefficients
in the \powheg\ MC sample are found to have negligible impact on the
measured cross sections.

The MC simulations used for modelling the underlying event and parton
shower processes are not explicitly studied here, but are only expected to
influence this measurement via the lepton isolation selection
efficiencies. Studies presented in reference~\cite{STDM-2012-20} indicate
that such effects are small. 

\subsubsection{PDF uncertainty}
As discussed in section \ref{sec:methodology}, the 
response matrix $\mathcal{M}$ also includes a small acceptance interpolation from
the measured region to the fiducial region. These acceptance
corrections differ in each of the three measurement channels
due to $\eta^{e,\mu}$ gaps in the detector. The corrections
are 5--10\% but can be larger in certain bins of the
triple-differential cross-section measurement. The PDF uncertainties
due to these acceptance corrections are estimated using the CT10 PDF
eigenvector set at 68\% confidence level.  They are found to be
small, with uncertainties on the order of 0.1\% or below for most
cross-section measurement bins in the electron channel. In the high
rapidity electron channel the uncertainty is also found to be small,
except at large $|\cos\theta^*|$ where it can reach 0.6\%. The
uncertainty evaluated in the muon channel is found to be about 0.5\%
at low $m_{\mu\mu}$, negligible for $m_{\mu\mu}$ at $m_Z$, and reaches
0.6\% for large $|\cos\theta^*|$ and large $|y_{\mu\mu}|$.

\subsubsection{Luminosity}
The uncertainty in the integrated luminosity is 1.9\%, which is
derived following the methodology detailed in
reference~\cite{DAPR-2013-01}. This is fully correlated across all
measurement bins and analysis channels.

\subsection{Summary of measurement uncertainties}

Tables~\ref{tab:central_elec_unc_detailed}--\ref{tab:muon_detailed} present
the contributions of the individual uncertainties discussed above for each
channel in selected analysis bins. The influence of the experimental systematic
uncertainties on the measurements of {\d3s} can be divided into three regions
of $m_{\ell\ell}$ -- below the resonance peak, on the peak region, and above
the resonance. In the electron channels, the largest measurement uncertainties
arise from background and efficiency correction uncertainties at low and high
$m_{\ell\ell}$. In the peak region the uncertainty is dominated by the energy
scale sources. The muon channel precision is limited by the background
uncertainty at low $m_{\ell\ell}$, and by both the momentum scale and
misalignment uncertainties in the peak region. At larger invariant mass the
uncertainties related to the muon reconstruction and isolation efficiency also
become important.

\begin{table}[p!]
\tiny
\centering
\scalebox{0.87}{
\setlength\tabcolsep{3pt}
\setlength\extrarowheight{2pt}
\begin{tabular}{C{0.6cm}C{0.8cm}C{0.9cm}C{1.1cm}L{0.5cm}L{0.45cm}L{0.45cm}L{0.45cm}L{0.45cm}L{0.45cm}L{0.45cm}L{0.45cm}L{0.45cm}L{0.45cm}L{0.45cm}L{0.45cm}L{0.45cm}L{0.45cm}L{0.45cm}L{0.45cm}}
\toprule
Bin
& $m_{ee}$
& $|y_{ee}|$
& $\cos\theta^*$
& $\delta^{\textrm{stat}}_{\textrm{unc}}$
& $\delta^{\textrm{sig}}_{\textrm{unc}}$
& $\delta^{\textrm{bkg}}_{\textrm{unc}}$
& $\delta^{\textrm{mj}}_{\textrm{unc}}$
& $\delta^{\textrm{bkg}}_{\textrm{cor}}$
& $\delta^{\textrm{mj}}_{\textrm{cor}}$
& $\delta^{\textrm{scl}}_{\textrm{cor}}$
& $\delta^{\textrm{res}}_{\textrm{cor}}$
& $\delta^{\textrm{rec}}_{\textrm{cor}}$
& $\delta^{\textrm{id}}_{\textrm{cor}}$
& $\delta^{\textrm{trig}}_{\textrm{cor}}$
& $\delta^{\textrm{qmid}}_{\textrm{cor}}$
& $\delta^{\textrm{kfac}}_{\textrm{cor}}$
& $\delta^{\textrm{zpt}}_{\textrm{cor}}$
& $\delta^{\textrm{pdf}}_{\textrm{cor}}$
& $\delta^{\textrm{tot}}$ \\
&	[GeV] & & & [\%] & [\%] & [\%] & [\%] & [\%] & [\%] & [\%] & [\%] & [\%] & [\%] & [\%] & [\%] & [\%] & [\%] & [\%] & [\%]\\
\noalign{\vskip 0.05cm}
\midrule
  $1$ &   $46, 66$ &   $0.0, 0.2$ &    $-1.0, -0.7$  & $6.7$ & $2.4$ & $3.4$ & $3.1$ & $1.9$ & $5.2$ & $0.5$ & $0.7$ & $0.5$ & $2.5$ & $0.7$ & $0.2$ & $0.0$ & $0.9$ & $0.2$ & $10.6$  \\ [0.25ex]
  $2$ &   $46, 66$ &   $0.0, 0.2$ &    $-0.7, -0.4$  & $2.3$ & $0.8$ & $1.2$ & $0.9$ & $1.1$ & $2.0$ & $0.2$ & $0.2$ & $0.5$ & $2.7$ & $0.9$ & $0.0$ & $0.0$ & $0.0$ & $0.1$ & $4.7$  \\ [0.25ex]
  $3$ &   $46, 66$ &   $0.0, 0.2$ &    $-0.4, \phantom{-}0.0$  & $1.4$ & $0.5$ & $0.9$ & $0.4$ & $0.9$ & $0.9$ & $0.3$ & $0.1$ & $0.3$ & $1.9$ & $0.3$ & $0.0$ & $0.0$ & $0.0$ & $0.0$ & $2.9$  \\ [0.25ex]
  $4$ &   $46, 66$ &   $0.0, 0.2$ &    $\phantom{+}0.0, +0.4$  & $1.4$ & $0.5$ & $0.8$ & $0.5$ & $0.9$ & $0.9$ & $0.3$ & $0.1$ & $0.3$ & $1.9$ & $0.3$ & $0.0$ & $0.0$ & $0.0$ & $0.1$ & $3.0$  \\ [0.25ex]
  $5$ &   $46, 66$ &   $0.0, 0.2$ &    $+0.4, +0.7$  & $2.2$ & $0.8$ & $0.9$ & $0.9$ & $1.1$ & $2.0$ & $0.2$ & $0.1$ & $0.5$ & $2.6$ & $0.8$ & $0.0$ & $0.0$ & $0.0$ & $0.1$ & $4.5$  \\ [0.25ex]
  $6$ &   $46, 66$ &   $0.0, 0.2$ &    $+0.7, +1.0$  & $6.7$ & $2.3$ & $4.8$ & $3.1$ & $1.8$ & $4.9$ & $0.9$ & $0.5$ & $0.5$ & $2.6$ & $0.7$ & $0.1$ & $0.0$ & $0.9$ & $0.2$ & $10.9$  \\ [0.25ex]
\midrule
  $79$ &   $66, 80$ &   $0.2, 0.4$ &    $-1.0, -0.7$  & $2.7$ & $1.3$ & $0.5$ & $0.7$ & $0.5$ & $1.6$ & $1.5$ & $1.1$ & $0.6$ & $3.7$ & $1.2$ & $0.1$ & $0.0$ & $0.3$ & $0.2$ & $5.6$  \\ [0.25ex]
  $80$ &   $66, 80$ &   $0.2, 0.4$ &    $-0.7, -0.4$  & $1.3$ & $0.6$ & $0.4$ & $0.3$ & $0.3$ & $0.3$ & $0.4$ & $0.4$ & $0.3$ & $1.7$ & $0.4$ & $0.1$ & $0.0$ & $0.0$ & $0.0$ & $2.5$  \\ [0.25ex]
  $81$ &   $66, 80$ &   $0.2, 0.4$ &    $-0.4, \phantom{-}0.0$  & $1.3$ & $0.4$ & $0.4$ & $0.3$ & $0.3$ & $0.1$ & $0.3$ & $0.1$ & $0.1$ & $0.7$ & $0.2$ & $0.0$ & $0.0$ & $0.0$ & $0.0$ & $1.6$  \\ [0.25ex]
  $82$ &   $66, 80$ &   $0.2, 0.4$ &    $\phantom{+}0.0, +0.4$  & $1.2$ & $0.5$ & $0.3$ & $0.4$ & $0.3$ & $0.1$ & $0.3$ & $0.1$ & $0.1$ & $0.7$ & $0.2$ & $0.1$ & $0.0$ & $0.0$ & $0.0$ & $1.7$  \\ [0.25ex]
  $83$ &   $66, 80$ &   $0.2, 0.4$ &    $+0.4, +0.7$  & $1.4$ & $0.6$ & $0.3$ & $0.3$ & $0.3$ & $0.3$ & $0.6$ & $0.2$ & $0.3$ & $1.7$ & $0.4$ & $0.1$ & $0.0$ & $0.1$ & $0.0$ & $2.6$  \\ [0.25ex]
  $84$ &   $66, 80$ &   $0.2, 0.4$ &    $+0.7, +1.0$  & $2.7$ & $1.4$ & $0.4$ & $0.7$ & $0.4$ & $1.6$ & $2.8$ & $1.0$ & $0.6$ & $3.8$ & $1.2$ & $0.2$ & $0.0$ & $0.3$ & $0.1$ & $6.1$  \\ [0.25ex]
\midrule
  $157$ &   $80, 91$ &   $0.4, 0.6$ &    $-1.0, -0.7$  & $0.6$ & $0.3$ & $0.0$ & $0.1$ & $0.0$ & $0.1$ & $1.4$ & $0.3$ & $0.3$ & $3.2$ & $0.4$ & $0.1$ & $0.0$ & $0.0$ & $0.1$ & $3.6$  \\ [0.25ex]
  $158$ &   $80, 91$ &   $0.4, 0.6$ &    $-0.7, -0.4$  & $0.4$ & $0.2$ & $0.0$ & $0.0$ & $0.0$ & $0.0$ & $1.0$ & $0.1$ & $0.1$ & $0.5$ & $0.2$ & $0.1$ & $0.0$ & $0.1$ & $0.0$ & $1.2$  \\ [0.25ex]
  $159$ &   $80, 91$ &   $0.4, 0.6$ &    $-0.4, \phantom{-}0.0$  & $0.4$ & $0.1$ & $0.0$ & $0.0$ & $0.0$ & $0.0$ & $1.0$ & $0.1$ & $0.0$ & $0.3$ & $0.1$ & $0.1$ & $0.0$ & $0.0$ & $0.0$ & $1.1$  \\ [0.25ex]
  $160$ &   $80, 91$ &   $0.4, 0.6$ &    $\phantom{+}0.0, +0.4$  & $0.4$ & $0.1$ & $0.0$ & $0.0$ & $0.0$ & $0.0$ & $1.0$ & $0.0$ & $0.0$ & $0.3$ & $0.1$ & $0.1$ & $0.0$ & $0.0$ & $0.0$ & $1.2$  \\ [0.25ex]
  $161$ &   $80, 91$ &   $0.4, 0.6$ &    $+0.4, +0.7$  & $0.4$ & $0.2$ & $0.0$ & $0.0$ & $0.0$ & $0.0$ & $1.0$ & $0.1$ & $0.1$ & $0.5$ & $0.2$ & $0.1$ & $0.0$ & $0.1$ & $0.0$ & $1.2$  \\ [0.25ex]
  $162$ &   $80, 91$ &   $0.4, 0.6$ &    $+0.7, +1.0$  & $0.6$ & $0.3$ & $0.0$ & $0.0$ & $0.0$ & $0.1$ & $1.6$ & $0.2$ & $0.3$ & $3.2$ & $0.4$ & $0.1$ & $0.0$ & $0.0$ & $0.1$ & $3.7$  \\ [0.25ex]
\midrule
  $235$ &   $\phantom{1}91, 102$ &   $0.6, 0.8$ &    $-1.0, -0.7$  & $0.5$ & $0.2$ & $0.0$ & $0.1$ & $0.0$ & $0.0$ & $2.1$ & $0.2$ & $0.3$ & $2.6$ & $0.5$ & $0.0$ & $0.0$ & $0.2$ & $0.0$ & $3.5$  \\ [0.25ex]
  $236$ &   $\phantom{1}91, 102$ &   $0.6, 0.8$ &    $-0.7, -0.4$  & $0.4$ & $0.2$ & $0.0$ & $0.0$ & $0.0$ & $0.0$ & $1.3$ & $0.0$ & $0.1$ & $0.5$ & $0.2$ & $0.0$ & $0.0$ & $0.1$ & $0.0$ & $1.5$  \\ [0.25ex]
  $237$ &   $\phantom{1}91, 102$ &   $0.6, 0.8$ &    $-0.4, \phantom{-}0.0$  & $0.4$ & $0.1$ & $0.0$ & $0.0$ & $0.0$ & $0.0$ & $1.0$ & $0.1$ & $0.0$ & $0.2$ & $0.1$ & $0.0$ & $0.0$ & $0.0$ & $0.0$ & $1.1$  \\ [0.25ex]
  $238$ &   $\phantom{1}91, 102$ &   $0.6, 0.8$ &    $\phantom{+}0.0, +0.4$  & $0.3$ & $0.1$ & $0.0$ & $0.0$ & $0.0$ & $0.0$ & $1.0$ & $0.0$ & $0.0$ & $0.2$ & $0.1$ & $0.0$ & $0.0$ & $0.0$ & $0.0$ & $1.1$  \\ [0.25ex]
  $239$ &   $\phantom{1}91, 102$ &   $0.6, 0.8$ &    $+0.4, +0.7$  & $0.4$ & $0.2$ & $0.0$ & $0.0$ & $0.0$ & $0.0$ & $1.2$ & $0.0$ & $0.1$ & $0.5$ & $0.2$ & $0.0$ & $0.0$ & $0.1$ & $0.0$ & $1.4$  \\ [0.25ex]
  $240$ &   $\phantom{1}91, 102$ &   $0.6, 0.8$ &    $+0.7, +1.0$  & $0.5$ & $0.2$ & $0.0$ & $0.1$ & $0.0$ & $0.1$ & $2.1$ & $0.1$ & $0.3$ & $2.6$ & $0.5$ & $0.0$ & $0.0$ & $0.2$ & $0.0$ & $3.4$  \\ [0.25ex]
\midrule
  $313$ &   $102, 116$ &   $0.8, 1.0$ &    $-1.0, -0.7$  & $2.8$ & $1.2$ & $0.6$ & $0.8$ & $0.5$ & $0.7$ & $2.1$ & $0.9$ & $0.2$ & $1.4$ & $0.3$ & $0.1$ & $0.0$ & $0.1$ & $0.0$ & $4.3$  \\ [0.25ex]
  $314$ &   $102, 116$ &   $0.8, 1.0$ &    $-0.7, -0.4$  & $2.6$ & $1.2$ & $0.2$ & $0.5$ & $0.2$ & $0.9$ & $2.3$ & $1.0$ & $0.0$ & $0.4$ & $0.2$ & $0.0$ & $0.0$ & $0.1$ & $0.1$ & $4.0$  \\ [0.25ex]
  $315$ &   $102, 116$ &   $0.8, 1.0$ &    $-0.4, \phantom{-}0.0$  & $2.0$ & $0.8$ & $1.6$ & $0.3$ & $0.2$ & $0.2$ & $1.0$ & $0.3$ & $0.1$ & $0.3$ & $0.1$ & $0.1$ & $0.0$ & $0.0$ & $0.0$ & $2.9$  \\ [0.25ex]
  $316$ &   $102, 116$ &   $0.8, 1.0$ &    $\phantom{+}0.0, +0.4$  & $1.8$ & $0.7$ & $0.1$ & $0.2$ & $0.2$ & $0.1$ & $0.9$ & $0.5$ & $0.1$ & $0.3$ & $0.1$ & $0.0$ & $0.0$ & $0.1$ & $0.1$ & $2.2$  \\ [0.25ex]
  $317$ &   $102, 116$ &   $0.8, 1.0$ &    $+0.4, +0.7$  & $2.3$ & $1.0$ & $0.5$ & $0.4$ & $0.2$ & $0.7$ & $1.7$ & $1.3$ & $0.0$ & $0.4$ & $0.2$ & $0.1$ & $0.0$ & $0.0$ & $0.1$ & $3.5$  \\ [0.25ex]
  $318$ &   $102, 116$ &   $0.8, 1.0$ &    $+0.7, +1.0$  & $2.3$ & $1.0$ & $0.2$ & $0.6$ & $0.3$ & $0.6$ & $2.1$ & $0.6$ & $0.2$ & $1.4$ & $0.3$ & $0.0$ & $0.0$ & $0.0$ & $0.1$ & $3.8$  \\ [0.25ex]
\midrule
  $391$ &   $116, 150$ &   $1.0, 1.2$ &    $-1.0, -0.7$  & $4.8$ & $1.0$ & $2.8$ & $1.8$ & $1.3$ & $5.1$ & $0.2$ & $0.4$ & $0.1$ & $0.4$ & $0.2$ & $0.1$ & $0.0$ & $0.0$ & $0.1$ & $8.0$  \\ [0.25ex]
  $392$ &   $116, 150$ &   $1.0, 1.2$ &    $-0.7, -0.4$  & $3.5$ & $0.9$ & $0.4$ & $0.7$ & $0.7$ & $0.6$ & $0.6$ & $0.1$ & $0.1$ & $0.3$ & $0.2$ & $0.1$ & $0.0$ & $0.2$ & $0.1$ & $3.9$  \\ [0.25ex]
  $393$ &   $116, 150$ &   $1.0, 1.2$ &    $-0.4, \phantom{-}0.0$  & $3.1$ & $0.8$ & $1.3$ & $0.4$ & $0.5$ & $0.8$ & $0.6$ & $0.1$ & $0.1$ & $0.4$ & $0.2$ & $0.1$ & $0.0$ & $0.0$ & $0.2$ & $3.7$  \\ [0.25ex]
  $394$ &   $116, 150$ &   $1.0, 1.2$ &    $\phantom{+}0.0, +0.4$  & $3.0$ & $0.8$ & $0.6$ & $0.5$ & $0.4$ & $0.9$ & $0.6$ & $0.2$ & $0.1$ & $0.4$ & $0.2$ & $0.1$ & $0.0$ & $0.1$ & $0.0$ & $3.5$  \\ [0.25ex]
  $395$ &   $116, 150$ &   $1.0, 1.2$ &    $+0.4, +0.7$  & $2.8$ & $0.7$ & $0.5$ & $0.5$ & $0.5$ & $0.4$ & $0.6$ & $0.4$ & $0.1$ & $0.3$ & $0.2$ & $0.1$ & $0.0$ & $0.1$ & $0.1$ & $3.2$  \\ [0.25ex]
  $396$ &   $116, 150$ &   $1.0, 1.2$ &    $+0.7, +1.0$  & $3.7$ & $0.8$ & $2.2$ & $1.1$ & $0.8$ & $3.4$ & $0.4$ & $0.2$ & $0.1$ & $0.4$ & $0.2$ & $0.1$ & $0.0$ & $0.0$ & $0.1$ & $5.7$  \\ [0.25ex]
\midrule
  $469$ &   $150, 200$ &   $1.2, 1.4$ &    $-1.0, -0.7$  & $11.9$ & $1.4$ & $2.0$ & $3.6$ & $2.2$ & $1.5$ & $0.4$ & $0.3$ & $0.1$ & $0.5$ & $0.3$ & $0.1$ & $0.0$ & $0.0$ & $0.2$ & $12.9$  \\ [0.25ex]
  $470$ &   $150, 200$ &   $1.2, 1.4$ &    $-0.7, -0.4$  & $6.6$ & $0.8$ & $1.0$ & $5.9$ & $1.6$ & $0.9$ & $0.9$ & $0.2$ & $0.1$ & $0.5$ & $0.3$ & $0.1$ & $0.0$ & $0.0$ & $0.1$ & $9.2$  \\ [0.25ex]
  $471$ &   $150, 200$ &   $1.2, 1.4$ &    $-0.4, \phantom{-}0.0$  & $6.6$ & $1.0$ & $3.1$ & $1.9$ & $1.0$ & $0.4$ & $1.0$ & $0.1$ & $0.2$ & $0.6$ & $0.3$ & $0.2$ & $0.0$ & $0.0$ & $0.1$ & $7.8$  \\ [0.25ex]
  $472$ &   $150, 200$ &   $1.2, 1.4$ &    $\phantom{+}0.0, +0.4$  & $5.3$ & $0.9$ & $0.8$ & $0.9$ & $0.6$ & $0.2$ & $0.6$ & $0.2$ & $0.2$ & $0.6$ & $0.3$ & $0.2$ & $0.0$ & $0.1$ & $0.0$ & $5.6$  \\ [0.25ex]
  $473$ &   $150, 200$ &   $1.2, 1.4$ &    $+0.4, +0.7$  & $4.4$ & $0.6$ & $0.5$ & $1.9$ & $0.7$ & $0.4$ & $0.9$ & $0.2$ & $0.1$ & $0.5$ & $0.3$ & $0.1$ & $0.0$ & $0.0$ & $0.0$ & $5.0$  \\ [0.25ex]
  $474$ &   $150, 200$ &   $1.2, 1.4$ &    $+0.7, +1.0$  & $7.6$ & $0.9$ & $1.1$ & $2.3$ & $1.1$ & $0.7$ & $0.3$ & $0.2$ & $0.1$ & $0.5$ & $0.3$ & $0.2$ & $0.0$ & $0.0$ & $0.1$ & $8.3$  \\ [0.25ex]
\bottomrule
\end{tabular}
}
\caption{Central rapidity electron channel uncertainties in selected
bins. All uncertainties quoted are in units of percent, relative to the
measured differential cross section. The uncertainties are separated
into those which are bin-to-bin correlated within a single channel
(marked ``cor'') and those which are uncorrelated (marked ``unc'').
The sources are the uncertainties arising from
the data sample size ($\delta^{\textrm{stat}}_{\textrm{unc}}$);
the signal MC sample size ($\delta^{\textrm{sig}}_{\textrm{unc}}$);
the sizes of the background MC samples ($\delta^{\textrm{bkg}}_{\textrm{unc}}$);
the statistical component of the multijet estimation ($\delta^{\textrm{mj}}_{\textrm{unc}}$);
the combined correlated (normalisation) component of all background MC samples ($\delta^{\textrm{bkg}}_{\textrm{cor}}$);
the multijet estimation ($\delta^{\textrm{mj}}_{\textrm{cor}}$);
the electron energy scale ($\delta^{\textrm{scl}}_{\textrm{cor}}$) and
resolution ($\delta^{\textrm{res}}_{\textrm{cor}}$);
the reconstruction ($\delta^{\textrm{rec}}_{\textrm{cor}}$),
identification ($\delta^{\textrm{id}}_{\textrm{cor}}$), and
trigger efficiencies ($\delta^{\textrm{trig}}_{\textrm{cor}}$);
the electron charge misidentification ($\delta^{\textrm{qmid}}_{\textrm{cor}}$);
the $K$-factors ($\delta^{\textrm{kfac}}_{\textrm{cor}}$);
the $Z$ boson $p_{\textrm{T}}$ modelling ($\delta^{\textrm{zpt}}_{\textrm{cor}}$);
the PDF variation ($\delta^{\textrm{pdf}}_{\textrm{cor}}$); and
the total measurement uncertainty ($\delta^{\textrm{tot}}$).
The luminosity uncertainty is not included in these tables.}
\label{tab:central_elec_unc_detailed}
\end{table}
\begin{table}[htp!]
\tiny
\centering
\scalebox{0.87}{
\setlength\tabcolsep{3pt}
\setlength\extrarowheight{2pt}
\begin{tabular}{C{0.6cm}C{0.8cm}C{0.9cm}C{1.1cm}L{0.5cm}L{0.48cm}L{0.48cm}L{0.48cm}L{0.48cm}L{0.48cm}L{0.48cm}L{0.48cm}L{0.48cm}L{0.48cm}L{0.48cm}L{0.48cm}L{0.48cm}L{0.48cm}L{0.48cm}L{0.48cm}L{0.48cm}L{0.48cm}L{0.48cm}L{0.48cm}}
\toprule
Bin
& $m_{ee}$
& $|y_{ee}|$
& $\cos\theta^*$
& $\delta^{\textrm{stat}}_{\textrm{unc}}$
& $\delta^{\textrm{sig}}_{\textrm{unc}}$
& $\delta^{\textrm{bkg}}_{\textrm{unc}}$
& $\delta^{\textrm{mj}}_{\textrm{unc}}$
& $\delta^{\textrm{bkg}}_{\textrm{cor}}$
& $\delta^{\textrm{mj}}_{\textrm{cor}}$
& $\delta^{\textrm{scl}}_{\textrm{cor}}$
& $\delta^{\textrm{res}}_{\textrm{cor}}$
& $\delta^{\textrm{fscl}}_{\textrm{cor}}$
& $\delta^{\textrm{fres}}_{\textrm{cor}}$
& $\delta^{\textrm{rec}}_{\textrm{cor}}$
& $\delta^{\textrm{id}}_{\textrm{cor}}$
& $\delta^{\textrm{trig}}_{\textrm{cor}}$
& $\delta^{\textrm{iso}}_{\textrm{cor}}$
& $\delta^{\textrm{fid}}_{\textrm{cor}}$
& $\delta^{\textrm{qmid}}_{\textrm{cor}}$
& $\delta^{\textrm{kfac}}_{\textrm{cor}}$
& $\delta^{\textrm{zpt}}_{\textrm{cor}}$
& $\delta^{\textrm{pdf}}_{\textrm{cor}}$
& $\delta^{\textrm{tot}}$ \\
&	[GeV] & & & [\%] & [\%] & [\%] & [\%] & [\%] & [\%] & [\%] & [\%] & [\%] & [\%] & [\%] & [\%] & [\%] & [\%] & [\%] & [\%] & [\%] & [\%] & [\%] & [\%]\\
\noalign{\vskip 0.05cm}
\midrule
  $1$ &   $66, 80$ &   $1.2, 1.6$ &    $-1.0, -0.7$  & $6.4$ & $3.0$ & $6.0$ & $4.5$ & $0.9$ & $11.5$ & $0.4$ & $0.6$ & $3.1$ & $2.1$ & $0.2$ & $0.8$ & $0.3$ & $0.0$ & $0.7$ & $0.0$ & $0.0$ & $0.8$ & $0.6$ & $16.0$  \\ [0.25ex]
  $2$ &   $66, 80$ &   $1.2, 1.6$ &    $-0.7, -0.4$  & $16.4$ & $8.7$ & $8.0$ & $9.9$ & $0.5$ & $11.4$ & $0.5$ & $1.2$ & $5.8$ & $2.5$ & $0.1$ & $0.2$ & $0.1$ & $0.0$ & $0.8$ & $0.0$ & $0.0$ & $0.8$ & $0.3$ & $26.0$  \\ [0.25ex]
  $3$ &   $66, 80$ &   $1.2, 1.6$ &    $-0.4, \phantom{-}0.0$  & $-$ & $-$ & $-$ & $-$ & $-$ & $-$ & $-$ & $-$ & $-$ & $-$ & $-$ & $-$ & $-$ & $-$ & $-$ & $-$ & $-$ & $-$ & $-$ & $-$  \\ [0.25ex]
  $4$ &   $66, 80$ &   $1.2, 1.6$ &    $\phantom{+}0.0, +0.4$  & $-$ & $-$ & $-$ & $-$ & $-$ & $-$ & $-$ & $-$ & $-$ & $-$ & $-$ & $-$ & $-$ & $-$ & $-$ & $-$ & $-$ & $-$ & $-$ & $-$  \\ [0.25ex]
  $5$ &   $66, 80$ &   $1.2, 1.6$ &    $+0.4, +0.7$  & $15.7$ & $8.0$ & $6.7$ & $7.9$ & $0.5$ & $10.7$ & $0.9$ & $0.8$ & $3.8$ & $5.5$ & $0.1$ & $0.1$ & $0.1$ & $0.0$ & $0.8$ & $0.0$ & $0.0$ & $1.6$ & $1.4$ & $24.1$  \\ [0.25ex]
  $6$ &   $66, 80$ &   $1.2, 1.6$ &    $+0.7, +1.0$  & $7.9$ & $3.3$ & $8.8$ & $5.8$ & $1.6$ & $15.3$ & $0.7$ & $0.7$ & $2.3$ & $2.9$ & $0.2$ & $0.8$ & $0.3$ & $0.0$ & $0.7$ & $0.0$ & $0.0$ & $0.9$ & $0.3$ & $20.9$  \\ [0.25ex]
\midrule
  $19$ &   $66, 80$ &   $2.4, 2.8$ &    $-1.0, -0.7$  & $3.4$ & $2.2$ & $1.4$ & $2.8$ & $0.3$ & $3.4$ & $2.5$ & $0.7$ & $4.3$ & $5.2$ & $0.2$ & $1.6$ & $0.4$ & $0.1$ & $1.4$ & $0.0$ & $0.0$ & $2.4$ & $0.2$ & $10.1$  \\ [0.25ex]
  $20$ &   $66, 80$ &   $2.4, 2.8$ &    $-0.7, -0.4$  & $2.2$ & $1.3$ & $0.8$ & $1.6$ & $0.3$ & $1.1$ & $1.2$ & $0.6$ & $3.1$ & $3.9$ & $0.1$ & $0.8$ & $0.2$ & $0.0$ & $1.3$ & $0.0$ & $0.0$ & $0.5$ & $0.1$ & $6.4$  \\ [0.25ex]
  $21$ &   $66, 80$ &   $2.4, 2.8$ &    $-0.4, \phantom{-}0.0$  & $2.3$ & $1.0$ & $0.8$ & $1.4$ & $0.2$ & $1.5$ & $0.4$ & $0.2$ & $0.9$ & $0.3$ & $0.1$ & $0.5$ & $0.2$ & $0.0$ & $0.8$ & $0.0$ & $0.0$ & $0.1$ & $0.0$ & $3.6$  \\ [0.25ex]
  $22$ &   $66, 80$ &   $2.4, 2.8$ &    $\phantom{+}0.0, +0.4$  & $2.8$ & $1.2$ & $1.5$ & $1.9$ & $0.4$ & $2.0$ & $0.4$ & $0.5$ & $1.3$ & $0.3$ & $0.1$ & $0.5$ & $0.2$ & $0.0$ & $0.7$ & $0.0$ & $0.0$ & $0.3$ & $0.1$ & $4.7$  \\ [0.25ex]
  $23$ &   $66, 80$ &   $2.4, 2.8$ &    $+0.4, +0.7$  & $2.7$ & $1.6$ & $1.3$ & $2.3$ & $0.4$ & $1.7$ & $1.6$ & $0.2$ & $4.0$ & $6.0$ & $0.1$ & $0.8$ & $0.2$ & $0.0$ & $1.4$ & $0.1$ & $0.0$ & $1.1$ & $0.2$ & $8.8$  \\ [0.25ex]
  $24$ &   $66, 80$ &   $2.4, 2.8$ &    $+0.7, +1.0$  & $4.2$ & $2.7$ & $3.4$ & $3.7$ & $0.7$ & $5.5$ & $2.8$ & $0.9$ & $4.9$ & $6.5$ & $0.2$ & $1.6$ & $0.4$ & $0.1$ & $1.4$ & $0.0$ & $0.0$ & $3.6$ & $0.3$ & $13.2$  \\ [0.25ex]
\midrule
  $73$ &   $\phantom{1}91, 102$ &   $2.0, 2.4$ &    $-1.0, -0.7$  & $0.9$ & $0.6$ & $0.2$ & $0.3$ & $0.0$ & $0.8$ & $0.8$ & $0.1$ & $1.9$ & $0.1$ & $0.2$ & $0.8$ & $0.2$ & $0.0$ & $1.2$ & $0.0$ & $0.0$ & $0.8$ & $0.1$ & $2.9$  \\ [0.25ex]
  $74$ &   $\phantom{1}91, 102$ &   $2.0, 2.4$ &    $-0.7, -0.4$  & $0.5$ & $0.3$ & $0.0$ & $0.2$ & $0.0$ & $0.7$ & $0.9$ & $0.1$ & $1.5$ & $0.2$ & $0.0$ & $0.4$ & $0.1$ & $0.0$ & $0.8$ & $0.0$ & $0.0$ & $0.1$ & $0.1$ & $2.1$  \\ [0.25ex]
  $75$ &   $\phantom{1}91, 102$ &   $2.0, 2.4$ &    $-0.4, \phantom{-}0.0$  & $0.7$ & $0.3$ & $0.1$ & $0.4$ & $0.0$ & $0.6$ & $0.6$ & $0.1$ & $1.7$ & $0.1$ & $0.0$ & $0.2$ & $0.1$ & $0.0$ & $0.7$ & $0.0$ & $0.0$ & $0.1$ & $0.0$ & $2.2$  \\ [0.25ex]
  $76$ &   $\phantom{1}91, 102$ &   $2.0, 2.4$ &    $\phantom{+}0.0, +0.4$  & $0.6$ & $0.3$ & $0.1$ & $0.4$ & $0.0$ & $0.5$ & $0.5$ & $0.1$ & $1.5$ & $0.1$ & $0.0$ & $0.2$ & $0.1$ & $0.0$ & $0.7$ & $0.0$ & $0.0$ & $0.1$ & $0.1$ & $2.0$  \\ [0.25ex]
  $77$ &   $\phantom{1}91, 102$ &   $2.0, 2.4$ &    $+0.4, +0.7$  & $0.5$ & $0.3$ & $0.1$ & $0.1$ & $0.0$ & $0.5$ & $0.9$ & $0.2$ & $1.3$ & $0.3$ & $0.0$ & $0.4$ & $0.1$ & $0.0$ & $0.8$ & $0.0$ & $0.0$ & $0.2$ & $0.1$ & $2.0$  \\ [0.25ex]
  $78$ &   $\phantom{1}91, 102$ &   $2.0, 2.4$ &    $+0.7, +1.0$  & $0.9$ & $0.5$ & $0.2$ & $0.3$ & $0.0$ & $0.3$ & $0.7$ & $0.2$ & $1.6$ & $0.2$ & $0.2$ & $0.7$ & $0.2$ & $0.0$ & $1.2$ & $0.0$ & $0.0$ & $0.8$ & $0.0$ & $2.6$  \\ [0.25ex]
\midrule
  $97$ &   $102, 116$ &   $1.6, 2.0$ &    $-1.0, -0.7$  & $3.8$ & $1.8$ & $2.0$ & $2.9$ & $0.7$ & $4.2$ & $0.6$ & $0.3$ & $2.4$ & $2.2$ & $0.1$ & $0.3$ & $0.1$ & $0.0$ & $0.8$ & $0.0$ & $0.0$ & $1.5$ & $0.1$ & $7.9$  \\ [0.25ex]
  $98$ &   $102, 116$ &   $1.6, 2.0$ &    $-0.7, -0.4$  & $4.4$ & $2.1$ & $2.0$ & $3.4$ & $0.3$ & $3.6$ & $1.2$ & $0.6$ & $2.1$ & $1.2$ & $0.0$ & $0.2$ & $0.0$ & $0.0$ & $0.7$ & $0.0$ & $0.0$ & $1.5$ & $0.2$ & $8.0$  \\ [0.25ex]
  $99$ &   $102, 116$ &   $1.6, 2.0$ &    $-0.4, \phantom{-}0.0$  & $-$ & $-$ & $-$ & $-$ & $-$ & $-$ & $-$ & $-$ & $-$ & $-$ & $-$ & $-$ & $-$ & $-$ & $-$ & $-$ & $-$ & $-$ & $-$ & $-$  \\ [0.25ex]
  $100$ &   $102, 116$ &   $1.6, 2.0$ &    $\phantom{+}0.0, +0.4$  & $-$ & $-$ & $-$ & $-$ & $-$ & $-$ & $-$ & $-$ & $-$ & $-$ & $-$ & $-$ & $-$ & $-$ & $-$ & $-$ & $-$ & $-$ & $-$ & $-$  \\ [0.25ex]
  $101$ &   $102, 116$ &   $1.6, 2.0$ &    $+0.4, +0.7$  & $3.3$ & $1.5$ & $1.6$ & $2.1$ & $0.2$ & $2.2$ & $1.0$ & $0.7$ & $1.7$ & $1.0$ & $0.0$ & $0.2$ & $0.0$ & $0.0$ & $0.7$ & $0.0$ & $0.0$ & $1.1$ & $0.1$ & $5.6$  \\ [0.25ex]
  $102$ &   $102, 116$ &   $1.6, 2.0$ &    $+0.7, +1.0$  & $2.6$ & $1.4$ & $1.3$ & $1.5$ & $0.3$ & $1.9$ & $0.3$ & $0.1$ & $2.1$ & $1.0$ & $0.1$ & $0.3$ & $0.1$ & $0.0$ & $0.8$ & $0.0$ & $0.0$ & $0.9$ & $0.2$ & $4.9$  \\ [0.25ex]
\midrule
  $109$ &   $102, 116$ &   $2.4, 2.8$ &    $-1.0, -0.7$  & $3.7$ & $2.2$ & $2.3$ & $3.4$ & $0.8$ & $6.2$ & $3.3$ & $1.2$ & $6.7$ & $6.6$ & $0.1$ & $0.6$ & $0.1$ & $0.0$ & $1.4$ & $0.0$ & $0.0$ & $3.3$ & $0.3$ & $13.7$  \\ [0.25ex]
  $110$ &   $102, 116$ &   $2.4, 2.8$ &    $-0.7, -0.4$  & $4.2$ & $2.3$ & $1.0$ & $3.7$ & $0.3$ & $3.3$ & $1.4$ & $1.2$ & $5.5$ & $4.2$ & $0.0$ & $0.2$ & $0.1$ & $0.0$ & $1.2$ & $0.0$ & $0.0$ & $2.0$ & $0.2$ & $10.2$  \\ [0.25ex]
  $111$ &   $102, 116$ &   $2.4, 2.8$ &    $-0.4, \phantom{-}0.0$  & $3.9$ & $1.9$ & $1.5$ & $4.5$ & $0.2$ & $4.6$ & $0.7$ & $0.9$ & $2.3$ & $1.2$ & $0.1$ & $0.3$ & $0.2$ & $0.0$ & $0.7$ & $0.0$ & $0.0$ & $0.9$ & $0.2$ & $8.5$  \\ [0.25ex]
  $112$ &   $102, 116$ &   $2.4, 2.8$ &    $\phantom{+}0.0, +0.4$  & $3.1$ & $1.5$ & $0.7$ & $2.9$ & $0.1$ & $3.2$ & $0.6$ & $0.4$ & $2.3$ & $1.3$ & $0.1$ & $0.3$ & $0.1$ & $0.0$ & $0.8$ & $0.0$ & $0.0$ & $0.9$ & $0.1$ & $6.3$  \\ [0.25ex]
  $113$ &   $102, 116$ &   $2.4, 2.8$ &    $+0.4, +0.7$  & $2.7$ & $1.6$ & $1.1$ & $1.7$ & $0.2$ & $1.6$ & $1.2$ & $0.8$ & $4.0$ & $2.1$ & $0.0$ & $0.2$ & $0.1$ & $0.0$ & $1.2$ & $0.0$ & $0.0$ & $1.4$ & $0.2$ & $6.5$  \\ [0.25ex]
  $114$ &   $102, 116$ &   $2.4, 2.8$ &    $+0.7, +1.0$  & $2.2$ & $1.4$ & $1.3$ & $1.5$ & $0.3$ & $2.4$ & $2.0$ & $0.8$ & $3.3$ & $3.2$ & $0.1$ & $0.6$ & $0.1$ & $0.0$ & $1.3$ & $0.0$ & $0.0$ & $2.2$ & $0.1$ & $7.0$  \\ [0.25ex]
\midrule
  $127$ &   $116, 150$ &   $1.6, 2.0$ &    $-1.0, -0.7$  & $8.4$ & $1.7$ & $8.7$ & $7.1$ & $2.9$ & $29.0$ & $0.2$ & $0.4$ & $1.8$ & $1.2$ & $0.0$ & $0.1$ & $0.0$ & $0.0$ & $0.6$ & $0.0$ & $0.0$ & $0.7$ & $0.2$ & $32.5$  \\ [0.25ex]
  $128$ &   $116, 150$ &   $1.6, 2.0$ &    $-0.7, -0.4$  & $7.6$ & $2.0$ & $4.2$ & $9.0$ & $1.3$ & $8.6$ & $0.6$ & $0.2$ & $0.3$ & $0.5$ & $0.0$ & $0.1$ & $0.0$ & $0.0$ & $0.6$ & $0.0$ & $0.0$ & $0.5$ & $0.2$ & $15.4$  \\ [0.25ex]
  $129$ &   $116, 150$ &   $1.6, 2.0$ &    $-0.4, \phantom{-}0.0$  & $-$ & $-$ & $-$ & $-$ & $-$ & $-$ & $-$ & $-$ & $-$ & $-$ & $-$ & $-$ & $-$ & $-$ & $-$ & $-$ & $-$ & $-$ & $-$ & $-$  \\ [0.25ex]
  $130$ &   $116, 150$ &   $1.6, 2.0$ &    $\phantom{+}0.0, +0.4$  & $-$ & $-$ & $-$ & $-$ & $-$ & $-$ & $-$ & $-$ & $-$ & $-$ & $-$ & $-$ & $-$ & $-$ & $-$ & $-$ & $-$ & $-$ & $-$ & $-$  \\ [0.25ex]
  $131$ &   $116, 150$ &   $1.6, 2.0$ &    $+0.4, +0.7$  & $4.4$ & $1.2$ & $3.1$ & $3.8$ & $0.5$ & $3.1$ & $0.2$ & $0.1$ & $0.3$ & $0.2$ & $0.0$ & $0.1$ & $0.0$ & $0.0$ & $0.6$ & $0.0$ & $0.0$ & $0.3$ & $0.1$ & $7.4$  \\ [0.25ex]
  $132$ &   $116, 150$ &   $1.6, 2.0$ &    $+0.7, +1.0$  & $3.9$ & $0.9$ & $5.5$ & $2.5$ & $1.2$ & $9.8$ & $0.2$ & $0.1$ & $0.9$ & $0.2$ & $0.0$ & $0.1$ & $0.0$ & $0.0$ & $0.7$ & $0.0$ & $0.0$ & $0.5$ & $0.1$ & $12.3$  \\ [0.25ex]
\midrule
  $139$ &   $116, 150$ &   $2.4, 2.8$ &    $-1.0, -0.7$  & $16.3$ & $2.9$ & $11.4$ & $14.0$ & $5.4$ & $29.3$ & $1.3$ & $0.5$ & $5.4$ & $1.7$ & $0.1$ & $0.3$ & $0.1$ & $0.0$ & $1.1$ & $0.1$ & $0.0$ & $1.3$ & $0.3$ & $39.1$  \\ [0.25ex]
  $140$ &   $116, 150$ &   $2.4, 2.8$ &    $-0.7, -0.4$  & $7.5$ & $3.0$ & $7.5$ & $7.3$ & $1.2$ & $10.7$ & $0.2$ & $0.2$ & $1.2$ & $1.4$ & $0.0$ & $0.2$ & $0.1$ & $0.0$ & $0.9$ & $0.0$ & $0.0$ & $1.6$ & $0.3$ & $17.2$  \\ [0.25ex]
  $141$ &   $116, 150$ &   $2.4, 2.8$ &    $-0.4, \phantom{-}0.0$  & $6.0$ & $1.7$ & $3.8$ & $5.6$ & $0.5$ & $6.8$ & $0.2$ & $0.1$ & $1.8$ & $0.5$ & $0.1$ & $0.4$ & $0.1$ & $0.0$ & $0.6$ & $0.1$ & $0.0$ & $0.9$ & $0.1$ & $11.6$  \\ [0.25ex]
  $142$ &   $116, 150$ &   $2.4, 2.8$ &    $\phantom{+}0.0, +0.4$  & $4.5$ & $1.4$ & $3.1$ & $3.2$ & $0.5$ & $3.4$ & $0.1$ & $0.5$ & $0.8$ & $0.2$ & $0.1$ & $0.4$ & $0.1$ & $0.0$ & $0.6$ & $0.0$ & $0.0$ & $0.5$ & $0.1$ & $7.4$  \\ [0.25ex]
  $143$ &   $116, 150$ &   $2.4, 2.8$ &    $+0.4, +0.7$  & $3.8$ & $1.4$ & $2.4$ & $2.4$ & $0.4$ & $3.3$ & $0.3$ & $0.3$ & $0.9$ & $0.7$ & $0.0$ & $0.2$ & $0.1$ & $0.0$ & $1.0$ & $0.0$ & $0.0$ & $0.9$ & $0.1$ & $6.5$  \\ [0.25ex]
  $144$ &   $116, 150$ &   $2.4, 2.8$ &    $+0.7, +1.0$  & $3.3$ & $1.0$ & $1.7$ & $2.0$ & $0.7$ & $3.8$ & $0.7$ & $0.2$ & $1.8$ & $0.6$ & $0.1$ & $0.3$ & $0.1$ & $0.0$ & $1.1$ & $0.0$ & $0.0$ & $0.2$ & $0.1$ & $6.3$  \\ [0.25ex]
\bottomrule
\end{tabular}
}
\caption{High rapidity electron channel uncertainties in selected bins.
All uncertainties quoted are in units of percent, relative to the measured
differential cross section. Bins with blank entries (``$-$'') are those that
have been omitted from the measurement due to a lack of expected
events. The uncertainties are separated
into those which are bin-to-bin correlated within a single channel
(marked ``cor'') and those which are uncorrelated (marked ``unc'').
The sources are the uncertainties arising from
the data sample size ($\delta^{\textrm{stat}}_{\textrm{unc}}$);
the signal MC sample size ($\delta^{\textrm{sig}}_{\textrm{unc}}$);
the sizes of the background MC samples ($\delta^{\textrm{bkg}}_{\textrm{unc}}$);
the statistical component of the multijet estimation ($\delta^{\textrm{mj}}_{\textrm{unc}}$);
the combined correlated (normalisation) component of all background MC samples ($\delta^{\textrm{bkg}}_{\textrm{cor}}$);
the multijet estimation ($\delta^{\textrm{mj}}_{\textrm{cor}}$);
the electron energy scale ($\delta^{\textrm{scl}}_{\textrm{cor}}$) and
resolution ($\delta^{\textrm{res}}_{\textrm{cor}}$);
the forward electron energy scale ($\delta^{\textrm{fscl}}_{\textrm{cor}}$) and
resolution ($\delta^{\textrm{fres}}_{\textrm{cor}}$);
the reconstruction ($\delta^{\textrm{rec}}_{\textrm{cor}}$),
identification ($\delta^{\textrm{id}}_{\textrm{cor}}$),
trigger ($\delta^{\textrm{trig}}_{\textrm{cor}}$),
isolation ($\delta^{\textrm{iso}}_{\textrm{cor}}$), and
forward identification efficiencies ($\delta^{\textrm{fid}}_{\textrm{cor}}$);
the electron charge misidentification ($\delta^{\textrm{qmid}}_{\textrm{cor}}$);
the $K$-factors ($\delta^{\textrm{kfac}}_{\textrm{cor}}$);
the $Z$ boson $p_{\textrm{T}}$ modelling ($\delta^{\textrm{zpt}}_{\textrm{cor}}$);
the PDF variation ($\delta^{\textrm{pdf}}_{\textrm{cor}}$); and
the total measurement uncertainty ($\delta^{\textrm{tot}}$).
The luminosity uncertainty is not included in these tables.}
\label{tab:high_rapidity_elec_unc_detailed}
\end{table}
\begin{table}[htp!]
\tiny
\centering
\scalebox{0.87}{
\setlength\tabcolsep{3pt}
\setlength\extrarowheight{2pt}
\begin{tabular}{C{0.6cm}C{0.8cm}C{0.9cm}C{1.1cm}L{0.5cm}L{0.45cm}L{0.45cm}L{0.45cm}L{0.45cm}L{0.45cm}L{0.45cm}L{0.45cm}L{0.45cm}L{0.45cm}L{0.45cm}L{0.45cm}L{0.45cm}L{0.45cm}L{0.45cm}}
\toprule
Bin
& $m_{\mu\mu}$
& $|y_{\mu\mu}|$
& $\cos\theta^*$
& $\delta^{\textrm{stat}}_{\textrm{unc}}$
& $\delta^{\textrm{sig}}_{\textrm{unc}}$
& $\delta^{\textrm{bkg}}_{\textrm{unc}}$
& $\delta^{\textrm{bkg}}_{\textrm{cor}}$
& $\delta^{\textrm{mj}}_{\textrm{cor}}$
& $\delta^{\textrm{scl}}_{\textrm{cor}}$
& $\delta^{\textrm{sag}}_{\textrm{cor}}$
& $\delta^{\textrm{res}}_{\textrm{cor}}$
& $\delta^{\textrm{rec}}_{\textrm{cor}}$
& $\delta^{\textrm{id}}_{\textrm{cor}}$
& $\delta^{\textrm{trig}}_{\textrm{cor}}$
& $\delta^{\textrm{kfac}}_{\textrm{cor}}$
& $\delta^{\textrm{zpt}}_{\textrm{cor}}$
& $\delta^{\textrm{pdf}}_{\textrm{cor}}$
& $\delta^{\textrm{tot}}$ \\
&	[GeV] & & & [\%] & [\%] & [\%] & [\%] & [\%] & [\%] & [\%] & [\%] & [\%] & [\%] & [\%] & [\%] & [\%] & [\%] & [\%]\\
\noalign{\vskip 0.05cm}
\midrule
1	&	$46, 66$	&	$0.0, 0.2$	&	$-1.0, -0.7$	&	5.4	&	2.0	&  2.1 	&	1.5	&	 0.5	&	0.2	&	0.5	&	0.6	&	0.3	&	0.3	&	0.7	&	0.0	&	0.5	&	0.3	&	6.6	\\
2	&	$46, 66$	&	$0.0, 0.2$	&	$-0.7, -0.4$	&	1.8	&	0.7	&  1.1 	&	1.2	&	 0.0	&	0.0	&	0.1	&	0.1	&	0.2	&	0.2	&	0.5	&	0.2	&	0.3	&	0.2	&	2.7	\\
3	&	$46, 66$	&	$0.0, 0.2$	&	$-0.4, \phantom{-}0.0$	&	1.5	&	0.6	&  0.8 	&	0.9	&	 0.5	&	0.0	&	0.1	&	0.0	&	0.5	&	0.4	&	0.0	&	0.2	&	0.4	&	0.2	&	2.3	\\
4	&	$46, 66$	&	$0.0, 0.2$	&	$\phantom{+}0.0, +0.4$	&	1.5	&	0.6	&  0.9 	&	0.9	&	 0.5	&	0.0	&	0.1	&	0.1	&	0.5	&	0.4	&	0.0	&	0.2	&	0.5	&	0.2	&	2.3	\\
5	&	$46, 66$	&	$0.0, 0.2$	&	$+0.4, +0.7$	&	1.9	&	0.6	&  1.2 	&	1.2	&	 0.0	&	0.1	&	0.1	&	0.4	&	0.2	&	0.2	&	0.5	&	0.2	&	0.3	&	0.2	&	2.8	\\
6	&	$46, 66$	&	$0.0, 0.2$	&	$+0.7, +1.0$	&	5.7	&	2.0	&  3.6 	&	1.8	&	 0.5	&	0.1	&	1.0	&	0.1	&	0.3	&	0.3	&	0.8	&	0.2	&	0.6	&	0.8	&	7.7	\\
\midrule                                                                                                                                             
79	&	$66, 80$	&	$0.2, 0.4$	&	$-1.0, -0.7$	&	2.3	&	1.1	&  0.5 	&	0.6	&	 0.7	&	0.1	&	0.7	&	0.4	&	0.2	&	0.2	&	0.3	&	0.0	&	0.0	&	0.1	&	3.0	\\
80	&	$66, 80$	&	$0.2, 0.4$	&	$-0.7, -0.4$	&	1.3	&	0.7	&  0.3 	&	0.4	&	 0.1	&	0.1	&	0.2	&	0.1	&	0.3	&	0.3	&	0.0	&	0.0	&	0.0	&	0.1	&	1.7	\\
81	&	$66, 80$	&	$0.2, 0.4$	&	$-0.4, \phantom{-}0.0$	&	1.4	&	0.7	&  0.4 	&	0.3	&	 0.2	&	0.1	&	0.2	&	0.2	&	0.4	&	0.4	&	0.0	&	0.1	&	0.1	&	0.3	&	1.8	\\
82	&	$66, 80$	&	$0.2, 0.4$	&	$\phantom{+}0.0, +0.4$	&	1.4	&	0.7	&  0.3 	&	0.3	&	 0.2	&	0.1	&	0.1	&	0.2	&	0.4	&	0.4	&	0.1	&	0.1	&	0.1	&	0.2	&	1.8	\\
83	&	$66, 80$	&	$0.2, 0.4$	&	$+0.4, +0.7$	&	1.4	&	0.7	&  0.4 	&	0.4	&	 0.1	&	0.1	&	0.2	&	0.2	&	0.3	&	0.3	&	0.0	&	0.1	&	0.1	&	0.1	&	1.8	\\
84	&	$66, 80$	&	$0.2, 0.4$	&	$+0.7, +1.0$	&	2.2	&	1.1	&  0.4 	&	0.6	&	 0.8	&	0.2	&	0.7	&	0.1	&	0.2	&	0.2	&	0.3	&	0.0	&	0.0	&	0.3	&	3.0	\\
\midrule                                                                                                                                             
157	&	$80, 91$	&	$0.4, 0.6$	&	$-1.0, -0.7$	&	0.4	&	0.2	&  0.0  &	0.0	&	 0.0	&	0.1	&	1.0	&	0.1	&	0.3	&	0.3	&	0.0	&	0.0	&	0.0	&	0.1	&	1.4	\\
158	&	$80, 91$	&	$0.4, 0.6$	&	$-0.7, -0.4$	&	0.4	&	0.2	&  0.0  &	0.0	&	 0.0	&	0.2	&	0.6	&	0.1	&	0.4	&	0.4	&	0.1	&	0.0	&	0.0	&	0.0	&	1.1	\\
159	&	$80, 91$	&	$0.4, 0.6$	&	$-0.4, \phantom{-}0.0$	&	0.3	&	0.1	&  0.0        &	0.0	&	 0.0	&	0.2	&	0.3	&	0.1	&	0.3	&	0.3	&	0.0	&	0.0	&	0.0	&	0.0	&	0.9	\\
160	&	$80, 91$	&	$0.4, 0.6$	&	$\phantom{+}0.0, +0.4$	&	0.3	&	0.1	&  0.0        &	0.0	&	 0.0	&	0.2	&	0.3	&	0.1	&	0.3	&	0.3	&	0.0	&	0.0	&	0.0	&	0.0	&	0.9	\\
161	&	$80, 91$	&	$0.4, 0.6$	&	$+0.4, +0.7$	&	0.4	&	0.2	&  0.0  &	0.0	&	 0.0	&	0.2	&	0.6	&	0.0	&	0.4	&	0.4	&	0.1	&	0.0	&	0.0	&	0.0	&	1.1	\\
162	&	$80, 91$	&	$0.4, 0.6$	&	$+0.7, +1.0$	&	0.4	&	0.2	&  0.0        &	0.0	&	 0.0	&	0.2	&	1.1	&	0.1	&	0.3	&	0.3	&	0.1	&	0.0	&	0.1	&	0.0	&	1.4	\\
\midrule                                                                                                                                             
235	&	$\phantom{1}91, 102$	&	$0.6, 0.8$	&	$-1.0, -0.7$	&	0.4	&	0.2	&  0.0        &	0.0	&	 0.0	&	0.1	&	0.5	&	0.0	&	0.3	&	0.3	&	0.1	&	0.0	&	0.1	&	0.0	&	1.0	\\
236	&	$\phantom{1}91, 102$	&	$0.6, 0.8$	&	$-0.7, -0.4$	&	0.3	&	0.2	&  0.0        &	0.0	&	 0.0	&	0.1	&	1.0	&	0.0	&	0.4	&	0.4	&	0.2	&	0.0	&	0.0	&	0.0	&	1.3	\\
237	&	$\phantom{1}91, 102$	&	$0.6, 0.8$	&	$-0.4, \phantom{-}0.0$	&	0.3	&	0.1	&  0.0        &	0.0	&	 0.0	&	0.1	&	0.3	&	0.0	&	0.2	&	0.2	&	0.0	&	0.0	&	0.0	&	0.0	&	0.8	\\
238	&	$\phantom{1}91, 102$	&	$0.6, 0.8$	&	$\phantom{+}0.0, +0.4$	&	0.3	&	0.1	&  0.0        &	0.0	&	 0.0	&	0.2	&	0.3	&	0.0	&	0.3	&	0.2	&	0.0	&	0.0	&	0.0	&	0.0	&	0.8	\\
239	&	$\phantom{1}91, 102$	&	$0.6, 0.8$	&	$+0.4, +0.7$	&	0.3	&	0.2	&  0.0        &	0.0	&	 0.0	&	0.2	&	1.0	&	0.0	&	0.4	&	0.4	&	0.1	&	0.0	&	0.0	&	0.0	&	1.3	\\
240	&	$\phantom{1}91, 102$	&	$0.6, 0.8$	&	$+0.7, +1.0$	&	0.4	&	0.2	&  0.0  &	0.0	&	 0.0	&	0.1	&	0.5	&	0.0	&	0.3	&	0.3	&	0.1	&	0.0	&	0.1	&	0.1	&	1.0	\\
\midrule                                                                                                                                             
313	&	$102, 116$	&	$0.8, 1.0$	&	$-1.0, -0.7$	&	2.1	&	1.0	&  0.1  &	0.4	&	 0.0	&	0.2	&	0.9	&	1.4	&	0.4	&	0.4	&	0.2	&	0.0	&	0.0	&	0.1	&	3.0	\\
314	&	$102, 116$	&	$0.8, 1.0$	&	$-0.7, -0.4$	&	1.8	&	0.8	&  0.0  &	0.2	&	 0.1	&	0.2	&	1.8	&	0.3	&	0.3	&	0.3	&	0.2	&	0.0	&	0.0	&	0.0	&	2.8	\\
315	&	$102, 116$	&	$0.8, 1.0$	&	$-0.4, \phantom{-}0.0$	&	1.7	&	0.7	&  0.0  &	0.1	&	 0.0	&	0.1	&	0.4	&	0.6	&	0.3	&	0.3	&	0.1	&	0.0	&	0.0	&	0.0	&	2.0	\\
316	&	$102, 116$	&	$0.8, 1.0$	&	$\phantom{+}0.0, +0.4$	&	1.6	&	0.6	&  0.0  &	0.1	&	 0.0	&	0.2	&	0.4	&	0.5	&	0.3	&	0.3	&	0.0	&	0.0	&	0.0	&	0.0	&	2.0	\\
317	&	$102, 116$	&	$0.8, 1.0$	&	$+0.4, +0.7$	&	1.6	&	0.7	&  0.0  &	0.2	&	 0.1	&	0.2	&	2.0	&	0.8	&	0.4	&	0.3	&	0.1	&	0.0	&	0.0	&	0.1	&	2.8	\\
318	&	$102, 116$	&	$0.8, 1.0$	&	$+0.7, +1.0$	&	2.0	&	0.9	&  0.1  &	0.3	&	 0.0	&	0.2	&	0.8	&	1.5	&	0.4	&	0.4	&	0.0	&	0.0	&	0.0	&	0.0	&	2.7	\\
\midrule                                                                                                                                             
391	&	$116, 150$	&	$1.0, 1.2$	&	$-1.0, -0.7$	&	4.1	&	1.2	&  0.3  &	1.3	&	 0.0	&	0.1	&	0.5	&	0.3	&	0.5	&	0.5	&	0.2	&	0.1	&	0.0	&	0.1	&	4.8	\\
392	&	$116, 150$	&	$1.0, 1.2$	&	$-0.7, -0.4$	&	2.9	&	0.7	&  0.2  &	0.7	&	 0.1	&	0.1	&	0.7	&	0.4	&	0.4	&	0.3	&	0.2	&	0.0	&	0.1	&	0.1	&	3.4	\\
393	&	$116, 150$	&	$1.0, 1.2$	&	$-0.4, \phantom{-}0.0$	&	2.5	&	0.6	&  0.1  &	0.5	&	 0.1	&	0.1	&	0.5	&	0.1	&	0.3	&	0.3	&	0.2	&	0.0	&	0.1	&	0.1	&	2.8	\\
394	&	$116, 150$	&	$1.0, 1.2$	&	$\phantom{+}0.0, +0.4$	&	2.2	&	0.6	&  0.1  &	0.4	&	 0.0	&	0.0	&	0.5	&	0.0	&	0.3	&	0.3	&	0.1	&	0.0	&	0.1	&	0.1	&	2.5	\\
395	&	$116, 150$	&	$1.0, 1.2$	&	$+0.4, +0.7$	&	2.3	&	0.6	&  0.2  &	0.5	&	 0.0	&	0.0	&	0.4	&	0.3	&	0.3	&	0.3	&	0.0	&	0.0	&	0.1	&	0.0	&	2.6	\\
396	&	$116, 150$	&	$1.0, 1.2$	&	$+0.7, +1.0$	&	3.2	&	0.9	&  0.3  &	0.7	&	 0.1	&	0.1	&	0.8	&	0.1	&	0.5	&	0.5	&	0.0	&	0.0	&	0.0	&	0.1	&	3.8	\\
\midrule                                                                                                                                             
469	&	$150, 200$	&	$1.2, 1.4$	&	$-1.0, -0.7$	&	11.1	&	1.5	&  1.2  &	2.9	&	 0.1	&	0.3	&	2.7	&	0.5	&	0.7	&	0.5	&	0.2	&	0.1	&	0.0	&	0.1	&	13.6	\\
470	&	$150, 200$	&	$1.2, 1.4$	&	$-0.7, -0.4$	&	5.6	&	0.8	&  0.5  &	1.4	&	 0.0	&	0.1	&	1.3	&	0.1	&	0.5	&	0.4	&	0.2	&	0.0	&	0.0	&	0.1	&	6.2	\\
471	&	$150, 200$	&	$1.2, 1.4$	&	$-0.4, \phantom{-}0.0$	&	4.6	&	0.6	&  0.3  &	0.9	&	 0.1	&	0.0	&	1.0	&	0.2	&	0.4	&	0.4	&	0.2	&	0.0	&	0.0	&	0.1	&	5.1	\\
472	&	$150, 200$	&	$1.2, 1.4$	&	$\phantom{+}0.0, +0.4$	&	4.1	&	0.5	&  0.2  &	0.7	&	 0.1	&	0.0	&	1.1	&	0.0	&	0.4	&	0.4	&	0.1	&	0.0	&	0.0	&	0.0	&	4.5	\\
473	&	$150, 200$	&	$1.2, 1.4$	&	$+0.4, +0.7$	&	4.0	&	0.5	&  0.2  &	0.8	&	 0.0	&	0.1	&	0.8	&	0.2	&	0.4	&	0.4	&	0.0	&	0.0	&	0.0	&	0.1	&	4.3	\\
474	&	$150, 200$	&	$1.2, 1.4$	&	$+0.7, +1.0$	&	6.6	&	0.9	&  0.5  &	1.2	&	 0.1	&	0.0	&	1.7	&	0.0	&	0.6	&	0.5	&	0.0	&	0.0	&	0.1	&	0.1	&	8.0	\\
\bottomrule
\end{tabular}
}
\caption{Central rapidity muon channel uncertainties in selected bins.
All uncertainties quoted are in units of percent, relative to the measured
differential cross section. The uncertainties are separated into those which
are bin-to-bin correlated within a single channel (marked ``cor'') and those
which are uncorrelated (marked ``unc''). The sources are the uncertainties arising from
the data sample size ($\delta^{\textrm{stat}}_{\textrm{unc}}$);
the signal MC sample size ($\delta^{\textrm{sig}}_{\textrm{unc}}$);
the sizes of the background MC samples ($\delta^{\textrm{bkg}}_{\textrm{unc}}$);
the combined correlated (normalisation) component of all background MC samples ($\delta^{\textrm{bkg}}_{\textrm{cor}}$);
the multijet estimation ($\delta^{\textrm{mj}}_{\textrm{cor}}$);
the muon momentum scale ($\delta^{\textrm{scl}}_{\textrm{cor}}$);
the sagitta bias corrections ($\delta^{\textrm{sag}}_{\textrm{cor}}$);
the muon momentum resolution ($\delta^{\textrm{res}}_{\textrm{cor}}$);
the reconstruction ($\delta^{\textrm{rec}}_{\textrm{cor}}$),
identification ($\delta^{\textrm{id}}_{\textrm{cor}}$), and
trigger efficiencies ($\delta^{\textrm{trig}}_{\textrm{cor}}$);
the $K$-factors ($\delta^{\textrm{kfac}}_{\textrm{cor}}$);
the $Z$ boson $p_{\textrm{T}}$ modelling ($\delta^{\textrm{zpt}}_{\textrm{cor}}$);
the PDF variation ($\delta^{\textrm{pdf}}_{\textrm{cor}}$); and the
total measurement uncertainty ($\delta^{\textrm{tot}}$). The
luminosity uncertainty is not included in these tables.}
\label{tab:muon_detailed}
\end{table}
\FloatBarrier

\clearpage

\FloatBarrier

\section{Results}
\label{sec:results}

In the two invariant mass bins in the region $80<m_{\ell\ell}<102$~GeV, the
measurement of {\d3s} in the central electron channel
achieves a total uncertainty (excluding the luminosity contribution)
of 1--2\% per bin. In the muon channel the precision is better than 1\%. In
both cases the measurement precision is dominated by the experimental
systematic uncertainties, compared to a data statistical uncertainty
of about 0.5\% per bin in this high-precision region. In the high
rapidity electron channel, the precision of the measurement reaches
2--3\% per bin, of which the statistical uncertainty is about 0.5\%.

The data tables provided in this paper contain compact summaries of
the measurement uncertainties; however, complete tables with the full
breakdown of all systematic uncertainties and their correlated components
are provided in HEPData~\cite{Hepdata,Maguire:2017ypu}. These complete
tables also include the correction factors used to translate the
unfolded measurements from the dressed-level to the Born-level as
discussed in section~\ref{sec:methodology}. 

\subsection{Combination of the central rapidity electron and muon channels}
\label{combination}
The central rapidity electron and muon measurement channels are defined
with a common fiducial region given in section~\ref{sec:methodology} and
therefore are combined to further reduce the experimental uncertainties. A
$\chi^2$-minimisation technique is used to combine
the cross sections~\cite{Glazov:2005rn,Aaron:2009bp,Aaron:2009aa}. This method
introduces a nuisance parameter for each systematic error source which
contributes to the total $\chi^2$. The sources of uncertainty
considered are discussed in section~\ref{sec:sys}.  Correlated sources
of uncertainty which are propagated with the pseudo-experiment or
bootstrap resampling methods can be represented in covariance matrix
form for each source. The covariance matrices are decomposed into
eigenvector representations as input to the $\chi^2$-minimisation
function. For each covariance matrix the eigenvectors are sorted
by the magnitude of their corresponding eigenvalues. The largest
of the eigenvalues are added in order of decreasing value until their
sum exceeds a certain fraction of the sum of all eigenvalues,
$f_{\textrm eig}$. At which point the correlation information for
the eigenvectors whose eigenvalues were not included in the sum
is ignored and the eigenvectors are added in quadrature to form a
diagonal uncorrelated uncertainty matrix. The resulting numbers of
nuisance parameters depends on the complexity of the correlation
pattern and on $f_{\textrm{eig}}$, for which values between 99\% and
20\% are chosen depending on the source.

This method of decomposition can 
accurately describe the full covariance matrix, and simultaneously
reduce the number of nuisance parameters. The method
preserves the total uncertainty and marginally enhances the
uncorrelated component of the uncertainty by construction. The
original and decomposed covariance matrices are compared and found to
agree well such  that the combined results are found to be 
stable in terms of $\chi^2$ and the central values and their uncertainties
when $f_{\textrm{eig}}$ is  varied around the chosen value in a wide range.

Bin-to-bin correlated sources of uncertainty which are also correlated
between the two measurement channels share common nuisance parameters,
and are listed in section~\ref{sec:sysCommon}.  In total, 275 nuisance
parameters are used in the procedure. The behaviour of the
uncertainties with respect to the combined cross-section values can lead to
non-Gaussian distributions of the nuisance parameters. For example,
sources related to the selection efficiencies are expected to be
proportional to the combined cross-section value, i.e. have
multiplicative behaviour; sources related to background subtraction
are expected to be independent of the combined cross section and
therefore have an additive behaviour. Finally, data statistical sources
are expected to be proportional to the square-root of the combined
cross section, and have Poisson-like behaviour even after unfolding.

The combination of the central electron and muon channels introduces shifts and constraints
to the nuisance parameters. These shifts are propagated to high rapidity electron channel
measurement but only have a small impact on this channel since it is
dominated by the forward calorimeter uncertainties. The combination of
the electron and muon channel cross-section measurements results in a
$\chi^2$ per degree of freedom (dof) of 489/451 ($p$-value of 10\%). The pulls
of the individual channel measurements to the combined data are found
to be Gaussian-distributed about zero with unit RMS. 
They do not indicate any trends as a function of the kinematic variables.
The pulls of the nuisance parameters are similarly found to be Gaussian-distributed
about zero with a somewhat larger width of 1.18. Only six nuisance
parameters have shifts exceeding three standard deviations, which are
sources related to the calibration of the electromagnetic calorimeter,
and the source describing the normalisation of the
$Z\rightarrow\tau\tau$ background MC sample. These particular sources have
negligible impact on the measurement.

\subsection{Compatibility tests and integrated measurements}
\label{sec:compatibilityandintegrated}

In the following subsections, the triple-differential cross sections measured in
each of the three channels are compared to one another. The compatibility of
the combined data with published ATLAS DY measurements made using the same
2012 dataset is briefly discussed. Moreover, the combined triple-differential cross
section is integrated to produce single- and double-differential cross sections which
are then compared to theoretical predictions.

\subsubsection{Compatibility of the central and high rapidity measurements}
\label{sec:compatibility}
The measurements performed in the central electron and muon channels are
compared with the high rapidity analysis to test for
compatibility. The measurements are made in two different fiducial
regions and therefore a common fiducial volume is defined within which
the comparison is made. This volume is chosen to be $66<m_{\ell\ell}<150$~{\GeV},
$p_{\textrm{T}}^{\ell}>20$~{\GeV}, and no requirement is made on the
pseudorapidity of the lepton. The comparison is performed in the overlapping
$|y_{\ell\ell}|$ bins of the central and high rapidity analyses. 

The corresponding acceptance corrections are obtained from the \powheg\
simulation for each individual measurement bin. Bins with
extrapolation factors smaller than 0.1 are excluded from this test,
since they correspond to very restricted regions of phase space. Such regions
are subject to large modelling uncertainties, in particular the uncertainty associated
with modelling the $Z$ boson transverse momentum. In each bin, the
sum of the extrapolation factors for the central and high rapidity
channels are found to be close to 80\%, indicating that the
two sets of measurements cover most of the phase space for
$66<m_{\ell\ell}<150$~{\GeV} and $p_{\textrm{T}}^{\ell}>20$~{\GeV}.
A second calculation of the extrapolation factors to the full phase space
(i.e. $p_{\textrm{T}}^{\ell}>0$~{\GeV}) has an uncertainty of 1.5\%. This is assumed
to be strongly anti-correlated between the factors for the central and
high rapidity channels since the sum of factors is close
to unity. Therefore, an additional 1\% anti-correlated uncertainty in
the extrapolation factors is used.

The uncertainties arising from electron efficiency corrections are
taken to be uncorrelated between the central and high rapidity electron
channels since they use different identification criteria and
triggers. The multijet uncertainty is also taken to be
uncorrelated. The $\chi^2$/dof of the compatibility test is found to
be 32/30 ($p$-value of 37\%) for the electron channel and 39/30
($p$-value of 13\%) for the muon channel.

\subsubsection{Compatibility with published data}
\label{sec:hmdy}
The cross-section measurements in the central electron and muon
channels partially overlap with published DY measurements from ATLAS
using the same data set. They are differential measurements of the $Z$
boson transverse momentum spectrum~\cite{STDM-2014-12} and of the
high-mass DY cross section for $m_{\ell\ell}>116$~{\GeV}~\cite{STDM-2014-06}.
The compatibility of the data presented here with these two published
measurements has been tested in identical fiducial regions, separately
for the electron and muon channels. The measurements are in good
agreement with each other.

The reader is referred to \cite{STDM-2014-12} where the most precise
measurements of integrated and $p_{\textrm{T}}$-differential $Z$
cross sections were made in the fiducial region
$p_{\textrm{T}}^{\ell}>20$~{\GeV} and $|\eta^{\ell}|<2.4$.

For cross sections differential in $m_{\ell\ell}$ and
$|y_{\ell\ell}|$ in the region $m_{\ell\ell} > 116$~{\GeV},
see the results presented in reference~\cite{STDM-2014-06}.
These measurements are given in the fiducial region of
$p_{\textrm{T}}^{\ell}>40,30$~{\GeV} for leading and
subleading leptons, and $|\eta^{\ell}|<2.5$. 
Note that the published cross sections include the
$\gamma\gamma\to\ell^+\ell^-$ process.

For cross sections measured in the region
$m_{\ell\ell}<116$~{\GeV} and differential in $m_{\ell\ell}$ and
$|y_{\ell\ell}|$, the data presented in this paper should
be used.

\subsubsection{Integrated cross sections}
\label{integXsec}
The combined measurements are integrated over the kinematic variables
$\cos\theta^*$ and $y_{\ell\ell}$ in order to determine the cross
section $\textrm{d}\sigma/\textrm{d}m_{\ell\ell}$. Similarly, the
integration is performed in $\cos\theta^*$ to determine the cross
section $\textrm{d}^2\sigma/\textrm{d}m_{\ell\ell}\textrm{d}|y_{\ell\ell}|$.
The integration is firstly performed for the electron and muon
channels separately to allow a $\chi^2$-test for compatibility of the
two channels. The measurements are simply summed
in the $e$ and $\mu$ channels for the bins in which
both electron and muon measurements are present. Statistical and uncorrelated
uncertainties are added in quadrature, whereas correlated systematic
uncertainties are propagated linearly. The compatibility tests return
$\chi^2$/dof~=~12.8/7 ($p$-value of 7.7\%) for the one-dimensional cross section, and
103/84 ($p$-value of 7.4\%) for the two-dimensional cross section.

\begin{figure}[htp!]
\centering
\includegraphics[width=0.6\textwidth]{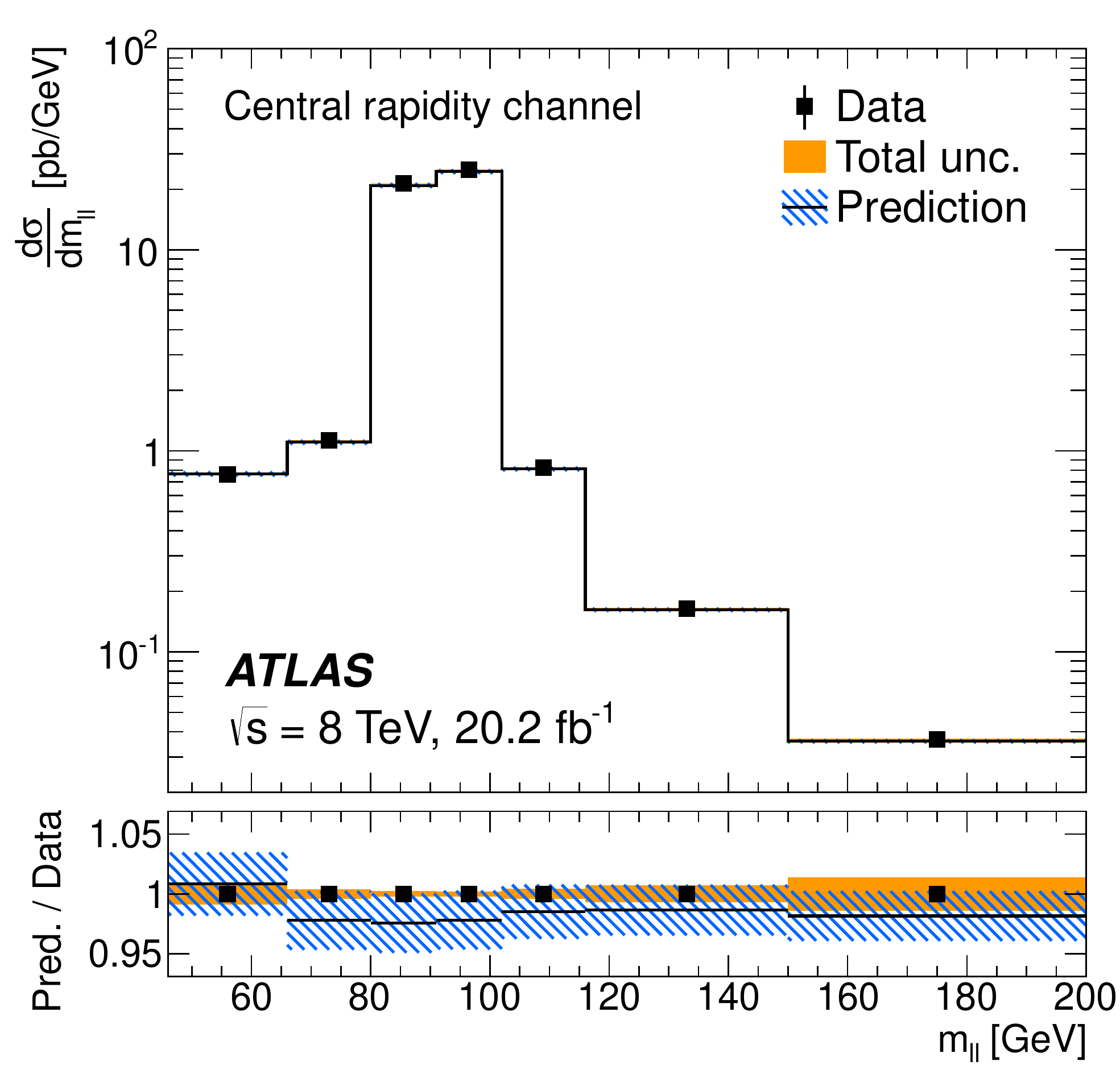}
\caption{The combined Born-level fiducial cross section
	$\textrm{d}\sigma/\textrm{d}m_{\ell\ell}$. The data are shown as
	solid markers and the prediction from \powheg\ including NNLO
	QCD and NLO EW $K$-factors is shown as the solid line. The lower
	panel shows the ratio of prediction to measurement. The inner error
	bars represent the data statistical uncertainty and the solid band
	shows the total experimental uncertainty. The contribution to the
	uncertainty from the luminosity measurement is excluded. The
	hatched band represents the statistical and PDF uncertainties in the
	prediction.}
\label{fig:1D_Combined_Truth}
\end{figure}

The integrated cross sections
$\textrm{d}\sigma/\textrm{d}m_{\ell\ell}$ and
$\textrm{d}^2\sigma/\textrm{d}m_{\ell\ell}\textrm{d}|y_{\ell\ell}|$
are determined from the combined Born-level fiducial
triple-differential cross sections. The one-dimensional result is
shown in figure~\ref{fig:1D_Combined_Truth}. The corresponding table
of measurements is given in table~\ref{tab:full_summary_m} located in
the appendix. The data shows that the combined Born-level fiducial cross section falls by
three orders of magnitude in the invariant mass region from the
resonant peak to 200~{\GeV}.
The data have an uncertainty of about 2\%, dominated by the luminosity 
uncertainty of $1.9\%$, while uncertainties from the experimental
systematic sources can be as low as $0.5\%$ for the peak region.
The statistical precision is $0.5\%$
or better, even for the highest invariant mass bin. The fiducial
measurements are well predicted by the NLO QCD and parton shower
simulation from \powheg\ partially corrected for NNLO QCD and NLO EW effects, and
scattering amplitude coefficients as described in
section~\ref{sec:MC}. The uncertainties in the predictions include those
arising from the sample size and the PDF variations. No
renormalization, factorisation and matching scale variation
uncertainties are included although they can be sizeable -- as large as 5\% for
NLO predictions. Except in the lowest mass bin, the predictions underestimate the cross section by about 1--2\%
(smaller than the luminosity uncertainty), as seen in the lower panel of
the figure which shows the ratio of prediction to the measurement.

The two-dimensional Born-level fiducial cross section,
$\textrm{d}^2\sigma/\textrm{d}m_{\ell\ell}\textrm{d}|y_{\ell\ell}|$,
is illustrated in figure~\ref{fig:2D_Combined_Truth} and listed in
table~\ref{tab:full_summary_ym} of the appendix. In each measured invariant
mass bin, the shape of the rapidity distribution shows a plateau at
small $|y_{\ell\ell}|$ leading to a broad shoulder followed by a cross
section falling to zero at the highest accessible
$|y_{\ell\ell}|$. The width of the plateau narrows with increasing
$m_{\ell\ell}$.  In the two high-precision $Z$-peak mass bins,
the measured cross-section values have a
total uncertainty (excluding the common luminosity uncertainty) of
0.4\% for $|y_{\ell\ell}|<1$ rising to 0.7\% at $|y_{\ell\ell}|=2.4$. At
high invariant mass, the statistical and experimental uncertainty
components contribute equally to the total measurement precision
in the plateau region, increasing from 0.5\% to 1.8\%. The
theoretical predictions replicate the features in the data well.
The lower panel of each figure shows the ratio of the prediction to
the measurement.  Here, in addition to overall rate difference already observed in the one-dimensional
distribution, a small tendency of the data to exceed the predictions at
the highest $|y_{\ell\ell}|$ can be seen in some of the mass bins.

\begin{figure}[htp!]
\centering
\includegraphics[width=0.32\textwidth]{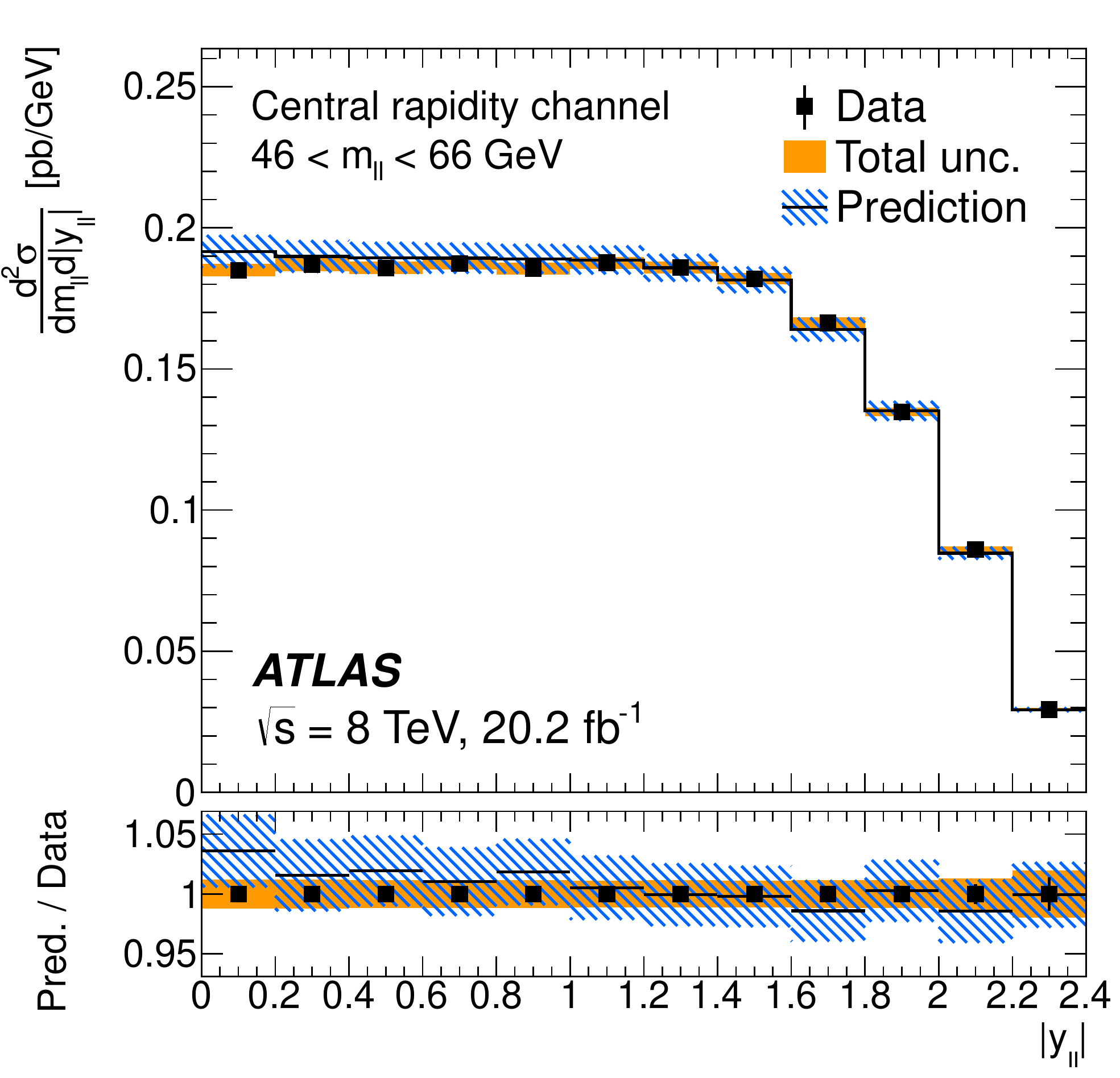}~
\includegraphics[width=0.32\textwidth]{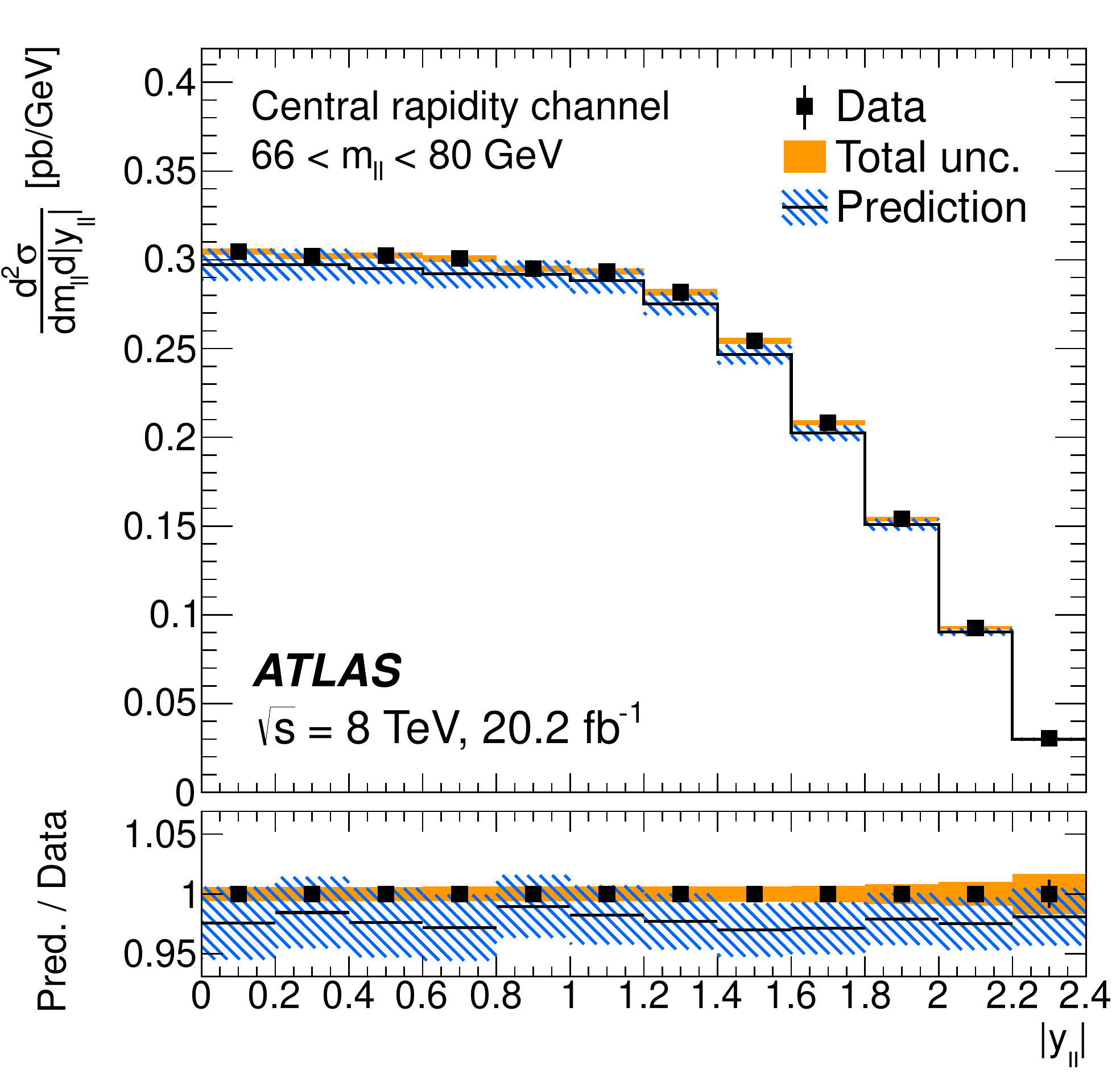}~
\includegraphics[width=0.32\textwidth]{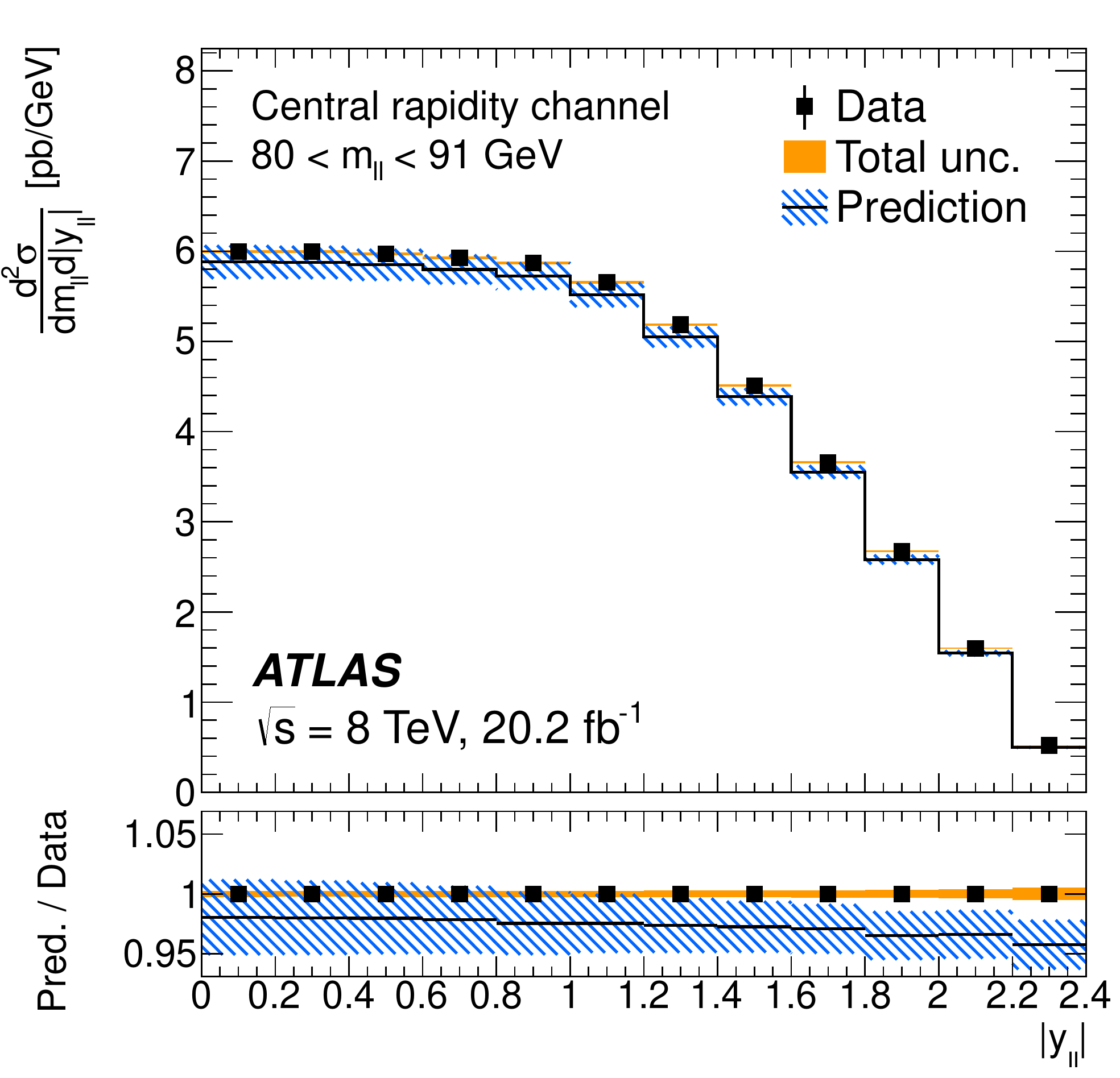}\\
\includegraphics[width=0.32\textwidth]{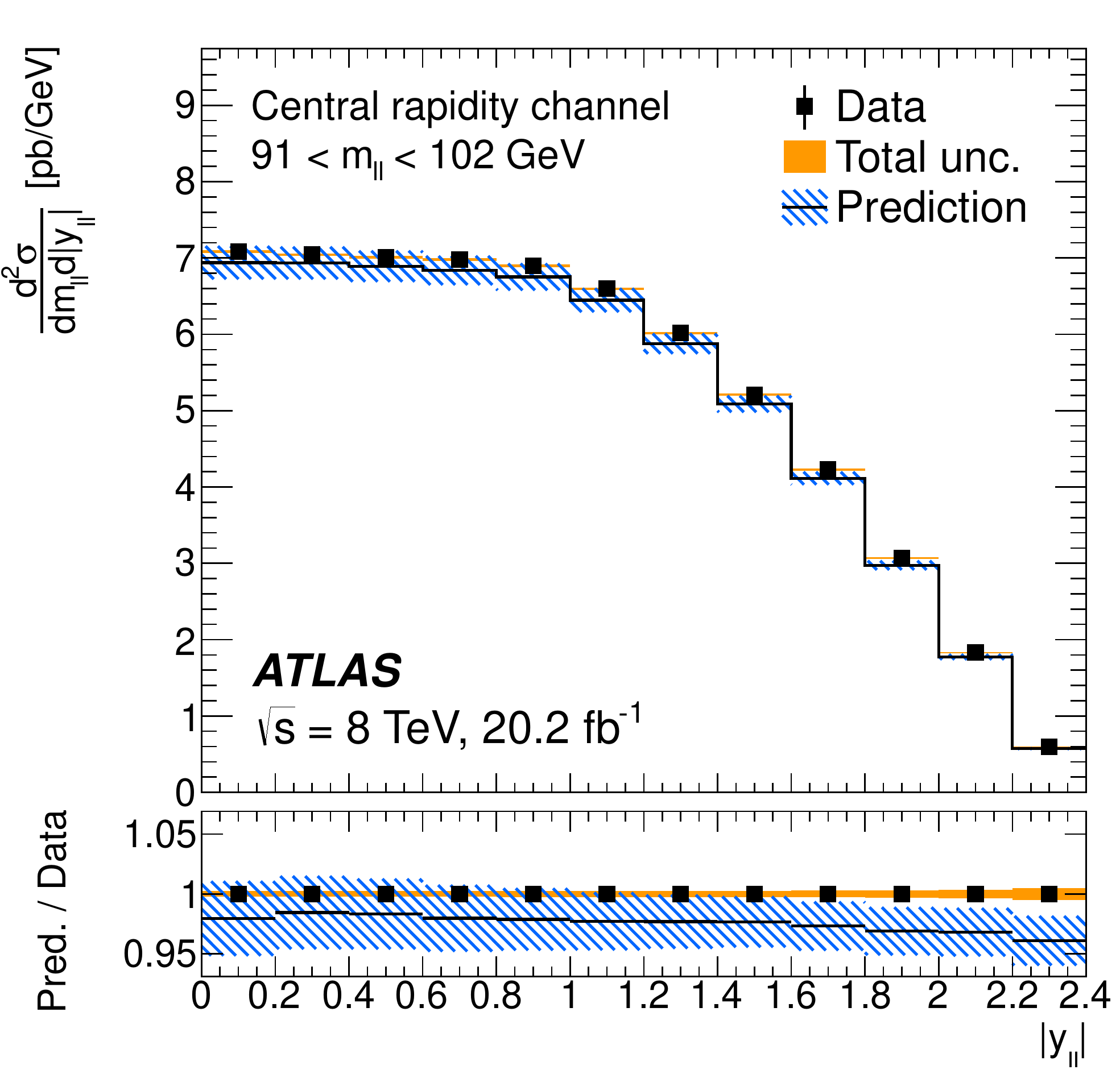}~
\includegraphics[width=0.32\textwidth]{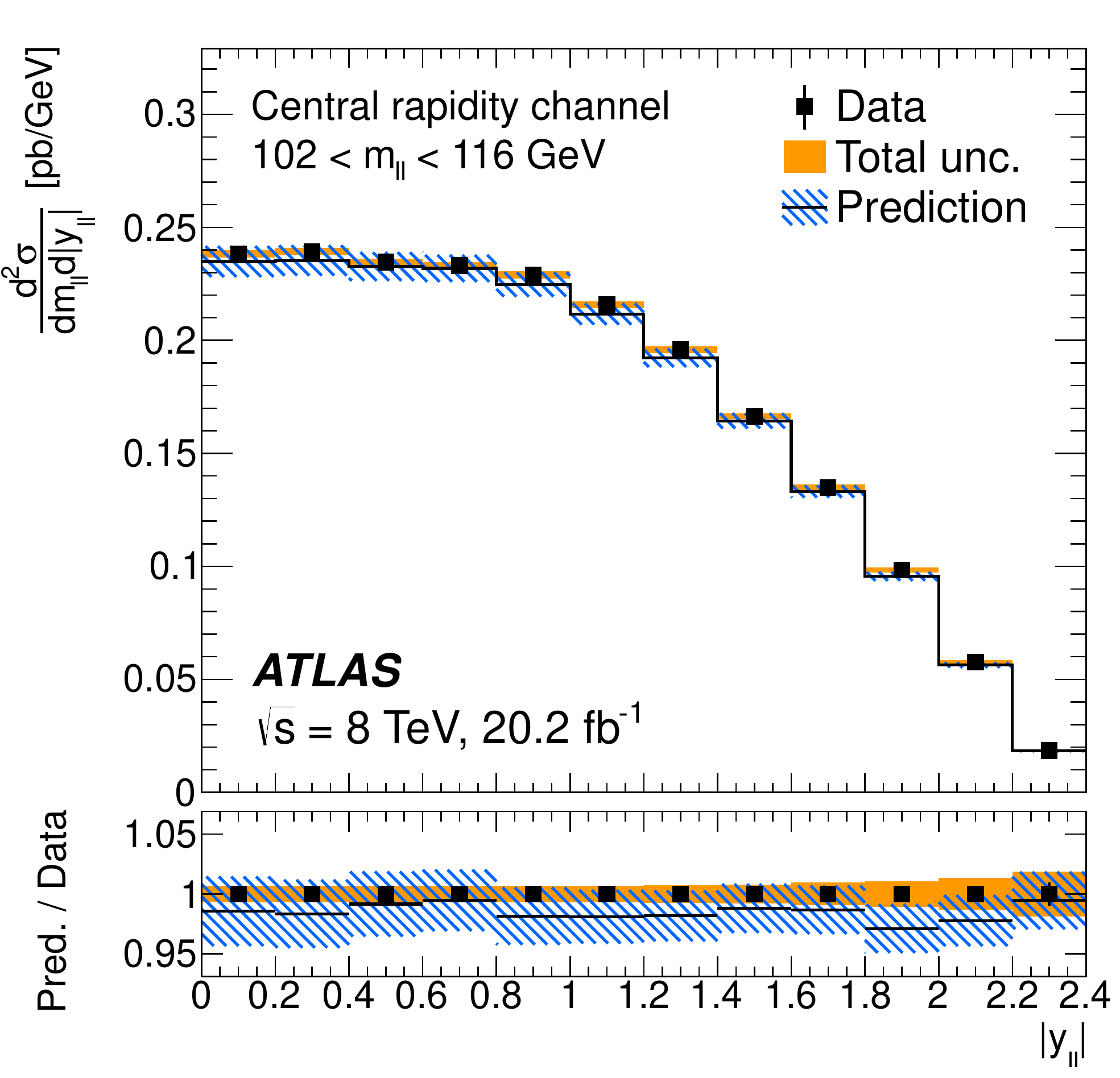}~
\includegraphics[width=0.32\textwidth]{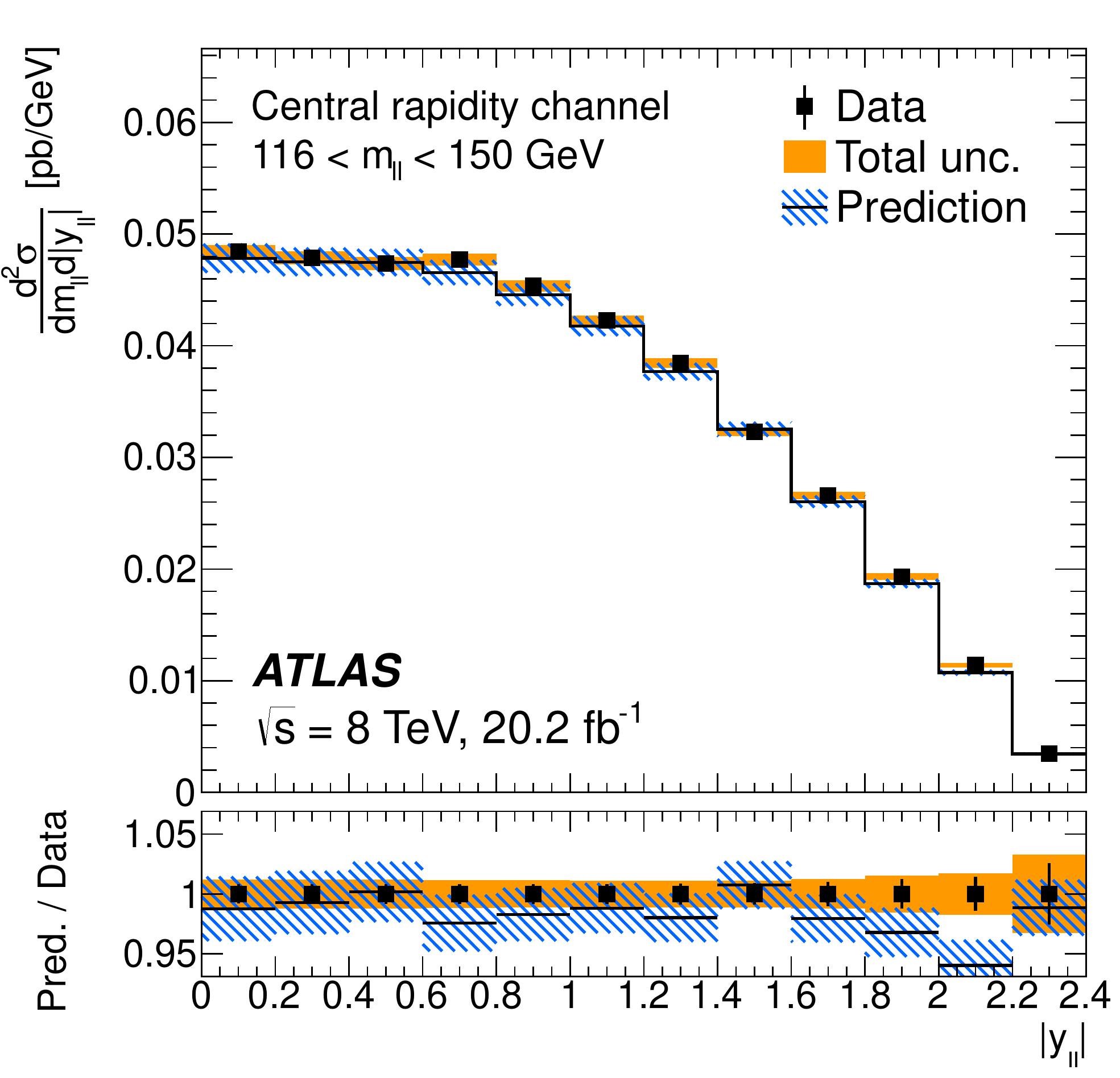}\\
\includegraphics[width=0.32\textwidth]{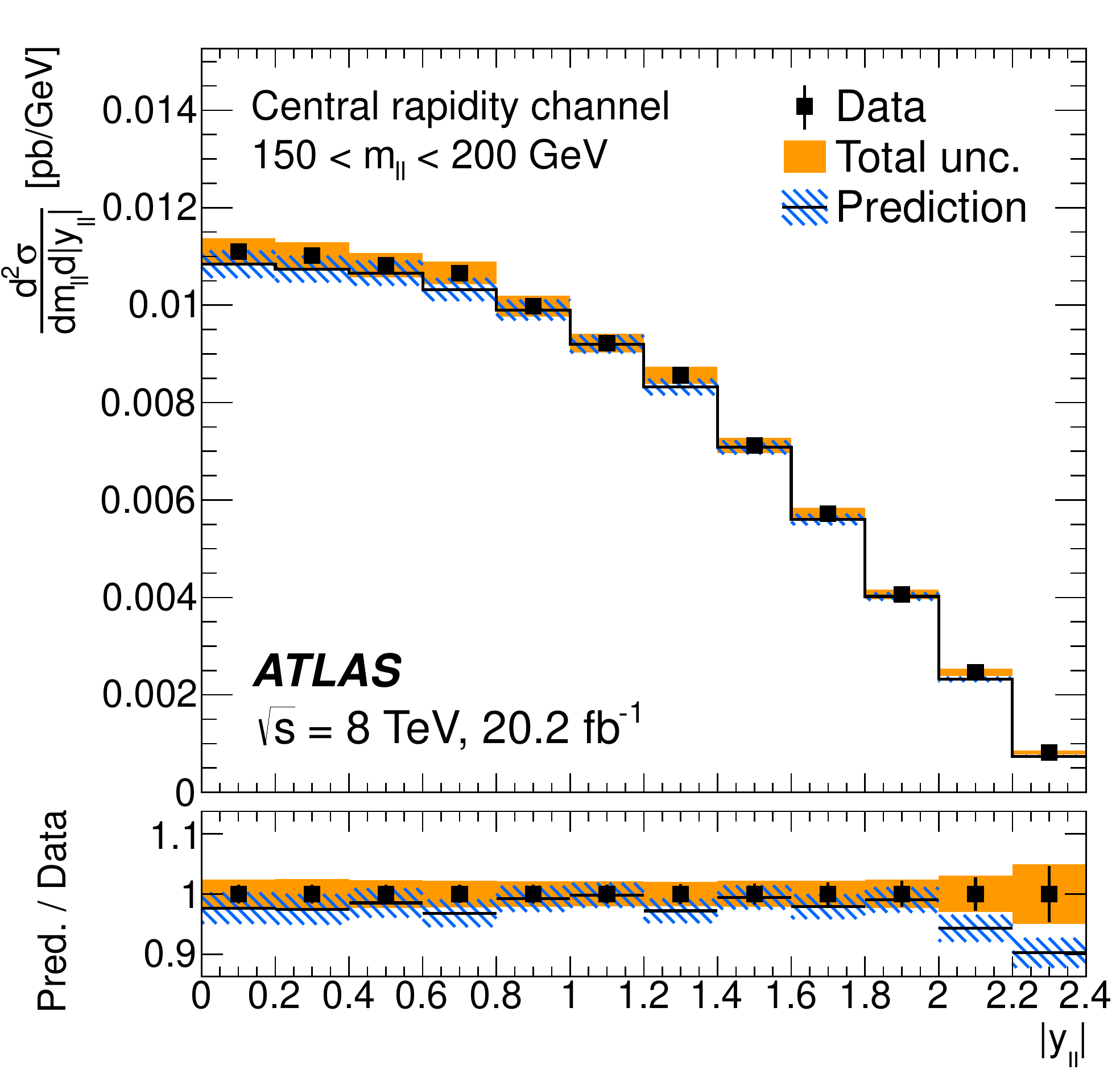}
\caption{The combined Born-level fiducial cross section
	$\textrm{d}^2\sigma/\textrm{d}m_{\ell\ell}\textrm{d}|y_{\ell\ell}|$
	in the seven invariant mass bins of the central measurements. The data are shown as
	solid markers and the prediction from \powheg\ including NNLO
	QCD and NLO EW $K$-factors is shown as the solid line. The lower
	panel shows the ratio of prediction to measurement. The inner error
	bars represent the data statistical uncertainty and the solid band
	shows the total experimental uncertainty. The contribution to the
	uncertainty from the luminosity measurement is excluded. The
	hatched band represents the statistical and PDF uncertainties in the
	prediction.}
\label{fig:2D_Combined_Truth}
\end{figure}
\FloatBarrier

\subsection{Triple-differential cross sections}
\label{Xsec3d}
\begin{figure}[htp!]
	\centering
	\includegraphics[width=0.49\textwidth]{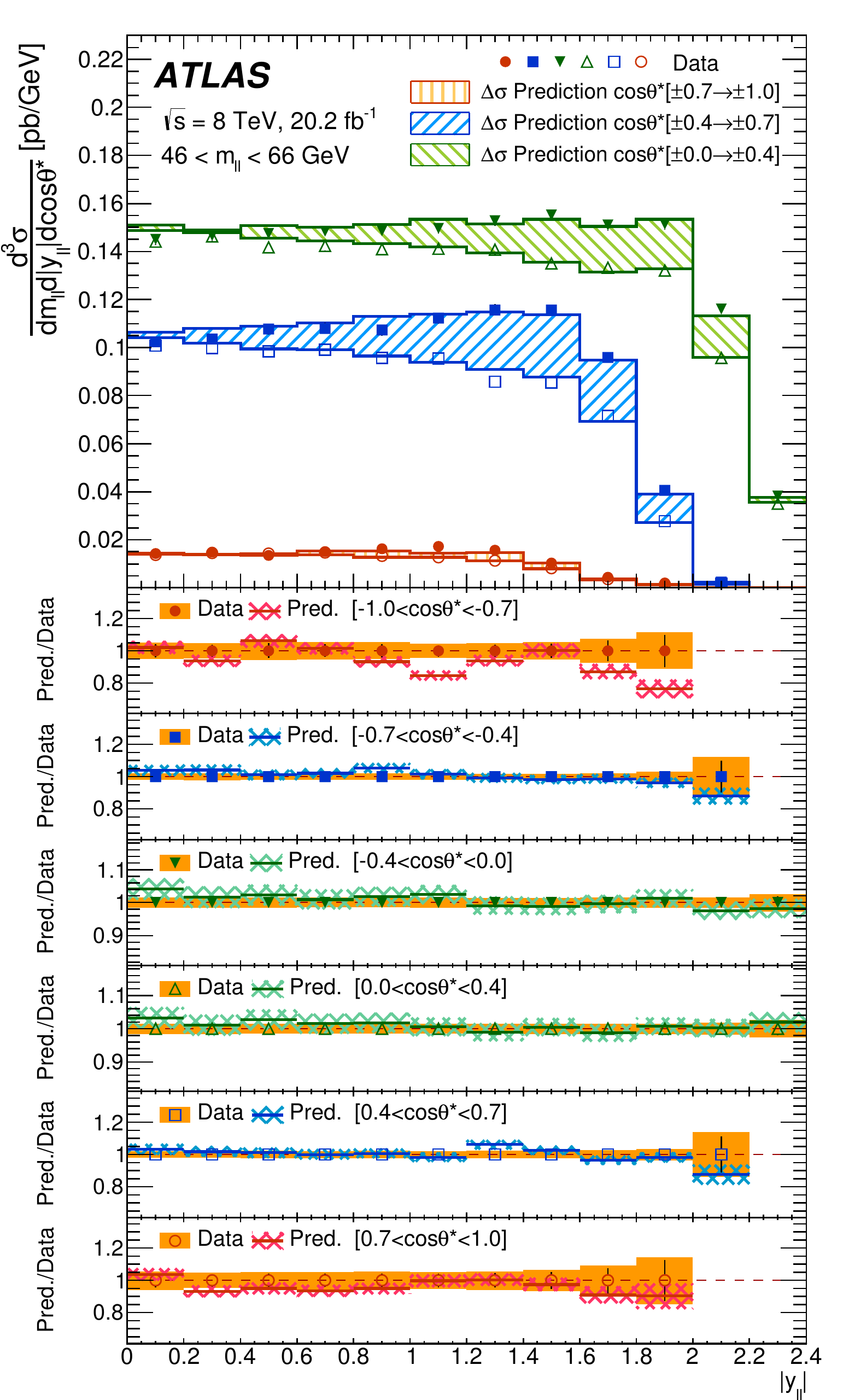}~
	\includegraphics[width=0.49\textwidth]{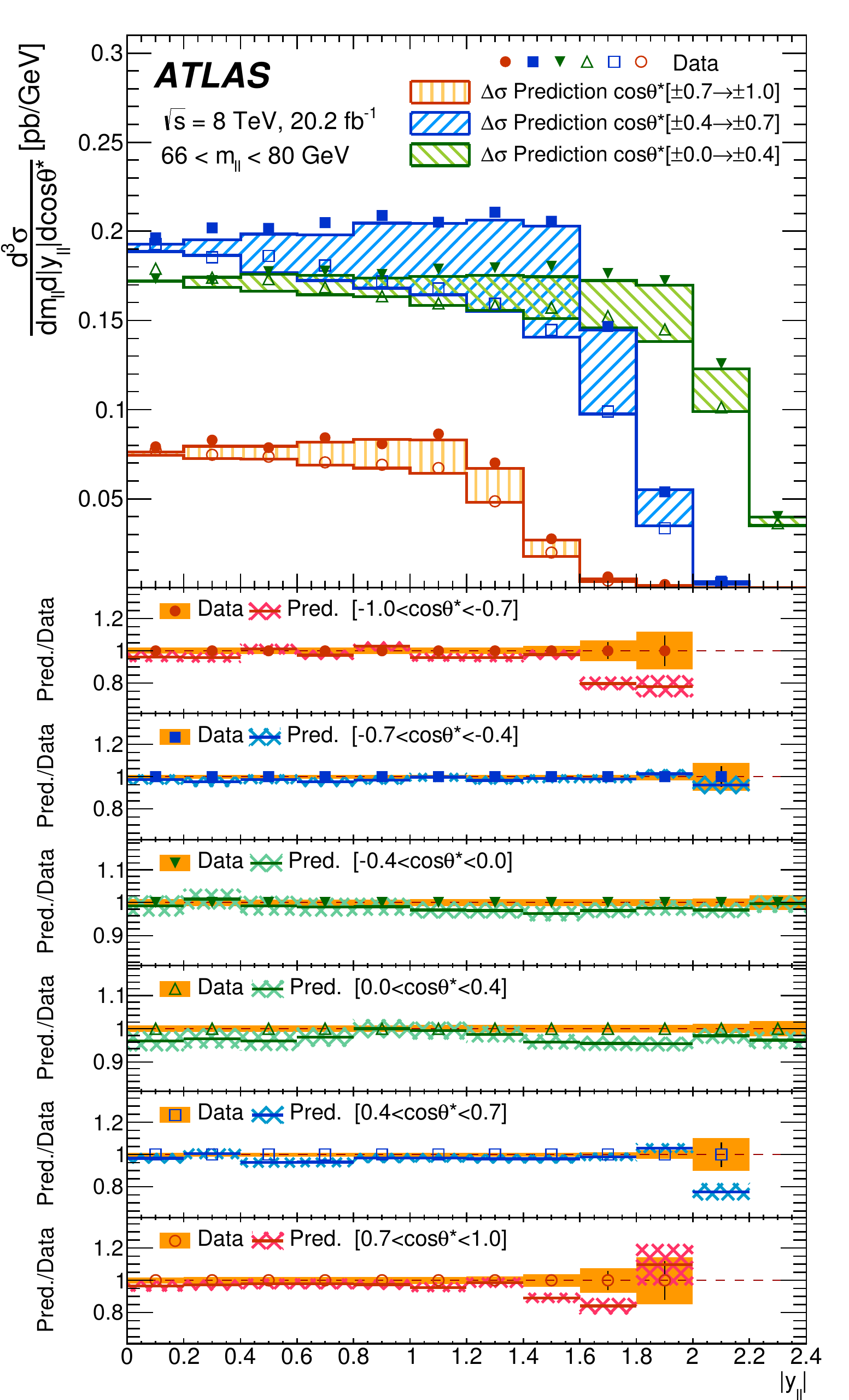}
	\caption{The combined Born-level fiducial cross sections $\textrm{d}^3\sigma$.
		The kinematic region shown is labelled in each plot. The data are shown as
		solid ($\cos \theta^*<0$) and open ($\cos \theta^*>0$) markers
		and the prediction from \powheg\ including NNLO QCD and NLO EW
		$K$-factors is shown as the solid line. The difference, $\Delta\sigma$,
		between the predicted cross sections in the two measurement bins at equal
		|$\cos\theta^*$| symmetric around $\cos\theta^* = 0$ is represented by the hatched shading.
		In each plot, the lower panel shows the ratio of prediction to measurement.
		The inner error bars represent the statistical uncertainty of the data and the
		solid band	shows the total experimental uncertainty. The contribution to the
		uncertainty from the luminosity measurement is excluded. The crosshatched
		band represents the statistical and PDF uncertainties in the prediction.}
	\label{l_comb_format_1}
\end{figure}

\begin{figure}[htp!]
	\centering
	\includegraphics[width=0.49\textwidth]{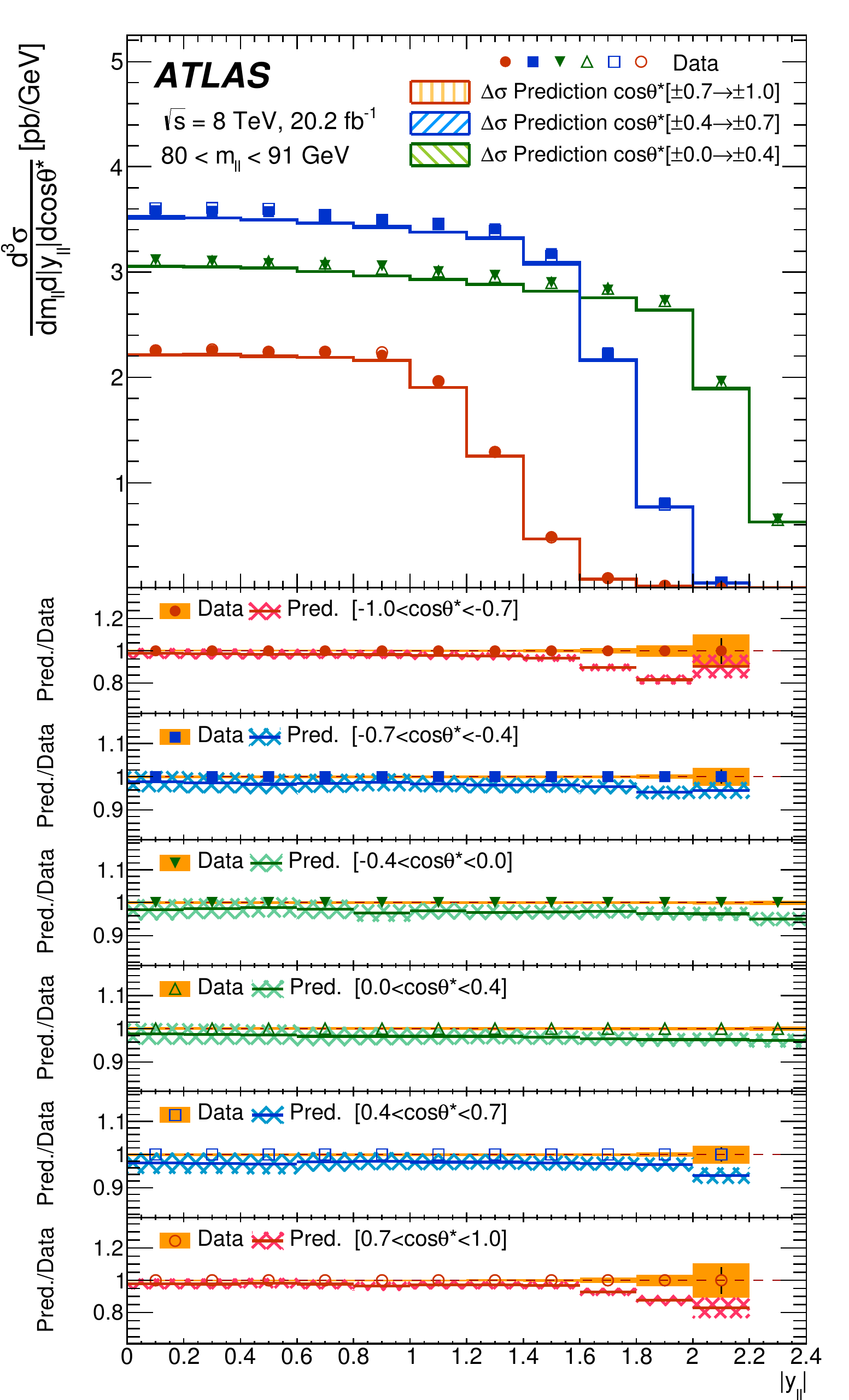}~
	\includegraphics[width=0.49\textwidth]{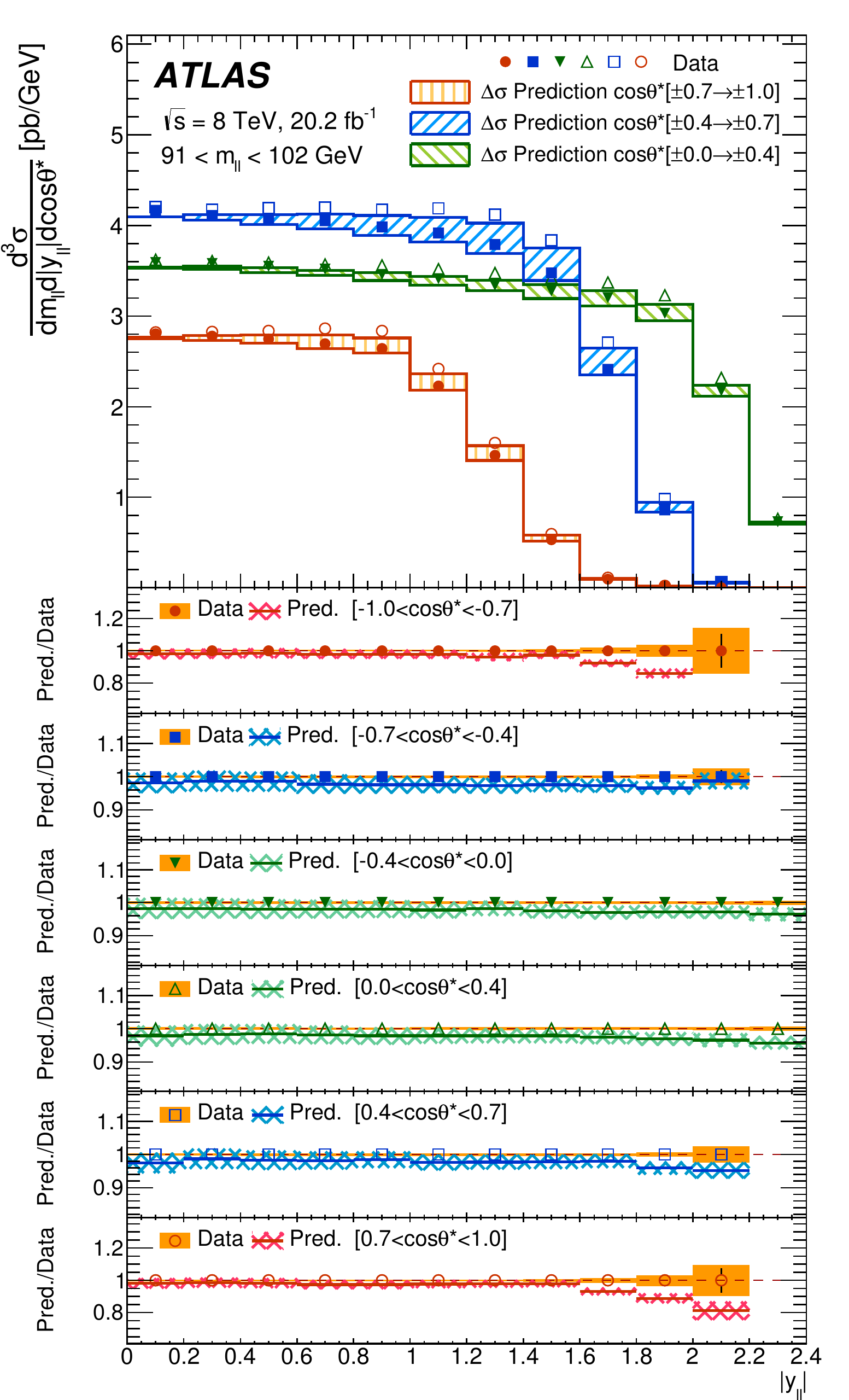}
	\caption{The combined Born-level fiducial cross sections $\textrm{d}^3\sigma$.
		The kinematic region shown is labelled in each plot. The data are shown as
		solid ($\cos \theta^*<0$) and open ($\cos \theta^*>0$) markers
		and the prediction from \powheg\ including NNLO QCD and NLO EW
		$K$-factors is shown as the solid line. The difference, $\Delta\sigma$,
		between the predicted cross sections in the two measurement bins at equal
		|$\cos\theta^*$| symmetric around $\cos\theta^* = 0$ is represented by the hatched shading.
		In each plot, the lower panel shows the ratio of prediction to measurement.
		The inner error bars represent the statistical uncertainty of the data and the
		solid band	shows the total experimental uncertainty. The contribution to the
		uncertainty from the luminosity measurement is excluded. The crosshatched
		band represents the statistical and PDF uncertainties in the prediction.}
	\label{l_comb_format_2}
\end{figure}

\begin{figure}[htp!]
	\centering
	\includegraphics[width=0.49\textwidth]{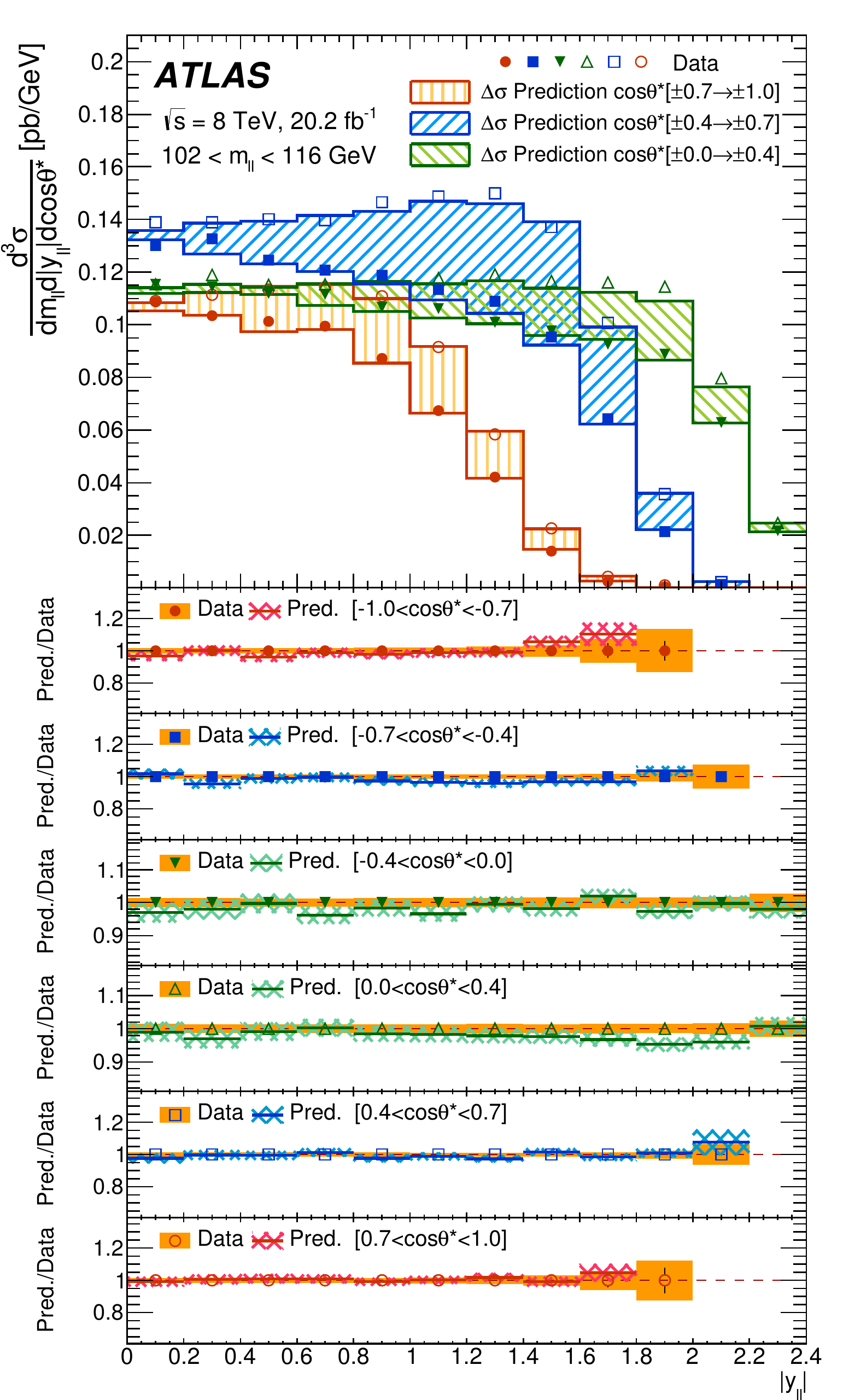}~
	\includegraphics[width=0.49\textwidth]{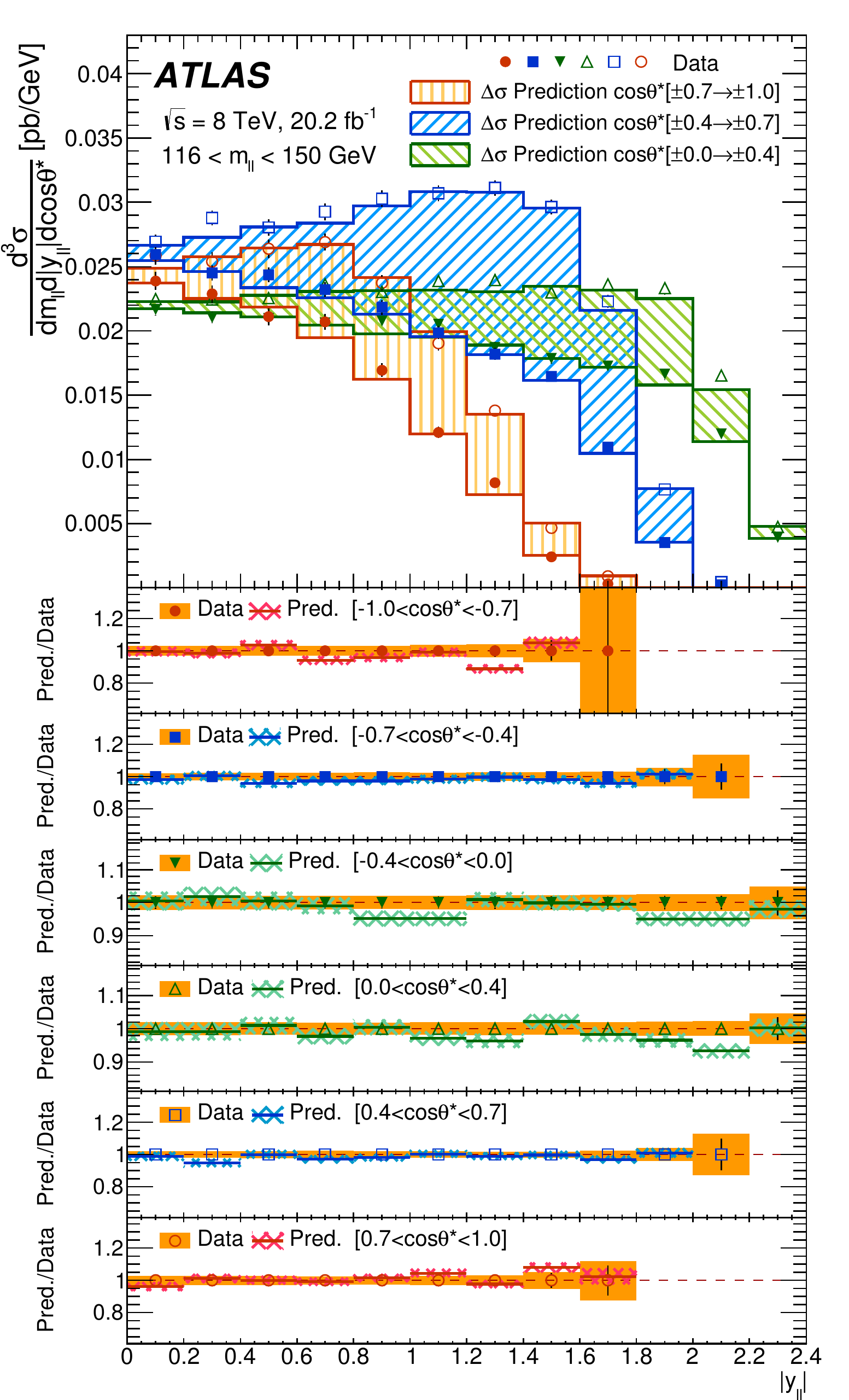}
	\caption{The combined Born-level fiducial cross sections $\textrm{d}^3\sigma$.
		The kinematic region shown is labelled in each plot. The data are shown as
		solid ($\cos \theta^*<0$) and open ($\cos \theta^*>0$) markers
		and the prediction from \powheg\ including NNLO QCD and NLO EW
		$K$-factors is shown as the solid line. The difference, $\Delta\sigma$,
		between the predicted cross sections in the two measurement bins at equal
		|$\cos\theta^*$| symmetric around $\cos\theta^* = 0$ is represented by the hatched shading.
		In each plot, the lower panel shows the ratio of prediction to measurement.
		The inner error bars represent the statistical uncertainty of the data and the
		solid band	shows the total experimental uncertainty. The contribution to the
		uncertainty from the luminosity measurement is excluded. The crosshatched
		band represents the statistical and PDF uncertainties in the prediction.}
	\label{l_comb_format_3}
\end{figure}

\begin{figure}[htp!]
	\centering
	\includegraphics[width=0.49\textwidth]{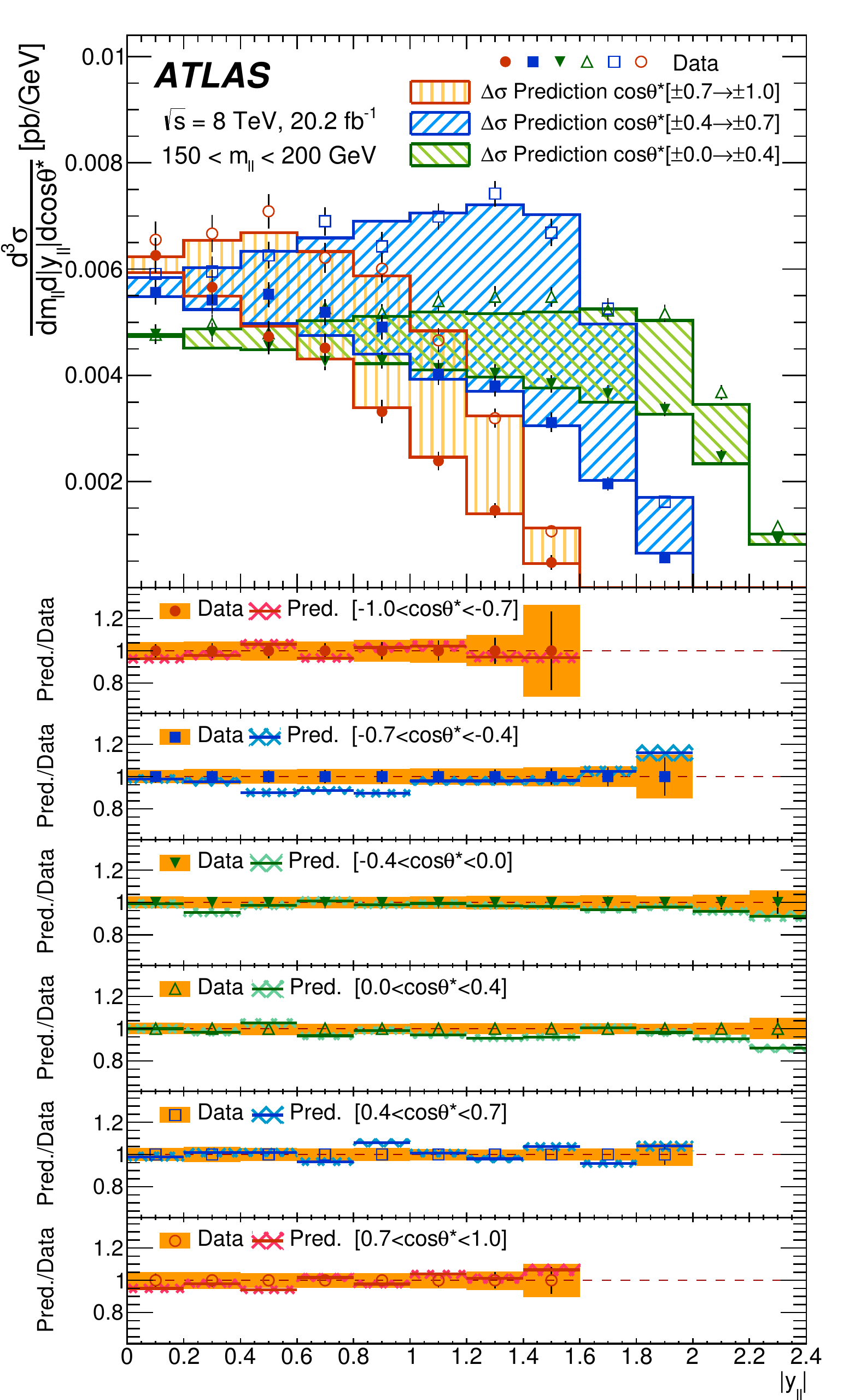}
	\caption{The combined Born-level fiducial cross sections $\textrm{d}^3\sigma$.
		The kinematic region shown is labelled in each plot. The data are shown as
		solid ($\cos \theta^*<0$) and open ($\cos \theta^*>0$) markers
		and the prediction from \powheg\ including NNLO QCD and NLO EW
		$K$-factors is shown as the solid line. The difference, $\Delta\sigma$,
		between the predicted cross sections in the two measurement bins at equal
		|$\cos\theta^*$| symmetric around $\cos\theta^* = 0$ is represented by the hatched shading.
		In each plot, the lower panel shows the ratio of prediction to measurement.
		The inner error bars represent the statistical uncertainty of the data and the
		solid band	shows the total experimental uncertainty. The contribution to the
		uncertainty from the luminosity measurement is excluded. The crosshatched
		band represents the statistical and PDF uncertainties in the prediction.}
	\label{l_comb_format_4}
\end{figure}

The combined triple-differential Born-level cross section is shown in
figures~\ref{l_comb_format_1}--\ref{l_comb_format_4}.  For each
invariant mass bin, the data are presented as a function of
$|y_{\ell\ell}|$, with each of the six $\cos\theta^*$ regions overlaid
in the main panel of the figures. The lower panels show in more
detail the ratio of the prediction to the data for each $\cos\theta^*$ bin in
turn. The statistical and total, excluding the contribution from the luminosity,
uncertainties in the data are shown in the ratio panels.

The accessible range of the $|y_{\ell\ell}|$ distribution is largest for the region
close to $\cos\theta^*\simeq 0$, and smallest at the extremes of $\cos\theta^*$.
In the lowest invariant mass bin, the cross-section measurements in $\cos\theta^*$
bins with the same absolute value, e.g. bins $-1.0 < \cos\theta^* < -0.7$ and
$+0.7 < \cos\theta^* <  +1.0$, are consistent with each other at low
$|y_{\ell\ell}|\simeq0$, but exhibit an asymmetry which increases with $|y_{\ell\ell}|$.
At large $|y_{\ell\ell}|$, the cross sections for $\cos\theta^*<0$ are up to 35\% larger
than the corresponding measurements at $\cos\theta^*>0$.  In the
$66<m_{\ell\ell}<80$~{\GeV} bin, all cross sections are larger, for large
$|\cos \theta^*|$ in particular, due to reduced influence of the  fiducial selection
on  $p^{\ell}_{\textrm{T}}$.

The next two invariant mass bins show the peak of the cross section
where the asymmetry is smallest. In fact, for $80<m_{\ell\ell}<91$~{\GeV}
the difference between $\cos\theta^*>0$ and $\cos\theta^*<0$ is close to
zero. The dramatic improvement in the overall precision of the
measurements in this region is also apparent. 
For the $91<m_{\ell\ell}<102$~{\GeV} region, the small
asymmetry is observed to change sign, yielding larger cross sections for
the $\cos\theta^*<0$ part of the phase space. This behaviour is 
expected from the interference effects between the $Z$ and $\gamma^*$
contributions to the scattering amplitudes. For bins of higher invariant mass
the asymmetry increases albeit with larger uncertainties due to
the limited statistical precision of the data. The combined measurement
is listed in table~\ref{tab:comb_summary} with its uncertainties. 

The predictions describe the data very well, as can be seen from the ratio panels,
apart from some bins at large $|y_{\ell\ell}|$ and $|\cos\theta^*|$. These bins correspond
to edges of the fiducial acceptance and may be affected by the $p_{\textrm{T},\ell\ell}$
modelling uncertainties which are not shown for the predictions.

\begin{figure}[htp!]
\centering
\includegraphics[width=0.48\textwidth]{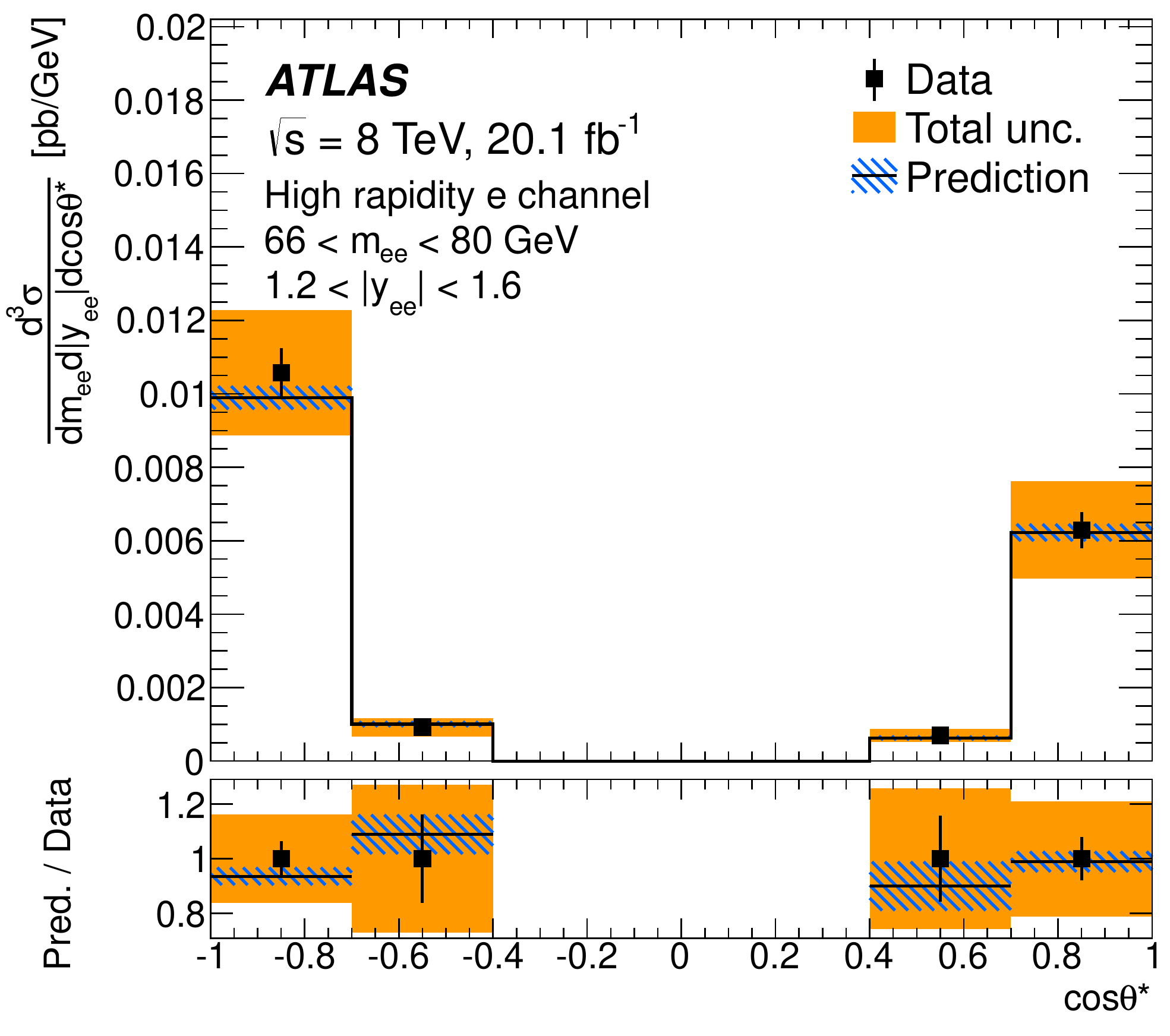}~
\includegraphics[width=0.48\textwidth]{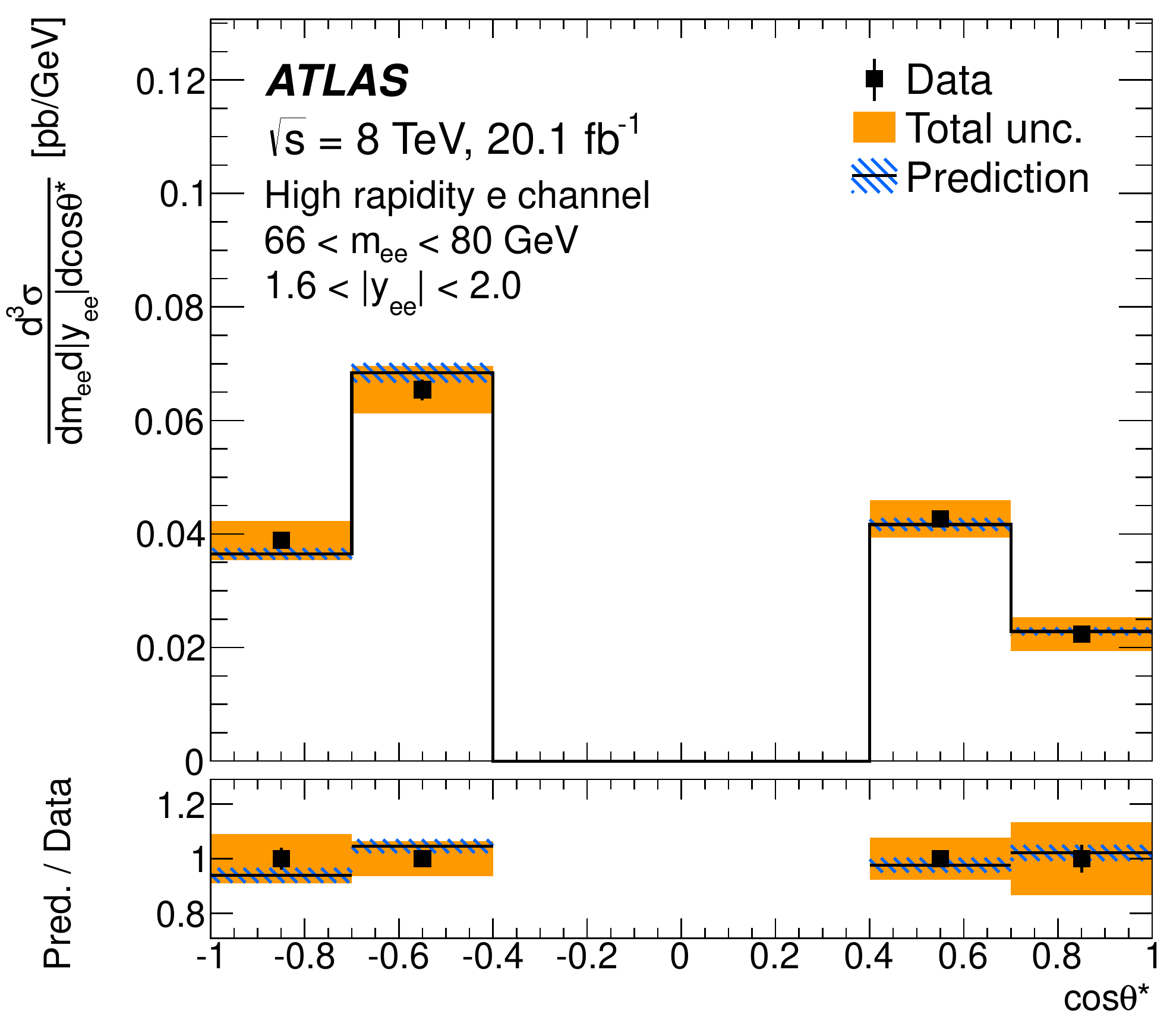}\\
\includegraphics[width=0.48\textwidth]{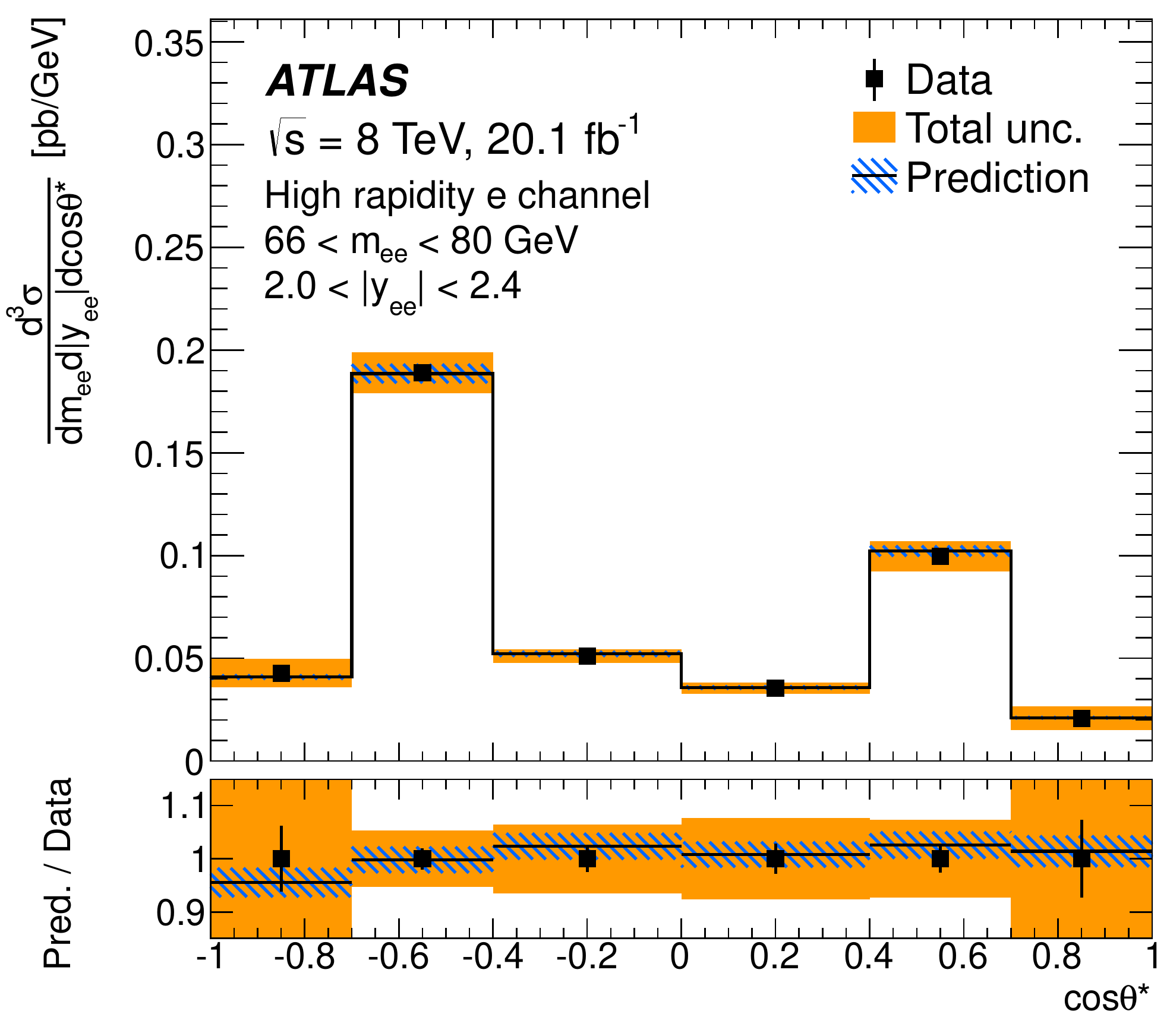}~
\includegraphics[width=0.48\textwidth]{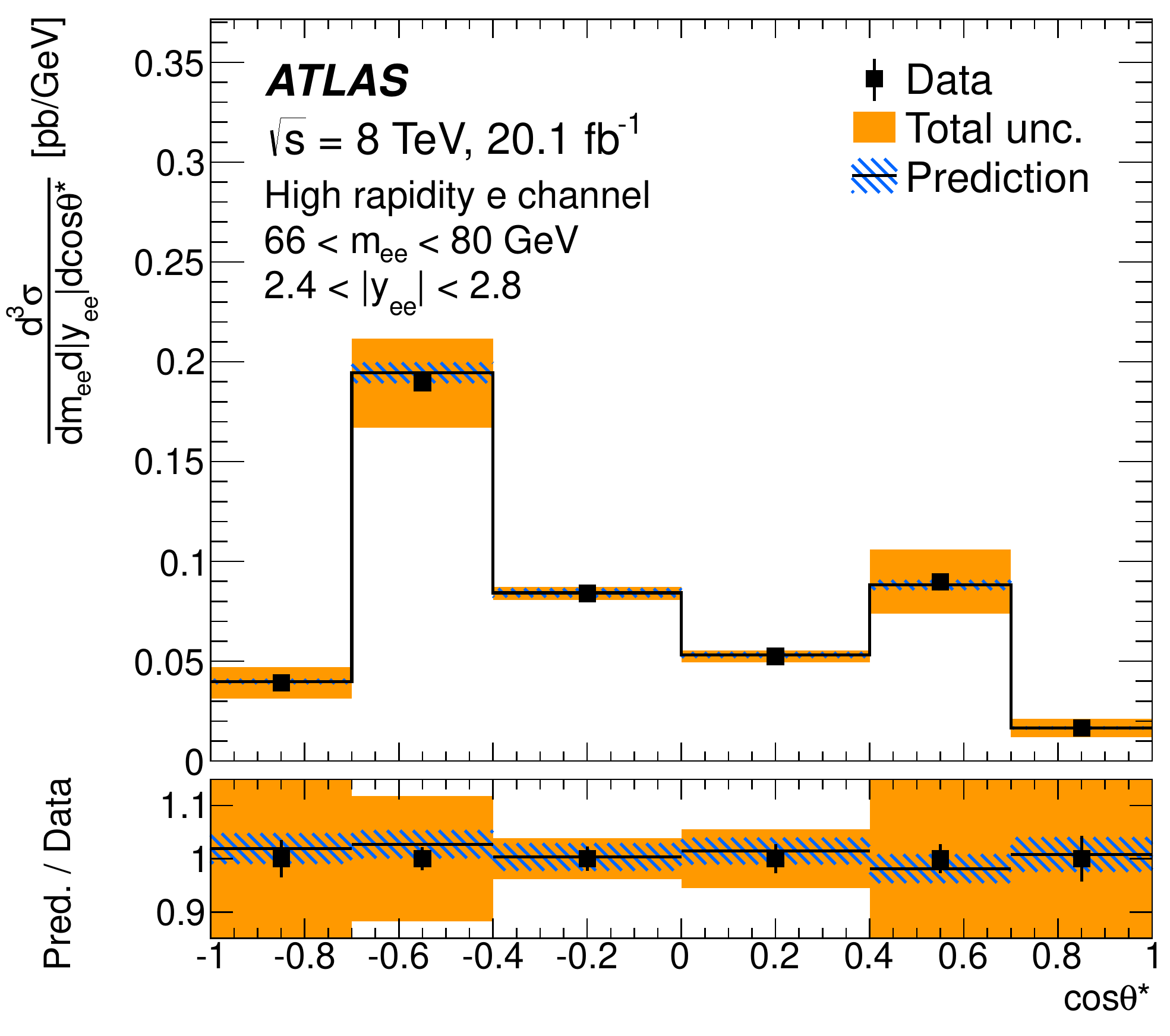}\\
\includegraphics[width=0.48\textwidth]{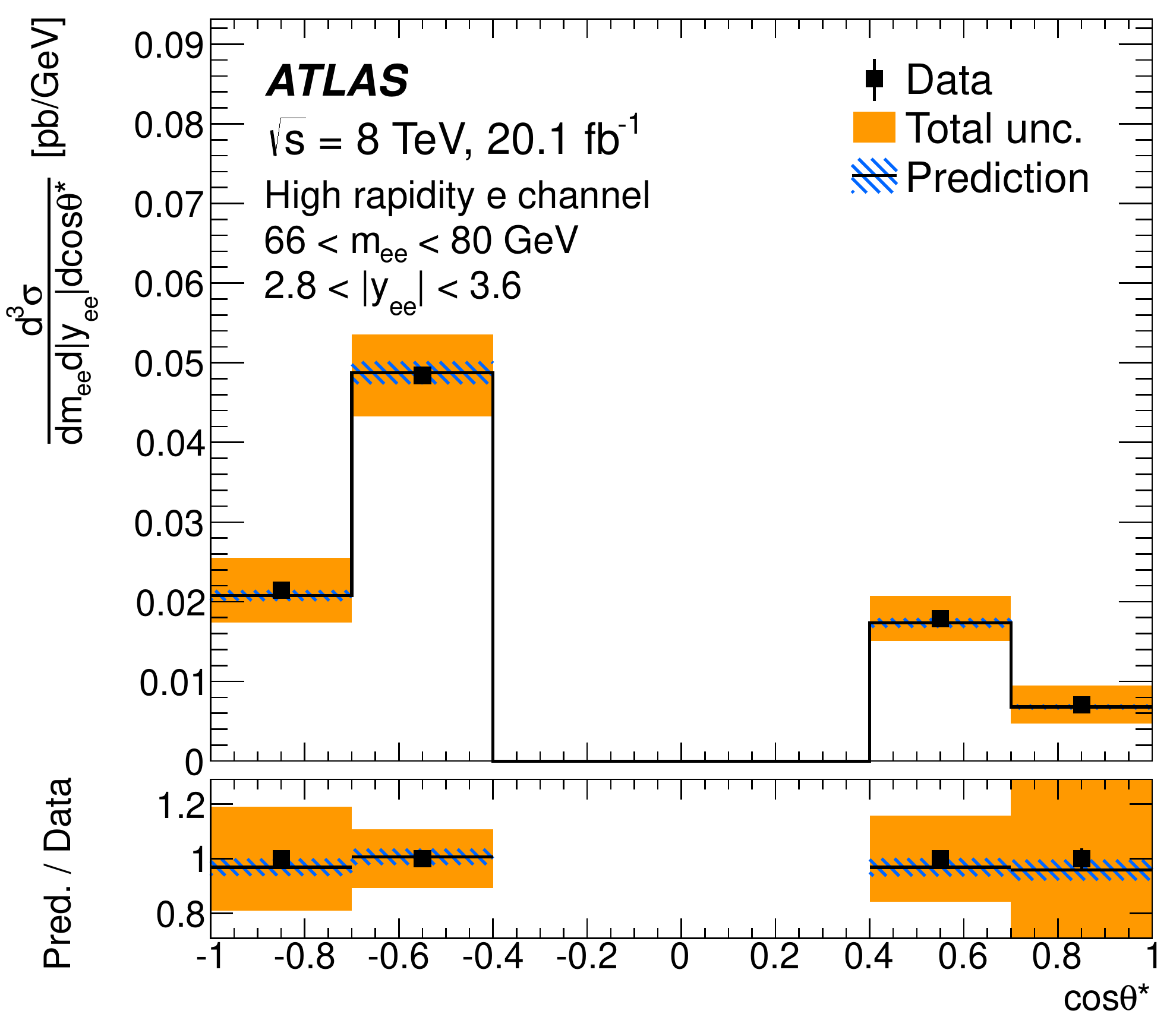}
\caption{The high rapidity electron channel Born-level fiducial cross section
		$\textrm{d}^3\sigma$. The kinematic region shown is labelled in each
		plot. The data are shown as solid markers and the prediction from
		\powheg\ including NNLO QCD and NLO EW $K$-factors is shown as
		the solid line. In each plot, the lower panel shows the ratio of prediction
		to measurement. The inner error bars represent the statistical uncertainty
		of the data and the solid band shows the total experimental uncertainty.
		The contribution from the uncertainty of the luminosity measurement is
		excluded. The hatched band represents the statistical and PDF uncertainties
		in the prediction.}
\label{fig:3D_ZCF_1}
\end{figure}

\begin{figure}[htp!]
\centering
\includegraphics[width=0.48\textwidth]{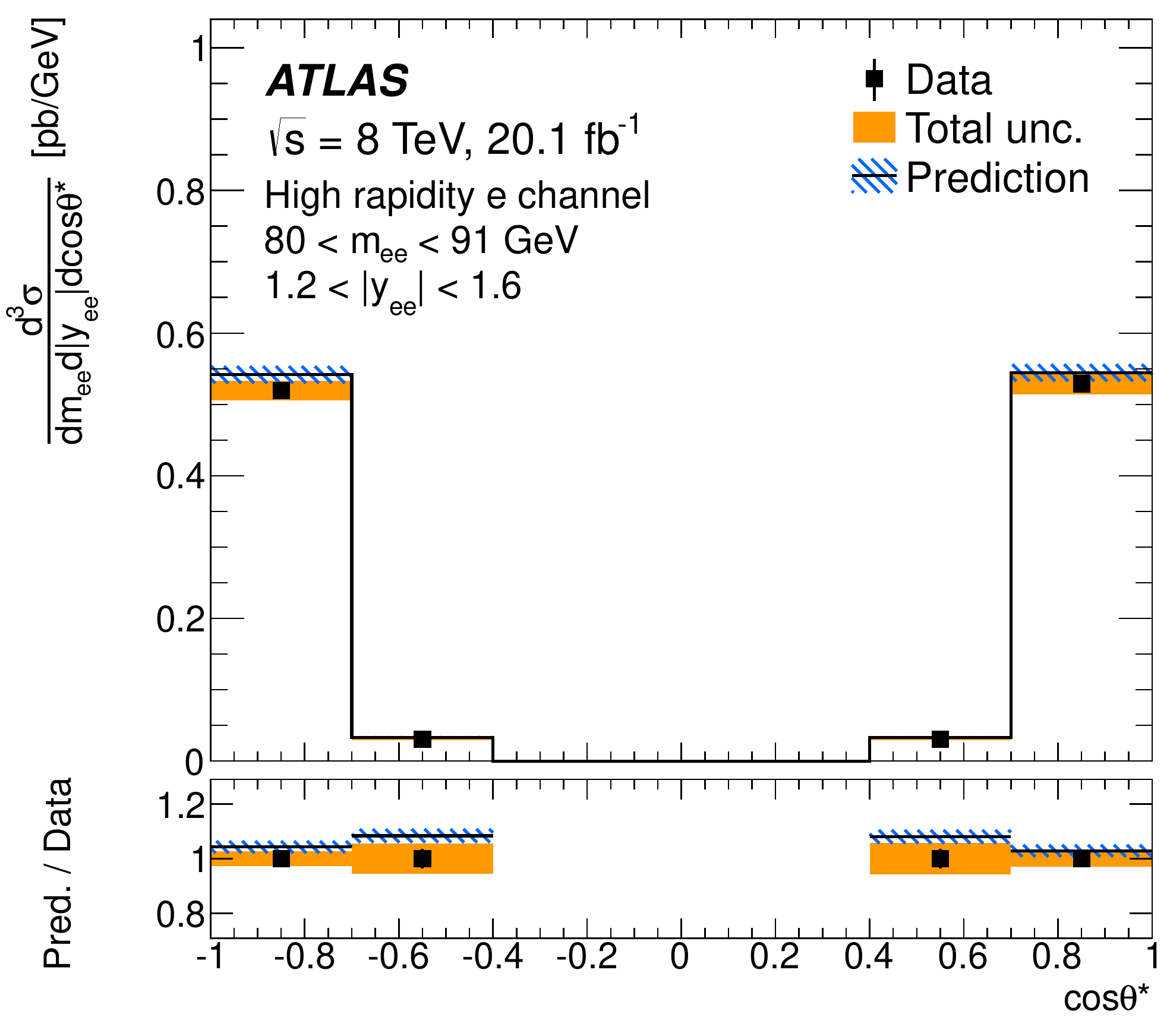}~
\includegraphics[width=0.48\textwidth]{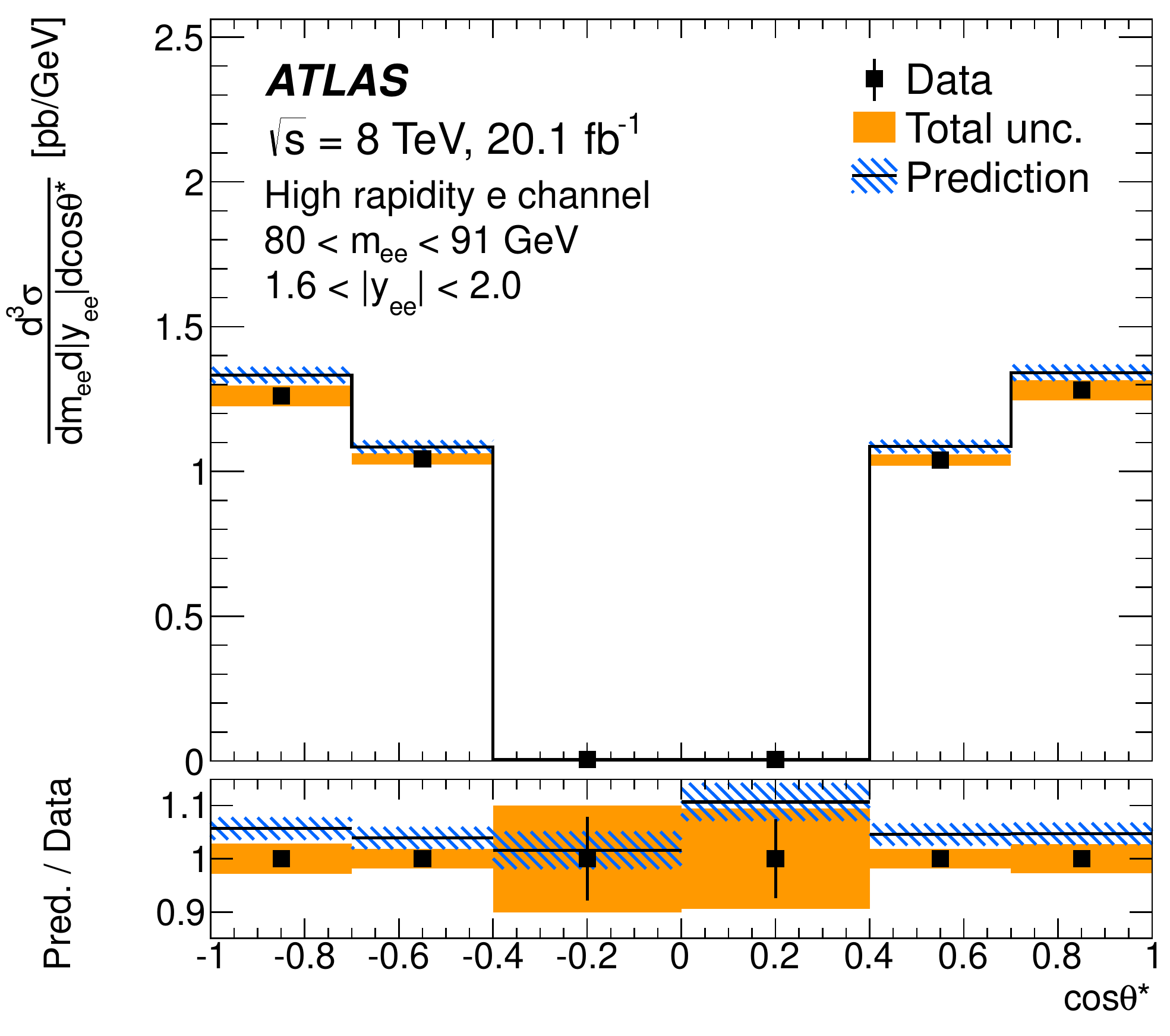}\\
\includegraphics[width=0.48\textwidth]{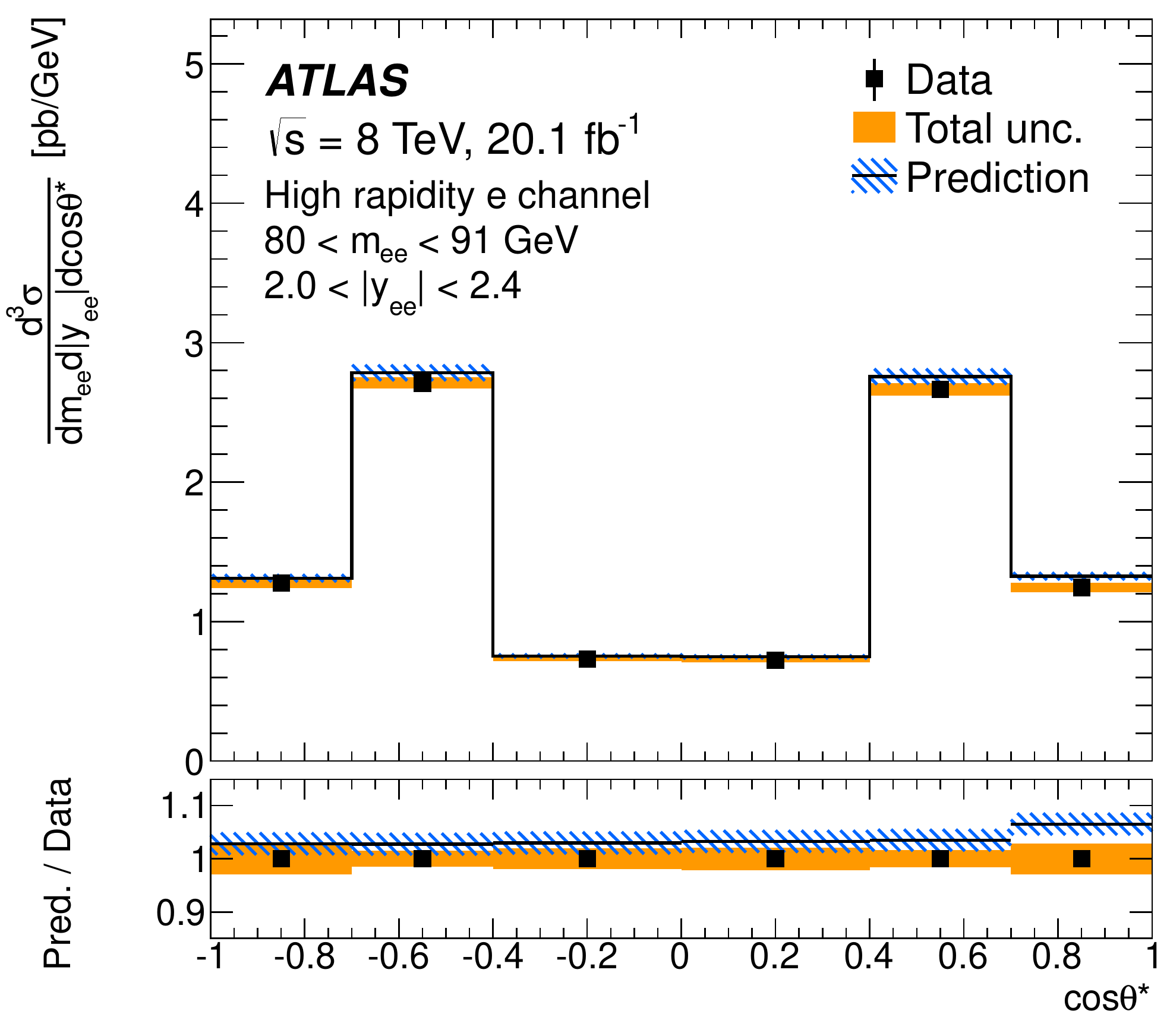}~
\includegraphics[width=0.48\textwidth]{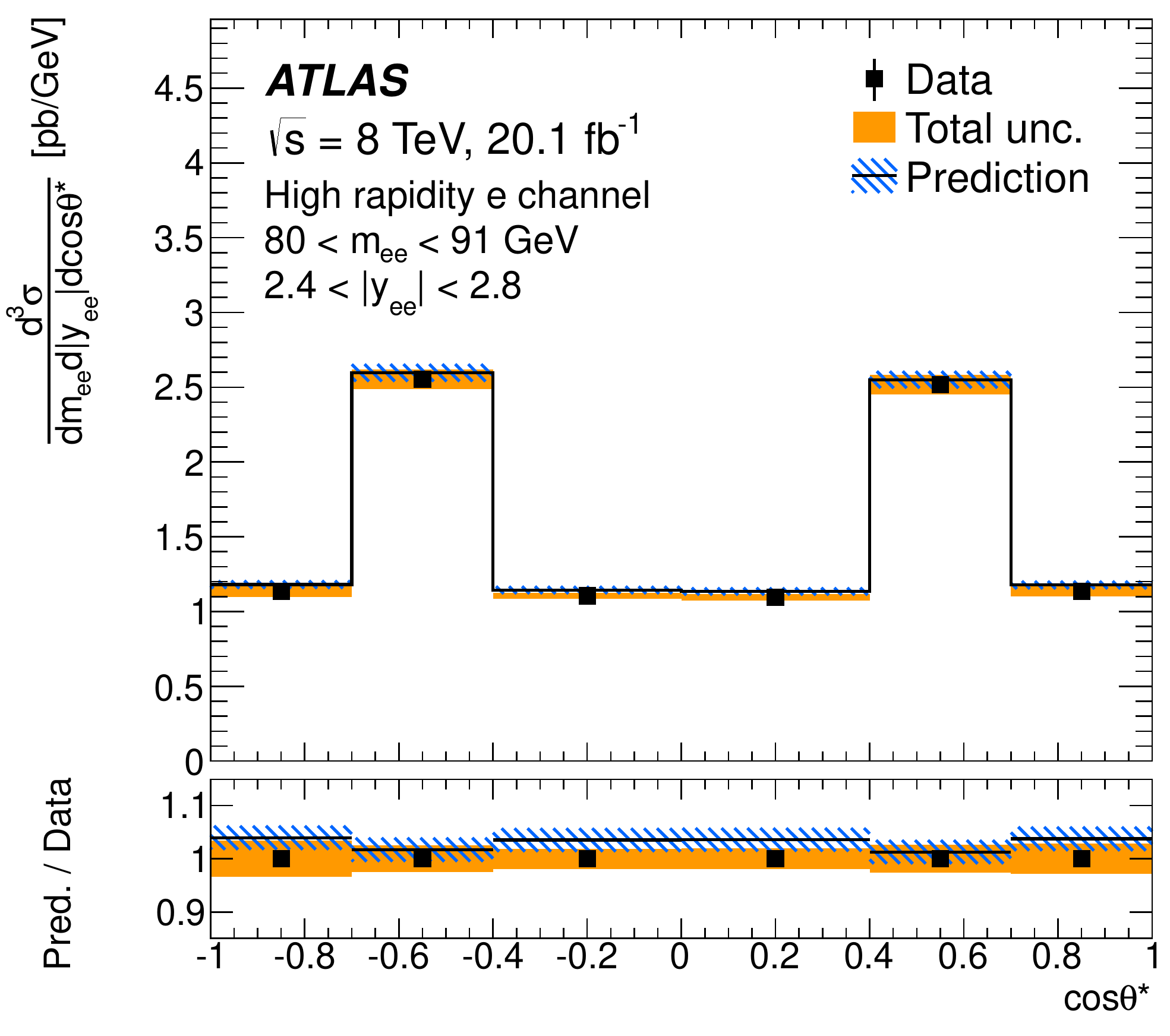}\\
\includegraphics[width=0.48\textwidth]{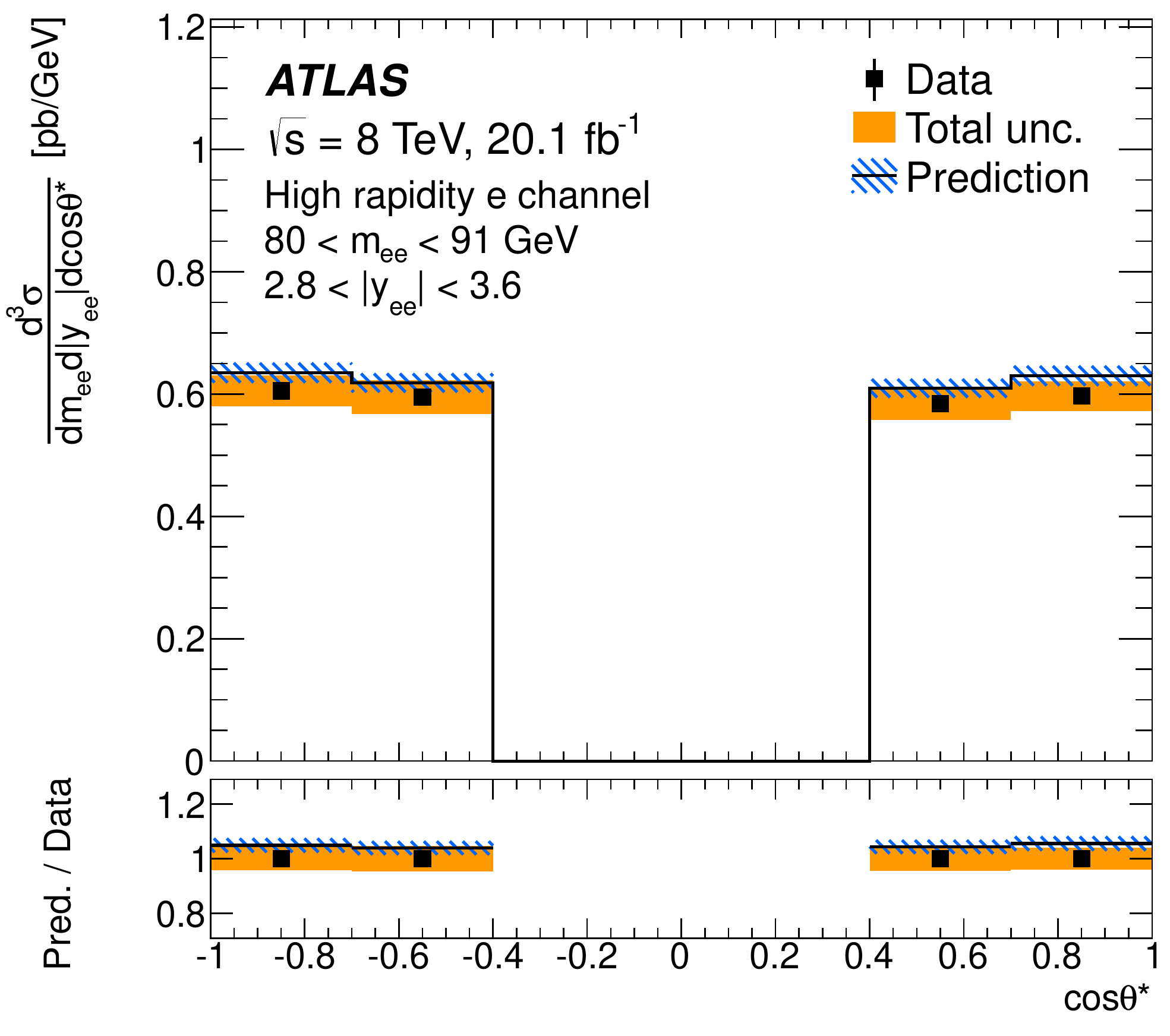}
\caption{The high rapidity electron channel Born-level fiducial cross section
		$\textrm{d}^3\sigma$. The kinematic region shown is labelled in each
		plot. The data are shown as solid markers and the prediction from
		\powheg\ including NNLO QCD and NLO EW $K$-factors is shown as
		the solid line. In each plot, the lower panel shows the ratio of prediction
		to measurement. The inner error bars represent the statistical uncertainty
		of the data and the solid band shows the total experimental uncertainty.
		The contribution from the uncertainty of the luminosity measurement is
		excluded. The hatched band represents the statistical and PDF uncertainties
		in the prediction.}
\label{fig:3D_ZCF_2}
\end{figure}

\begin{figure}[htp!]
\centering
\includegraphics[width=0.48\textwidth]{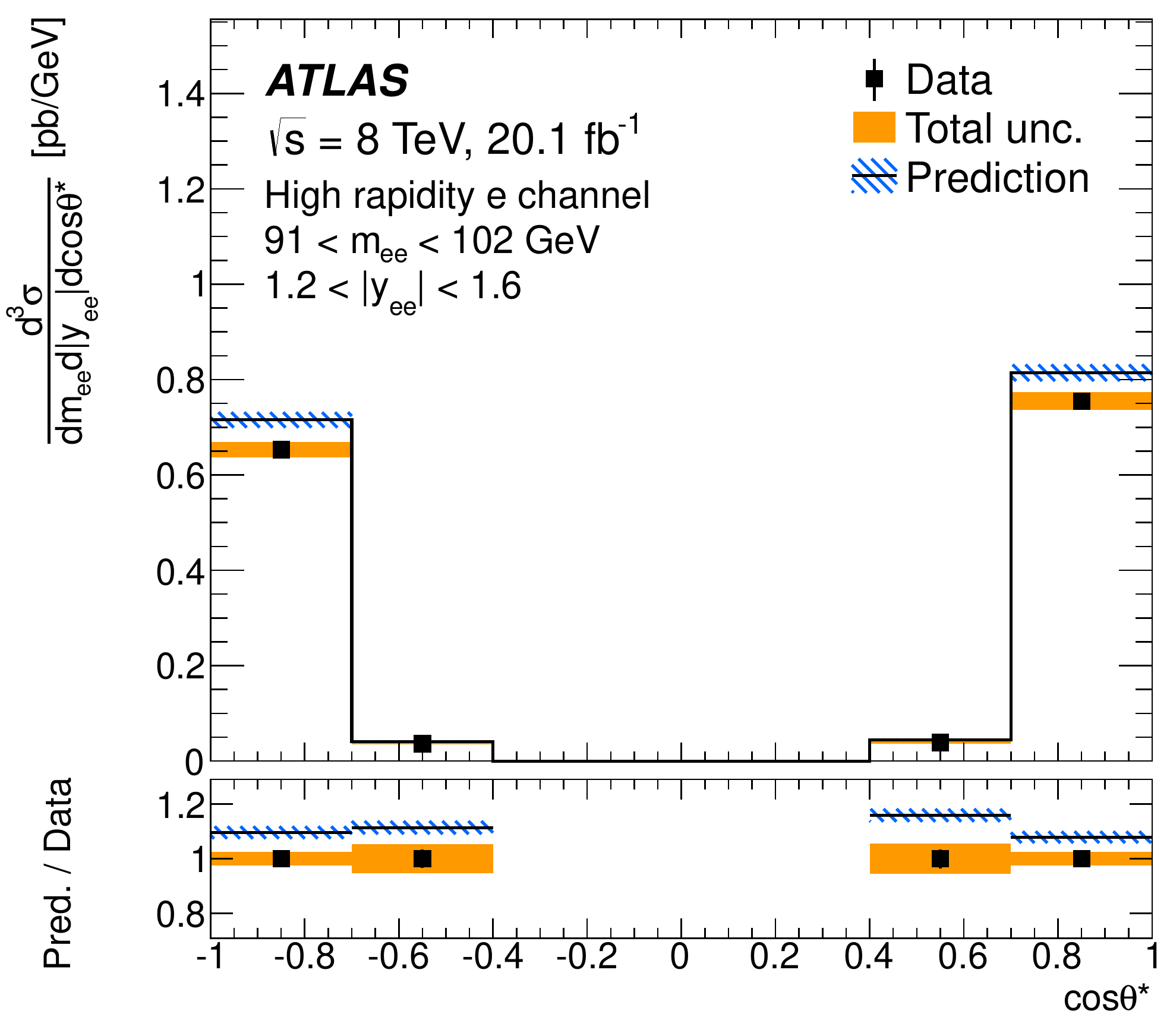}~
\includegraphics[width=0.48\textwidth]{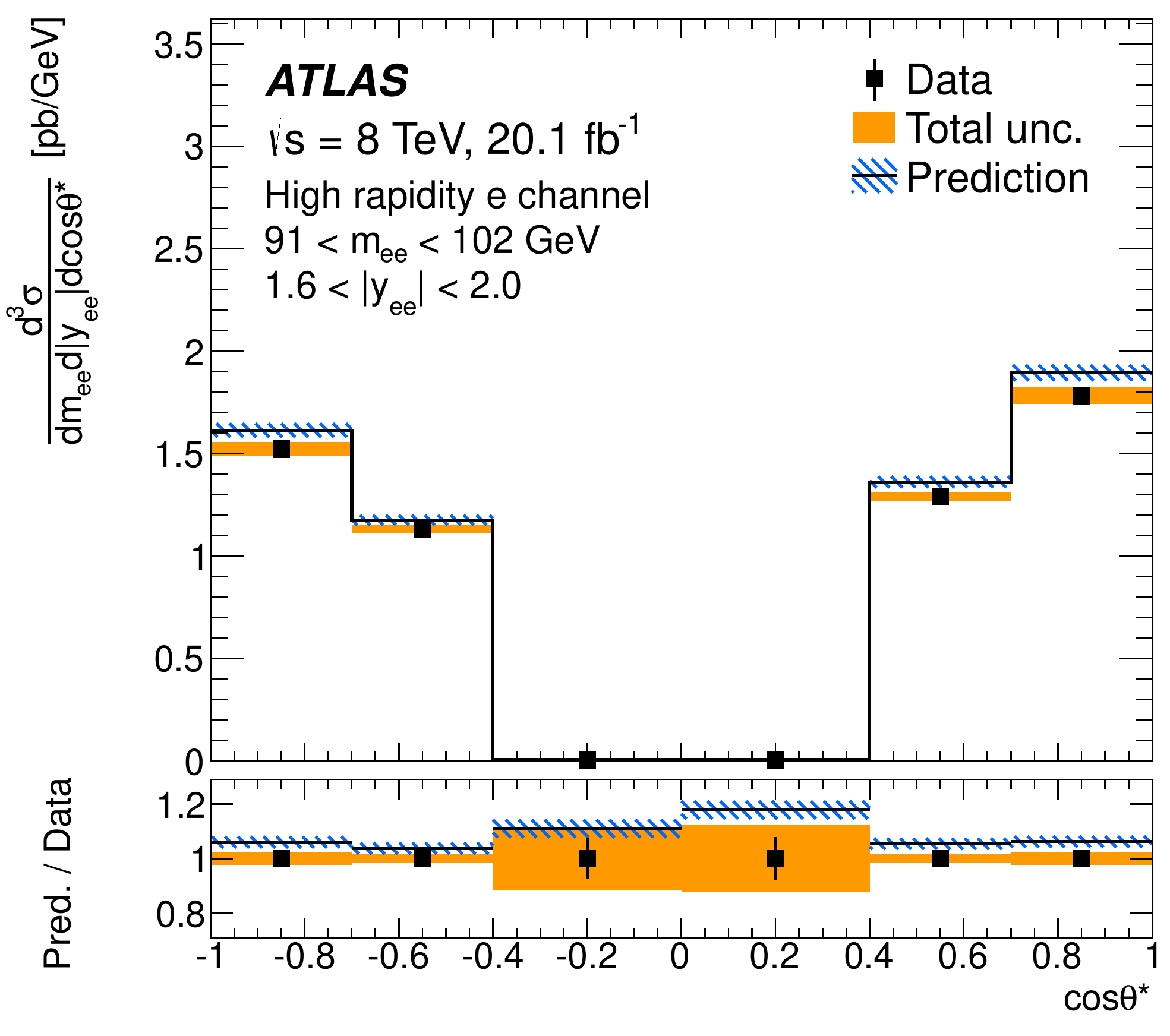}\\
\includegraphics[width=0.48\textwidth]{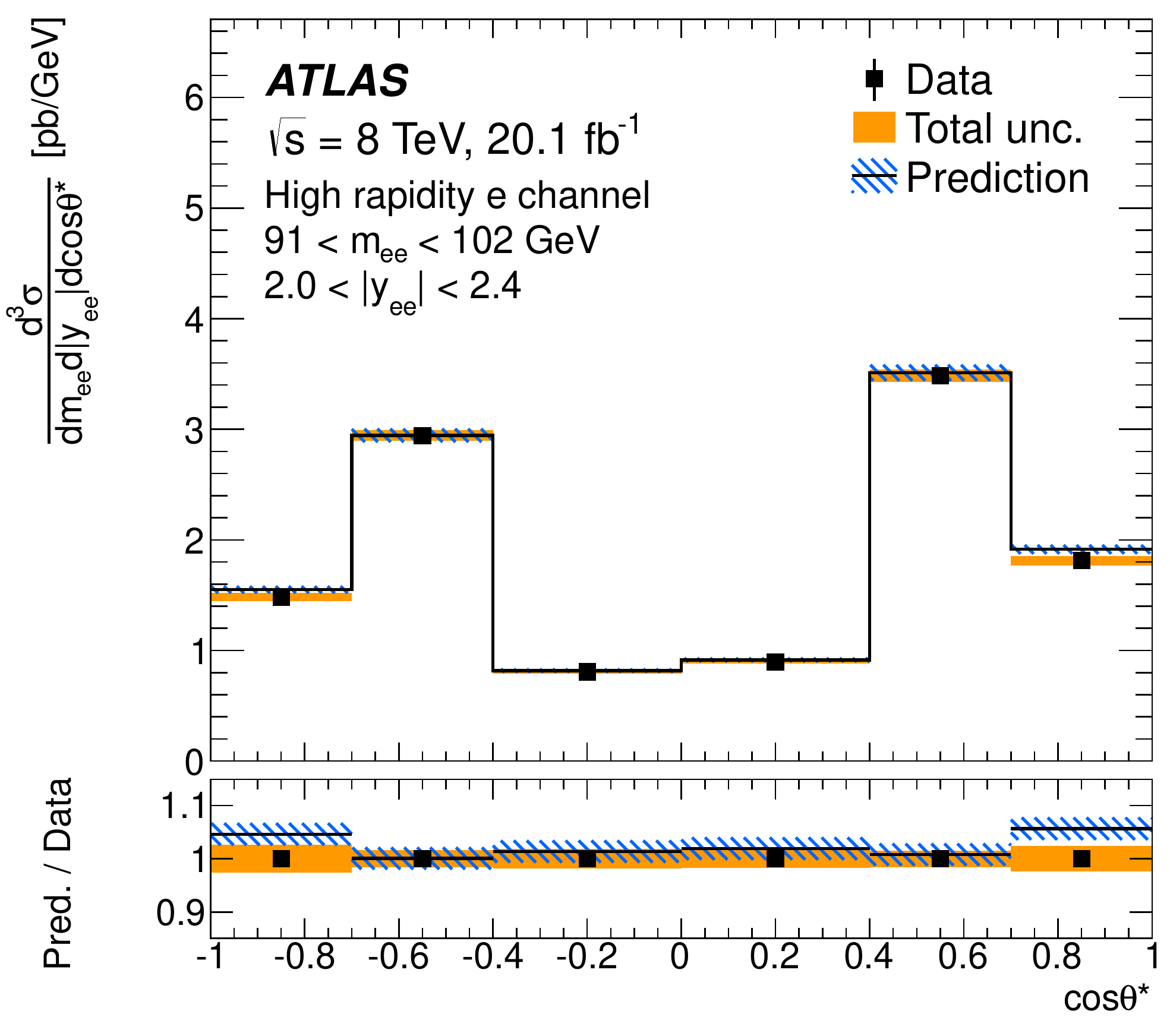}~
\includegraphics[width=0.48\textwidth]{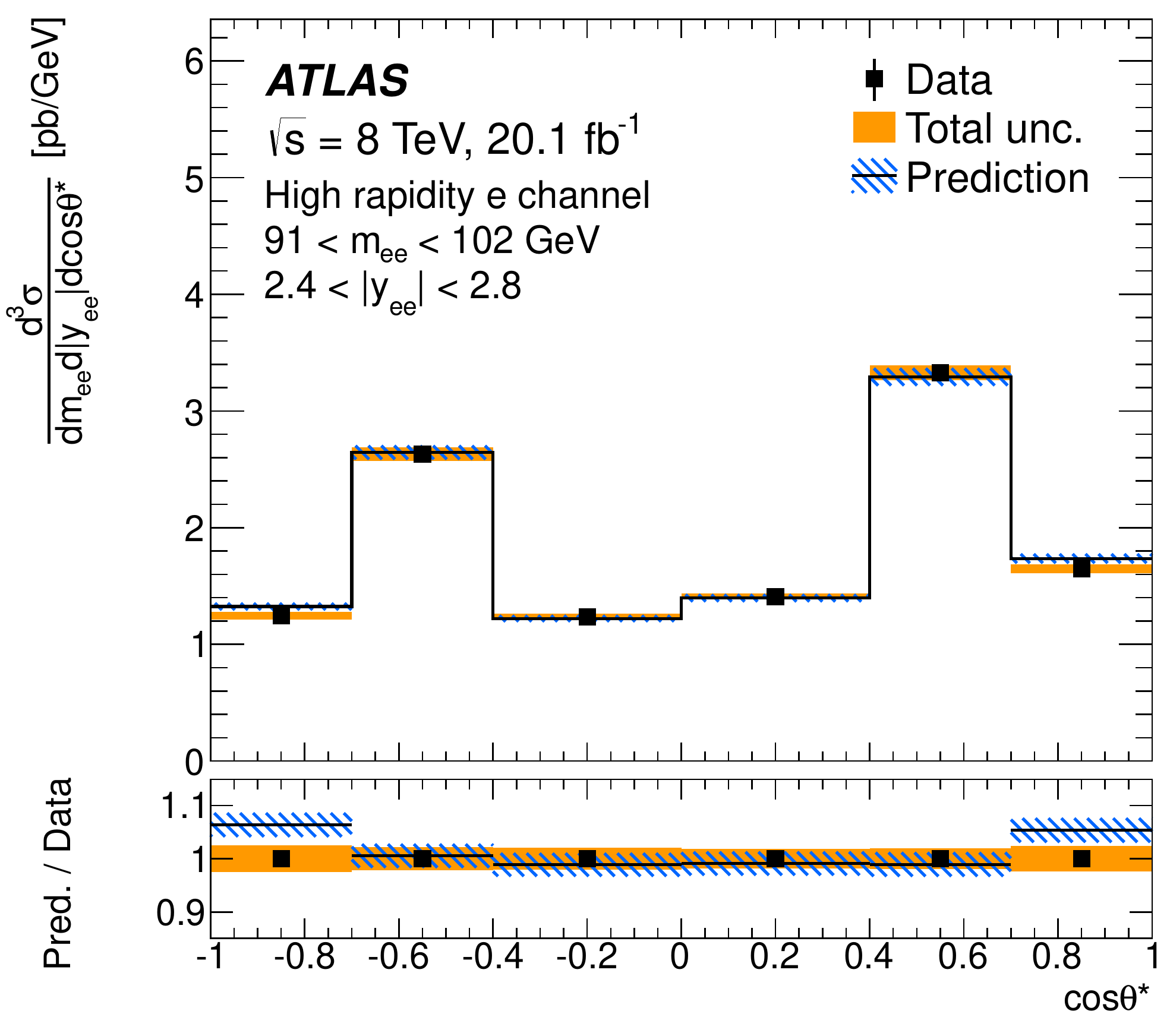}\\
\includegraphics[width=0.48\textwidth]{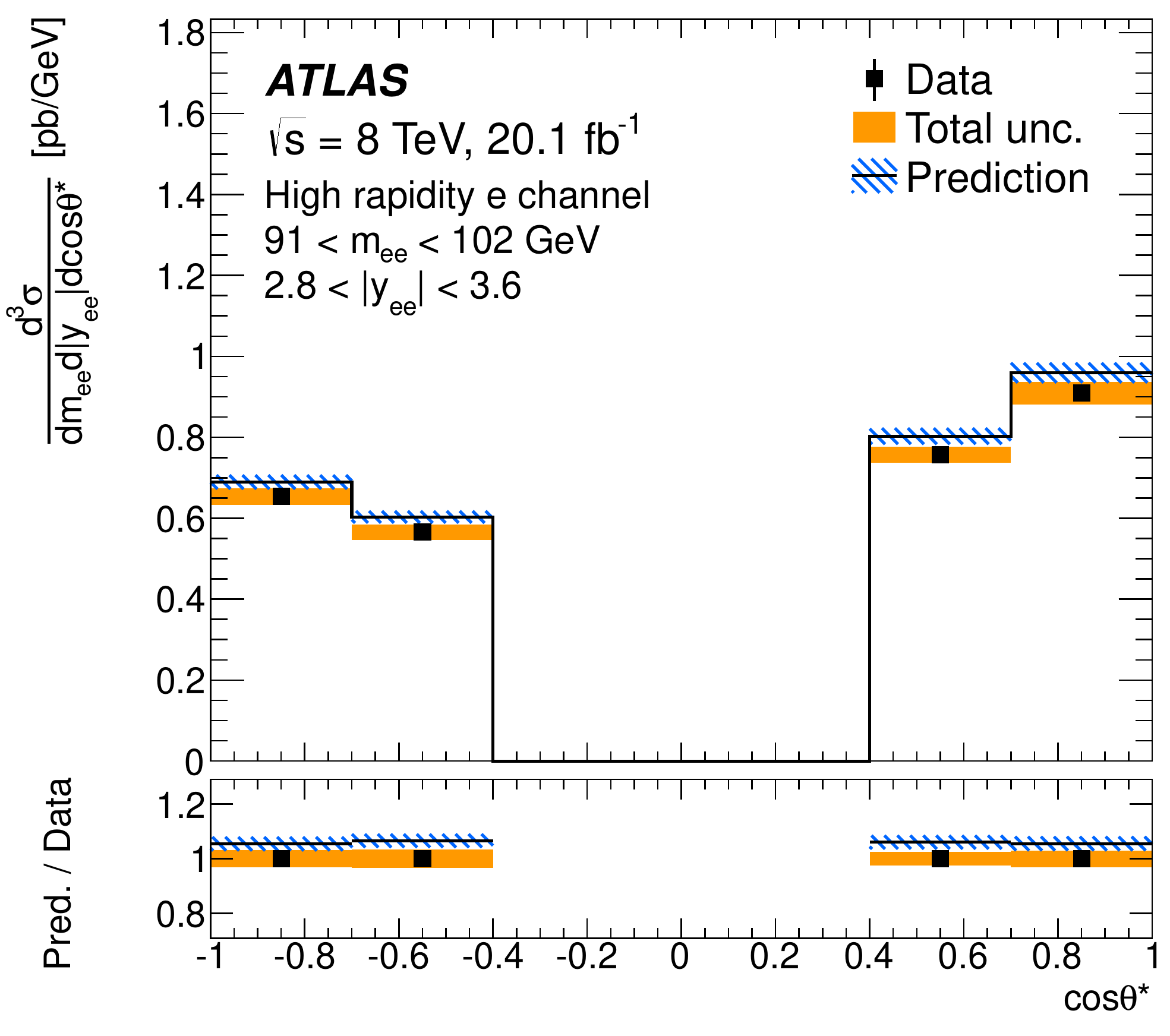}
\caption{The high rapidity electron channel Born-level fiducial cross section
		$\textrm{d}^3\sigma$. The kinematic region shown is labelled in each
		plot. The data are shown as solid markers and the prediction from
		\powheg\ including NNLO QCD and NLO EW $K$-factors is shown as
		the solid line. In each plot, the lower panel shows the ratio of prediction
		to measurement. The inner error bars represent the statistical uncertainty
		of the data and the solid band shows the total experimental uncertainty.
		The contribution from the uncertainty of the luminosity measurement is
		excluded. The hatched band represents the statistical and PDF uncertainties
		in the prediction.}
\label{fig:3D_ZCF_3}
\end{figure}

\begin{figure}[htp!]
\centering
\includegraphics[width=0.48\textwidth]{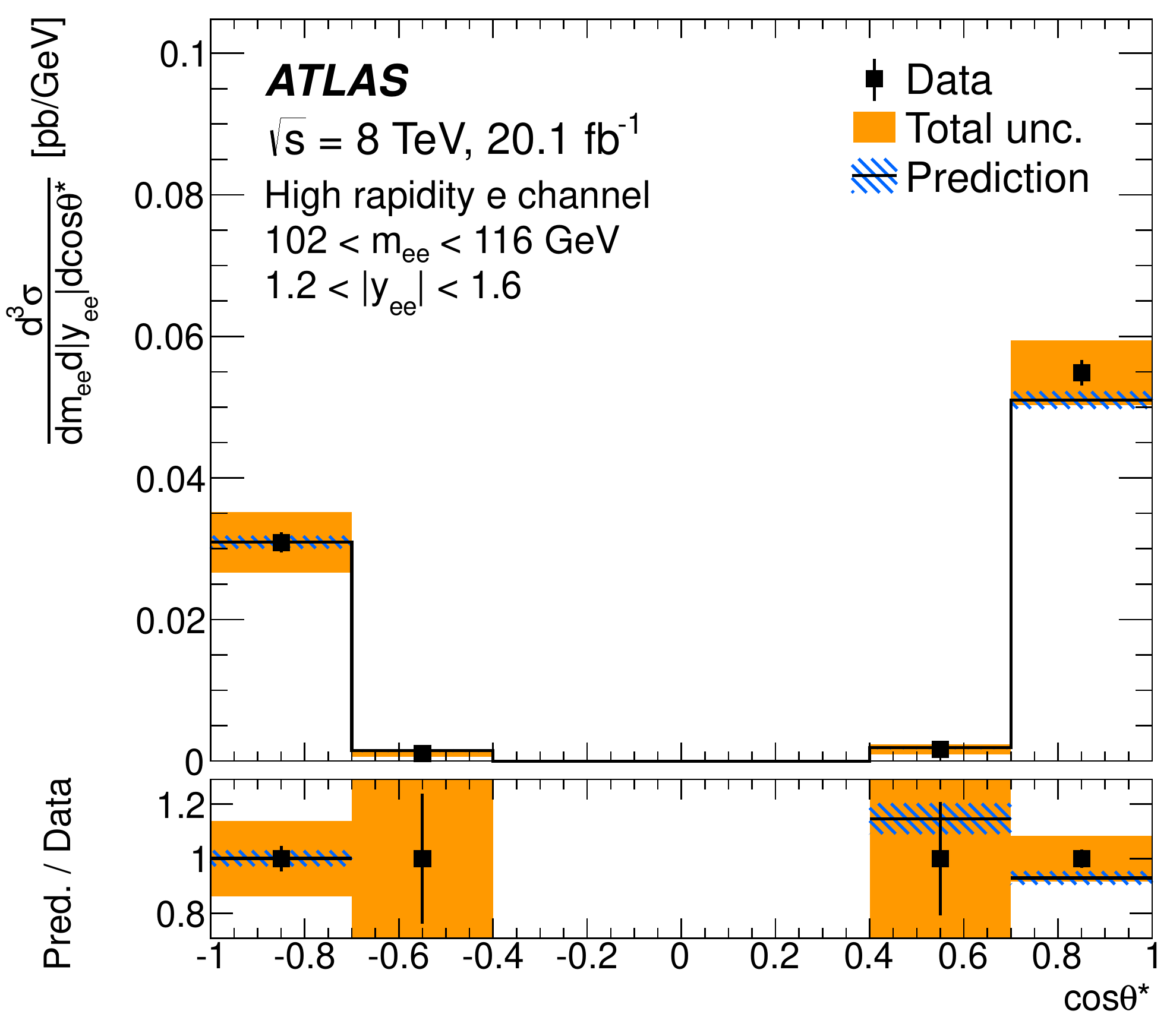}~
\includegraphics[width=0.48\textwidth]{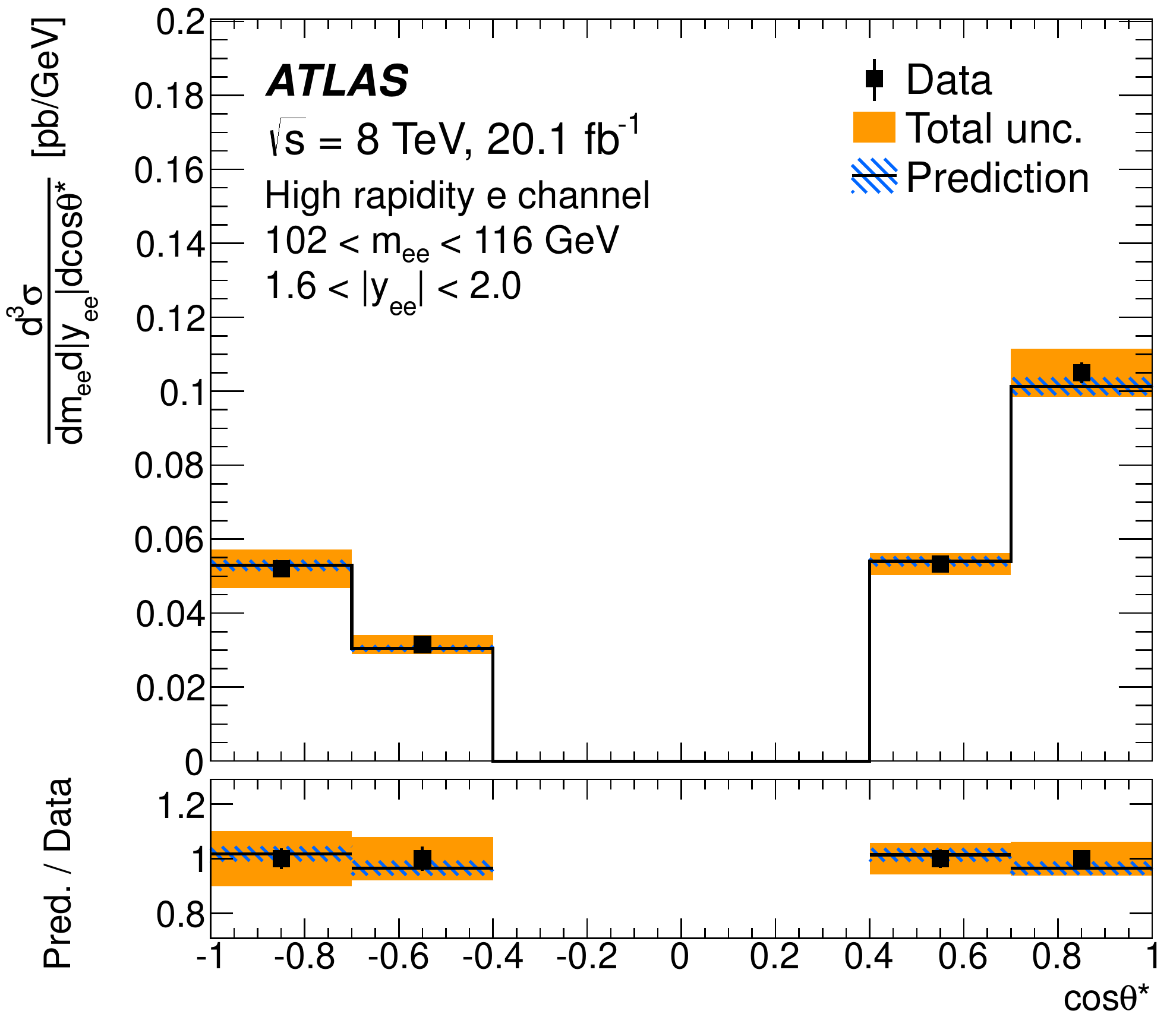}\\
\includegraphics[width=0.48\textwidth]{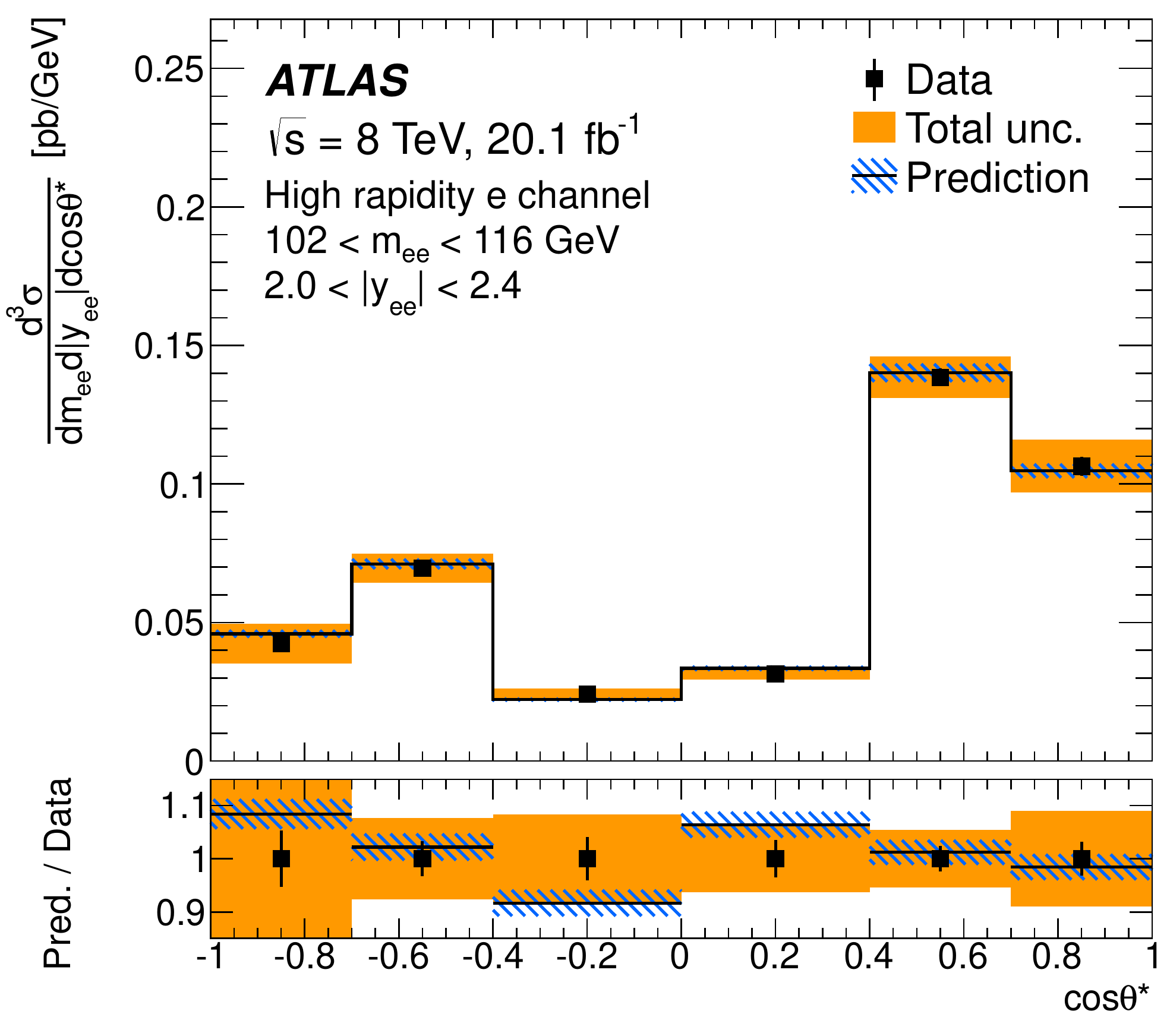}~
\includegraphics[width=0.48\textwidth]{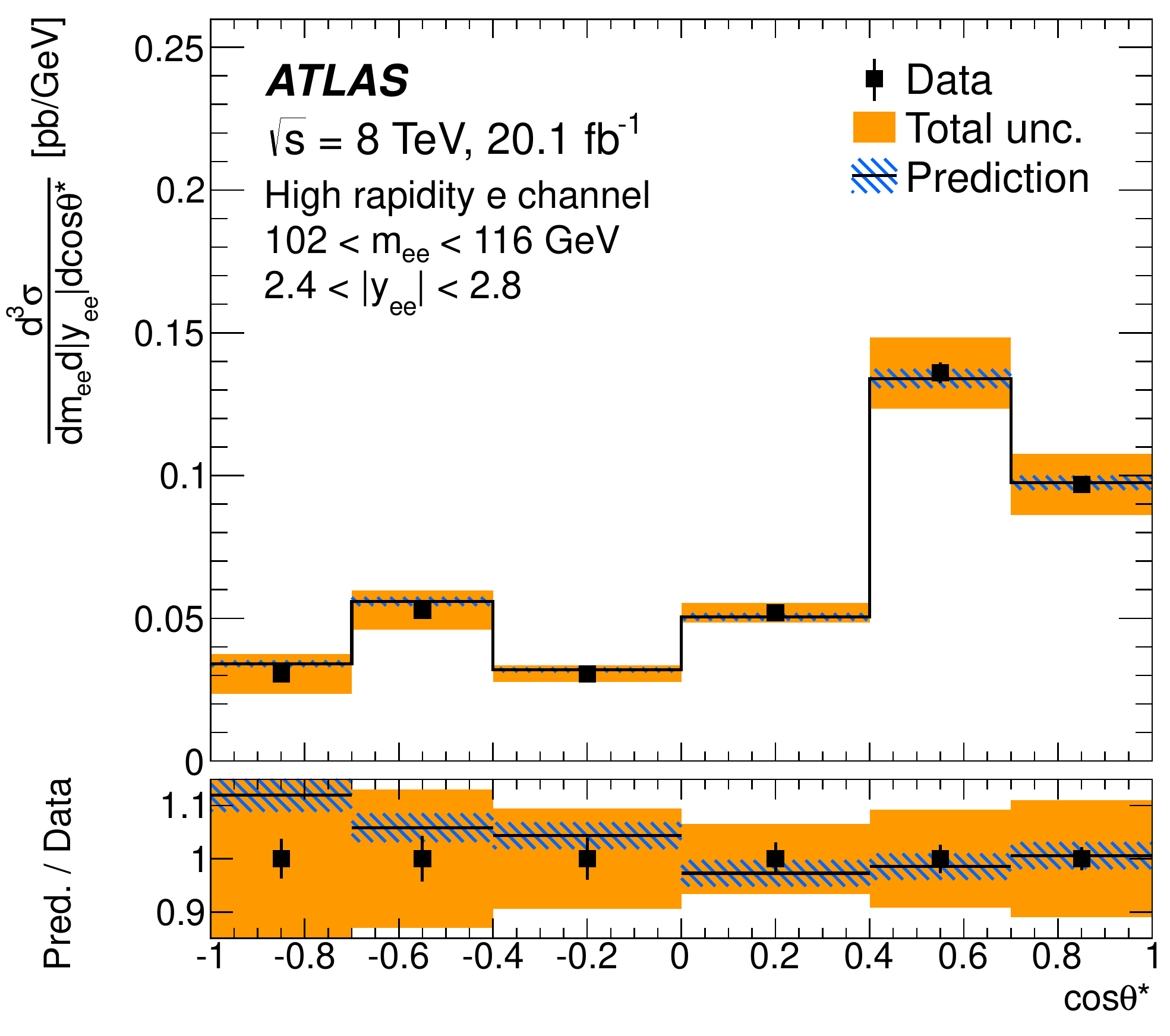}\\
\includegraphics[width=0.48\textwidth]{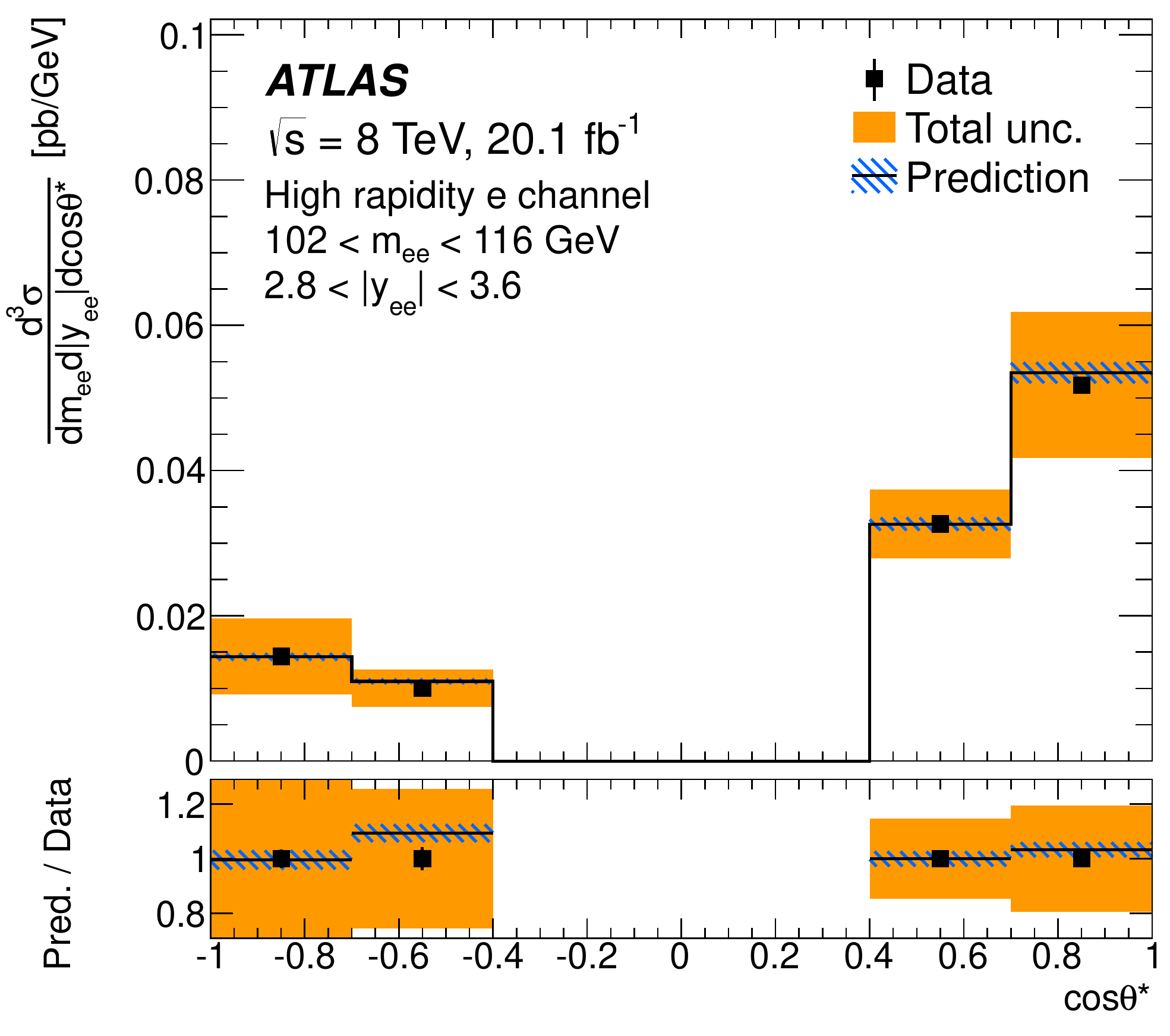}
\caption{The high rapidity electron channel Born-level fiducial cross section
		$\textrm{d}^3\sigma$. The kinematic region shown is labelled in each
		plot. The data are shown as solid markers and the prediction from
		\powheg\ including NNLO QCD and NLO EW $K$-factors is shown as
		the solid line. In each plot, the lower panel shows the ratio of prediction
		to measurement. The inner error bars represent the statistical uncertainty
		of the data and the solid band shows the total experimental uncertainty.
		The contribution from the uncertainty of the luminosity measurement is
		excluded. The hatched band represents the statistical and PDF uncertainties
		in the prediction.}
\label{fig:3D_ZCF_4}
\end{figure}

\begin{figure}[htp!]
\centering
\includegraphics[width=0.48\textwidth]{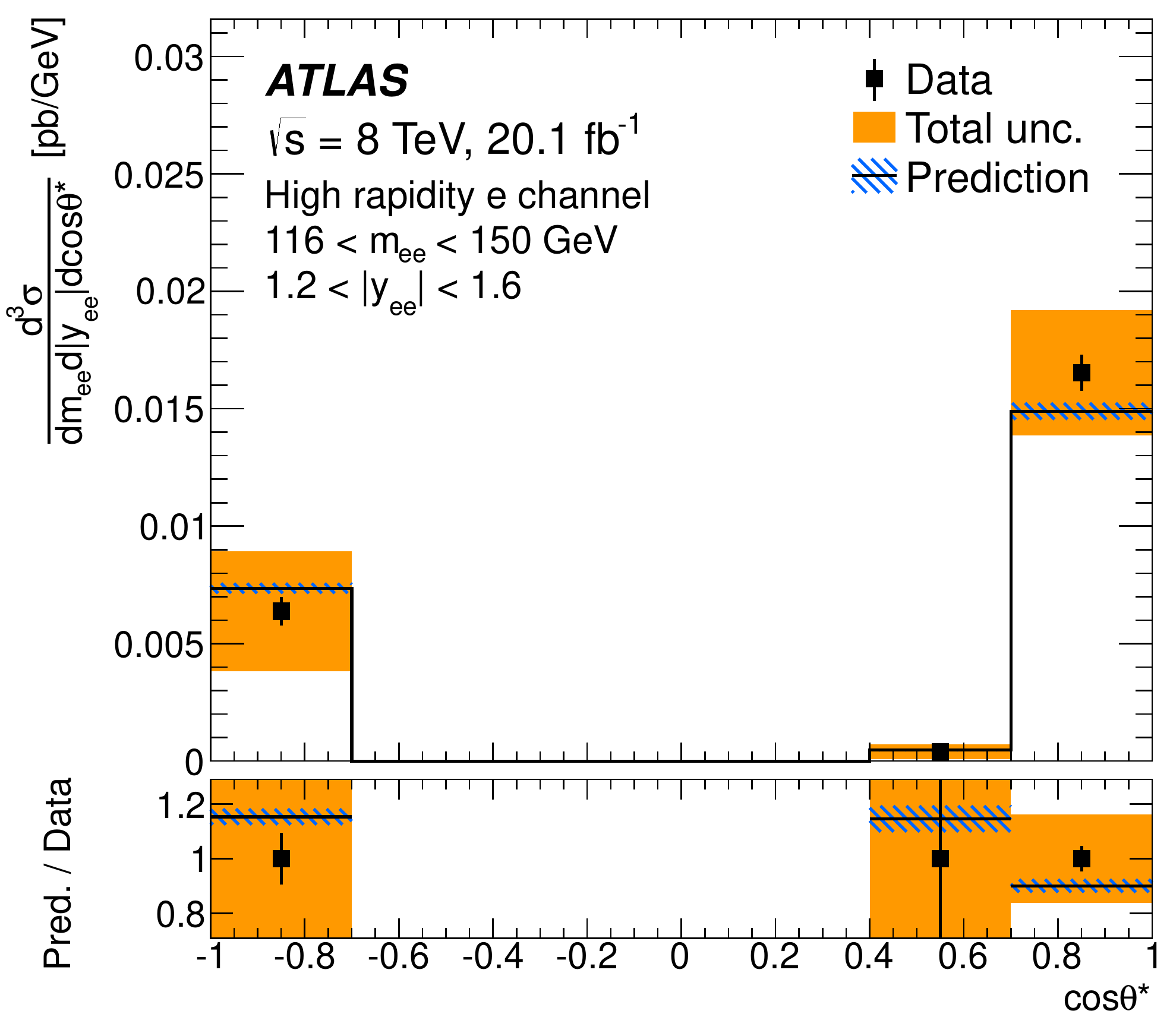}~
\includegraphics[width=0.48\textwidth]{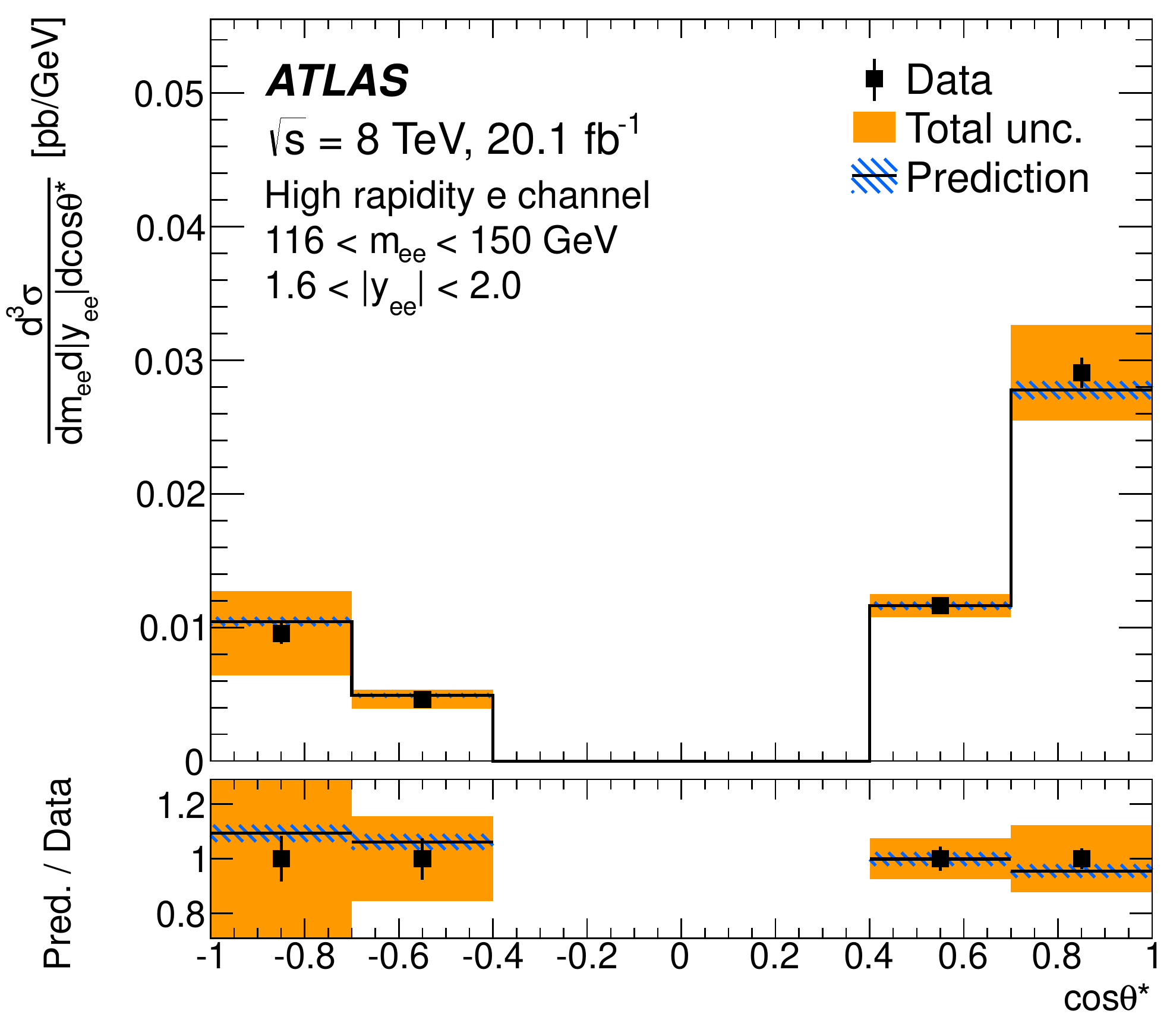}\\
\includegraphics[width=0.48\textwidth]{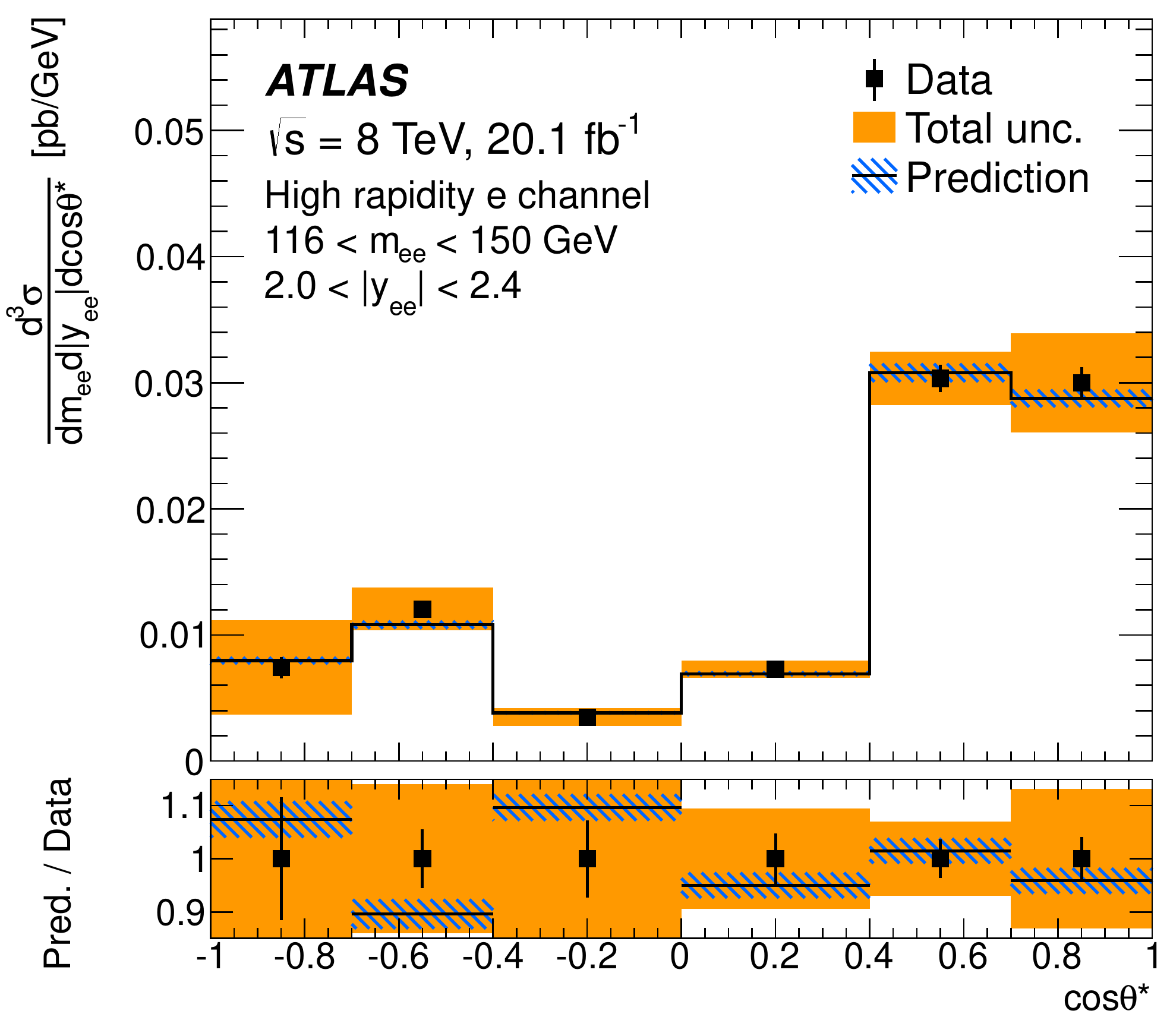}~
\includegraphics[width=0.48\textwidth]{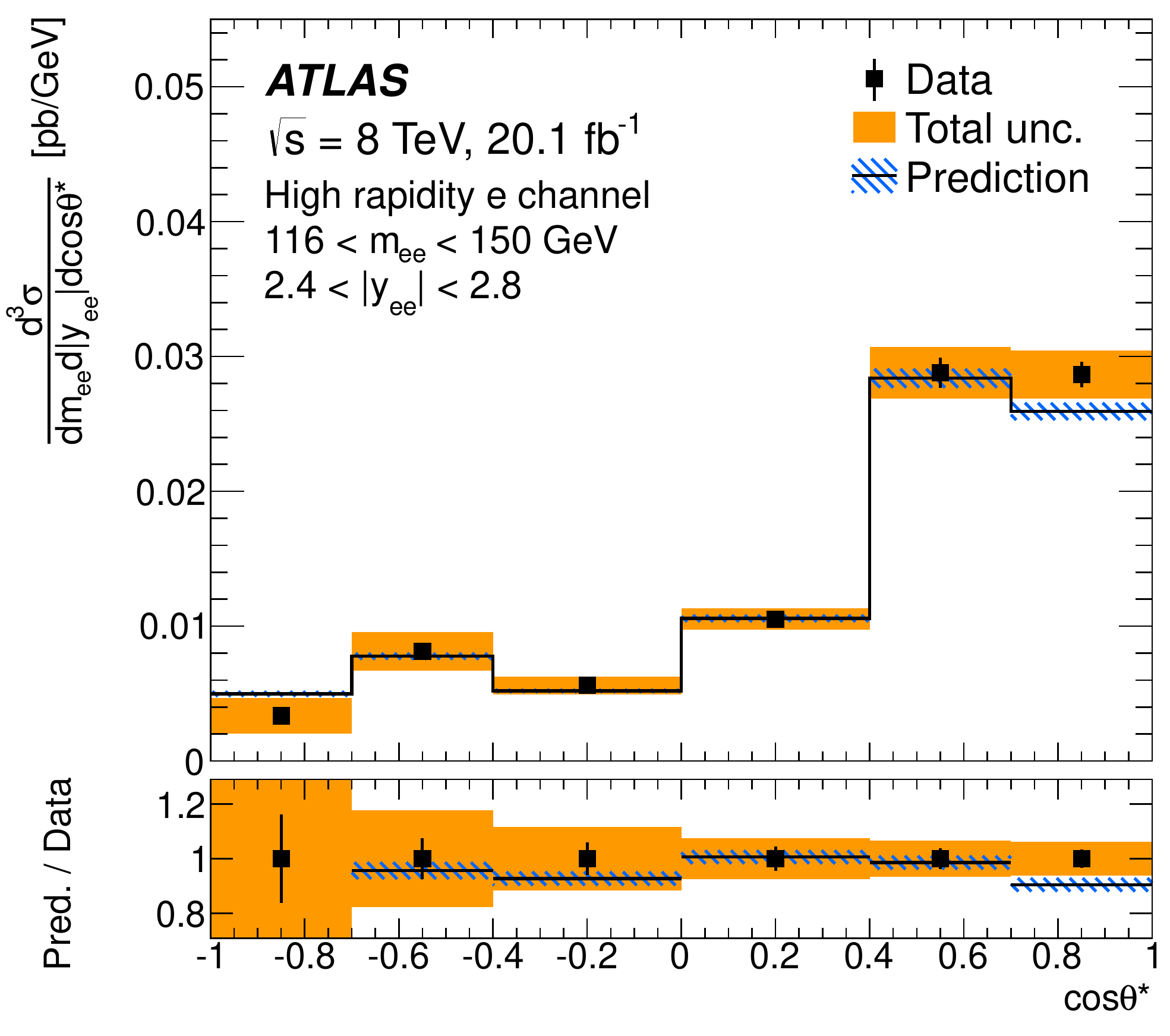}\\
\includegraphics[width=0.48\textwidth]{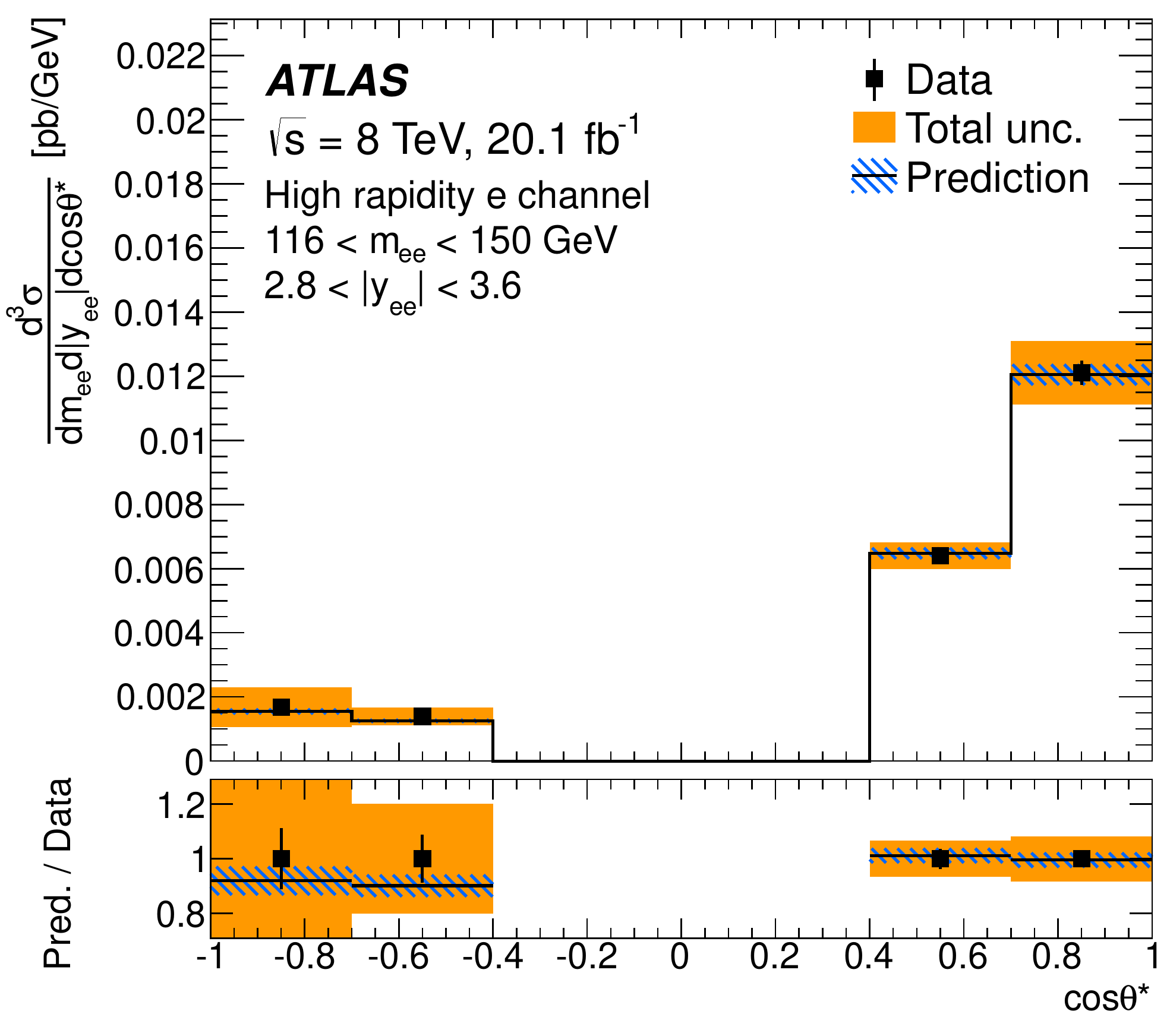}
\caption{The high rapidity electron channel Born-level fiducial cross section
		$\textrm{d}^3\sigma$. The kinematic region shown is labelled in each
		plot. The data are shown as solid markers and the prediction from
		\powheg\ including NNLO QCD and NLO EW $K$-factors is shown as
		the solid line. In each plot, the lower panel shows the ratio of prediction
		to measurement. The inner error bars represent the statistical uncertainty
		of the data and the solid band shows the total experimental uncertainty.
		The contribution from the uncertainty of the luminosity measurement is
		excluded. The hatched band represents the statistical and PDF uncertainties
		in the prediction.}
\label{fig:3D_ZCF_5}
\end{figure}

In figures~\ref{fig:3D_ZCF_1}--\ref{fig:3D_ZCF_5} the measured triple
differential Born-level cross section for the high rapidity electron
channel analysis is presented as a function of $\cos\theta^*$.  In
this channel the region of small $|\cos\theta^*|$ is experimentally
accessible only for moderate values of rapidity, i.e.
$|y_{\ell\ell}|\simeq$~2.0--2.8. Nevertheless the same features of the
cross section are observed: the cross sections are largest for the
region $m_{\ell\ell}\sim m_Z$; an asymmetry in the $\cos\theta^*$
spectrum is observed with larger cross sections at negative
$\cos\theta^*$ for $m_{\ell\ell}<m_Z$, and larger cross sections at
positive $\cos\theta^*$ for $m_{\ell\ell}>m_Z$; the magnitude of the
asymmetry is smallest for $80<m_{\ell\ell}<91$~{\GeV} and increases
with $m_{\ell\ell}$. The triple-differential measurement is listed in
table~\ref{tab:zcf_summary} with its uncertainties. 

\FloatBarrier

\subsection{Forward-backward asymmetry}
\label{afb}
The effect of parity violation in $Z$ boson decays is more
clearly visible in the forward-backward asymmetry, $A_{\textrm{FB}}$,
derived from the cross-section measurements of {\d3s}. The combined
Born-level cross sections are used to determine $A_{\textrm{FB}}$ in
the region $0 < |y_{\ell\ell}|< 2.4$ by summing the measurement
bins for $\cos\theta^*>0$ and for $\cos\theta^*<0$ and calculating the
asymmetry according to equation~(\ref{eq:afb}).

The uncorrelated uncertainty in $A_{\textrm{FB}}$ is determined using
standard error propagation. The correlated uncertainty is determined
for each source in turn by coherently shifting {\d3s} by the
associated correlated uncertainty and calculating the difference to
the nominal value of $A_{\textrm{FB}}$. Finally, the total uncertainty
in $A_{\textrm{FB}}$ is taken as the sum in quadrature of the correlated
and uncorrelated components. The uncertainties in $A_{\textrm{FB}}$
are significantly reduced, especially the correlated uncertainties
such as the electron energy scale and resolution. The total
uncertainty is dominated by the data statistical uncertainty
everywhere. An experimental uncertainty of $1\times 10^{-3}$ is reached
for the combined measurement, and $4\times 10^{-3}$ for the high
rapidity electron channel measurement. 
In the high-precision region of $80<m_{\ell\ell}<102$~{\GeV} the largest
systematic uncertainty contributions are from the MC
sample size (which are a factor two smaller than the data statistical
uncertainty) and the lepton scale contributions, which are an order of
magnitude smaller. At low $m_{\ell\ell}$ the uncorrelated and
statistical contributions from the background sources are also of
comparable size. Summary tables of these
measurements are given in tables~\ref{tab:comb_afb} and
~\ref{tab:zcf_afb} in the appendix.

\begin{figure}[htp!]
\centering
\includegraphics[width=0.95\textwidth]{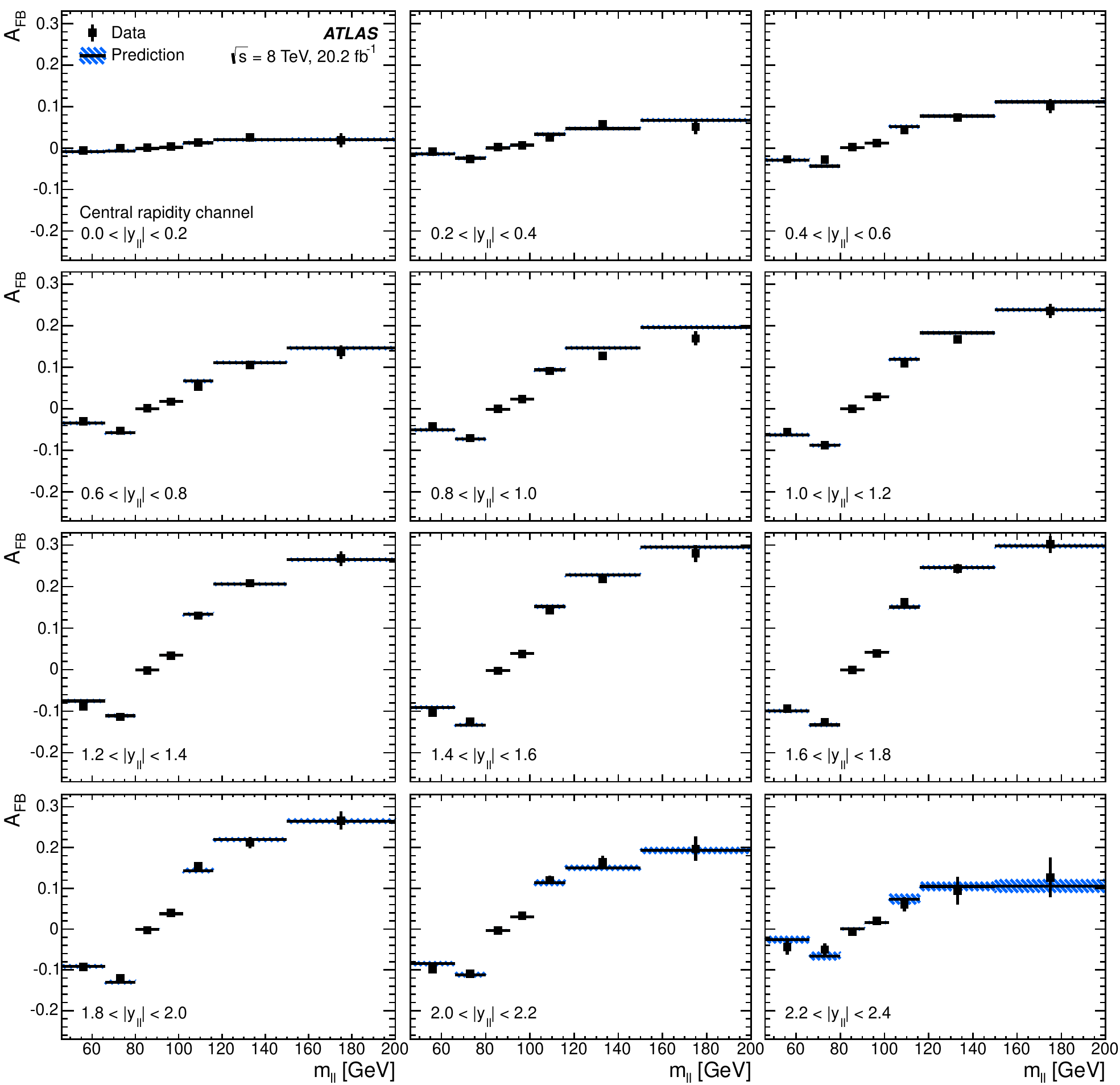}
\caption{Forward-backward asymmetry, $A_{\textrm{FB}}$, determined from the combined Born-level
		fiducial cross section. The kinematic region shown is labelled in
		each plot. The data are shown as solid markers and the error bars
		represent the total experimental uncertainty. The prediction from
		\powheg\ including NNLO QCD and NLO EW $K$-factors is shown
		as the solid line and the hatched band represents the statistical and
		PDF uncertainties in the prediction.}
\label{fig:2D_Combined_Truth_AFB}
\end{figure}

The measurements of $A_{\textrm{FB}}$ are shown in
figure~\ref{fig:2D_Combined_Truth_AFB} for the combined data. The data
are compared to a Born-level prediction from \powheg\ including
$K$-factors for NNLO QCD and NLO EW corrections. The value of
$\sin^2\theta^{\textrm{eff}}_{\textrm{lept}}$ used in the simulation
is 0.23113~\cite{Olive:2016xmw}. The measured asymmetry is found to
generally increase with $m_{\ell\ell}$ from a negative to a positive
asymmetry which is close to zero near
$m_{\ell\ell}=m_Z$. The magnitude of $A_{\textrm{FB}}$ is smallest for
$|y_{\ell\ell}|=0$ and increases to a maximum in the region
$1.0<|y_{\ell\ell}|<2.0$, before decreasing at larger rapidity. This is
expected from the effect of dilution and the unknown direction of the
incident $q$ on an event-by-event basis. At larger $|y_{\ell\ell}|$,
and hence larger $x$, the influence of the higher-momentum valence $u$-
and $d$-quarks becomes increasingly apparent through the longitudinal
boost in the valence direction. This allows a correct determination of
the $q$ direction to be made on average and is well modelled by the
\powheg\ prediction. At even larger $|y_{\ell\ell}|$ in the combined
measurements the maximum of $|A_{\textrm{FB}}|$ decreases again due to
the limited acceptance of the detector in $\eta^{e,\mu}$.

\begin{figure}[p!]
\centering
\includegraphics[width=0.95\textwidth]{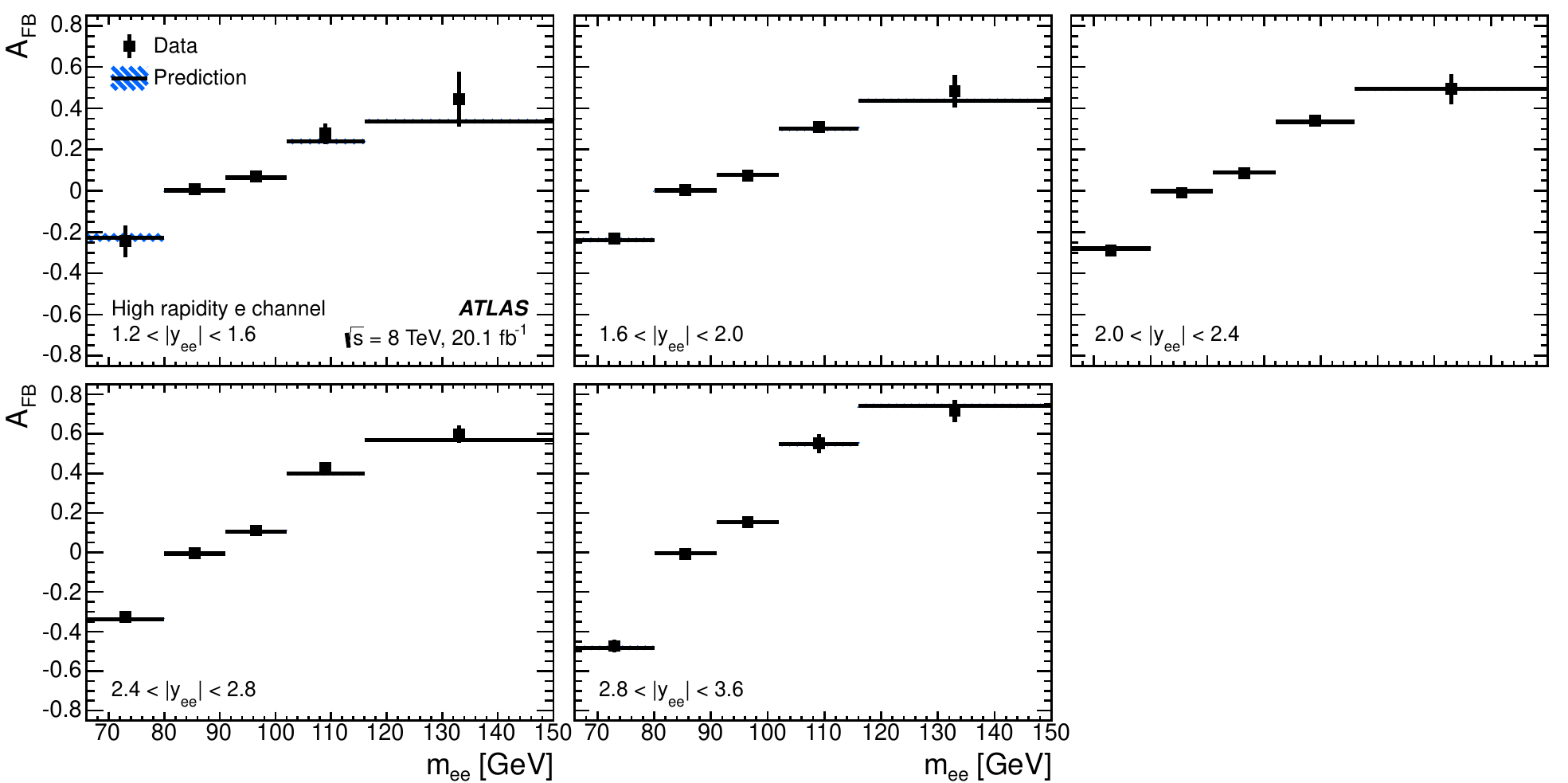}
\caption{Forward-backward asymmetry, $A_{\textrm{FB}}$, determined from the high rapidity electron Born-level
		fiducial cross section. The kinematic region shown is labelled in
		each plot. The data are shown as solid markers and the error bars
		represent the total experimental uncertainty. The prediction from
		\powheg\ including NNLO QCD and NLO EW $K$-factors is shown
		as the solid line and the hatched band represents the statistical and
		PDF uncertainties in the prediction.}
\label{fig:2D_ZCF_Truth_AFB}
\end{figure}

The measurements of $A_{\textrm{FB}}$ in the high rapidity electron
channel analysis, which is expected to be more sensitive to the asymmetry,
are presented in figure~\ref{fig:2D_ZCF_Truth_AFB}.
Qualitatively, the asymmetry shows behaviour similar to that seen in
the combined measurement: the asymmetry increases with $m_{ee}$
and values of $|A_{\textrm{FB}}|$ reaching 0.7 are observed at the
highest $|y_{ee}|$ where the influence of dilution is smallest. As was
the case in the combined measurement, the high rapidity $A_{\textrm{FB}}$
measurement is well-described by the \powheg\ prediction.

\FloatBarrier

\FloatBarrier

\section{Conclusion}
\label{sec:conclusion}

The triple-differential Drell--Yan production cross section
${\textrm{d}}^3\sigma/{\textrm{d}}m_{\ell\ell}{\textrm{d}}|y_{\ell\ell}|{\textrm{d}}\cos\theta^*$
is measured in the range $46<m_{\ell\ell}<200$~{\GeV} and $|y_{\ell\ell}|<2.4$
for electron and muon pairs. The measurements are extended to high rapidity
in the electron channel up to $|y_{ee}|=3.6$ in the mass range $66<m_{\ell\ell}<150$~{\GeV}.
The analysis uses $20.2$~fb$^{-1}$ of $pp$ collision data at $\sqrt{s}=8$~{\TeV}
collected in 2012 by the ATLAS detector at the LHC. The central rapidity measurement
channels are combined taking into account the systematic uncertainty correlations. Their
combination achieves an experimental precision of better than $0.5\%$, excluding the
overall uncertainty in the luminosity measurement of $1.9\%$.

The combined cross sections are integrated to produce the single- and
double-differential cross sections ${\textrm{d}}\sigma/{\textrm{d}}m_{\ell\ell}$
and ${\textrm{d}}^2\sigma/{\textrm{d}}m_{\ell\ell}{\textrm{d}}|y_{\ell\ell}|$.
The fiducial cross sections are compared to a theoretical prediction calculated
using \powheg\ at NLO with matched leading-logarithm parton showers. The
calculation is approximately corrected for NNLO QCD effects and for additional
higher-order electroweak effects applied as a function of $m_{\ell\ell}$. The
single- and double-differential measurements are well described by the prediction.
Having applied corrections to the scattering amplitude coefficients in \powheg\,
the prediction also provides a good description of the triple-differential
measurements.

The measured cross sections are used to determine the forward-backward
asymmetry $A_{\textrm{FB}}$ as a function of dilepton invariant mass and
rapidity. The \powheg\ predictions enhanced with NNLO QCD and NLO EW
$K$-factors describe the observed behaviour of $A_{\textrm{FB}}$ well. 
\FloatBarrier

\clearpage
\section*{Acknowledgements}

We thank CERN for the very successful operation of the LHC, as well as the
support staff from our institutions without whom ATLAS could not be
operated efficiently.

We acknowledge the support of ANPCyT, Argentina; YerPhI, Armenia; ARC, Australia; BMWFW and FWF, Austria; ANAS, Azerbaijan; SSTC, Belarus; CNPq and FAPESP, Brazil; NSERC, NRC and CFI, Canada; CERN; CONICYT, Chile; CAS, MOST and NSFC, China; COLCIENCIAS, Colombia; MSMT CR, MPO CR and VSC CR, Czech Republic; DNRF and DNSRC, Denmark; IN2P3-CNRS, CEA-DRF/IRFU, France; SRNSF, Georgia; BMBF, HGF, and MPG, Germany; GSRT, Greece; RGC, Hong Kong SAR, China; ISF, I-CORE and Benoziyo Center, Israel; INFN, Italy; MEXT and JSPS, Japan; CNRST, Morocco; NWO, Netherlands; RCN, Norway; MNiSW and NCN, Poland; FCT, Portugal; MNE/IFA, Romania; MES of Russia and NRC KI, Russian Federation; JINR; MESTD, Serbia; MSSR, Slovakia; ARRS and MIZ\v{S}, Slovenia; DST/NRF, South Africa; MINECO, Spain; SRC and Wallenberg Foundation, Sweden; SERI, SNSF and Cantons of Bern and Geneva, Switzerland; MOST, Taiwan; TAEK, Turkey; STFC, United Kingdom; DOE and NSF, United States of America. In addition, individual groups and members have received support from BCKDF, the Canada Council, CANARIE, CRC, Compute Canada, FQRNT, and the Ontario Innovation Trust, Canada; EPLANET, ERC, ERDF, FP7, Horizon 2020 and Marie Sk{\l}odowska-Curie Actions, European Union; Investissements d'Avenir Labex and Idex, ANR, R{\'e}gion Auvergne and Fondation Partager le Savoir, France; DFG and AvH Foundation, Germany; Herakleitos, Thales and Aristeia programmes co-financed by EU-ESF and the Greek NSRF; BSF, GIF and Minerva, Israel; BRF, Norway; CERCA Programme Generalitat de Catalunya, Generalitat Valenciana, Spain; the Royal Society and Leverhulme Trust, United Kingdom.

The crucial computing support from all WLCG partners is acknowledged gratefully, in particular from CERN, the ATLAS Tier-1 facilities at TRIUMF (Canada), NDGF (Denmark, Norway, Sweden), CC-IN2P3 (France), KIT/GridKA (Germany), INFN-CNAF (Italy), NL-T1 (Netherlands), PIC (Spain), ASGC (Taiwan), RAL (UK) and BNL (USA), the Tier-2 facilities worldwide and large non-WLCG resource providers. Major contributors of computing resources are listed in reference~\cite{ATL-GEN-PUB-2016-002}.

\FloatBarrier

\clearpage
\appendix
\section*{Data Tables}
\addcontentsline{toc}{part}{Data Tables}
\renewcommand{\thesubsection}{\Alph{subsection}}
Summary tables of
${\textrm{d}}^3\sigma/{\textrm{d}}m_{\ell\ell}{\textrm{d}}|y_{\ell\ell}|{\textrm{d}}\cos\theta^*$
cross sections and $A_{\textrm{FB}}$ are given in this appendix.
Tables containing the complete breakdown of systematic
uncertainties are available in HEPData~\cite{Hepdata,Maguire:2017ypu}.

\subsection{Integrated cross-section tables}
\label{tab:integ}
\begin{table}[htp!]
\scriptsize
\centering
\scalebox{1}{
\setlength\tabcolsep{3pt}
\setlength\extrarowheight{2pt}
\begin{tabular}{C{1.5cm}C{2cm}C{1.1cm}C{1.1cm}C{1.1cm}C{1.1cm}}
\toprule
$m_{\ell\ell}$
& $\mathrm{d}\sigma/\mathrm{d}m_{\ell\ell}$
& $\delta^{\textrm {stat}}$
& $\delta^{\textrm {syst}}_{\textrm {unc}}$
& $\delta^{\textrm {syst}}_{\textrm {cor}}$
& $\delta^{\textrm {total}}$ \\
$[$GeV$]$ & [pb/GeV] & [\%] & [\%] & [\%] & [\%] \\
\noalign{\vskip 0.05cm}
\cmidrule(lr){1-6}
  $46, 66$ &     $7.61\times 10^{-1}$ &    $0.2$ &    $0.1$ &    $0.9$ &    $0.9$ \\
  $66, 80$ &     $1.13$ &    $0.1$ &    $0.1$ &    $0.4$ &    $0.4$ \\
  $80, 91$ &     $21.4$ &    $0.0$ &    $0.0$ &    $0.2$ &    $0.2$ \\
  $\phantom{1}91, 102$ &     $25.0$ &    $0.0$ &    $0.0$ &    $0.2$ &    $0.2$ \\
  $102, 116$ &     $8.25\times 10^{-1}$ &    $0.2$ &    $0.1$ &    $0.4$ &    $0.4$ \\
  $116, 150$ &     $1.64\times 10^{-1}$ &    $0.3$ &    $0.1$ &    $0.7$ &    $0.7$ \\
  $150, 200$ &     $3.66\times 10^{-2}$ &    $0.5$ &    $0.2$ &    $1.3$ &    $1.4$ \\
\bottomrule
\end{tabular}
}
\caption{The combined Born-level single-differential cross section
${\textrm{d}}\sigma/{\textrm{d}}m_{\ell\ell}$. The measurements are
listed together with the statistical ($\delta^{\textrm{stat}}$),
uncorrelated systematic ($\delta^{\textrm {syst}}_{\textrm {unc}}$),
correlated systematic ($\delta^{\textrm{syst}}_{\textrm{cor}}$),
and total ($\delta^{\textrm{total}}$) uncertainties. The luminosity
uncertainty of 1.9\% is not shown and not included in the overall
systematic and total uncertainties.}
\label{tab:full_summary_m}
\end{table}

\begin{table}[htp!]
\tiny
\centering
\scalebox{1}{
\setlength\tabcolsep{3pt}
\setlength\extrarowheight{2pt}
\begin{tabular}{C{1.05cm}C{1cm}C{1.75cm}C{0.65cm}C{0.55cm}C{0.55cm}C{0.55cm} @{\hskip 20pt} C{1.05cm}C{1cm}C{1.75cm}C{0.65cm}C{0.55cm}C{0.55cm}C{0.55cm}}
\toprule
$m_{\ell\ell}$
& $|y_{\ell\ell}|$
& $\textrm{d}^2\sigma/\textrm{d}m_{\ell\ell}\textrm{d}|y_{\ell\ell}|$
& $\delta^{\textrm {stat}}$
& $\delta^{\textrm {syst}}_{\textrm {unc}}$
& $\delta^{\textrm {syst}}_{\textrm {cor}}$
& $\delta^{\textrm {total}}$
& $m_{ee}$
& $|y_{\ell\ell}|$
& $\textrm{d}^2\sigma/\textrm{d}m_{\ell\ell}\textrm{d}|y_{\ell\ell}|$
& $\delta^{\textrm {stat}}$
& $\delta^{\textrm {syst}}_{\textrm {unc}}$
& $\delta^{\textrm {syst}}_{\textrm {cor}}$
& $\delta^{\textrm {total}}$ \\
$[$GeV$]$ & & [pb/GeV] & [\%] & [\%] & [\%] & [\%] & $[$GeV$]$ & & [pb/GeV] & [\%] & [\%] & [\%] & [\%] \\
\noalign{\vskip 0.05cm}
\cmidrule(lr{22pt}){1-7}
\cmidrule(l{2.2pt}r){8-14}
  $46, 66$ &   $0.0, 0.2$ &     $1.85\times 10^{-1}$ &    $0.6$ &    $0.4$ &    $1.0$ &    $1.2$ &   $46, 66$ &   $1.2, 1.4$ &     $1.86\times 10^{-1}$ &    $0.6$ &    $0.4$ &    $0.9$ &    $1.1$ \\
  $46, 66$ &   $0.2, 0.4$ &     $1.87\times 10^{-1}$ &    $0.6$ &    $0.5$ &    $1.0$ &    $1.2$ &   $46, 66$ &   $1.4, 1.6$ &     $1.82\times 10^{-1}$ &    $0.6$ &    $0.4$ &    $0.9$ &    $1.1$ \\
  $46, 66$ &   $0.4, 0.6$ &     $1.86\times 10^{-1}$ &    $0.6$ &    $0.4$ &    $0.9$ &    $1.2$ &   $46, 66$ &   $1.6, 1.8$ &     $1.66\times 10^{-1}$ &    $0.6$ &    $0.5$ &    $0.9$ &    $1.2$ \\
  $46, 66$ &   $0.6, 0.8$ &     $1.87\times 10^{-1}$ &    $0.6$ &    $0.4$ &    $0.9$ &    $1.2$ &   $46, 66$ &   $1.8, 2.0$ &     $1.35\times 10^{-1}$ &    $0.7$ &    $0.5$ &    $0.8$ &    $1.2$ \\
  $46, 66$ &   $0.8, 1.0$ &     $1.86\times 10^{-1}$ &    $0.6$ &    $0.4$ &    $0.9$ &    $1.2$ &   $46, 66$ &   $2.0, 2.2$ &     $8.60\times 10^{-2}$ &    $0.8$ &    $0.6$ &    $0.8$ &    $1.3$ \\
  $46, 66$ &   $1.0, 1.2$ &     $1.88\times 10^{-1}$ &    $0.6$ &    $0.4$ &    $0.9$ &    $1.1$ &   $46, 66$ &   $2.2, 2.4$ &     $2.93\times 10^{-2}$ &    $1.4$ &    $1.1$ &    $0.9$ &    $2.0$ \\
\cmidrule(lr{22pt}){1-7}
\cmidrule(l{2.2pt}r){8-14}
  $66, 80$ &   $0.0, 0.2$ &     $3.05\times 10^{-1}$ &    $0.4$ &    $0.2$ &    $0.4$ &    $0.6$ &   $66, 80$ &   $1.2, 1.4$ &     $2.82\times 10^{-1}$ &    $0.4$ &    $0.2$ &    $0.4$ &    $0.6$ \\
  $66, 80$ &   $0.2, 0.4$ &     $3.02\times 10^{-1}$ &    $0.4$ &    $0.2$ &    $0.4$ &    $0.6$ &   $66, 80$ &   $1.4, 1.6$ &     $2.54\times 10^{-1}$ &    $0.5$ &    $0.3$ &    $0.4$ &    $0.6$ \\
  $66, 80$ &   $0.4, 0.6$ &     $3.02\times 10^{-1}$ &    $0.4$ &    $0.2$ &    $0.4$ &    $0.6$ &   $66, 80$ &   $1.6, 1.8$ &     $2.08\times 10^{-1}$ &    $0.5$ &    $0.3$ &    $0.4$ &    $0.7$ \\
  $66, 80$ &   $0.6, 0.8$ &     $3.01\times 10^{-1}$ &    $0.4$ &    $0.2$ &    $0.4$ &    $0.6$ &   $66, 80$ &   $1.8, 2.0$ &     $1.54\times 10^{-1}$ &    $0.6$ &    $0.3$ &    $0.5$ &    $0.8$ \\
  $66, 80$ &   $0.8, 1.0$ &     $2.95\times 10^{-1}$ &    $0.4$ &    $0.2$ &    $0.4$ &    $0.6$ &   $66, 80$ &   $2.0, 2.2$ &     $9.27\times 10^{-2}$ &    $0.7$ &    $0.4$ &    $0.6$ &    $1.0$ \\
  $66, 80$ &   $1.0, 1.2$ &     $2.93\times 10^{-1}$ &    $0.4$ &    $0.2$ &    $0.4$ &    $0.6$ &   $66, 80$ &   $2.2, 2.4$ &     $3.05\times 10^{-2}$ &    $1.2$ &    $0.7$ &    $0.9$ &    $1.7$ \\
\cmidrule(lr{22pt}){1-7}
\cmidrule(l{2.2pt}r){8-14}
  $80, 91$ &   $0.0, 0.2$ &     $6.00$ &    $0.1$ &    $0.0$ &    $0.2$ &    $0.2$ &   $80, 91$ &   $1.2, 1.4$ &     $5.19$ &    $0.1$ &    $0.1$ &    $0.2$ &    $0.3$ \\
  $80, 91$ &   $0.2, 0.4$ &     $6.00$ &    $0.1$ &    $0.0$ &    $0.2$ &    $0.2$ &   $80, 91$ &   $1.4, 1.6$ &     $4.51$ &    $0.1$ &    $0.1$ &    $0.2$ &    $0.3$ \\
  $80, 91$ &   $0.4, 0.6$ &     $5.97$ &    $0.1$ &    $0.1$ &    $0.2$ &    $0.2$ &   $80, 91$ &   $1.6, 1.8$ &     $3.66$ &    $0.1$ &    $0.1$ &    $0.3$ &    $0.3$ \\
  $80, 91$ &   $0.6, 0.8$ &     $5.93$ &    $0.1$ &    $0.0$ &    $0.2$ &    $0.3$ &   $80, 91$ &   $1.8, 2.0$ &     $2.67$ &    $0.1$ &    $0.1$ &    $0.3$ &    $0.3$ \\
  $80, 91$ &   $0.8, 1.0$ &     $5.87$ &    $0.1$ &    $0.1$ &    $0.2$ &    $0.3$ &   $80, 91$ &   $2.0, 2.2$ &     $1.60$ &    $0.2$ &    $0.1$ &    $0.3$ &    $0.4$ \\
  $80, 91$ &   $1.0, 1.2$ &     $5.66$ &    $0.1$ &    $0.1$ &    $0.2$ &    $0.3$ &   $80, 91$ &   $2.2, 2.4$ &     $5.20\times 10^{-1}$ &    $0.3$ &    $0.2$ &    $0.4$ &    $0.5$ \\
\cmidrule(lr{22pt}){1-7}
\cmidrule(l{2.2pt}r){8-14}
  $\phantom{1}91, 102$ &   $0.0, 0.2$ &     $7.08$ &    $0.1$ &    $0.1$ &    $0.2$ &    $0.2$ &   $\phantom{1}91, 102$ &   $1.2, 1.4$ &     $6.02$ &    $0.1$ &    $0.0$ &    $0.2$ &    $0.3$ \\
  $\phantom{1}91, 102$ &   $0.2, 0.4$ &     $7.04$ &    $0.1$ &    $0.1$ &    $0.2$ &    $0.2$ &   $\phantom{1}91, 102$ &   $1.4, 1.6$ &     $5.21$ &    $0.1$ &    $0.1$ &    $0.2$ &    $0.3$ \\
  $\phantom{1}91, 102$ &   $0.4, 0.6$ &     $7.01$ &    $0.1$ &    $0.1$ &    $0.2$ &    $0.2$ &   $\phantom{1}91, 102$ &   $1.6, 1.8$ &     $4.23$ &    $0.1$ &    $0.1$ &    $0.3$ &    $0.3$ \\
  $\phantom{1}91, 102$ &   $0.6, 0.8$ &     $6.98$ &    $0.1$ &    $0.0$ &    $0.2$ &    $0.2$ &   $\phantom{1}91, 102$ &   $1.8, 2.0$ &     $3.07$ &    $0.2$ &    $0.1$ &    $0.3$ &    $0.3$ \\
  $\phantom{1}91, 102$ &   $0.8, 1.0$ &     $6.90$ &    $0.1$ &    $0.0$ &    $0.2$ &    $0.2$ &   $\phantom{1}91, 102$ &   $2.0, 2.2$ &     $1.83$ &    $0.2$ &    $0.1$ &    $0.3$ &    $0.4$ \\
  $\phantom{1}91, 102$ &   $1.0, 1.2$ &     $6.60$ &    $0.1$ &    $0.1$ &    $0.2$ &    $0.3$ &   $\phantom{1}91, 102$ &   $2.2, 2.4$ &     $5.96\times 10^{-1}$ &    $0.3$ &    $0.2$ &    $0.4$ &    $0.5$ \\
\cmidrule(lr{22pt}){1-7}
\cmidrule(l{2.2pt}r){8-14}
  $102, 116$ &   $0.0, 0.2$ &     $2.38\times 10^{-1}$ &    $0.5$ &    $0.2$ &    $0.3$ &    $0.7$ &   $102, 116$ &   $1.2, 1.4$ &     $1.96\times 10^{-1}$ &    $0.5$ &    $0.3$ &    $0.5$ &    $0.7$ \\
  $102, 116$ &   $0.2, 0.4$ &     $2.39\times 10^{-1}$ &    $0.5$ &    $0.2$ &    $0.4$ &    $0.7$ &   $102, 116$ &   $1.4, 1.6$ &     $1.66\times 10^{-1}$ &    $0.5$ &    $0.3$ &    $0.5$ &    $0.8$ \\
  $102, 116$ &   $0.4, 0.6$ &     $2.35\times 10^{-1}$ &    $0.5$ &    $0.2$ &    $0.4$ &    $0.7$ &   $102, 116$ &   $1.6, 1.8$ &     $1.35\times 10^{-1}$ &    $0.6$ &    $0.4$ &    $0.7$ &    $1.0$ \\
  $102, 116$ &   $0.6, 0.8$ &     $2.33\times 10^{-1}$ &    $0.5$ &    $0.3$ &    $0.4$ &    $0.7$ &   $102, 116$ &   $1.8, 2.0$ &     $9.84\times 10^{-2}$ &    $0.6$ &    $0.4$ &    $0.8$ &    $1.1$ \\
  $102, 116$ &   $0.8, 1.0$ &     $2.29\times 10^{-1}$ &    $0.5$ &    $0.3$ &    $0.4$ &    $0.7$ &   $102, 116$ &   $2.0, 2.2$ &     $5.76\times 10^{-2}$ &    $0.7$ &    $0.5$ &    $1.0$ &    $1.3$ \\
  $102, 116$ &   $1.0, 1.2$ &     $2.16\times 10^{-1}$ &    $0.5$ &    $0.3$ &    $0.4$ &    $0.7$ &   $102, 116$ &   $2.2, 2.4$ &     $1.85\times 10^{-2}$ &    $1.0$ &    $0.9$ &    $1.3$ &    $1.9$ \\
\cmidrule(lr{22pt}){1-7}
\cmidrule(l{2.2pt}r){8-14}
  $116, 150$ &   $0.0, 0.2$ &     $4.84\times 10^{-2}$ &    $0.8$ &    $0.3$ &    $0.8$ &    $1.2$ &   $116, 150$ &   $1.2, 1.4$ &     $3.84\times 10^{-2}$ &    $0.9$ &    $0.4$ &    $0.6$ &    $1.1$ \\
  $116, 150$ &   $0.2, 0.4$ &     $4.79\times 10^{-2}$ &    $0.8$ &    $0.3$ &    $0.8$ &    $1.2$ &   $116, 150$ &   $1.4, 1.6$ &     $3.23\times 10^{-2}$ &    $0.9$ &    $0.4$ &    $0.5$ &    $1.1$ \\
  $116, 150$ &   $0.4, 0.6$ &     $4.74\times 10^{-2}$ &    $0.8$ &    $0.3$ &    $0.8$ &    $1.2$ &   $116, 150$ &   $1.6, 1.8$ &     $2.66\times 10^{-2}$ &    $1.0$ &    $0.5$ &    $0.5$ &    $1.2$ \\
  $116, 150$ &   $0.6, 0.8$ &     $4.77\times 10^{-2}$ &    $0.8$ &    $0.3$ &    $0.8$ &    $1.2$ &   $116, 150$ &   $1.8, 2.0$ &     $1.93\times 10^{-2}$ &    $1.2$ &    $0.7$ &    $0.6$ &    $1.5$ \\
  $116, 150$ &   $0.8, 1.0$ &     $4.54\times 10^{-2}$ &    $0.8$ &    $0.3$ &    $0.7$ &    $1.1$ &   $116, 150$ &   $2.0, 2.2$ &     $1.14\times 10^{-2}$ &    $1.4$ &    $0.7$ &    $0.7$ &    $1.7$ \\
  $116, 150$ &   $1.0, 1.2$ &     $4.23\times 10^{-2}$ &    $0.8$ &    $0.4$ &    $0.6$ &    $1.1$ &   $116, 150$ &   $2.2, 2.4$ &     $3.48\times 10^{-3}$ &    $2.6$ &    $1.7$ &    $1.2$ &    $3.3$ \\
\cmidrule(lr{22pt}){1-7}
\cmidrule(l{2.2pt}r){8-14}
  $150, 200$ &   $0.0, 0.2$ &     $1.11\times 10^{-2}$ &    $1.6$ &    $0.6$ &    $1.8$ &    $2.4$ &   $150, 200$ &   $1.2, 1.4$ &     $8.56\times 10^{-3}$ &    $1.6$ &    $0.6$ &    $1.0$ &    $2.0$ \\
  $150, 200$ &   $0.2, 0.4$ &     $1.10\times 10^{-2}$ &    $1.5$ &    $0.7$ &    $1.8$ &    $2.4$ &   $150, 200$ &   $1.4, 1.6$ &     $7.12\times 10^{-3}$ &    $1.8$ &    $0.9$ &    $0.9$ &    $2.2$ \\
  $150, 200$ &   $0.4, 0.6$ &     $1.08\times 10^{-2}$ &    $1.5$ &    $0.6$ &    $1.7$ &    $2.3$ &   $150, 200$ &   $1.6, 1.8$ &     $5.72\times 10^{-3}$ &    $1.9$ &    $0.7$ &    $0.8$ &    $2.2$ \\
  $150, 200$ &   $0.6, 0.8$ &     $1.07\times 10^{-2}$ &    $1.5$ &    $0.5$ &    $1.5$ &    $2.2$ &   $150, 200$ &   $1.8, 2.0$ &     $4.06\times 10^{-3}$ &    $2.2$ &    $0.7$ &    $0.7$ &    $2.4$ \\
  $150, 200$ &   $0.8, 1.0$ &     $9.98\times 10^{-3}$ &    $1.6$ &    $0.5$ &    $1.3$ &    $2.1$ &   $150, 200$ &   $2.0, 2.2$ &     $2.46\times 10^{-3}$ &    $2.8$ &    $1.0$ &    $0.7$ &    $3.0$ \\
  $150, 200$ &   $1.0, 1.2$ &     $9.22\times 10^{-3}$ &    $1.6$ &    $0.6$ &    $1.2$ &    $2.1$ &   $150, 200$ &   $2.2, 2.4$ &     $8.20\times 10^{-4}$ &    $4.7$ &    $1.3$ &    $1.0$ &    $5.0$ \\
\bottomrule
\end{tabular}
}
\caption{The combined Born-level double-differential cross section
${\textrm{d}}^2\sigma/{\textrm{d}}m_{\ell\ell}{\textrm{d}}|y_{\ell\ell}|$.
The measurements are listed together with the statistical ($\delta^{\textrm{stat}}$),
uncorrelated systematic ($\delta^{\textrm {syst}}_{\textrm {unc}}$),
correlated systematic ($\delta^{\textrm{syst}}_{\textrm{cor}}$),
and total ($\delta^{\textrm{total}}$) uncertainties. The luminosity
uncertainty of 1.9\% is not shown and not included in the overall
systematic and total uncertainties.}
\label{tab:full_summary_ym}
\end{table}

\FloatBarrier

\subsection{Triple-differential cross-section tables}
\label{tab:3d}
\begin{table}[htp!]
\tiny
\centering
\scalebox{0.75}{
\setlength\tabcolsep{3pt}
\setlength\extrarowheight{2pt}
\begin{tabular}{C{0.6cm}C{0.9cm}C{0.9cm}C{1.1cm}C{1.6cm}L{0.5cm}L{0.45cm}L{0.45cm}L{0.45cm} @{\hskip 20pt} C{0.6cm}C{0.9cm}C{0.9cm}C{1.1cm}C{1.6cm}L{0.5cm}L{0.45cm}L{0.45cm}L{0.45cm}}
\toprule
Bin
& $m_{\ell\ell}$
& $|y_{\ell\ell}|$
& $\cos\theta^*$
& $\textrm{d}^3\sigma$
& $\delta^{\textrm {stat}}$
& $\delta^{\textrm {syst}}_{\textrm {unc}}$
& $\delta^{\textrm {syst}}_{\textrm {cor}}$
& $\delta^{\textrm {total}}$
& Bin
& $m_{\ell\ell}$
& $|y_{\ell\ell}|$
& $\cos\theta^*$
& $\textrm{d}^3\sigma$
& $\delta^{\textrm {stat}}$
& $\delta^{\textrm {syst}}_{\textrm {unc}}$
& $\delta^{\textrm {syst}}_{\textrm {cor}}$
& $\delta^{\textrm {total}}$ \\
&	[GeV] & & & [pb/GeV] & [\%] & [\%] & [\%] & [\%] & & [GeV] & & & [pb/GeV] & [\%] & [\%] & [\%] & [\%]\\
\noalign{\vskip 0.05cm}
\cmidrule(lr{19.5pt}){1-9}
\cmidrule(r{3pt}l){10-18}
  $1$ &   $46, 66$ &   $0.0, 0.2$ &   $-1.0, -0.7$ &     $1.42\times 10^{-2}$ &    $4.0$ &    $2.5$ &    $1.8$ &    $5.1$ &   $6$ &   $46, 66$ &   $0.0, 0.2$ &   $+0.7, +1.0$ &     $1.37\times 10^{-2}$ &    $4.2$ &    $3.4$ &    $2.0$ &    $5.7$ \\
  $2$ &   $46, 66$ &   $0.0, 0.2$ &   $-0.7, -0.4$ &     $1.02\times 10^{-1}$ &    $1.4$ &    $1.1$ &    $1.2$ &    $2.1$ &   $5$ &   $46, 66$ &   $0.0, 0.2$ &   $+0.4, +0.7$ &     $1.01\times 10^{-1}$ &    $1.4$ &    $1.1$ &    $1.2$ &    $2.1$ \\
  $3$ &   $46, 66$ &   $0.0, 0.2$ &   $-0.4, \phantom{-}0.0$ &     $1.45\times 10^{-1}$ &    $1.0$ &    $0.8$ &    $0.9$ &    $1.5$ &   $4$ &   $46, 66$ &   $0.0, 0.2$ &   $\phantom{+}0.0, +0.4$ &     $1.44\times 10^{-1}$ &    $1.0$ &    $0.8$ &    $0.9$ &    $1.6$ \\
\cmidrule(lr{19.5pt}){1-9}
\cmidrule(r{3pt}l){10-18}
  $7$ &   $46, 66$ &   $0.2, 0.4$ &   $-1.0, -0.7$ &     $1.48\times 10^{-2}$ &    $3.9$ &    $2.1$ &    $1.8$ &    $4.8$ &   $12$ &   $46, 66$ &   $0.2, 0.4$ &   $+0.7, +1.0$ &     $1.46\times 10^{-2}$ &    $3.9$ &    $1.9$ &    $1.8$ &    $4.7$ \\
\bottomrule
\end{tabular}
}
\end{table}

\begin{table}[htp!]
\tiny
\centering
\scalebox{0.75}{
\setlength\tabcolsep{3pt}
\setlength\extrarowheight{2pt}
\begin{tabular}{C{0.6cm}C{0.9cm}C{0.9cm}C{1.1cm}C{1.6cm}L{0.5cm}L{0.45cm}L{0.45cm}L{0.45cm} @{\hskip 20pt} C{0.6cm}C{0.9cm}C{0.9cm}C{1.1cm}C{1.6cm}L{0.5cm}L{0.45cm}L{0.45cm}L{0.45cm}}
\toprule
Bin
& $m_{\ell\ell}$
& $|y_{\ell\ell}|$
& $\cos\theta^*$
& $\textrm{d}^3\sigma$
& $\delta^{\textrm {stat}}$
& $\delta^{\textrm {syst}}_{\textrm {unc}}$
& $\delta^{\textrm {syst}}_{\textrm {cor}}$
& $\delta^{\textrm {total}}$
& Bin
& $m_{\ell\ell}$
& $|y_{\ell\ell}|$
& $\cos\theta^*$
& $\textrm{d}^3\sigma$
& $\delta^{\textrm {stat}}$
& $\delta^{\textrm {syst}}_{\textrm {unc}}$
& $\delta^{\textrm {syst}}_{\textrm {cor}}$
& $\delta^{\textrm {total}}$ \\
&	[GeV] & & & [pb/GeV] & [\%] & [\%] & [\%] & [\%] & & [GeV] & & & [pb/GeV] & [\%] & [\%] & [\%] & [\%]\\
\noalign{\vskip 0.05cm}
\cmidrule(lr{19.5pt}){1-9}
\cmidrule(r{3pt}l){10-18}
  $145$ &   $\phantom{1}80, \phantom{1}91$ &   $0.0, 0.2$ &   $-1.0, -0.7$ &     $2.25$ &    $0.3$ &    $0.2$ &    $0.3$ &    $0.5$ &   $150$ &   $\phantom{1}80, \phantom{1}91$ &   $0.0, 0.2$ &   $+0.7, +1.0$ &     $2.26$ &    $0.3$ &    $0.2$ &    $0.3$ &    $0.5$ \\
  $146$ &   $\phantom{1}80, \phantom{1}91$ &   $0.0, 0.2$ &   $-0.7, -0.4$ &     $3.58$ &    $0.3$ &    $0.1$ &    $0.2$ &    $0.3$ &   $149$ &   $\phantom{1}80, \phantom{1}91$ &   $0.0, 0.2$ &   $+0.4, +0.7$ &     $3.60$ &    $0.3$ &    $0.1$ &    $0.2$ &    $0.3$ \\
  $147$ &   $\phantom{1}80, \phantom{1}91$ &   $0.0, 0.2$ &   $-0.4, \phantom{-}0.0$ &     $3.12$ &    $0.3$ &    $0.1$ &    $0.2$ &    $0.4$ &   $148$ &   $\phantom{1}80, \phantom{1}91$ &   $0.0, 0.2$ &   $\phantom{+}0.0, +0.4$ &     $3.10$ &    $0.3$ &    $0.1$ &    $0.2$ &    $0.4$ \\
\cmidrule(lr{19.5pt}){1-9}
\cmidrule(r{3pt}l){10-18}
  $151$ &   $\phantom{1}80, \phantom{1}91$ &   $0.2, 0.4$ &   $-1.0, -0.7$ &     $2.26$ &    $0.3$ &    $0.2$ &    $0.3$ &    $0.5$ &   $156$ &   $\phantom{1}80, \phantom{1}91$ &   $0.2, 0.4$ &   $+0.7, +1.0$ &     $2.27$ &    $0.3$ &    $0.2$ &    $0.3$ &    $0.5$ \\
\bottomrule
\end{tabular}
}
\end{table}

\begin{table}[htp!]
\tiny
\centering
\scalebox{0.75}{
\setlength\tabcolsep{3pt}
\setlength\extrarowheight{2pt}
\begin{tabular}{C{0.6cm}C{0.9cm}C{0.9cm}C{1.1cm}C{1.6cm}L{0.5cm}L{0.45cm}L{0.45cm}L{0.45cm} @{\hskip 20pt} C{0.6cm}C{0.9cm}C{0.9cm}C{1.1cm}C{1.6cm}L{0.5cm}L{0.45cm}L{0.45cm}L{0.45cm}}
\toprule
Bin
& $m_{\ell\ell}$
& $|y_{\ell\ell}|$
& $\cos\theta^*$
& $\textrm{d}^3\sigma$
& $\delta^{\textrm {stat}}$
& $\delta^{\textrm {syst}}_{\textrm {unc}}$
& $\delta^{\textrm {syst}}_{\textrm {cor}}$
& $\delta^{\textrm {total}}$
& Bin
& $m_{\ell\ell}$
& $|y_{\ell\ell}|$
& $\cos\theta^*$
& $\textrm{d}^3\sigma$
& $\delta^{\textrm {stat}}$
& $\delta^{\textrm {syst}}_{\textrm {unc}}$
& $\delta^{\textrm {syst}}_{\textrm {cor}}$
& $\delta^{\textrm {total}}$ \\
&	[GeV] & & & [pb/GeV] & [\%] & [\%] & [\%] & [\%] & & [GeV] & & & [pb/GeV] & [\%] & [\%] & [\%] & [\%]\\
\noalign{\vskip 0.05cm}
\cmidrule(lr{19.5pt}){1-9}
\cmidrule(r{3pt}l){10-18}
  $289$ &   $102, 116$ &   $0.0, 0.2$ &   $-1.0, -0.7$ &     $1.09\times 10^{-1}$ &    $1.4$ &    $0.8$ &    $0.6$ &    $1.7$ &   $294$ &   $102, 116$ &   $0.0, 0.2$ &   $+0.7, +1.0$ &     $1.09\times 10^{-1}$ &    $1.4$ &    $0.8$ &    $0.6$ &    $1.7$ \\
  $290$ &   $102, 116$ &   $0.0, 0.2$ &   $-0.7, -0.4$ &     $1.30\times 10^{-1}$ &    $1.2$ &    $0.6$ &    $0.4$ &    $1.4$ &   $293$ &   $102, 116$ &   $0.0, 0.2$ &   $+0.4, +0.7$ &     $1.39\times 10^{-1}$ &    $1.2$ &    $0.5$ &    $0.3$ &    $1.4$ \\
  $291$ &   $102, 116$ &   $0.0, 0.2$ &   $-0.4, \phantom{-}0.0$ &     $1.15\times 10^{-1}$ &    $1.2$ &    $0.5$ &    $0.3$ &    $1.4$ &   $292$ &   $102, 116$ &   $0.0, 0.2$ &   $\phantom{+}0.0, +0.4$ &     $1.15\times 10^{-1}$ &    $1.2$ &    $0.5$ &    $0.4$ &    $1.4$ \\
\cmidrule(lr{19.5pt}){1-9}
\cmidrule(r{3pt}l){10-18}
  $295$ &   $102, 116$ &   $0.2, 0.4$ &   $-1.0, -0.7$ &     $1.03\times 10^{-1}$ &    $1.5$ &    $0.8$ &    $0.6$ &    $1.8$ &   $300$ &   $102, 116$ &   $0.2, 0.4$ &   $+0.7, +1.0$ &     $1.11\times 10^{-1}$ &    $1.5$ &    $0.8$ &    $0.6$ &    $1.7$ \\
\bottomrule
\end{tabular}
}
\end{table}

\begin{table}[htp!]
\tiny
\centering
\scalebox{0.75}{
\setlength\tabcolsep{3pt}
\setlength\extrarowheight{2pt}
\begin{tabular}{C{0.6cm}C{0.9cm}C{0.9cm}C{1.1cm}C{1.6cm}L{0.5cm}L{0.45cm}L{0.45cm}L{0.45cm} @{\hskip 20pt} C{0.6cm}C{0.9cm}C{0.9cm}C{1.1cm}C{1.6cm}L{0.5cm}L{0.45cm}L{0.45cm}L{0.45cm}}
\toprule
Bin
& $m_{\ell\ell}$
& $|y_{\ell\ell}|$
& $\cos\theta^*$
& $\textrm{d}^3\sigma$
& $\delta^{\textrm {stat}}$
& $\delta^{\textrm {syst}}_{\textrm {unc}}$
& $\delta^{\textrm {syst}}_{\textrm {cor}}$
& $\delta^{\textrm {total}}$
& Bin
& $m_{\ell\ell}$
& $|y_{\ell\ell}|$
& $\cos\theta^*$
& $\textrm{d}^3\sigma$
& $\delta^{\textrm {stat}}$
& $\delta^{\textrm {syst}}_{\textrm {unc}}$
& $\delta^{\textrm {syst}}_{\textrm {cor}}$
& $\delta^{\textrm {total}}$ \\
&	[GeV] & & & [pb/GeV] & [\%] & [\%] & [\%] & [\%] & & [GeV] & & & [pb/GeV] & [\%] & [\%] & [\%] & [\%]\\
\noalign{\vskip 0.05cm}
\cmidrule(lr{19.5pt}){1-9}
\cmidrule(r{3pt}l){10-18}
  $433$ &   $150, 200$ &   $0.0, 0.2$ &   $-1.0, -0.7$ &     $6.26\times 10^{-3}$ &    $4.2$ &    $1.6$ &    $2.8$ &    $5.3$ &   $438$ &   $150, 200$ &   $0.0, 0.2$ &   $+0.7, +1.0$ &     $6.55\times 10^{-3}$ &    $4.3$ &    $1.7$ &    $2.8$ &    $5.4$ \\
  $434$ &   $150, 200$ &   $0.0, 0.2$ &   $-0.7, -0.4$ &     $5.56\times 10^{-3}$ &    $3.7$ &    $1.2$ &    $1.6$ &    $4.2$ &   $437$ &   $150, 200$ &   $0.0, 0.2$ &   $+0.4, +0.7$ &     $5.91\times 10^{-3}$ &    $3.6$ &    $1.3$ &    $1.6$ &    $4.1$ \\
  $435$ &   $150, 200$ &   $0.0, 0.2$ &   $-0.4, \phantom{-}0.0$ &     $4.77\times 10^{-3}$ &    $3.4$ &    $1.3$ &    $1.2$ &    $3.8$ &   $436$ &   $150, 200$ &   $0.0, 0.2$ &   $\phantom{+}0.0, +0.4$ &     $4.76\times 10^{-3}$ &    $3.5$ &    $0.7$ &    $1.1$ &    $3.7$ \\
\cmidrule(lr{19.5pt}){1-9}
\cmidrule(r{3pt}l){10-18}
  $439$ &   $150, 200$ &   $0.2, 0.4$ &   $-1.0, -0.7$ &     $5.66\times 10^{-3}$ &    $4.6$ &    $1.7$ &    $3.2$ &    $5.8$ &   $444$ &   $150, 200$ &   $0.2, 0.4$ &   $+0.7, +1.0$ &     $6.67\times 10^{-3}$ &    $4.0$ &    $2.0$ &    $2.7$ &    $5.2$ \\
\bottomrule
\end{tabular}
}
\caption{The combined Born-level triple-differential cross section
${\textrm{d}}^3\sigma/{\textrm{d}}m_{\ell\ell}{\textrm{d}}|y_{\ell\ell}|{\textrm{d}}\cos\theta^*$.
The measurements are listed together with the statistical ($\delta^{\textrm{stat}}$),
uncorrelated systematic ($\delta^{\textrm {syst}}_{\textrm {unc}}$),
correlated systematic ($\delta^{\textrm{syst}}_{\textrm{cor}}$),
and total ($\delta^{\textrm{total}}$) uncertainties. The luminosity
uncertainty of 1.9\% is not shown and not included in the overall
systematic and total uncertainties.}
\label{tab:comb_summary}
\end{table}
\begin{table}[htp!]
\tiny
\centering
\scalebox{0.85}{
\setlength\tabcolsep{3pt}
\setlength\extrarowheight{2pt}
\begin{tabular}{C{0.6cm}C{0.9cm}C{0.9cm}C{1.1cm}C{1.6cm}L{0.5cm}L{0.45cm}L{0.45cm}L{0.45cm} @{\hskip 20pt} C{0.6cm}C{0.9cm}C{0.9cm}C{1.1cm}C{1.6cm}L{0.5cm}L{0.45cm}L{0.45cm}L{0.45cm}}
\toprule
Bin
& $m_{ee}$
& $|y_{ee}|$
& $\cos\theta^*$
& $\textrm{d}^3\sigma$
& $\delta^{\textrm {stat}}$
& $\delta^{\textrm {syst}}_{\textrm {unc}}$
& $\delta^{\textrm {syst}}_{\textrm {cor}}$
& $\delta^{\textrm {total}}$
& Bin
& $m_{ee}$
& $|y_{ee}|$
& $\cos\theta^*$
& $\textrm{d}^3\sigma$
& $\delta^{\textrm {stat}}$
& $\delta^{\textrm {syst}}_{\textrm {unc}}$
& $\delta^{\textrm {syst}}_{\textrm {cor}}$
& $\delta^{\textrm {total}}$ \\
&	[GeV] & & & [pb/GeV] & [\%] & [\%] & [\%] & [\%] & & [GeV] & & & [pb/GeV] & [\%] & [\%] & [\%] & [\%]\\
\noalign{\vskip 0.05cm}
\cmidrule(lr{19.5pt}){1-9}
\cmidrule(r{3pt}l){10-18}
  $1$ &   $66, 80$ &   $1.2, 1.6$ &   $-1.0, -0.7$ &     $1.06\times 10^{-2}$ &    $6.4$ &    $8.1$ &    $12.4$ &    $16.1$ &   $6$ &   $66, 80$ &   $1.2, 1.6$ &   $+0.7, +1.0$ &     $6.29\times 10^{-3}$ &    $7.8$ &    $11.0$ &    $16.1$ &    $21.0$ \\
  $2$ &   $66, 80$ &   $1.2, 1.6$ &   $-0.7, -0.4$ &     $9.24\times 10^{-4}$ &    $16.2$ &    $15.4$ &    $15.3$ &    $27.1$ &   $5$ &   $66, 80$ &   $1.2, 1.6$ &   $+0.4, +0.7$ &     $6.97\times 10^{-4}$ &    $15.7$ &    $13.1$ &    $15.5$ &    $25.7$ \\
  $3$ &   $66, 80$ &   $1.2, 1.6$ &   $-0.4, 0.0$ & $-$ & $-$ & $-$ & $-$ & $-$ &   $4$ &   $66, 80$ &   $1.2, 1.6$ &   $\phantom{+}0.0, +0.4$ & $-$ & $-$ & $-$ & $-$ & $-$ \\
\cmidrule(lr{19.5pt}){1-9}
\cmidrule(r{3pt}l){10-18}
  $7$ &   $66, 80$ &   $1.6, 2.0$ &   $-1.0, -0.7$ &     $3.89\times 10^{-2}$ &    $3.9$ &    $3.8$ &    $7.0$ &    $8.9$ &   $12$ &   $66, 80$ &   $1.6, 2.0$ &   $+0.7, +1.0$ &     $2.24\times 10^{-2}$ &    $5.1$ &    $6.7$ &    $10.4$ &    $13.3$ \\
  $8$ &   $66, 80$ &   $1.6, 2.0$ &   $-0.7, -0.4$ &     $6.54\times 10^{-2}$ &    $2.8$ &    $2.6$ &    $5.1$ &    $6.4$ &   $11$ &   $66, 80$ &   $1.6, 2.0$ &   $+0.4, +0.7$ &     $4.27\times 10^{-2}$ &    $3.2$ &    $3.5$ &    $6.1$ &    $7.8$ \\
  $9$ &   $66, 80$ &   $1.6, 2.0$ &   $-0.4, \phantom{-}0.0$ & $-$ & $-$ & $-$ & $-$ & $-$ &   $10$ &   $66, 80$ &   $1.6, 2.0$ &   $\phantom{+}0.0, +0.4$ & $-$ & $-$ & $-$ & $-$ & $-$ \\
\cmidrule(lr{19.5pt}){1-9}
\cmidrule(r{3pt}l){10-18}
  $13$ &   $66, 80$ &   $2.0, 2.4$ &   $-1.0, -0.7$ &     $4.28\times 10^{-2}$ &    $6.2$ &    $6.6$ &    $13.1$ &    $16.0$ &   $18$ &   $66, 80$ &   $2.0, 2.4$ &   $+0.7, +1.0$ &     $2.08\times 10^{-2}$ &    $7.3$ &    $8.6$ &    $25.4$ &    $27.8$ \\
  $14$ &   $66, 80$ &   $2.0, 2.4$ &   $-0.7, -0.4$ &     $1.89\times 10^{-1}$ &    $2.0$ &    $2.1$ &    $4.4$ &    $5.2$ &   $17$ &   $66, 80$ &   $2.0, 2.4$ &   $+0.4, +0.7$ &     $9.97\times 10^{-2}$ &    $2.6$ &    $3.4$ &    $6.0$ &    $7.3$ \\
  $15$ &   $66, 80$ &   $2.0, 2.4$ &   $-0.4, \phantom{-}0.0$ &     $5.10\times 10^{-2}$ &    $2.5$ &    $2.5$ &    $5.4$ &    $6.5$ &   $16$ &   $66, 80$ &   $2.0, 2.4$ &   $\phantom{+}0.0, +0.4$ &     $3.55\times 10^{-2}$ &    $2.8$ &    $3.2$ &    $6.3$ &    $7.6$ \\
\cmidrule(lr{19.5pt}){1-9}
\cmidrule(r{3pt}l){10-18}
  $19$ &   $66, 80$ &   $2.4, 2.8$ &   $-1.0, -0.7$ &     $3.91\times 10^{-2}$ &    $3.5$ &    $3.8$ &    $19.5$ &    $20.2$ &   $24$ &   $66, 80$ &   $2.4, 2.8$ &   $+0.7, +1.0$ &     $1.65\times 10^{-2}$ &    $4.3$ &    $5.7$ &    $26.9$ &    $27.9$ \\
  $20$ &   $66, 80$ &   $2.4, 2.8$ &   $-0.7, -0.4$ &     $1.89\times 10^{-1}$ &    $2.2$ &    $2.2$ &    $11.3$ &    $11.8$ &   $23$ &   $66, 80$ &   $2.4, 2.8$ &   $+0.4, +0.7$ &     $8.98\times 10^{-2}$ &    $2.7$ &    $3.1$ &    $17.5$ &    $18.0$ \\
  $21$ &   $66, 80$ &   $2.4, 2.8$ &   $-0.4, \phantom{-}0.0$ &     $8.40\times 10^{-2}$ &    $2.3$ &    $1.9$ &    $2.4$ &    $3.8$ &   $22$ &   $66, 80$ &   $2.4, 2.8$ &   $\phantom{+}0.0, +0.4$ &     $5.24\times 10^{-2}$ &    $2.8$ &    $2.7$ &    $3.9$ &    $5.5$ \\
\cmidrule(lr{19.5pt}){1-9}
\cmidrule(r{3pt}l){10-18}
  $25$ &   $66, 80$ &   $2.8, 3.6$ &   $-1.0, -0.7$ &     $2.14\times 10^{-2}$ &    $3.0$ &    $3.1$ &    $18.4$ &    $18.9$ &   $30$ &   $66, 80$ &   $2.8, 3.6$ &   $+0.7, +1.0$ &     $7.09\times 10^{-3}$ &    $3.8$ &    $7.4$ &    $32.7$ &    $33.8$ \\
  $26$ &   $66, 80$ &   $2.8, 3.6$ &   $-0.7, -0.4$ &     $4.84\times 10^{-2}$ &    $2.1$ &    $1.9$ &    $10.3$ &    $10.7$ &   $29$ &   $66, 80$ &   $2.8, 3.6$ &   $+0.4, +0.7$ &     $1.79\times 10^{-2}$ &    $2.8$ &    $3.2$ &    $15.2$ &    $15.7$ \\
  $27$ &   $66, 80$ &   $2.8, 3.6$ &   $-0.4, \phantom{-}0.0$ & $-$ & $-$ & $-$ & $-$ & $-$ &   $28$ &   $66, 80$ &   $2.8, 3.6$ &   $\phantom{+}0.0, +0.4$ & $-$ & $-$ & $-$ & $-$ & $-$ \\
\cmidrule(lr{19.5pt}){1-9}
\cmidrule(r{3pt}l){10-18}
  $31$ &   $80, 91$ &   $1.2, 1.6$ &   $-1.0, -0.7$ &     $5.20\times 10^{-1}$ &    $0.9$ &    $0.7$ &    $2.4$ &    $2.7$ &   $36$ &   $80, 91$ &   $1.2, 1.6$ &   $+0.7, +1.0$ &     $5.29\times 10^{-1}$ &    $0.9$ &    $0.7$ &    $2.6$ &    $2.8$ \\
  $32$ &   $80, 91$ &   $1.2, 1.6$ &   $-0.7, -0.4$ &     $3.07\times 10^{-2}$ &    $3.6$ &    $2.6$ &    $3.4$ &    $5.6$ &   $35$ &   $80, 91$ &   $1.2, 1.6$ &   $+0.4, +0.7$ &     $3.07\times 10^{-2}$ &    $3.5$ &    $2.8$ &    $3.6$ &    $5.8$ \\
  $33$ &   $80, 91$ &   $1.2, 1.6$ &   $-0.4, \phantom{-}0.0$ & $-$ & $-$ & $-$ & $-$ & $-$ &   $34$ &   $80, 91$ &   $1.2, 1.6$ &   $\phantom{+}0.0, +0.4$ & $-$ & $-$ & $-$ & $-$ & $-$ \\
\cmidrule(lr{19.5pt}){1-9}
\cmidrule(r{3pt}l){10-18}
  $37$ &   $80, 91$ &   $1.6, 2.0$ &   $-1.0, -0.7$ &     $1.26$ &    $0.7$ &    $0.4$ &    $2.7$ &    $2.8$ &   $42$ &   $80, 91$ &   $1.6, 2.0$ &   $+0.7, +1.0$ &     $1.28$ &    $0.6$ &    $0.4$ &    $2.6$ &    $2.7$ \\
  $38$ &   $80, 91$ &   $1.6, 2.0$ &   $-0.7, -0.4$ &     $1.04$ &    $0.7$ &    $0.5$ &    $1.7$ &    $1.9$ &   $41$ &   $80, 91$ &   $1.6, 2.0$ &   $+0.4, +0.7$ &     $1.04$ &    $0.7$ &    $0.5$ &    $1.7$ &    $1.9$ \\
  $39$ &   $80, 91$ &   $1.6, 2.0$ &   $-0.4, \phantom{-}0.0$ &     $5.59\times 10^{-3}$ &    $7.9$ &    $4.4$ &    $4.4$ &    $10.0$ &   $40$ &   $80, 91$ &   $1.6, 2.0$ &   $\phantom{+}0.0, +0.4$ &     $5.05\times 10^{-3}$ &    $7.4$ &    $4.6$ &    $3.5$ &    $9.4$ \\
\cmidrule(lr{19.5pt}){1-9}
\cmidrule(r{3pt}l){10-18}
  $43$ &   $80, 91$ &   $2.0, 2.4$ &   $-1.0, -0.7$ &     $1.28$ &    $0.9$ &    $0.8$ &    $2.7$ &    $2.9$ &   $48$ &   $80, 91$ &   $2.0, 2.4$ &   $+0.7, +1.0$ &     $1.24$ &    $0.9$ &    $0.7$ &    $2.6$ &    $2.9$ \\
  $44$ &   $80, 91$ &   $2.0, 2.4$ &   $-0.7, -0.4$ &     $2.71$ &    $0.5$ &    $0.4$ &    $1.4$ &    $1.5$ &   $47$ &   $80, 91$ &   $2.0, 2.4$ &   $+0.4, +0.7$ &     $2.66$ &    $0.5$ &    $0.4$ &    $1.5$ &    $1.6$ \\
  $45$ &   $80, 91$ &   $2.0, 2.4$ &   $-0.4, \phantom{-}0.0$ &     $7.32\times 10^{-1}$ &    $0.6$ &    $0.6$ &    $1.8$ &    $2.0$ &   $46$ &   $80, 91$ &   $2.0, 2.4$ &   $\phantom{+}0.0, +0.4$ &     $7.23\times 10^{-1}$ &    $0.6$ &    $0.6$ &    $1.9$ &    $2.1$ \\
\cmidrule(lr{19.5pt}){1-9}
\cmidrule(r{3pt}l){10-18}
  $49$ &   $80, 91$ &   $2.4, 2.8$ &   $-1.0, -0.7$ &     $1.14$ &    $0.7$ &    $0.6$ &    $3.2$ &    $3.4$ &   $54$ &   $80, 91$ &   $2.4, 2.8$ &   $+0.7, +1.0$ &     $1.14$ &    $0.6$ &    $0.6$ &    $2.7$ &    $2.9$ \\
  $50$ &   $80, 91$ &   $2.4, 2.8$ &   $-0.7, -0.4$ &     $2.55$ &    $0.6$ &    $0.5$ &    $2.4$ &    $2.5$ &   $53$ &   $80, 91$ &   $2.4, 2.8$ &   $+0.4, +0.7$ &     $2.52$ &    $0.6$ &    $0.5$ &    $2.5$ &    $2.6$ \\
  $51$ &   $80, 91$ &   $2.4, 2.8$ &   $-0.4, \phantom{-}0.0$ &     $1.10$ &    $0.6$ &    $0.5$ &    $1.7$ &    $1.9$ &   $52$ &   $80, 91$ &   $2.4, 2.8$ &   $\phantom{+}0.0, +0.4$ &     $1.10$ &    $0.6$ &    $0.5$ &    $1.7$ &    $1.9$ \\
\cmidrule(lr{19.5pt}){1-9}
\cmidrule(r{3pt}l){10-18}
  $55$ &   $80, 91$ &   $2.8, 3.6$ &   $-1.0, -0.7$ &     $6.05\times 10^{-1}$ &    $0.6$ &    $0.6$ &    $4.1$ &    $4.2$ &   $60$ &   $80, 91$ &   $2.8, 3.6$ &   $+0.7, +1.0$ &     $5.97\times 10^{-1}$ &    $0.6$ &    $0.6$ &    $4.0$ &    $4.1$ \\
  $56$ &   $80, 91$ &   $2.8, 3.6$ &   $-0.7, -0.4$ &     $5.95\times 10^{-1}$ &    $0.5$ &    $0.5$ &    $4.6$ &    $4.6$ &   $59$ &   $80, 91$ &   $2.8, 3.6$ &   $+0.4, +0.7$ &     $5.84\times 10^{-1}$ &    $0.5$ &    $0.5$ &    $4.5$ &    $4.5$ \\
  $57$ &   $80, 91$ &   $2.8, 3.6$ &   $-0.4, \phantom{-}0.0$ & $-$ & $-$ & $-$ & $-$ & $-$ &   $58$ &   $80, 91$ &   $2.8, 3.6$ &   $\phantom{+}0.0, +0.4$ & $-$ & $-$ & $-$ & $-$ & $-$ \\
\cmidrule(lr{19.5pt}){1-9}
\cmidrule(r{3pt}l){10-18}
  $61$ &   $\phantom{1}91, 102$ &   $1.2, 1.6$ &   $-1.0, -0.7$ &     $6.53\times 10^{-1}$ &    $0.8$ &    $0.6$ &    $2.2$ &    $2.4$ &   $66$ &   $\phantom{1}91, 102$ &   $1.2, 1.6$ &   $+0.7, +1.0$ &     $7.55\times 10^{-1}$ &    $0.8$ &    $0.6$ &    $2.3$ &    $2.5$ \\
  $62$ &   $\phantom{1}91, 102$ &   $1.2, 1.6$ &   $-0.7, -0.4$ &     $3.64\times 10^{-2}$ &    $3.5$ &    $2.8$ &    $2.7$ &    $5.2$ &   $65$ &   $\phantom{1}91, 102$ &   $1.2, 1.6$ &   $+0.4, +0.7$ &     $3.88\times 10^{-2}$ &    $3.5$ &    $3.0$ &    $3.2$ &    $5.6$ \\
  $63$ &   $\phantom{1}91, 102$ &   $1.2, 1.6$ &   $-0.4, \phantom{-}0.0$ & $-$ & $-$ & $-$ & $-$ & $-$ &   $64$ &   $\phantom{1}91, 102$ &   $1.2, 1.6$ &   $\phantom{+}0.0, +0.4$ & $-$ & $-$ & $-$ & $-$ & $-$ \\
\cmidrule(lr{19.5pt}){1-9}
\cmidrule(r{3pt}l){10-18}
  $67$ &   $\phantom{1}91, 102$ &   $1.6, 2.0$ &   $-1.0, -0.7$ &     $1.52$ &    $0.6$ &    $0.4$ &    $2.2$ &    $2.3$ &   $72$ &   $\phantom{1}91, 102$ &   $1.6, 2.0$ &   $+0.7, +1.0$ &     $1.78$ &    $0.6$ &    $0.4$ &    $2.1$ &    $2.3$ \\
  $68$ &   $\phantom{1}91, 102$ &   $1.6, 2.0$ &   $-0.7, -0.4$ &     $1.13$ &    $0.7$ &    $0.5$ &    $1.5$ &    $1.7$ &   $71$ &   $\phantom{1}91, 102$ &   $1.6, 2.0$ &   $+0.4, +0.7$ &     $1.29$ &    $0.7$ &    $0.4$ &    $1.4$ &    $1.6$ \\
  $69$ &   $\phantom{1}91, 102$ &   $1.6, 2.0$ &   $-0.4, \phantom{-}0.0$ &     $6.15\times 10^{-3}$ &    $7.6$ &    $6.1$ &    $6.4$ &    $11.6$ &   $70$ &   $\phantom{1}91, 102$ &   $1.6, 2.0$ &   $\phantom{+}0.0, +0.4$ &     $5.73\times 10^{-3}$ &    $8.0$ &    $6.7$ &    $6.3$ &    $12.2$ \\
\cmidrule(lr{19.5pt}){1-9}
\cmidrule(r{3pt}l){10-18}
  $73$ &   $\phantom{1}91, 102$ &   $2.0, 2.4$ &   $-1.0, -0.7$ &     $1.48$ &    $0.9$ &    $0.7$ &    $2.3$ &    $2.6$ &   $78$ &   $\phantom{1}91, 102$ &   $2.0, 2.4$ &   $+0.7, +1.0$ &     $1.81$ &    $0.8$ &    $0.6$ &    $2.2$ &    $2.4$ \\
  $74$ &   $\phantom{1}91, 102$ &   $2.0, 2.4$ &   $-0.7, -0.4$ &     $2.94$ &    $0.5$ &    $0.3$ &    $1.6$ &    $1.7$ &   $77$ &   $\phantom{1}91, 102$ &   $2.0, 2.4$ &   $+0.4, +0.7$ &     $3.48$ &    $0.4$ &    $0.3$ &    $1.5$ &    $1.6$ \\
  $75$ &   $\phantom{1}91, 102$ &   $2.0, 2.4$ &   $-0.4, \phantom{-}0.0$ &     $8.06\times 10^{-1}$ &    $0.7$ &    $0.6$ &    $1.6$ &    $1.9$ &   $76$ &   $\phantom{1}91, 102$ &   $2.0, 2.4$ &   $\phantom{+}0.0, +0.4$ &     $8.97\times 10^{-1}$ &    $0.6$ &    $0.6$ &    $1.5$ &    $1.7$ \\
\cmidrule(lr{19.5pt}){1-9}
\cmidrule(r{3pt}l){10-18}
  $79$ &   $\phantom{1}91, 102$ &   $2.4, 2.8$ &   $-1.0, -0.7$ &     $1.25$ &    $0.7$ &    $0.6$ &    $2.3$ &    $2.5$ &   $84$ &   $\phantom{1}91, 102$ &   $2.4, 2.8$ &   $+0.7, +1.0$ &     $1.65$ &    $0.6$ &    $0.5$ &    $2.3$ &    $2.4$ \\
  $80$ &   $\phantom{1}91, 102$ &   $2.4, 2.8$ &   $-0.7, -0.4$ &     $2.63$ &    $0.6$ &    $0.5$ &    $2.0$ &    $2.1$ &   $83$ &   $\phantom{1}91, 102$ &   $2.4, 2.8$ &   $+0.4, +0.7$ &     $3.33$ &    $0.5$ &    $0.5$ &    $1.8$ &    $2.0$ \\
  $81$ &   $\phantom{1}91, 102$ &   $2.4, 2.8$ &   $-0.4, \phantom{-}0.0$ &     $1.24$ &    $0.6$ &    $1.0$ &    $1.7$ &    $2.1$ &   $82$ &   $\phantom{1}91, 102$ &   $2.4, 2.8$ &   $\phantom{+}0.0, +0.4$ &     $1.41$ &    $0.6$ &    $0.8$ &    $1.5$ &    $1.8$ \\
\cmidrule(lr{19.5pt}){1-9}
\cmidrule(r{3pt}l){10-18}
  $85$ &   $\phantom{1}91, 102$ &   $2.8, 3.6$ &   $-1.0, -0.7$ &     $6.54\times 10^{-1}$ &    $0.6$ &    $0.6$ &    $3.0$ &    $3.1$ &   $90$ &   $\phantom{1}91, 102$ &   $2.8, 3.6$ &   $+0.7, +1.0$ &     $9.09\times 10^{-1}$ &    $0.5$ &    $0.6$ &    $2.9$ &    $3.0$ \\
  $86$ &   $\phantom{1}91, 102$ &   $2.8, 3.6$ &   $-0.7, -0.4$ &     $5.66\times 10^{-1}$ &    $0.6$ &    $0.5$ &    $3.3$ &    $3.4$ &   $89$ &   $\phantom{1}91, 102$ &   $2.8, 3.6$ &   $+0.4, +0.7$ &     $7.57\times 10^{-1}$ &    $0.5$ &    $0.4$ &    $2.5$ &    $2.6$ \\
  $87$ &   $\phantom{1}91, 102$ &   $2.8, 3.6$ &   $-0.4, \phantom{-}0.0$ & $-$ & $-$ & $-$ & $-$ & $-$ &   $88$ &   $\phantom{1}91, 102$ &   $2.8, 3.6$ &   $\phantom{+}0.0, +0.4$ & $-$ & $-$ & $-$ & $-$ & $-$ \\
\cmidrule(lr{19.5pt}){1-9}
\cmidrule(r{3pt}l){10-18}
  $91$ &   $102, 116$ &   $1.2, 1.6$ &   $-1.0, -0.7$ &     $3.09\times 10^{-2}$ &    $4.6$ &    $6.3$ &    $11.4$ &    $13.8$ &   $96$ &   $102, 116$ &   $1.2, 1.6$ &   $+0.7, +1.0$ &     $5.49\times 10^{-2}$ &    $3.3$ &    $4.0$ &    $6.5$ &    $8.3$ \\
  $92$ &   $102, 116$ &   $1.2, 1.6$ &   $-0.7, -0.4$ &     $1.06\times 10^{-3}$ &    $23.7$ &    $22.9$ &    $33.4$ &    $46.9$ &   $95$ &   $102, 116$ &   $1.2, 1.6$ &   $+0.4, +0.7$ &     $1.64\times 10^{-3}$ &    $20.6$ &    $29.7$ &    $24.2$ &    $43.5$ \\
  $93$ &   $102, 116$ &   $1.2, 1.6$ &   $-0.4, \phantom{-}0.0$ & $-$ & $-$ & $-$ & $-$ & $-$ &   $94$ &   $102, 116$ &   $1.2, 1.6$ &   $\phantom{+}0.0, +0.4$ & $-$ & $-$ & $-$ & $-$ & $-$ \\
\cmidrule(lr{19.5pt}){1-9}
\cmidrule(r{3pt}l){10-18}
  $97$ &   $102, 116$ &   $1.6, 2.0$ &   $-1.0, -0.7$ &     $5.20\times 10^{-2}$ &    $3.8$ &    $4.0$ &    $8.5$ &    $10.1$ &   $102$ &   $102, 116$ &   $1.6, 2.0$ &   $+0.7, +1.0$ &     $1.05\times 10^{-1}$ &    $2.6$ &    $2.4$ &    $5.0$ &    $6.2$ \\
  $98$ &   $102, 116$ &   $1.6, 2.0$ &   $-0.7, -0.4$ &     $3.15\times 10^{-2}$ &    $4.4$ &    $4.4$ &    $4.9$ &    $8.0$ &   $101$ &   $102, 116$ &   $1.6, 2.0$ &   $+0.4, +0.7$ &     $5.32\times 10^{-2}$ &    $3.3$ &    $3.0$ &    $3.5$ &    $5.7$ \\
  $99$ &   $102, 116$ &   $1.6, 2.0$ &   $-0.4, \phantom{-}0.0$ & $-$ & $-$ & $-$ & $-$ & $-$ &   $100$ &   $102, 116$ &   $1.6, 2.0$ &   $\phantom{+}0.0, +0.4$ & $-$ & $-$ & $-$ & $-$ & $-$ \\
\cmidrule(lr{19.5pt}){1-9}
\cmidrule(r{3pt}l){10-18}
  $103$ &   $102, 116$ &   $2.0, 2.4$ &   $-1.0, -0.7$ &     $4.24\times 10^{-2}$ &    $5.3$ &    $5.0$ &    $15.5$ &    $17.1$ &   $108$ &   $102, 116$ &   $2.0, 2.4$ &   $+0.7, +1.0$ &     $1.06\times 10^{-1}$ &    $3.2$ &    $3.1$ &    $7.8$ &    $9.0$ \\
  $104$ &   $102, 116$ &   $2.0, 2.4$ &   $-0.7, -0.4$ &     $6.96\times 10^{-2}$ &    $3.2$ &    $3.5$ &    $5.9$ &    $7.6$ &   $107$ &   $102, 116$ &   $2.0, 2.4$ &   $+0.4, +0.7$ &     $1.38\times 10^{-1}$ &    $2.4$ &    $2.2$ &    $4.4$ &    $5.4$ \\
  $105$ &   $102, 116$ &   $2.0, 2.4$ &   $-0.4, \phantom{-}0.0$ &     $2.42\times 10^{-2}$ &    $4.0$ &    $4.4$ &    $5.9$ &    $8.4$ &   $106$ &   $102, 116$ &   $2.0, 2.4$ &   $\phantom{+}0.0, +0.4$ &     $3.15\times 10^{-2}$ &    $3.5$ &    $3.5$ &    $4.0$ &    $6.3$ \\
\cmidrule(lr{19.5pt}){1-9}
\cmidrule(r{3pt}l){10-18}
  $109$ &   $102, 116$ &   $2.4, 2.8$ &   $-1.0, -0.7$ &     $3.05\times 10^{-2}$ &    $3.8$ &    $4.6$ &    $22.0$ &    $22.8$ &   $114$ &   $102, 116$ &   $2.4, 2.8$ &   $+0.7, +1.0$ &     $9.69\times 10^{-2}$ &    $2.2$ &    $2.4$ &    $10.5$ &    $11.0$ \\
  $110$ &   $102, 116$ &   $2.4, 2.8$ &   $-0.7, -0.4$ &     $5.28\times 10^{-2}$ &    $4.3$ &    $4.5$ &    $11.4$ &    $12.9$ &   $113$ &   $102, 116$ &   $2.4, 2.8$ &   $+0.4, +0.7$ &     $1.36\times 10^{-1}$ &    $2.7$ &    $2.6$ &    $8.4$ &    $9.2$ \\
  $111$ &   $102, 116$ &   $2.4, 2.8$ &   $-0.4, \phantom{-}0.0$ &     $3.06\times 10^{-2}$ &    $4.0$ &    $5.1$ &    $6.9$ &    $9.5$ &   $112$ &   $102, 116$ &   $2.4, 2.8$ &   $\phantom{+}0.0, +0.4$ &     $5.19\times 10^{-2}$ &    $3.1$ &    $3.3$ &    $4.8$ &    $6.6$ \\
\cmidrule(lr{19.5pt}){1-9}
\cmidrule(r{3pt}l){10-18}
  $115$ &   $102, 116$ &   $2.8, 3.6$ &   $-1.0, -0.7$ &     $1.44\times 10^{-2}$ &    $3.4$ &    $3.9$ &    $35.9$ &    $36.3$ &   $120$ &   $102, 116$ &   $2.8, 3.6$ &   $+0.7, +1.0$ &     $5.18\times 10^{-2}$ &    $2.0$ &    $1.8$ &    $19.3$ &    $19.4$ \\
  $116$ &   $102, 116$ &   $2.8, 3.6$ &   $-0.7, -0.4$ &     $1.00\times 10^{-2}$ &    $4.2$ &    $4.6$ &    $24.8$ &    $25.5$ &   $119$ &   $102, 116$ &   $2.8, 3.6$ &   $+0.4, +0.7$ &     $3.27\times 10^{-2}$ &    $2.4$ &    $2.1$ &    $14.2$ &    $14.6$ \\
  $117$ &   $102, 116$ &   $2.8, 3.6$ &   $-0.4, \phantom{-}0.0$ & $-$ & $-$ & $-$ & $-$ & $-$ &   $118$ &   $102, 116$ &   $2.8, 3.6$ &   $\phantom{+}0.0, +0.4$ & $-$ & $-$ & $-$ & $-$ & $-$ \\
\bottomrule
\end{tabular}
}
\end{table}

\begin{table}[htp!]
\tiny
\centering
\scalebox{0.85}{
\setlength\tabcolsep{3pt}
\setlength\extrarowheight{2pt}
\begin{tabular}{C{0.6cm}C{0.9cm}C{0.9cm}C{1.1cm}C{1.6cm}L{0.5cm}L{0.45cm}L{0.45cm}L{0.45cm} @{\hskip 20pt} C{0.6cm}C{0.9cm}C{0.9cm}C{1.1cm}C{1.6cm}L{0.5cm}L{0.45cm}L{0.45cm}L{0.45cm}}
\toprule
Bin
& $m_{ee}$
& $|y_{ee}|$
& $\cos\theta^*$
& $\textrm{d}^3\sigma$
& $\delta^{\textrm {stat}}$
& $\delta^{\textrm {syst}}_{\textrm {unc}}$
& $\delta^{\textrm {syst}}_{\textrm {cor}}$
& $\delta^{\textrm {total}}$
& Bin
& $m_{ee}$
& $|y_{ee}|$
& $\cos\theta^*$
& $\textrm{d}^3\sigma$
& $\delta^{\textrm {stat}}$
& $\delta^{\textrm {syst}}_{\textrm {unc}}$
& $\delta^{\textrm {syst}}_{\textrm {cor}}$
& $\delta^{\textrm {total}}$ \\
&	[GeV] & & & [pb/GeV] & [\%] & [\%] & [\%] & [\%] & & [GeV] & & & [pb/GeV] & [\%] & [\%] & [\%] & [\%]\\
\noalign{\vskip 0.05cm}
\cmidrule(lr{19.5pt}){1-9}
\cmidrule(r{3pt}l){10-18}
  $121$ &   $116, 150$ &   $1.2, 1.6$ &   $-1.0, -0.7$ &     $6.38\times 10^{-3}$ &    $9.4$ &    $13.8$ &    $36.2$ &    $39.9$ &   $126$ &   $116, 150$ &   $1.2, 1.6$ &   $+0.7, +1.0$ &     $1.65\times 10^{-2}$ &    $4.7$ &    $5.6$ &    $14.4$ &    $16.1$ \\
  $122$ &   $116, 150$ &   $1.2, 1.6$ &   $-0.7, -0.4$ & $-$ & $-$ & $-$ & $-$ & $-$ &   $125$ &   $116, 150$ &   $1.2, 1.6$ &   $+0.4, +0.7$ &     $4.05\times 10^{-4}$ &    $33.3$ &    $55.4$ &    $43.5$ &    $78.0$ \\
  $123$ &   $116, 150$ &   $1.2, 1.6$ &   $-0.4, \phantom{-}0.0$ & $-$ & $-$ & $-$ & $-$ & $-$ &   $124$ &   $116, 150$ &   $1.2, 1.6$ &   $\phantom{+}0.0, +0.4$ & $-$ & $-$ & $-$ & $-$ & $-$ \\
\cmidrule(lr{19.5pt}){1-9}
\cmidrule(r{3pt}l){10-18}
  $127$ &   $116, 150$ &   $1.6, 2.0$ &   $-1.0, -0.7$ &     $9.56\times 10^{-3}$ &    $8.4$ &    $11.4$ &    $29.7$ &    $32.9$ &   $132$ &   $116, 150$ &   $1.6, 2.0$ &   $+0.7, +1.0$ &     $2.91\times 10^{-2}$ &    $3.9$ &    $6.2$ &    $9.9$ &    $12.3$ \\
  $128$ &   $116, 150$ &   $1.6, 2.0$ &   $-0.7, -0.4$ &     $4.62\times 10^{-3}$ &    $7.6$ &    $10.1$ &    $9.1$ &    $15.6$ &   $131$ &   $116, 150$ &   $1.6, 2.0$ &   $+0.4, +0.7$ &     $1.16\times 10^{-2}$ &    $4.4$ &    $5.1$ &    $3.3$ &    $7.5$ \\
  $129$ &   $116, 150$ &   $1.6, 2.0$ &   $-0.4, \phantom{-}0.0$ & $-$ & $-$ & $-$ & $-$ & $-$ &   $130$ &   $116, 150$ &   $1.6, 2.0$ &   $\phantom{+}0.0, +0.4$ & $-$ & $-$ & $-$ & $-$ & $-$ \\
\cmidrule(lr{19.5pt}){1-9}
\cmidrule(r{3pt}l){10-18}
  $133$ &   $116, 150$ &   $2.0, 2.4$ &   $-1.0, -0.7$ &     $7.42\times 10^{-3}$ &    $11.5$ &    $16.3$ &    $46.1$ &    $50.3$ &   $138$ &   $116, 150$ &   $2.0, 2.4$ &   $+0.7, +1.0$ &     $3.00\times 10^{-2}$ &    $4.1$ &    $5.2$ &    $11.3$ &    $13.1$ \\
  $134$ &   $116, 150$ &   $2.0, 2.4$ &   $-0.7, -0.4$ &     $1.21\times 10^{-2}$ &    $5.5$ &    $6.4$ &    $11.2$ &    $14.0$ &   $137$ &   $116, 150$ &   $2.0, 2.4$ &   $+0.4, +0.7$ &     $3.03\times 10^{-2}$ &    $3.6$ &    $3.8$ &    $4.7$ &    $7.0$ \\
  $135$ &   $116, 150$ &   $2.0, 2.4$ &   $-0.4, \phantom{-}0.0$ &     $3.48\times 10^{-3}$ &    $7.2$ &    $12.2$ &    $14.1$ &    $20.1$ &   $136$ &   $116, 150$ &   $2.0, 2.4$ &   $\phantom{+}0.0, +0.4$ &     $7.29\times 10^{-3}$ &    $4.7$ &    $5.9$ &    $5.6$ &    $9.4$ \\
\cmidrule(lr{19.5pt}){1-9}
\cmidrule(r{3pt}l){10-18}
  $139$ &   $116, 150$ &   $2.4, 2.8$ &   $-1.0, -0.7$ &     $3.36\times 10^{-3}$ &    $16.2$ &    $18.3$ &    $30.4$ &    $39.0$ &   $144$ &   $116, 150$ &   $2.4, 2.8$ &   $+0.7, +1.0$ &     $2.87\times 10^{-2}$ &    $3.3$ &    $3.0$ &    $4.3$ &    $6.2$ \\
  $140$ &   $116, 150$ &   $2.4, 2.8$ &   $-0.7, -0.4$ &     $8.13\times 10^{-3}$ &    $7.6$ &    $10.9$ &    $11.6$ &    $17.6$ &   $143$ &   $116, 150$ &   $2.4, 2.8$ &   $+0.4, +0.7$ &     $2.88\times 10^{-2}$ &    $3.8$ &    $3.8$ &    $3.9$ &    $6.7$ \\
  $141$ &   $116, 150$ &   $2.4, 2.8$ &   $-0.4, \phantom{-}0.0$ &     $5.61\times 10^{-3}$ &    $5.9$ &    $7.0$ &    $7.1$ &    $11.6$ &   $142$ &   $116, 150$ &   $2.4, 2.8$ &   $\phantom{+}0.0, +0.4$ &     $1.05\times 10^{-2}$ &    $4.5$ &    $4.7$ &    $3.8$ &    $7.5$ \\
\cmidrule(lr{19.5pt}){1-9}
\cmidrule(r{3pt}l){10-18}
  $145$ &   $116, 150$ &   $2.8, 3.6$ &   $-1.0, -0.7$ &     $1.68\times 10^{-3}$ &    $11.1$ &    $13.6$ &    $32.3$ &    $36.8$ &   $150$ &   $116, 150$ &   $2.8, 3.6$ &   $+0.7, +1.0$ &     $1.21\times 10^{-2}$ &    $3.1$ &    $3.6$ &    $6.8$ &    $8.3$ \\
  $146$ &   $116, 150$ &   $2.8, 3.6$ &   $-0.7, -0.4$ &     $1.39\times 10^{-3}$ &    $8.8$ &    $11.4$ &    $14.1$ &    $20.1$ &   $149$ &   $116, 150$ &   $2.8, 3.6$ &   $+0.4, +0.7$ &     $6.41\times 10^{-3}$ &    $3.7$ &    $4.0$ &    $3.8$ &    $6.6$ \\
  $147$ &   $116, 150$ &   $2.8, 3.6$ &   $-0.4, \phantom{-}0.0$ & $-$ & $-$ & $-$ & $-$ & $-$ &   $148$ &   $116, 150$ &   $2.8, 3.6$ &   $\phantom{+}0.0, +0.4$ & $-$ & $-$ & $-$ & $-$ & $-$ \\
\bottomrule
\end{tabular}
}
\caption{The high rapidity electron channel Born-level triple-differential cross section
${\textrm{d}}^3\sigma/{\textrm{d}}m_{\ell\ell}{\textrm{d}}|y_{\ell\ell}|{\textrm{d}}\cos\theta^*$.
The measurements are listed together with the statistical ($\delta^{\textrm{stat}}$),
uncorrelated systematic ($\delta^{\textrm {syst}}_{\textrm {unc}}$),
correlated systematic ($\delta^{\textrm{syst}}_{\textrm{cor}}$),
and total ($\delta^{\textrm{total}}$) uncertainties. The luminosity
uncertainty of 1.9\% is not shown and not included in the overall
systematic and total uncertainties.}
\label{tab:zcf_summary}
\end{table}
\FloatBarrier

\subsection{Forward-backward asymmetry tables}
\label{tab:afb}
\begin{table}[htp!]
\tiny
\centering
\scalebox{0.8}{
\setlength\tabcolsep{2.5pt}
\setlength\extrarowheight{2.5pt}
\begin{tabular}{C{1.1cm}C{1.1cm}C{1.6cm}C{1.15cm}C{1.15cm}C{1.15cm}C{1.15cm} @{\hskip 15pt} C{1.1cm}C{1.1cm}C{1.6cm}C{1.15cm}C{1.15cm}C{1.15cm}C{1.15cm}}
\toprule
$|y_{\ell\ell}|$
& $m_{\ell\ell}$
& $A_{\textrm {FB}}$
& $\Delta^{\textrm {stat}}$
& $\Delta^{\textrm {syst}}_{\textrm {unc}}$
& $\Delta^{\textrm {syst}}_{\textrm {cor}}$
& $\Delta^{\textrm {total}}$
& $|y_{\ell\ell}|$
& $m_{\ell\ell}$
& $A_{\textrm {FB}}$
& $\Delta^{\textrm {stat}}$
& $\Delta^{\textrm {syst}}_{\textrm {unc}}$
& $\Delta^{\textrm {syst}}_{\textrm {cor}}$
& $\Delta^{\textrm {total}}$ \\
&	[GeV] & & & & & & & [GeV] & & & & & \\
\noalign{\vskip 0.05cm}
\cmidrule(lr{17.7pt}){1-7}
\cmidrule(l{2.5pt}r){8-14}
  $0.0, 0.2$ &   $46, 66$ &     $-5.97\times 10^{-3}$ &    $5.6\times 10^{-3}$ &    $4.5\times 10^{-3}$ &    $7.2\times 10^{-4}$ &    $7.3\times 10^{-3}$ &   $0.2, 0.4$ &   $46, 66$ &     $-8.12\times 10^{-3}$ &    $5.7\times 10^{-3}$ &    $4.7\times 10^{-3}$ &    $8.7\times 10^{-4}$ &    $7.4\times 10^{-3}$ \\
  $0.0, 0.2$ &   $66, 80$ &     $-2.48\times 10^{-4}$ &    $4.2\times 10^{-3}$ &    $2.3\times 10^{-3}$ &    $7.5\times 10^{-4}$ &    $4.8\times 10^{-3}$ &   $0.2, 0.4$ &   $66, 80$ &     $-2.61\times 10^{-2}$ &    $4.2\times 10^{-3}$ &    $2.3\times 10^{-3}$ &    $6.7\times 10^{-4}$ &    $4.9\times 10^{-3}$ \\
  $0.0, 0.2$ &   $80, 91$ & $\phantom{-}8.26\times 10^{-4}$ &    $1.1\times 10^{-3}$ &    $5.6\times 10^{-4}$ &    $3.5\times 10^{-4}$ &    $1.3\times 10^{-3}$ &   $0.2, 0.4$ &   $80, 91$ & $\phantom{-}1.85\times 10^{-3}$ &    $1.1\times 10^{-3}$ &    $5.5\times 10^{-4}$ &    $5.5\times 10^{-4}$ &    $1.4\times 10^{-3}$ \\
  $0.0, 0.2$ &   $\phantom{1}91, 102$ & $\phantom{-}3.78\times 10^{-3}$ &    $1.1\times 10^{-3}$ &    $5.1\times 10^{-4}$ &    $3.2\times 10^{-4}$ &    $1.2\times 10^{-3}$ &   $0.2, 0.4$ &   $\phantom{1}91, 102$ & $\phantom{-}6.49\times 10^{-3}$ &    $1.1\times 10^{-3}$ &    $5.0\times 10^{-4}$ &    $5.1\times 10^{-4}$ &    $1.3\times 10^{-3}$ \\
  $0.0, 0.2$ &   $102, 116$ & $\phantom{-}1.29\times 10^{-2}$ &    $5.3\times 10^{-3}$ &    $2.5\times 10^{-3}$ &    $9.1\times 10^{-4}$ &    $5.9\times 10^{-3}$ &   $0.2, 0.4$ &   $102, 116$ & $\phantom{-}2.59\times 10^{-2}$ &    $5.3\times 10^{-3}$ &    $2.5\times 10^{-3}$ &    $1.1\times 10^{-3}$ &    $5.9\times 10^{-3}$ \\
  $0.0, 0.2$ &   $116, 150$ & $\phantom{-}2.60\times 10^{-2}$ &    $8.3\times 10^{-3}$ &    $3.3\times 10^{-3}$ &    $1.3\times 10^{-3}$ &    $9.0\times 10^{-3}$ &   $0.2, 0.4$ &   $116, 150$ & $\phantom{-}5.69\times 10^{-2}$ &    $8.4\times 10^{-3}$ &    $3.6\times 10^{-3}$ &    $1.5\times 10^{-3}$ &    $9.2\times 10^{-3}$ \\
  $0.0, 0.2$ &   $150, 200$ & $\phantom{-}1.88\times 10^{-2}$ &    $1.6\times 10^{-2}$ &    $5.8\times 10^{-3}$ &    $2.9\times 10^{-3}$ &    $1.7\times 10^{-2}$ &   $0.2, 0.4$ &   $150, 200$ & $\phantom{-}5.11\times 10^{-2}$ &    $1.6\times 10^{-2}$ &    $6.9\times 10^{-3}$ &    $4.2\times 10^{-3}$ &    $1.8\times 10^{-2}$ \\
\cmidrule(lr{17.7pt}){1-7}
\cmidrule(l{2.5pt}r){8-14}
  $0.4, 0.6$ &   $46, 66$ &     $-2.71\times 10^{-2}$ &    $5.9\times 10^{-3}$ &    $4.5\times 10^{-3}$ &    $9.1\times 10^{-4}$ &    $7.5\times 10^{-3}$ &   $0.6, 0.8$ &   $46, 66$ &     $-2.96\times 10^{-2}$ &    $5.7\times 10^{-3}$ &    $4.3\times 10^{-3}$ &    $9.5\times 10^{-4}$ &    $7.2\times 10^{-3}$ \\
  $0.4, 0.6$ &   $66, 80$ &     $-2.76\times 10^{-2}$ &    $4.3\times 10^{-3}$ &    $2.3\times 10^{-3}$ &    $7.6\times 10^{-4}$ &    $4.9\times 10^{-3}$ &   $0.6, 0.8$ &   $66, 80$ &     $-5.27\times 10^{-2}$ &    $4.4\times 10^{-3}$ &    $2.5\times 10^{-3}$ &    $9.2\times 10^{-4}$ &    $5.1\times 10^{-3}$ \\
  $0.4, 0.6$ &   $80, 91$ & $\phantom{-}2.61\times 10^{-3}$ &    $1.1\times 10^{-3}$ &    $5.5\times 10^{-4}$ &    $6.0\times 10^{-4}$ &    $1.4\times 10^{-3}$ &   $0.6, 0.8$ &   $80, 91$ & $\phantom{-}1.48\times 10^{-3}$ &    $1.1\times 10^{-3}$ &    $5.6\times 10^{-4}$ &    $6.6\times 10^{-4}$ &    $1.4\times 10^{-3}$ \\
  $0.4, 0.6$ &   $\phantom{1}91, 102$ & $\phantom{-}1.23\times 10^{-2}$ &    $1.1\times 10^{-3}$ &    $5.2\times 10^{-4}$ &    $5.7\times 10^{-4}$ &    $1.3\times 10^{-3}$ &   $0.6, 0.8$ &   $\phantom{1}91, 102$ & $\phantom{-}1.74\times 10^{-2}$ &    $1.1\times 10^{-3}$ &    $5.1\times 10^{-4}$ &    $6.4\times 10^{-4}$ &    $1.4\times 10^{-3}$ \\
  $0.4, 0.6$ &   $102, 116$ & $\phantom{-}4.39\times 10^{-2}$ &    $5.2\times 10^{-3}$ &    $2.6\times 10^{-3}$ &    $1.2\times 10^{-3}$ &    $6.0\times 10^{-3}$ &   $0.6, 0.8$ &   $102, 116$ & $\phantom{-}5.37\times 10^{-2}$ &    $5.2\times 10^{-3}$ &    $2.7\times 10^{-3}$ &    $1.7\times 10^{-3}$ &    $6.1\times 10^{-3}$ \\
  $0.4, 0.6$ &   $116, 150$ & $\phantom{-}7.39\times 10^{-2}$ &    $8.5\times 10^{-3}$ &    $3.3\times 10^{-3}$ &    $1.7\times 10^{-3}$ &    $9.3\times 10^{-3}$ &   $0.6, 0.8$ &   $116, 150$ & $\phantom{-}1.06\times 10^{-1}$ &    $8.3\times 10^{-3}$ &    $3.1\times 10^{-3}$ &    $1.8\times 10^{-3}$ &    $9.1\times 10^{-3}$ \\
  $0.4, 0.6$ &   $150, 200$ & $\phantom{-}1.01\times 10^{-1}$ &    $1.6\times 10^{-2}$ &    $5.8\times 10^{-3}$ &    $2.7\times 10^{-3}$ &    $1.7\times 10^{-2}$ &   $0.6, 0.8$ &   $150, 200$ & $\phantom{-}1.37\times 10^{-1}$ &    $1.6\times 10^{-2}$ &    $5.3\times 10^{-3}$ &    $3.2\times 10^{-3}$ &    $1.7\times 10^{-2}$ \\
\cmidrule(lr{17.7pt}){1-7}
\cmidrule(l{2.5pt}r){8-14}
  $0.8, 1.0$ &   $46, 66$ &     $-4.28\times 10^{-2}$ &    $5.8\times 10^{-3}$ &    $4.4\times 10^{-3}$ &    $8.9\times 10^{-4}$ &    $7.3\times 10^{-3}$ &   $1.0, 1.2$ &   $46, 66$ &     $-5.60\times 10^{-2}$ &    $5.7\times 10^{-3}$ &    $4.2\times 10^{-3}$ &    $1.2\times 10^{-3}$ &    $7.2\times 10^{-3}$ \\
  $0.8, 1.0$ &   $66, 80$ &     $-7.03\times 10^{-2}$ &    $4.4\times 10^{-3}$ &    $2.4\times 10^{-3}$ &    $9.0\times 10^{-4}$ &    $5.1\times 10^{-3}$ &   $1.0, 1.2$ &   $66, 80$ &     $-8.76\times 10^{-2}$ &    $4.3\times 10^{-3}$ &    $2.4\times 10^{-3}$ &    $9.5\times 10^{-4}$ &    $5.0\times 10^{-3}$ \\
  $0.8, 1.0$ &   $80, 91$ &     $-7.42\times 10^{-5}$ &    $1.1\times 10^{-3}$ &    $5.7\times 10^{-4}$ &    $7.5\times 10^{-4}$ &    $1.5\times 10^{-3}$ &   $1.0, 1.2$ &   $80, 91$ &     $\phantom{-}8.90\times 10^{-5}$ &    $1.1\times 10^{-3}$ &    $5.7\times 10^{-4}$ &    $7.8\times 10^{-4}$ &    $1.5\times 10^{-3}$ \\
  $0.8, 1.0$ &   $\phantom{1}91, 102$ & $\phantom{-}2.33\times 10^{-2}$ &    $1.1\times 10^{-3}$ &    $5.2\times 10^{-4}$ &    $6.9\times 10^{-4}$ &    $1.4\times 10^{-3}$ &   $1.0, 1.2$ &   $\phantom{1}91, 102$ & $\phantom{-}2.88\times 10^{-2}$ &    $1.1\times 10^{-3}$ &    $5.1\times 10^{-4}$ &    $7.4\times 10^{-4}$ &    $1.4\times 10^{-3}$ \\
  $0.8, 1.0$ &   $102, 116$ & $\phantom{-}9.15\times 10^{-2}$ &    $5.2\times 10^{-3}$ &    $2.8\times 10^{-3}$ &    $1.6\times 10^{-3}$ &    $6.1\times 10^{-3}$ &   $1.0, 1.2$ &   $102, 116$ & $\phantom{-}1.10\times 10^{-1}$ &    $5.1\times 10^{-3}$ &    $2.7\times 10^{-3}$ &    $1.7\times 10^{-3}$ &    $6.0\times 10^{-3}$ \\
  $0.8, 1.0$ &   $116, 150$ & $\phantom{-}1.28\times 10^{-1}$ &    $8.4\times 10^{-3}$ &    $3.4\times 10^{-3}$ &    $1.8\times 10^{-3}$ &    $9.2\times 10^{-3}$ &   $1.0, 1.2$ &   $116, 150$ & $\phantom{-}1.68\times 10^{-1}$ &    $8.4\times 10^{-3}$ &    $3.6\times 10^{-3}$ &    $2.0\times 10^{-3}$ &    $9.3\times 10^{-3}$ \\
  $0.8, 1.0$ &   $150, 200$ & $\phantom{-}1.69\times 10^{-1}$ &    $1.6\times 10^{-2}$ &    $5.3\times 10^{-3}$ &    $4.0\times 10^{-3}$ &    $1.7\times 10^{-2}$ &   $1.0, 1.2$ &   $150, 200$ & $\phantom{-}2.36\times 10^{-1}$ &    $1.6\times 10^{-2}$ &    $5.9\times 10^{-3}$ &    $4.2\times 10^{-3}$ &    $1.7\times 10^{-2}$ \\
\cmidrule(lr{17.7pt}){1-7}
\cmidrule(l{2.5pt}r){8-14}
  $1.2, 1.4$ &   $46, 66$ &     $-8.88\times 10^{-2}$ &    $5.7\times 10^{-3}$ &    $4.2\times 10^{-3}$ &    $1.4\times 10^{-3}$ &    $7.2\times 10^{-3}$ &   $1.4, 1.6$ &   $46, 66$ &     $-1.03\times 10^{-1}$ &    $5.8\times 10^{-3}$ &    $4.3\times 10^{-3}$ &    $1.7\times 10^{-3}$ &    $7.5\times 10^{-3}$ \\
  $1.2, 1.4$ &   $66, 80$ &     $-1.14\times 10^{-1}$ &    $4.2\times 10^{-3}$ &    $2.4\times 10^{-3}$ &    $1.1\times 10^{-3}$ &    $5.0\times 10^{-3}$ &   $1.4, 1.6$ &   $66, 80$ &     $-1.25\times 10^{-1}$ &    $4.4\times 10^{-3}$ &    $2.6\times 10^{-3}$ &    $1.3\times 10^{-3}$ &    $5.3\times 10^{-3}$ \\
  $1.2, 1.4$ &   $80, 91$ &     $-2.23\times 10^{-3}$ &    $1.1\times 10^{-3}$ &    $5.7\times 10^{-4}$ &    $8.7\times 10^{-4}$ &    $1.5\times 10^{-3}$ &   $1.4, 1.6$ &   $80, 91$ &     $-2.59\times 10^{-3}$ &    $1.2\times 10^{-3}$ &    $6.2\times 10^{-4}$ &    $9.2\times 10^{-4}$ &    $1.6\times 10^{-3}$ \\
  $1.2, 1.4$ &   $\phantom{1}91, 102$ & $\phantom{-}3.38\times 10^{-2}$ &    $1.1\times 10^{-3}$ &    $5.2\times 10^{-4}$ &    $8.5\times 10^{-4}$ &    $1.5\times 10^{-3}$ &   $1.4, 1.6$ &   $\phantom{1}91, 102$ & $\phantom{-}3.77\times 10^{-2}$ &    $1.1\times 10^{-3}$ &    $5.7\times 10^{-4}$ &    $8.9\times 10^{-4}$ &    $1.6\times 10^{-3}$ \\
  $1.2, 1.4$ &   $102, 116$ & $\phantom{-}1.30\times 10^{-1}$ &    $5.0\times 10^{-3}$ &    $2.8\times 10^{-3}$ &    $2.0\times 10^{-3}$ &    $6.1\times 10^{-3}$ &   $1.4, 1.6$ &   $102, 116$ & $\phantom{-}1.44\times 10^{-1}$ &    $5.1\times 10^{-3}$ &    $3.0\times 10^{-3}$ &    $2.2\times 10^{-3}$ &    $6.3\times 10^{-3}$ \\
  $1.2, 1.4$ &   $116, 150$ & $\phantom{-}2.09\times 10^{-1}$ &    $8.5\times 10^{-3}$ &    $3.7\times 10^{-3}$ &    $2.6\times 10^{-3}$ &    $9.6\times 10^{-3}$ &   $1.4, 1.6$ &   $116, 150$ & $\phantom{-}2.19\times 10^{-1}$ &    $8.9\times 10^{-3}$ &    $3.9\times 10^{-3}$ &    $2.8\times 10^{-3}$ &    $1.0\times 10^{-2}$ \\
  $1.2, 1.4$ &   $150, 200$ & $\phantom{-}2.68\times 10^{-1}$ &    $1.6\times 10^{-2}$ &    $6.6\times 10^{-3}$ &    $4.3\times 10^{-3}$ &    $1.8\times 10^{-2}$ &   $1.4, 1.6$ &   $150, 200$ & $\phantom{-}2.80\times 10^{-1}$ &    $1.8\times 10^{-2}$ &    $9.0\times 10^{-3}$ &    $4.5\times 10^{-3}$ &    $2.1\times 10^{-2}$ \\
\cmidrule(lr{17.7pt}){1-7}
\cmidrule(l{2.5pt}r){8-14}
  $1.6, 1.8$ &   $46, 66$ &     $-9.35\times 10^{-2}$ &    $6.1\times 10^{-3}$ &    $4.6\times 10^{-3}$ &    $1.4\times 10^{-3}$ &    $7.8\times 10^{-3}$ &   $1.8, 2.0$ &   $46, 66$ &     $-9.23\times 10^{-2}$ &    $6.7\times 10^{-3}$ &    $4.9\times 10^{-3}$ &    $1.5\times 10^{-3}$ &    $8.5\times 10^{-3}$ \\
  $1.6, 1.8$ &   $66, 80$ &     $-1.26\times 10^{-1}$ &    $4.9\times 10^{-3}$ &    $2.8\times 10^{-3}$ &    $1.5\times 10^{-3}$ &    $5.8\times 10^{-3}$ &   $1.8, 2.0$ &   $66, 80$ &     $-1.20\times 10^{-1}$ &    $5.6\times 10^{-3}$ &    $3.3\times 10^{-3}$ &    $2.0\times 10^{-3}$ &    $6.8\times 10^{-3}$ \\
  $1.6, 1.8$ &   $80, 91$ &     $-7.79\times 10^{-4}$ &    $1.3\times 10^{-3}$ &    $7.0\times 10^{-4}$ &    $1.0\times 10^{-3}$ &    $1.8\times 10^{-3}$ &   $1.8, 2.0$ &   $80, 91$ &     $-2.65\times 10^{-3}$ &    $1.5\times 10^{-3}$ &    $8.1\times 10^{-4}$ &    $1.1\times 10^{-3}$ &    $2.0\times 10^{-3}$ \\
  $1.6, 1.8$ &   $\phantom{1}91, 102$ & $\phantom{-}3.96\times 10^{-2}$ &    $1.3\times 10^{-3}$ &    $6.3\times 10^{-4}$ &    $9.6\times 10^{-4}$ &    $1.7\times 10^{-3}$ &   $1.8, 2.0$ &   $\phantom{1}91, 102$ & $\phantom{-}3.97\times 10^{-2}$ &    $1.5\times 10^{-3}$ &    $7.4\times 10^{-4}$ &    $1.1\times 10^{-3}$ &    $2.0\times 10^{-3}$ \\
  $1.6, 1.8$ &   $102, 116$ & $\phantom{-}1.63\times 10^{-1}$ &    $5.3\times 10^{-3}$ &    $3.6\times 10^{-3}$ &    $2.8\times 10^{-3}$ &    $7.0\times 10^{-3}$ &   $1.8, 2.0$ &   $102, 116$ & $\phantom{-}1.54\times 10^{-1}$ &    $5.6\times 10^{-3}$ &    $3.9\times 10^{-3}$ &    $3.5\times 10^{-3}$ &    $7.7\times 10^{-3}$ \\
  $1.6, 1.8$ &   $116, 150$ & $\phantom{-}2.43\times 10^{-1}$ &    $1.0\times 10^{-2}$ &    $4.6\times 10^{-3}$ &    $2.6\times 10^{-3}$ &    $1.2\times 10^{-2}$ &   $1.8, 2.0$ &   $116, 150$ & $\phantom{-}2.12\times 10^{-1}$ &    $1.2\times 10^{-2}$ &    $6.8\times 10^{-3}$ &    $3.9\times 10^{-3}$ &    $1.4\times 10^{-2}$ \\
  $1.6, 1.8$ &   $150, 200$ & $\phantom{-}3.02\times 10^{-1}$ &    $1.8\times 10^{-2}$ &    $6.8\times 10^{-3}$ &    $5.6\times 10^{-3}$ &    $2.0\times 10^{-2}$ &   $1.8, 2.0$ &   $150, 200$ & $\phantom{-}2.66\times 10^{-1}$ &    $2.1\times 10^{-2}$ &    $6.3\times 10^{-3}$ &    $4.2\times 10^{-3}$ &    $2.3\times 10^{-2}$ \\
\cmidrule(lr{17.7pt}){1-7}
\cmidrule(l{2.5pt}r){8-14}
  $2.0, 2.2$ &   $46, 66$ &     $-9.82\times 10^{-2}$ &    $8.3\times 10^{-3}$ &    $6.2\times 10^{-3}$ &    $2.6\times 10^{-3}$ &    $1.1\times 10^{-2}$ &   $2.2, 2.4$ &   $46, 66$ &     $-4.44\times 10^{-2}$ &    $1.4\times 10^{-2}$ &    $1.1\times 10^{-2}$ &    $3.1\times 10^{-3}$ &    $1.8\times 10^{-2}$ \\
  $2.0, 2.2$ &   $66, 80$ &     $-1.10\times 10^{-1}$ &    $6.9\times 10^{-3}$ &    $4.2\times 10^{-3}$ &    $3.0\times 10^{-3}$ &    $8.6\times 10^{-3}$ &   $2.2, 2.4$ &   $66, 80$ &     $-5.03\times 10^{-2}$ &    $1.2\times 10^{-2}$ &    $7.5\times 10^{-3}$ &    $6.1\times 10^{-3}$ &    $1.5\times 10^{-2}$ \\
  $2.0, 2.2$ &   $80, 91$ &     $-3.33\times 10^{-3}$ &    $1.9\times 10^{-3}$ &    $1.1\times 10^{-3}$ &    $1.5\times 10^{-3}$ &    $2.7\times 10^{-3}$ &   $2.2, 2.4$ &   $80, 91$ &     $-6.44\times 10^{-3}$ &    $3.1\times 10^{-3}$ &    $1.9\times 10^{-3}$ &    $2.1\times 10^{-3}$ &    $4.2\times 10^{-3}$ \\
  $2.0, 2.2$ &   $\phantom{1}91, 102$ & $\phantom{-}3.33\times 10^{-2}$ &    $1.9\times 10^{-3}$ &    $9.8\times 10^{-4}$ &    $1.5\times 10^{-3}$ &    $2.6\times 10^{-3}$ &   $2.2, 2.4$ &   $\phantom{1}91, 102$ & $\phantom{-}1.99\times 10^{-2}$ &    $3.1\times 10^{-3}$ &    $1.6\times 10^{-3}$ &    $2.3\times 10^{-3}$ &    $4.2\times 10^{-3}$ \\
  $2.0, 2.2$ &   $102, 116$ & $\phantom{-}1.20\times 10^{-1}$ &    $6.5\times 10^{-3}$ &    $5.1\times 10^{-3}$ &    $6.3\times 10^{-3}$ &    $1.0\times 10^{-2}$ &   $2.2, 2.4$ &   $102, 116$ & $\phantom{-}6.06\times 10^{-2}$ &    $9.7\times 10^{-3}$ &    $9.3\times 10^{-3}$ &    $1.2\times 10^{-2}$ &    $1.8\times 10^{-2}$ \\
  $2.0, 2.2$ &   $116, 150$ & $\phantom{-}1.64\times 10^{-1}$ &    $1.4\times 10^{-2}$ &    $6.6\times 10^{-3}$ &    $3.9\times 10^{-3}$ &    $1.6\times 10^{-2}$ &   $2.2, 2.4$ &   $116, 150$ & $\phantom{-}9.40\times 10^{-2}$ &    $2.6\times 10^{-2}$ &    $1.6\times 10^{-2}$ &    $1.5\times 10^{-2}$ &    $3.4\times 10^{-2}$ \\
  $2.0, 2.2$ &   $150, 200$ & $\phantom{-}1.97\times 10^{-1}$ &    $2.7\times 10^{-2}$ &    $1.0\times 10^{-2}$ &    $7.7\times 10^{-3}$ &    $3.0\times 10^{-2}$ &   $2.2, 2.4$ &   $150, 200$ & $\phantom{-}1.27\times 10^{-1}$ &    $4.6\times 10^{-2}$ &    $1.3\times 10^{-2}$ &    $1.0\times 10^{-2}$ &    $4.9\times 10^{-2}$ \\
\bottomrule
\end{tabular}
}
\caption{The asymmetry $A_{\textrm{FB}}$ determined from the combined
triple-differential cross-section measurements.
The measurements are listed together with the statistical ($\Delta^{\textrm{stat}}$),
uncorrelated systematic ($\Delta^{\textrm {syst}}_{\textrm {unc}}$),
correlated systematic ($\Delta^{\textrm{syst}}_{\textrm{cor}}$),
and total ($\Delta^{\textrm{total}}$) uncertainties. }
\label{tab:comb_afb}
\end{table}
\begin{table}[htp!]
\tiny
\centering
\scalebox{1}{
\setlength\tabcolsep{2.5pt}
\setlength\extrarowheight{2.5pt}
\begin{tabular}{C{1.1cm}C{1.1cm}C{1.75cm}C{1.2cm}C{1.2cm}C{1.2cm}C{1.2cm}}
\toprule
$|y_{ee}|$
& $m_{ee}$
& $A_{\textrm {FB}}$
& $\Delta^{\textrm {stat}}$
& $\Delta^{\textrm {syst}}_{\textrm {unc}}$
& $\Delta^{\textrm {syst}}_{\textrm {cor}}$
& $\Delta^{\textrm {total}}$ \\
&	[GeV] & & & & &\\
\noalign{\vskip 0.05cm}
\cmidrule(lr){1-7}
  $1.2, 1.6$ &   $66, 80$ &     $-2.44\times 10^{-1}$ &    $4.4\times 10^{-2}$ &    $5.9\times 10^{-2}$ &    $2.5\times 10^{-2}$ &    $7.8\times 10^{-2}$ \\
  $1.2, 1.6$ &   $80, 91$ & $\phantom{-}8.57\times 10^{-3}$ &    $6.2\times 10^{-3}$ &    $4.6\times 10^{-3}$ &    $3.6\times 10^{-3}$ &    $8.5\times 10^{-3}$ \\
  $1.2, 1.6$ &   $\phantom{1}91, 102$ & $\phantom{-}7.03\times 10^{-2}$ &    $5.7\times 10^{-3}$ &    $4.1\times 10^{-3}$ &    $4.9\times 10^{-3}$ &    $8.6\times 10^{-3}$ \\
  $1.2, 1.6$ &   $102, 116$ & $\phantom{-}2.78\times 10^{-1}$ &    $2.6\times 10^{-2}$ &    $3.4\times 10^{-2}$ &    $2.6\times 10^{-2}$ &    $5.0\times 10^{-2}$ \\
  $1.2, 1.6$ &   $116, 150$ & $\phantom{-}4.43\times 10^{-1}$ &    $4.2\times 10^{-2}$ &    $6.0\times 10^{-2}$ &    $1.1\times 10^{-1}$ &    $1.3\times 10^{-1}$ \\
\cmidrule(lr){1-7}
  $1.6, 2.0$ &   $66, 80$ &     $-2.32\times 10^{-1}$ &    $1.7\times 10^{-2}$ &    $1.9\times 10^{-2}$ &    $1.1\times 10^{-2}$ &    $2.7\times 10^{-2}$ \\
  $1.6, 2.0$ &   $80, 91$ & $\phantom{-}3.08\times 10^{-3}$ &    $3.3\times 10^{-3}$ &    $2.3\times 10^{-3}$ &    $2.5\times 10^{-3}$ &    $4.7\times 10^{-3}$ \\
  $1.6, 2.0$ &   $\phantom{1}91, 102$ & $\phantom{-}7.30\times 10^{-2}$ &    $3.2\times 10^{-3}$ &    $2.1\times 10^{-3}$ &    $1.8\times 10^{-3}$ &    $4.2\times 10^{-3}$ \\
  $1.6, 2.0$ &   $102, 116$ & $\phantom{-}3.09\times 10^{-1}$ &    $1.6\times 10^{-2}$ &    $1.6\times 10^{-2}$ &    $1.3\times 10^{-2}$ &    $2.6\times 10^{-2}$ \\
  $1.6, 2.0$ &   $116, 150$ & $\phantom{-}4.83\times 10^{-1}$ &    $2.6\times 10^{-2}$ &    $3.7\times 10^{-2}$ &    $6.5\times 10^{-2}$ &    $7.9\times 10^{-2}$ \\
\cmidrule(lr){1-7}
  $2.0, 2.4$ &   $66, 80$ &     $-2.89\times 10^{-1}$ &    $1.2\times 10^{-2}$ &    $1.4\times 10^{-2}$ &    $1.3\times 10^{-2}$ &    $2.3\times 10^{-2}$ \\
  $2.0, 2.4$ &   $80, 91$ &     $-9.15\times 10^{-3}$ &    $2.8\times 10^{-3}$ &    $2.1\times 10^{-3}$ &    $1.7\times 10^{-3}$ &    $3.9\times 10^{-3}$ \\
  $2.0, 2.4$ &   $\phantom{1}91, 102$ & $\phantom{-}8.43\times 10^{-2}$ &    $2.7\times 10^{-3}$ &    $1.9\times 10^{-3}$ &    $2.7\times 10^{-3}$ &    $4.3\times 10^{-3}$ \\
  $2.0, 2.4$ &   $102, 116$ & $\phantom{-}3.40\times 10^{-1}$ &    $1.3\times 10^{-2}$ &    $1.3\times 10^{-2}$ &    $1.6\times 10^{-2}$ &    $2.5\times 10^{-2}$ \\
  $2.0, 2.4$ &   $116, 150$ & $\phantom{-}4.93\times 10^{-1}$ &    $2.1\times 10^{-2}$ &    $2.7\times 10^{-2}$ &    $6.5\times 10^{-2}$ &    $7.3\times 10^{-2}$ \\
\cmidrule(lr){1-7}
  $2.4, 2.8$ &   $66, 80$ &     $-3.26\times 10^{-1}$ &    $1.1\times 10^{-2}$ &    $1.1\times 10^{-2}$ &    $1.7\times 10^{-2}$ &    $2.3\times 10^{-2}$ \\
  $2.4, 2.8$ &   $80, 91$ &     $-4.68\times 10^{-3}$ &    $2.6\times 10^{-3}$ &    $2.2\times 10^{-3}$ &    $2.4\times 10^{-3}$ &    $4.2\times 10^{-3}$ \\
  $2.4, 2.8$ &   $\phantom{1}91, 102$ & $\phantom{-}1.11\times 10^{-1}$ &    $2.6\times 10^{-3}$ &    $2.5\times 10^{-3}$ &    $2.1\times 10^{-3}$ &    $4.1\times 10^{-3}$ \\
  $2.4, 2.8$ &   $102, 116$ & $\phantom{-}4.29\times 10^{-1}$ &    $1.2\times 10^{-2}$ &    $1.3\times 10^{-2}$ &    $1.8\times 10^{-2}$ &    $2.6\times 10^{-2}$ \\
  $2.4, 2.8$ &   $116, 150$ & $\phantom{-}5.98\times 10^{-1}$ &    $1.8\times 10^{-2}$ &    $2.3\times 10^{-2}$ &    $3.3\times 10^{-2}$ &    $4.4\times 10^{-2}$ \\
\cmidrule(lr){1-7}
  $2.8, 3.6$ &   $66, 80$ &     $-4.73\times 10^{-1}$ &    $1.1\times 10^{-2}$ &    $1.4\times 10^{-2}$ &    $2.7\times 10^{-2}$ &    $3.2\times 10^{-2}$ \\
  $2.8, 3.6$ &   $80, 91$ &     $-8.07\times 10^{-3}$ &    $2.8\times 10^{-3}$ &    $2.7\times 10^{-3}$ &    $2.3\times 10^{-3}$ &    $4.5\times 10^{-3}$ \\
  $2.8, 3.6$ &   $\phantom{1}91, 102$ & $\phantom{-}1.55\times 10^{-1}$ &    $2.7\times 10^{-3}$ &    $2.7\times 10^{-3}$ &    $5.0\times 10^{-3}$ &    $6.2\times 10^{-3}$ \\
  $2.8, 3.6$ &   $102, 116$ & $\phantom{-}5.51\times 10^{-1}$ &    $1.1\times 10^{-2}$ &    $1.1\times 10^{-2}$ &    $4.5\times 10^{-2}$ &    $4.8\times 10^{-2}$ \\
  $2.8, 3.6$ &   $116, 150$ & $\phantom{-}7.15\times 10^{-1}$ &    $1.9\times 10^{-2}$ &    $2.3\times 10^{-2}$ &    $4.8\times 10^{-2}$ &    $5.7\times 10^{-2}$ \\
\bottomrule
\end{tabular}
}
\caption{The asymmetry $A_{\textrm{FB}}$ determined from the high rapidity electron channel
triple-differential cross-section measurement.
The measurement is listed together with the statistical ($\Delta^{\textrm{stat}}$),
uncorrelated systematic ($\Delta^{\textrm {syst}}_{\textrm {unc}}$),
correlated systematic ($\Delta^{\textrm{syst}}_{\textrm{cor}}$),
and total ($\Delta^{\textrm{total}}$) uncertainties. }
\label{tab:zcf_afb}
\end{table}
\FloatBarrier

\clearpage
\printbibliography

\clearpage
\begin{flushleft}
{\Large The ATLAS Collaboration}

\bigskip

M.~Aaboud$^\textrm{\scriptsize 137d}$,
G.~Aad$^\textrm{\scriptsize 88}$,
B.~Abbott$^\textrm{\scriptsize 115}$,
O.~Abdinov$^\textrm{\scriptsize 12}$$^{,*}$,
B.~Abeloos$^\textrm{\scriptsize 119}$,
S.H.~Abidi$^\textrm{\scriptsize 161}$,
O.S.~AbouZeid$^\textrm{\scriptsize 139}$,
N.L.~Abraham$^\textrm{\scriptsize 151}$,
H.~Abramowicz$^\textrm{\scriptsize 155}$,
H.~Abreu$^\textrm{\scriptsize 154}$
% ...
\bigskip
\\
% ...
$^{12}$ Institute of Physics, Azerbaijan Academy of Sciences, Baku, Azerbaijan\\
% ...
$^{88}$ CPPM, Aix-Marseille Universit{\'e} and CNRS/IN2P3, Marseille, France\\
% ...
$^{115}$ Homer L. Dodge Department of Physics and Astronomy, University of Oklahoma, Norman OK, United States of America\\
% ...
$^{119}$ LAL, Univ. Paris-Sud, CNRS/IN2P3, Universit{\'e} Paris-Saclay, Orsay, France\\
% ...
$^{137}$ $^{(a)}$ Facult{\'e} des Sciences Ain Chock, R{\'e}seau Universitaire de Physique des Hautes Energies - Universit{\'e} Hassan II, Casablanca; $^{(b)}$ Centre National de l'Energie des Sciences Techniques Nucleaires, Rabat; $^{(c)}$ Facult{\'e} des Sciences Semlalia, Universit{\'e} Cadi Ayyad, LPHEA-Marrakech; $^{(d)}$ Facult{\'e} des Sciences, Universit{\'e} Mohamed Premier and LPTPM, Oujda; $^{(e)}$ Facult{\'e} des sciences, Universit{\'e} Mohammed V, Rabat, Morocco\\
% ...
$^{139}$ Santa Cruz Institute for Particle Physics, University of California Santa Cruz, Santa Cruz CA, United States of America\\
% ...
$^{151}$ Department of Physics and Astronomy, University of Sussex, Brighton, United Kingdom\\
% ...
$^{154}$ Department of Physics, Technion: Israel Institute of Technology, Haifa, Israel\\
$^{155}$ Raymond and Beverly Sackler School of Physics and Astronomy, Tel Aviv University, Tel Aviv, Israel\\
% ...
$^{161}$ Department of Physics, University of Toronto, Toronto ON, Canada\\
% ...
$^{a}$ Also at Department of Physics, King's College London, London, United Kingdom\\
$^{b}$ Also at Institute of Physics, Azerbaijan Academy of Sciences, Baku, Azerbaijan\\
$^{c}$ Also at Novosibirsk State University, Novosibirsk, Russia\\
$^{d}$ Also at TRIUMF, Vancouver BC, Canada\\
$^{e}$ Also at Department of Physics {\&} Astronomy, University of Louisville, Louisville, KY, United States of America\\
% ...
$^{*}$ Deceased
\end{flushleft}


\end{document}